\documentclass[useAMS,usenatbib]{mn2e}
\usepackage{epsfig}
\usepackage{graphicx}
\usepackage{times}
\usepackage{color}

\def\mo{M$_\odot$}

\def\cm3{cm$^{-3}$}
\def\kms{km~s$^{-1}$}

\def\lsun{L$_{\odot}$}
\def\rsun{R$_{\odot}$}

\def\msun{M$_{\odot}$}

\def\one{{\,\sc i}}
\def\two{{\,\sc ii}}

\def\four{{\,\sc iv}}

\def\beq{\begin{equation}}
\def\eeq{\end{equation}}
\def\rstar{$R_{\ast}$}
\def\zsun{Z$_{\odot}$}
\def\lesssim{\mathrel{\hbox{\rlap{\hbox{\lower4pt\hbox{$\sim$}}}\hbox{$<$}}}}
\def\gtrsim{\mathrel{\hbox{\rlap{\hbox{\lower4pt\hbox{$\sim$}}}\hbox{$>$}}}}

\newcommand{\opar}{\hbox{$^{\scriptsize \rm  o}$}}
\newcommand{\oparsub}[1]{\hbox{$^{\scriptsize \rm  o}_{#1}$}}

\def\cmfgen{{\sc cmfgen}}
\def\mesa{{\sc mesa}}
\def\v1d{{\sc v1d}}

\newcommand{\iso}[2]{\ensuremath{^{#1}\rm{#2}}}

\def\apj{ApJ}
\def\apjs{ApJS}
\def\apjl{ApJL}
\def\aap{A\&A}
\def\araa{ARA\&A}

\def\aaps{A\&AS}
\def\apss{Ap\&SS}
\def\mnras{MNRAS}
\def\nat{Nature}

\def\solphys{Sol.~Phys.}

\def\gray{$\gamma$-ray}
\def\grays{$\gamma$-rays}
\def\isoni{$^{56}{\rm Ni}$}
\def\isoco{$^{56}{\rm Co}$}
\def\isofe{$^{56}{\rm Fe}$}

\title[Simulations of PISNe]{Radiative Properties of Pair-instability Supernova Explosions}
\author[Luc Dessart et al.]{\vspace{0.3cm} Luc Dessart,$^{1,2}$\thanks{email: Luc.Dessart@oamp.fr}
Roni Waldman,$^{3}$ Eli Livne,$^3$ D. John Hillier,$^{4}$ St\'ephane Blondin,$^{1,5}$ \\
 $^1$:  Aix Marseille Universit\'e, CNRS, LAM (Laboratoire d'Astrophysique de Marseille), UMR\,7326, 13388, Marseille, France \\
 $^2$: TAPIR, Mail code 350-17, California Institute of Technology, Pasadena, CA 91125, USA \\
  $^3$: Racah Institute of Physics, The Hebrew University, Jerusalem, 91104, Israel \\
  $^4$: Department of Physics and Astronomy, University of Pittsburgh, 3941 O'Hara Street, Pittsburgh, PA 15260, USA \\
  $^5$: Centre de Physique des Particules de Marseille (CPPM), Aix-Marseille Universit\'e, CNRS/IN2P3, Marseille, France}

\date{Accepted 2012 October 22.  Received 2012 October 19; in original form 2012 August 14}

\pagerange{\pageref{firstpage}--\pageref{lastpage}} \pubyear{2012}

\begin{document}

\maketitle

\label{firstpage}

\begin{abstract}
We present non-LTE time-dependent radiative-transfer simulations of pair-instability supernovae (PISNe)
stemming from red-supergiant (RSG), blue-supergiant (BSG) and Wolf-Rayet (WR) star rotation-free progenitors
born in the mass range 160--230\,\msun, at 10$^{-4}$\,\zsun.
Although subject to uncertainties in convection and stellar mass-loss rates,
our initial conditions come from physically-consistent models that treat evolution from the main-sequence,
the onset of the pair-production instability, and the explosion phase.
With our set of input models characterized by large \isoni\ and ejecta masses, and large kinetic energies,
we recover qualitatively the Type II-Plateau, II-peculiar, and Ib/c light-curve morphologies, although they have
larger peak bolometric luminosities ($\sim 10^{9}$ to $10^{10}$\,\lsun) and a longer duration ($\sim$\,200\,d).
We discuss the spectral properties for each model during the photospheric and nebular phases, including
Balmer lines in II-P and II-pec at early times, the dominance of lines from intermediate-mass-elements (IMEs)
near the bolometric maximum, and the strengthening of metal line blanketing thereafter.
Having similar He-core properties, all models exhibit similar post-peak spectra that are strongly blanketed by Fe\two\
and Fe\one\ lines, characterized by red colors, and that arise from  photospheres/ejecta with a temperature
of $\lesssim$\,4000\,K. Combined with the modest line widths after bolometric peak, these properties contrast
with those of  known superluminous SNe suggesting that PISNe are yet to be discovered. Being reddish, PISNe
will be difficult to observe at high redshift except when they stem from RSG explosions, in which case
they could be used as metallicity probes and distance indicators.
\end{abstract}

\begin{keywords} radiation hydrodynamics -- stars: atmospheres -- stars:
supernovae - stars: evolution -- stars: supernovae: individual: 2007bi, 2006gy
\end{keywords}


\section{Introduction}

About a decade ago, numerical simulations of star formation revealed that the first generation of stars may have
 been super-massive \citep{bromm_etal_02}, holding great promise to discover them, not as stars, but rather
through their catastrophic explosions.  \citet{barkat_etal_67} and \citet{rakavy_shaviv_67} proposed that for stars
sufficiently massive the production of e$^{-}$e$^{+}$ pairs from \gray\  annihilation would lead successively to an
implosion, a thermonuclear runaway, and an explosion leaving no remnant behind. These events are termed pair-instability
supernovae (PISNe). In the coming decade, the opportunity to identify PISNe and study them in detail may be realized
with new observatories such as LSST and JWST \citep{scannapieco_etal_05,hummel_etal_11}. However, there is considerable
uncertainty and speculation concerning the first generation of massive stellar  objects, impacting our understanding of their
formation, their evolution, and their explosion.

Recent simulations for the formation of the first stars has significantly altered the original picture.  The initial
mass function seems no longer to be heavily biased towards super-massive stars. Instead,  it stretches
down to lower masses, even very low masses \citep{caffau_etal_12},  perhaps exhibiting a bimodal distribution
\citep{stacy_etal_11,stacy_etal_12,clark_etal_11}. Binaries should also be created  \citep{stacy_etal_10}, although
the binary fraction is largely unknown. Simulations including feedback, in particular in the form of UV radiation, suggest
that the circumstellar disk feeding mass into the primordial star is evaporated soon after the onset of steady
thermonuclear burning. This feedback leads to the truncation of mass accretion and sets an upper mass limit
for the newly-formed star of $\sim$\,30\,\msun\ \citep{hosokawa_etal_11}. Theoretically, primordial star formation
does not seem to differ  significantly from present-day star formation. Paradoxically, it fails to explain the
existence in the Local Universe of super-massive stars like $\eta$ Car  \citep{etacar_rev_97,hillier_etal_01},
massive stars in the Galactic centre \citep{figer_etal_02,martins_etal_08}, or the spectacular R136 nursery
\citep{crowther_etal_10}. Since super-massive stars in the Local Universe systematically belong to dense star
clusters, they may form not in isolation but instead dynamically and chaotically from merger
events \cite[e.g.,][]{BB05_mergers,Vanbeveren09_mergers,pan_etal_12b}.

More generally, the nature of superluminous SNe remains perplexing. The interaction between two separate shells,
which converts kinetic energy to thermal and radiative energy, is one way to boost the luminosity of a SN, because standard
SNe typically have a hundred times more kinetic energy than time-integrated luminosities.
Another way of producing a bright display is by depositing magnetar radiation into the SN ejecta
\citep{maeda_etal_07,kasen_bildsten_10,woosley_10}. The energy lost by the magnetar leads
 to the spin down of the magnetar which eventually quenches the magnetar radiation.
 For a dipole field, the spin-down time scale is $t_{\rm sp} \sim 4.8 B^{-2}_{15} P^2_{10}$\,d, where
 $B_{\rm 15}$ is the magnetic strength in units of 10$^{15}$\,G and $P_{\rm 10}$ is the rotation period $P$ in
 units of 10\,ms. Depending on the magnetar parameters, its spin-down timescale may be comparable to the
decay timescale of \isoni,  making it an attractive substitute for models that require a larger-than-average production of \isoni\  for powering luminous SNe.  While a PISN may synthesize a maximum of 57\,\msun\ of \isoni\ \citep{HW02}, it is
unclear how much a ``standard" core-collapse SN can produce. This reflects the poorly known core properties at
the time of death, and the viability of the neutrino and magneto-rotational explosion mechanisms.  The collapsar
model is one instance where copious amounts of \isoni\ may be produced \citep{woosley_93}, but the required
core compactness to form such collapsars seem far more rarely matched by stellar evolution models,
favoring instead the formation of proto-magnetars  \citep{dessart_etal_12b}.

Restricting our focus to very low metallicity environments the ansatz of the current
study is that  super-massive stars do form, populating the mass range 140--260\,\msun.
Provided they do not lose much mass during the hydrogen-burning stage, they form a helium core with a mass in the range
64-133\,\msun\ that should experience the pair-production instability and lead to
a thermonuclear explosion that completely disrupts the star \citep{HW02,waldman_08}.
How often this  occurs is
completely speculative as it depends on the unknown formation rate of the progenitors as well as their evolution. A major component of this uncertainty is mass loss since it determines whether the star will explode as an extended blue-supergiant (BSG), as a very extended red-supergiant (RSG) star, or instead as a very compact Wolf-Rayet (WR) star. This, in turn, impacts the supernova (SN) light curve morphology and thus the potential detectability and reliable identification of these events \citep{scannapieco_etal_05,kasen_etal_11}.
As we discuss here, a H-rich PISN from a RSG star at 10$^{-4}$\zsun\  leaves little ambiguity about the origin of
the event while connecting a superluminous Type Ic event with a H-deficient PISN is more challenging.
It is unclear today if we have seen either H-rich or H-deficient PISNe \citep{dessart_etal_12d}.

Mass-loss in metal poor massive stars is highly uncertain. \citet{baraffe_etal_01} found that  metal-poor massive
stars are pulsationally stable and, based on this analysis, should die with their hydrogen envelope. The metallicity
dependence of radiation driving in massive-star winds suggests that this other form of mass loss, critical for OB
and WR stars at solar metallicity,  should be largely inhibited at very low metallicity \citep{cak}. As is quite typical
for proto-stellar collapse in star formation, the collapsed core is endowed with a large amount of
angular momentum at birth \citep{stacy_etal_11}, which may affect its evolution in a number of ways
\citep{HMM04_evol_rot,hirschi_07,heger_woosley_10,CW12,yoon_etal_12}.
For example, by increasing the oxygen core mass, fast rotation can lower the minimum main-sequence
mass needed to encounter the pair-production instability \citep{CW12}, thereby
affecting the \isoni-mass to ejecta-mass ratio and the resulting PISN
radiation properties \citep{dessart_etal_12d}.
High luminosity poses a severe
strain on the hydrostatic equilibrium of stars \citep{joss_etal_73}, especially near break-up rotation speeds
\citep{maeder_meynet_00}, establishing ideal conditions for centrifugally-driven mass loss (a disk) and/or a
continuum-driven wind \citep{owocki_etal_04}. In this context, radiation- and/or rotation-driven mass loss may in
fact compete with the episodic pulsations expected to occur in stars with a helium-core mass in the range
40--63\,\msun\ \citep{HW02,woosley_etal_07}.

Currently, the most promising PISN candidate  is SN2007bi \citep{galyam_etal_09,Y10_full}, although
this association raises several issues. The observations capture only the light-curve peak and beyond,
making the explosion time uncertain, compromising the inference of the ejecta mass and \isoni\ mass.
Furthermore, SN2007bi took place in an environment of 0.2--0.3\,\zsun; a hydrogen-deficient PISN at such a
high metallicity seems difficult to accommodate with stellar evolution
\citep{langer_etal_07}.\footnote{The lack of spatial resolution prevents the accurate inference
of the metallicity at the actual site of SN\,2007bi, so one cannot exclude the possibility that the
actual primordial cloud from which the progenitor of SN\,2007bi formed
may be of lower metallicity than the inferred value  of 0.2--0.3\,\zsun. In that case, a Type Ic PISN
may not be excluded.} The nebular spectral analysis of
\citet{galyam_etal_09} suggests a large ejecta mass, i.e., much larger than in standard core-collapse SNe,
but the accuracy of this inferrence does not unambiguously support a PISN ---
within the errors these masses are compatible with a progenitor that would not
experience the pair-production instability.

 The alternative core-collapse scenario of \citet{moriya_etal_10} for SN2007bi requires extreme
properties, which may be acceptable given the scarcity of these events, but it is also in contradiction
with the environmental metallicity. Their proposed model yields a 36\,B
($1\mathrm{B} \equiv 1 \mathrm{Bethe} = 10^{51}$\,erg) ejecta kinetic energy, which, combined
with the binding energy of the progenitor, requires the remarkable extraction of $\lesssim$\,50\% of the gravitational energy
of a typical neutron star.

\citet{kasen_bildsten_10} and \citet{kasen_etal_11} have fueled this controversy further by obtaining equally
satisfactory fits to the SN 2007bi bolometric light curve using either a magnetar model or a PISN model. In contrast,
in a recent study  based primarily on the simulations presented here \citep{dessart_etal_12d}, we emphasized that PISN
models systematically have red colors and narrow lines after light-curve peak -- properties that are in strong
contradiction with the observations of SN\,2007bi. \citet{dessart_etal_12d}
suggest that delayed energy injection, as from a  magnetar, into a lower mass ejecta would be more amenable
to producing both blue colors and broad emission lines typical of such superluminous SNe.

An independent problem with PISNe, if they were to be the first chemical nurseries in the Universe, is that their
distinctive nucleosynthetic yields \citep{HW02,umeda_nomoto_02,chieffi_limongi_04} are in apparent
contradiction with those inferred from extremely metal poor stars \citep{cayrel_etal_04,cohen_etal_08}. Such abundance
studies suggest that the metals arose from  ``standard'' core-collapse explosion of stars of
moderate mass \citep{heger_woosley_10}, although a contrived scenario involving a combination of SN with
strong fallback and highly-energetic SNe was proposed by \citet{nomoto_etal_06}.

In this work, we perform a quantitative study of PISN explosions arising from RSG, BSG, and WR star
progenitors, investigating their photometric and spectroscopic
signatures, and documenting in more details the results presented by \citet{kasen_etal_11} and \citet{pan_etal_12a}.
We allow for non-LTE at all times and include non-thermal processes,
which are not just important for exciting helium atoms in SNe Ib \citep{lucy_91} -- they directly
affect all species by driving level populations  away from LTE. They also indirectly affect the thermodynamic
state of the gas by channeling a fraction of the decay energy into excitation and ionization of ions and atoms
rather than into heat. We compute the full
time-dependent radiative transfer within the non-LTE framework, allowing the computation of the spectral
evolution from the photospheric to the nebular phase. Multi-band light curves are obtained by integrating the
resulting spectra over specified bandpasses. This unique approach allows to compute light curves and spectra
simultaneously and with the same level of complexity. It also allows us to start from physically-consistent ejecta
resulting from stellar evolution and explosion modelling. This detailed study aims at providing the key
spectroscopic signatures of the three possible types of PISN explosions, with the hope of identifying the critical
signatures that would help lifting the ambiguities surrounding the association of PISNe with events like SN\,
2007bi \citep{dessart_etal_12d}.

In the next section, we review the numerical setup for our calculations, including the pre-SN evolution from
the main sequence (Sect.~\ref{sect_presn}), the pair-instability explosion mechanism (Sect.~\ref{sect_expl}), and
the setup for the radiative-transfer calculations presented in this work (Sect.~\ref{sect_setup_cmfgen}). We then
describe the bolometric light curve of our PISN models, and compare them with those arising from ``standard"
core-collapse SN explosions of 15-25\,\msun\ RSG, BSG, and WR stars evolved as single or binary stars
(Sect.~\ref{sect_lc}). In Sect.~\ref{sect_phot} we connect these light-curve properties to those at the photosphere.
We then describe in detail the spectral evolution for each group of progenitors during the PISN photospheric
and nebular phases (Sect.~\ref{sect_spec_evol}).
We compare our simulations to PISN candidates SN\,2007bi and SN\,2006gy in Sect.~\ref{sect_comp_obs};
additional information, in particular the proposition that the SN 2007bi light curve is more
compatible with magnetar power, can be found in \citet{dessart_etal_12d}. We finally advocate the potential use
of PISN explosions as a diagnostic of the environmental metallicity in Sect.~\ref{sect_z_indic}. We summarize our
results and discuss their implications in Sect.~\ref{sect_disc}. To limit the length of the main body of the paper,
we provide additional results in an appendix (provided as online material only),
covering in more detail the evolution of the ejecta properties (e.g.,
ionization state and optical depth;  Appendix~A),  the atomic data and model atoms (Appendix~B),
and the dependency of our results on model atoms (Appendix~C).
We finally provide tabulated values for the luminosity and magnitudes for all time sequences
(Appendix~D).


\section{Pre-SN evolution, explosion, and model setup}
\label{sect_prep}

   The work presented in this paper was produced in several independent steps.
   First, a large grid of massive-star progenitors
   was evolved from the main sequence until all models underwent collapse due to the pressure deficit following
   pair production. These simulations were then remapped into a 1D radiation-hydrodynamical code,
   with allowance for (explosive) nuclear burning and radiation transport,
   and followed through the explosion phase for a few years. A detailed discussion of the results from
   these two steps will be presented in a separate paper (Waldman et al., in prep.). Here, we focus on a subset
   of four PISN models from this large grid of models, and specifically
   arising from the explosion of a RSG star, two BSG stars, and a bare He core. This paper discusses
   the evolution of these ejecta and their radiation taking into account both non-LTE and time-dependent effects.
   In Sect.~\ref{sect_presn}--\ref{sect_expl}, we present briefly the first two steps and then
   present the setup for the radiative-transfer calculations in Sect.~\ref{sect_setup_cmfgen}.
   Additional information is provided in the Appendix.

\subsection{pre-SN evolution}
\label{sect_presn}

Using the 1-D stellar-evolution code \mesa\ \citep{paxton_etal_11}, we perform calculations
of super-massive stars from the main-sequence until the onset of the pair-production instability.
We focus on progenitors with main-sequence masses between 160 and 230\,\msun, adopt a metallicity of
10$^{-4}$\,\zsun\  and neglect rotation.
In this paper, we focus on a small sample of PISN progenitors that synthesize a large amount of
\isoni\ suitable to produce a superluminous SN. We thus ignore the less massive progenitors that
encounter the pair-production instability but produce low explosion energies with a small \isoni\ mass (see, e.g.,
\citealt{kasen_etal_11}).
We treat convection by the well known mixing length theory (MLT)
using the Schwarzschild criterion and a mixing length parameter of 1.6, and ignore convective overshoot.
During core hydrogen and helium burning the ``basic'' 8-species nuclear reaction network of \mesa\ is used,
including the following isotopes: \iso{1}H, \iso{3,4}He, \iso{12}C, \iso{14}N, \iso{16}O, \iso{20}Ne, \iso{24}Mg.
From the onset of core carbon burning, the nuclear reaction network is expanded.
For the He100 model (see details below) the $\alpha$-chain elements are added, while for the other models,
in order to account for more nuclei that may manifest in the resulting spectra,
we use a more elaborate nuclear reaction network provided with  \mesa.
In addition to the $\alpha$-chain elements, it includes
the intermediate elements linking those through $(\alpha, p)(p,\gamma)$ reactions,
namely: n, p, \iso{3,4}He, \iso{12}C, \iso{13,14}N, \iso{16}O, \iso{19}F, \iso{20}Ne,
\iso{23}Na, \iso{24}Mg, \iso{27}Al, \iso{28}Si, \iso{31}P, \iso{32}S, \iso{35}Cl,
\iso{36}Ar, \iso{39}K, \iso{40}Ca, \iso{43}Sc, \iso{44}Ti, \iso{47}V, \iso{48}Cr,
\iso{51}Mn, \iso{52,54,56}Fe, \iso{55,56}Co, \iso{56}Ni. Pair creation is treated in the equation
of state used in \mesa, as well as in \v1d\ (see below and, e.g., \citealt{cox_giuli_68}).

A crucial ingredient in the models is mass-loss. Apart from the He100 model, which was evolved without mass loss, we use
two standard mass-loss prescriptions that are functions of the effective temperature, $T_{\rm eff}$, of the stellar model.
For cool stars, defined by $T_{\rm eff} < 10^4$\,K, we use the formula of \citet{deJager_etal_88},
multiplied by the metallicity-dependent factor $(Z/Z_\odot)^{0.5}$ \citep{kudritzki_etal_87}.
For hotter stars, we use the formula of \citet{vink_etal_01}, which contains a $(Z/Z_\odot)^{\alpha}$ metallicity
dependence, where $0.64<\alpha<0.69$ depending on the effective temperature.
Using these mass-loss prescriptions, the pair-instability is encountered when the star is a RSG. Our representative
model for this scenario is R190. None of our models, evolved from the main sequence,
lose enough mass to become a WR star, defined by
$X_{\rm surface}<0.4$ and  $T_{\rm eff} > 10^4$\,K. This is expected -- at the low metallicity of
10$^{-4}$\,\zsun\ adopted here, radiation-driven winds will be significantly inhibited during the
object's life as an O star.

To allow for uncertainties in mass loss (as well as for giant-eruptions, such as that associated with $\eta$ Car,
which are not covered by the above procedures), we perform additional simulations.
To mimic the possibility of a stronger mass loss, we artificially truncate the hydrogen
envelope of the RSG progenitor immediately before explosion, reducing the total mass and surface radius,
and producing a BSG progenitor (models B190 and B210). To mimic an even greater mass-loss rate by means
not understood today (although mass-transfer in a binary may be an option), we also evolve objects
directly from a bare helium core (model He100). We summarize the properties for these four models
in Table~\ref{tab_pre_pisn}. In our nomenclature, RX, BX, and HeX refer to explosions as RSG, BSG, and
He (WR) stars, where X is the initial mass.

The evolution is generally followed using \mesa's hydrostatic
mode. When an hydrodynamic instability is encountered, which occurs invariably
at central carbon exhaustion, the code is automatically switched to the
hydrodynamic mode, and the run is continued until \iso{20}Ne is exhausted
at the stellar center. At this time, before \iso{16}O burning turns on,
the profile is remapped into the 1D radiation hydrodynamical code V1D \citep{livne_93}.
We find that the nucleosynthetic
outcome is insensitive to the remapping epoch between hydrodynamic instability
and neon exhaustion, but does change for later remapping epochs, mainly due to
differences in the reaction networks used in the two codes.

\begin{figure*}
\epsfig{file=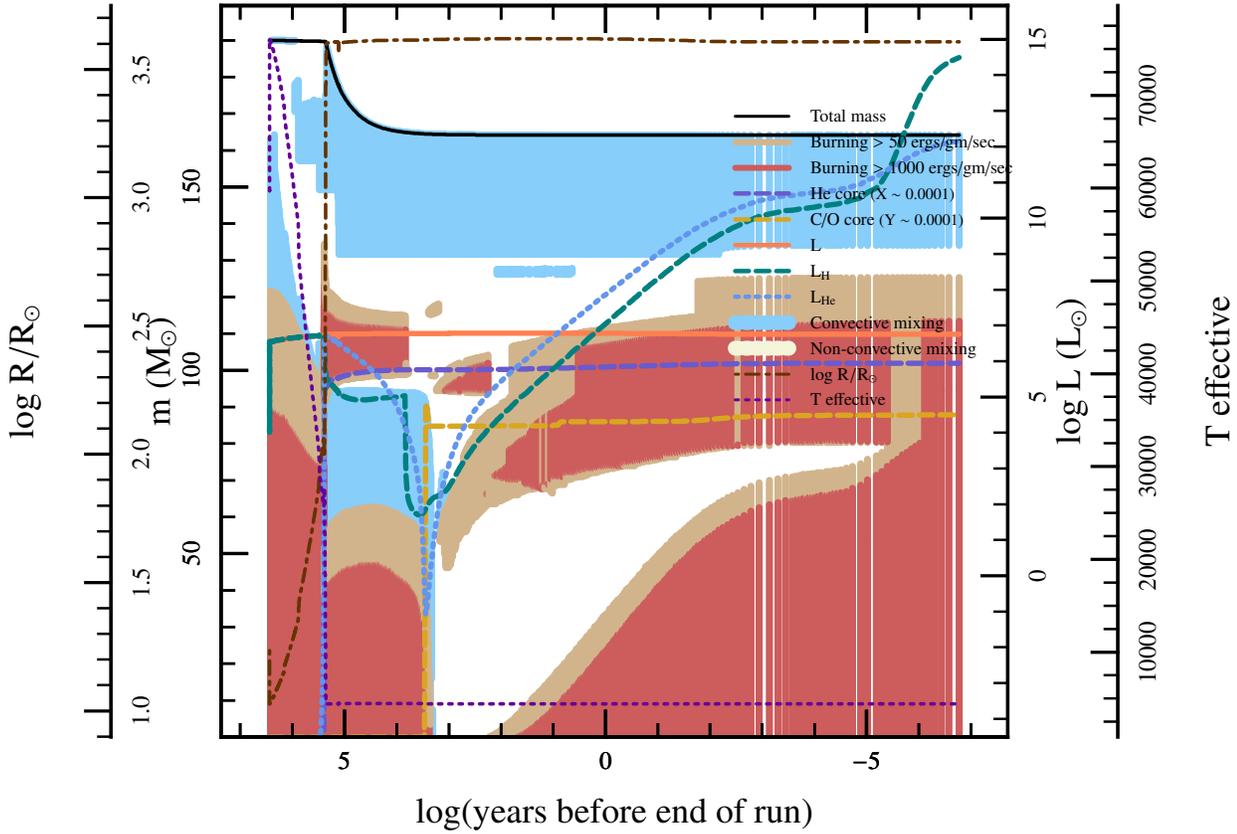,width=18cm}
\caption{
Kippenhahn plot of the 190\,\msun\ RSG model (i.e., R190), showing burning and
convective regions, core sizes, mass, luminosity, radius and effective temperature
as a function of the remaining time before the onset of the pair-production instability.
The ``luminosities" $L_{\rm H}$ and $L_{\rm He}$ correspond to the nuclear-energy generation rates
associated with H and He burning --- there is also a strong neutrino luminosity (not shown here).
Convective mixing refers to convective regions {\it per se}, while non-convective mixing corresponds to
regions where semi-convection, overshoot and/or thermohaline mixing occur.
\label{fig:kip}
}
\end{figure*}

Our models typically reside on the main sequence for about 2.5\,Myr,
gradually cooling from an O-star of about 80,000\,K and $\sim$\,10\,\rsun, finally
settling by the end of core helium burning, which lasts about 0.2-0.3\,Myr,
as a K-type supergiant of about 4300\,K and 4000--5000\,\rsun. Throughout its evolution,
the luminosity of the star does not experience dramatic changes --  it is 3--4$\times$10$^6$\,\lsun,  on the ZAMS,
gradually growing by about a factor of 1.5 towards the end
of the main sequence, and remaining virtually constant through the rest of the evolution.
The star burns hydrogen
and helium in shells for a period on the order of 10$^4$\,yr, while its core contracts, until carbon
is ignited at the center at $T_{\rm c}\sim$\,10$^9$\,K, $\rho_{\rm c}\sim$\,2$\times$\,10$^4$\,g\,cm$^{-3}$.
Since the carbon mass fraction is relatively low (on the order of 0.06) the energy release cannot prevent
the continued contraction of the core. Central carbon is exhausted in less than a year. An
hydrodynamic instability, which  typically starts at
$T_{\rm c}\sim$\,1.2$\times$\ 10$^9$\,K and  $\rho_{\rm c}\sim$\,6$\times$10$^4$g\,cm$^{-3}$, then occurs
and this is followed by central neon exhaustion that takes several hours to be achieved. At this stage
 $T_{\rm c}\sim$\,2$\times$\,10$^9$\,K and  $\rho_{\rm c}\sim$\,4$\times$10$^5$\,g\,cm$^{-3}$,
while infalling velocities typically reach several 100\,\kms.
The main properties of the models at the onset of the hydrodynamic instability are given in Table~
\ref{tab_pre_pisn}. We also show a Kippenhahn diagram for the RSG model R190 that illustrates
the evolution of the internal structure and surface properties prior to the pair-production instability
(Fig.~\ref{fig:kip}).

With our small grid, we thus cover the 3 different stellar types under which massive stars may explode, i.e.,
as RSG, BSG, or WR stars, resulting in explosions as SNe II-P, II-pec, and Ib/c. The only configuration
not considered here is SN IIn, which could stem from interaction between consecutively ejected shells
following pair instability pulses \citep{WBH07_puls_intsab}.

\begin{table}
\begin{center}
\caption[]{Summary of the composition, in solar masses, for each model immediately before the onset of the pair-instability
explosion.
\label{tab_pre_pisn}}.
\begin{tabular}{l@{\hspace{1.6mm}}c@{\hspace{1.6mm}}c@{\hspace{1.6mm}}c@{\hspace{1.6mm}}
c@{\hspace{1.6mm}}c@{\hspace{1.6mm}}c@{\hspace{1.6mm}}c@{\hspace{1.6mm}}
c@{\hspace{1.6mm}}c@{\hspace{1.6mm}}c@{\hspace{1.6mm}}}
\hline
  Model & M$_{\rm i}$ & M$_{\rm f}$ & \iso{1}H & \iso{4}He & \iso{12}C & \iso{16}O &
\iso{20}Ne & \iso{24}Mg & \iso{27}Al & \iso{28}Si \\
\hline
 R190 & 190.0 & 164.1 & 23.6 & 46.1 & 5.3 & 78.4 & 8.6 & 2.2 & 0.06 & 0.05  \\
 B190 & 190.0 & 133.9 & 5.7 & 33.8 & 5.3 & 78.4 & 8.6 & 2.2 & 0.06 & 0.05  \\
 B210 & 210.0 & 146.7 & 3.7 & 30.8 & 5.9 & 92.5 & 10.8 & 3.0 & 0.08 & 0.06  \\
 He100 & 100.0 & 100.0 & 0.0 & 8.5 & 5.6 & 75.4 & 8.1 & 2.1 & 0.00 & 0.07 \\
\hline
\end{tabular}
\end{center}
\end{table}


\begin{table}
\begin{center}
\caption[]{Summary of the main properties for the four hydrodynamical input models B190, R190, B210, and
He100 explored in this study. All \cmfgen\ models are run assuming either local energy deposition,
or non-local energy deposition (model name is then appended by ``NL'').
\label{tab_sum}
}
\begin{tabular}{l@{\hspace{1.6mm}}c@{\hspace{1.6mm}}c@{\hspace{1.6mm}}c@{\hspace{1.6mm}}
c@{\hspace{1.6mm}}c@{\hspace{1.6mm}}c@{\hspace{1.6mm}}c@{\hspace{1.6mm}}}
\hline
Model    &  Type & $M_{\rm i}$  & $M_{\rm f}$ &    \rstar       &   E$_{\rm kin}$    &    $M_{\rm ejecta}$  &
$M_{^{56}{\rm Ni}}$    \\
               & &   [\msun]         & [\msun]         &    [\rsun]    &   [B]                       &             [\msun]           &
[\msun]   \\
               \hline
B190(NL)             & BSG   & 190.0    & 133.9    &   186      &   34.5     & 133.9   & 2.99 \\
R190(NL)             &RSG  & 190.0    & 164.1    &   4044   &  33.2     &  164.1   & 2.63  \\
B210(NL)             &BSG  &  210.0    & 146.7   &    146     &  65.9     &   146.7  &  21.3 \\
He100(NL)$^a$  & WNE & 100.0    & 100.0   &     1         &    37.6   &    100.0  & 5.02 \\
He100K$^b$       & WNE  & 100.0    & 100.0   &     ...           &    40.9    & 100.0    & 5.00 \\
\hline
\end{tabular}
\end{center}
\flushleft
{\bf Notes:}
For all stellar-evolution models, the environmental metallicity is 10$^{-4}$\,\zsun.
Type stands for the progenitor stellar type.
$^a$: The He-star model is run with two different sets of model atoms: He100 uses the same model
atoms as B190, R190, and B210. He100ionI is identical in all ejecta and model parameters as He100 except that
it is modeled with the additional neutral species Mg\one, Si\one, S\one, and Ca\one\ (Appendix~C).
$^b$: Corresponding properties of the He100 model of \citet{kasen_etal_11}, which we name He100K to avoid
confusion.
\end{table}

\begin{table*}
\begin{center}
\caption[]{Summary of the chemical composition in ejecta models B190, R190, B210, and He100, immediately
after explosive nucleosynthesis stops and prior to decay of unstable isotopes (specifically \isoni). We limit the
table entries to H and He, CNO elements, the main IMEs, and finally Fe, Co, and Ni. The total mass of \isoni\
synthesized in the explosion is given in Table~\ref{tab_sum}. The top half of the table shows the cumulative
yields, and the bottom half the mass fractions at the progenitor surface. The \isoni\ surface mass fraction is zero
for all models. All models assume an original metallicity of 10$^{-4}$\,\zsun.
Numbers in parenthesis correspond to powers of ten.
\label{tab_presn}}
\begin{tabular}{l@{\hspace{1.2mm}}c@{\hspace{1.3mm}}c@{\hspace{1.3mm}}c@{\hspace{1.3mm}}
c@{\hspace{1.3mm}}
c@{\hspace{1.3mm}}c@{\hspace{1.3mm}}c@{\hspace{1.3mm}}c@{\hspace{1.3mm}}c@{\hspace{1.3mm}}
c@{\hspace{1.3mm}}
c@{\hspace{1.3mm}}c@{\hspace{1.3mm}}c@{\hspace{1.3mm}}c@{\hspace{1.3mm}}c@{\hspace{1.3mm}}}
\hline
Model      &$M_{\rm H}$&$M_{\rm He}$ &$M_{\rm C}$ &$M_{\rm N}$ &$M_{\rm O}$&$M_{\rm Ne}$&$M_{\rm
Mg}$&$M_{\rm Si}$
&$M_{\rm S}$&$M_{\rm Ar}$&$M_{\rm Ca}$&$M_{\rm Fe}$&$M_{\rm Co}$ &$M_{\rm Ni}$  \\
           & [\msun]  & [\msun]   & [\msun]   & [\msun]   & [\msun]  & [\msun]   & [\msun]    & [\msun]   & [\msun]   &
[\msun]   & [\msun]
           & [\msun]   & [\msun]   & [\msun]    \\
\hline
R190  & 2.36(1)  &  4.53(1)   & 1.06(0)   & 5.06(-3)   & 4.81(1)  &  2.03(0)   &  2.87(0)  &  1.96(1)  &
1.30(1)   & 2.69(0)
&  2.36(0)  &  1.81(-1)  &  5.09(-3)  &  2.64(0)     \\
B190  & 5.72(0)  &  3.30(1)   & 1.05(0)   & 5.56(-3)   & 4.81(1)  &  1.99(0)  &    2.81(0)  &  1.96(1)  &
1.31(1)   & 2.72(0)
&  2.39(0)  &  2.03(-1)  &  2.53(-3)  &  3.01(0)     \\
B210  & 3.72(0)  &  3.04(1)   & 6.73(-1)   & 7.35(-3)   & 3.96(1)  &  1.43(0)    &  2.49(0)  &  2.34(1)  &
1.58(1)   & 3.52(0)
&  3.19(0)  &  4.35(-1)  &  6.15(-4)  &  2.15(1)     \\
He100 & 5.22(-7)  &  7.91(0)   & 1.36(0)   & 5.05(-3)   & 4.36(1)  &  2.11(0)   &  2.52(0)  &  1.94(1)  &
1.27(1)   & 2.54(0)
&  2.41(0)  &  2.67(-1)  &  6.27(-3)  &  5.04(0)     \\
\hline
Model       &$X_{\rm H}$  &$X_{\rm He}$ &$X_{\rm C}$ &$X_{\rm N}$ &$X_{\rm O}$&$X_{\rm Ne}$&$X_{\rm
Mg}$&$X_{\rm Si}$
&$X_{\rm S}$&$X_{\rm Ar}$&$X_{\rm Ca}$&$X_{\rm Fe}$&$X_{\rm Co}$&$X_{\rm Ni}$  \\
\hline
R190    & 5.93(-1)  & 4.07(-1)   & 1.23(-7)   & 1.10(-6)  &  3.89(-7)  &  1.74(-7)   & 6.47(-8)  &  6.99(-8)
& 3.65(-8)
& 1.02(-8)  &  6.44(-9)  &  1.40(-7)  &  3.43(-10)  &  7.32(-9)     \\
B190    & 5.93(-1)  & 4.07(-1)   & 1.23(-7)   & 1.10(-6)  &  3.89(-7)  &  1.74(-7)   & 6.47(-8)  &  6.99(-8)
& 3.65(-8)
& 1.02(-8)  &  6.44(-9)  &  1.40(-7) &  3.43(-10)  &  7.32(-9)      \\
B210    & 5.78(-1) & 4.22(-1)   & 1.14(-7)   & 1.14(-6)  &  3.48(-7)  &  1.74(-7)    & 6.47(-8)  &  6.99(-8)
& 3.65(-8)
& 1.02(-8)  &  6.44(-9)  &  1.40(-7)  &  3.43(-10)  &  7.32(-9)    \\
He100 & 1.00(-15)  & 1.00(0)   & 3.37(-7)   & 9.87(-8)  &  9.16(-7)  &  1.74(-7)      & 2.76(-7)  &  6.99(-8)
& 3.65(-8)
& 1.02(-8)  &  6.44(-9)  &  1.41(-7)  &  3.43(-10)  &  7.32(-9)      \\
\hline
\end{tabular}
\end{center}
\end{table*}

\begin{figure*}
\vspace{0.5cm}
\epsfig{file=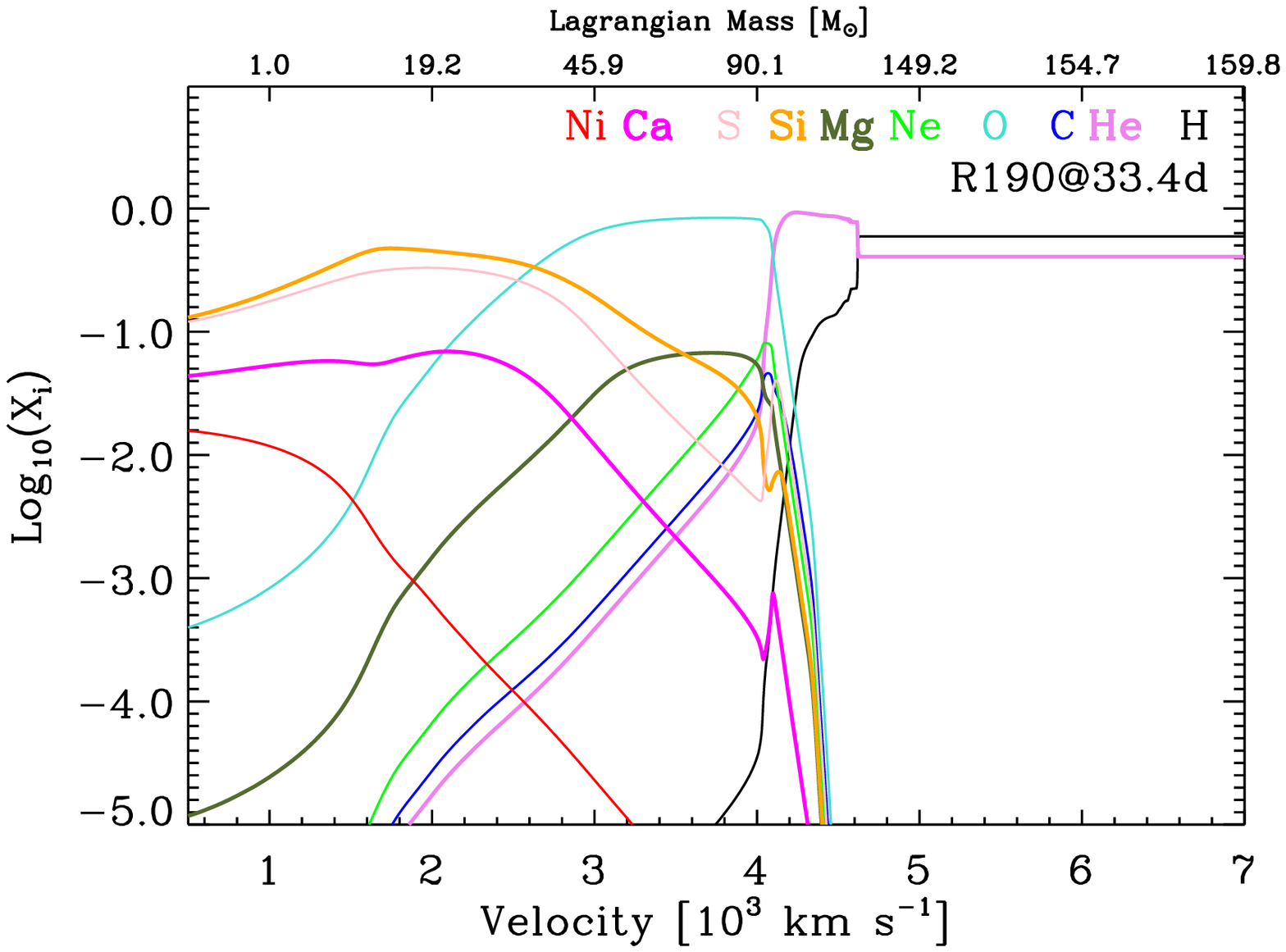,width=8.75cm}
\epsfig{file=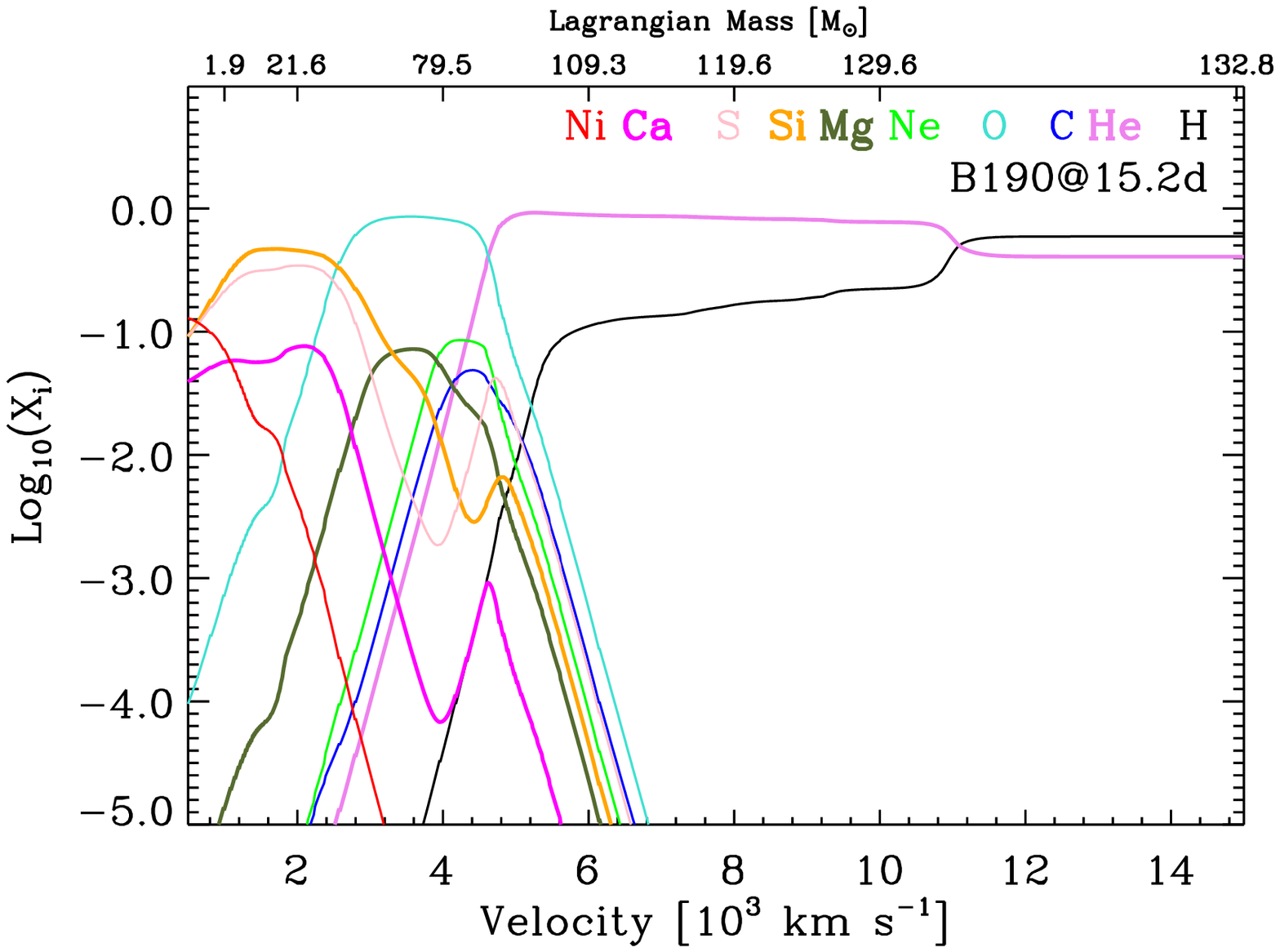,width=8.75cm}
\epsfig{file=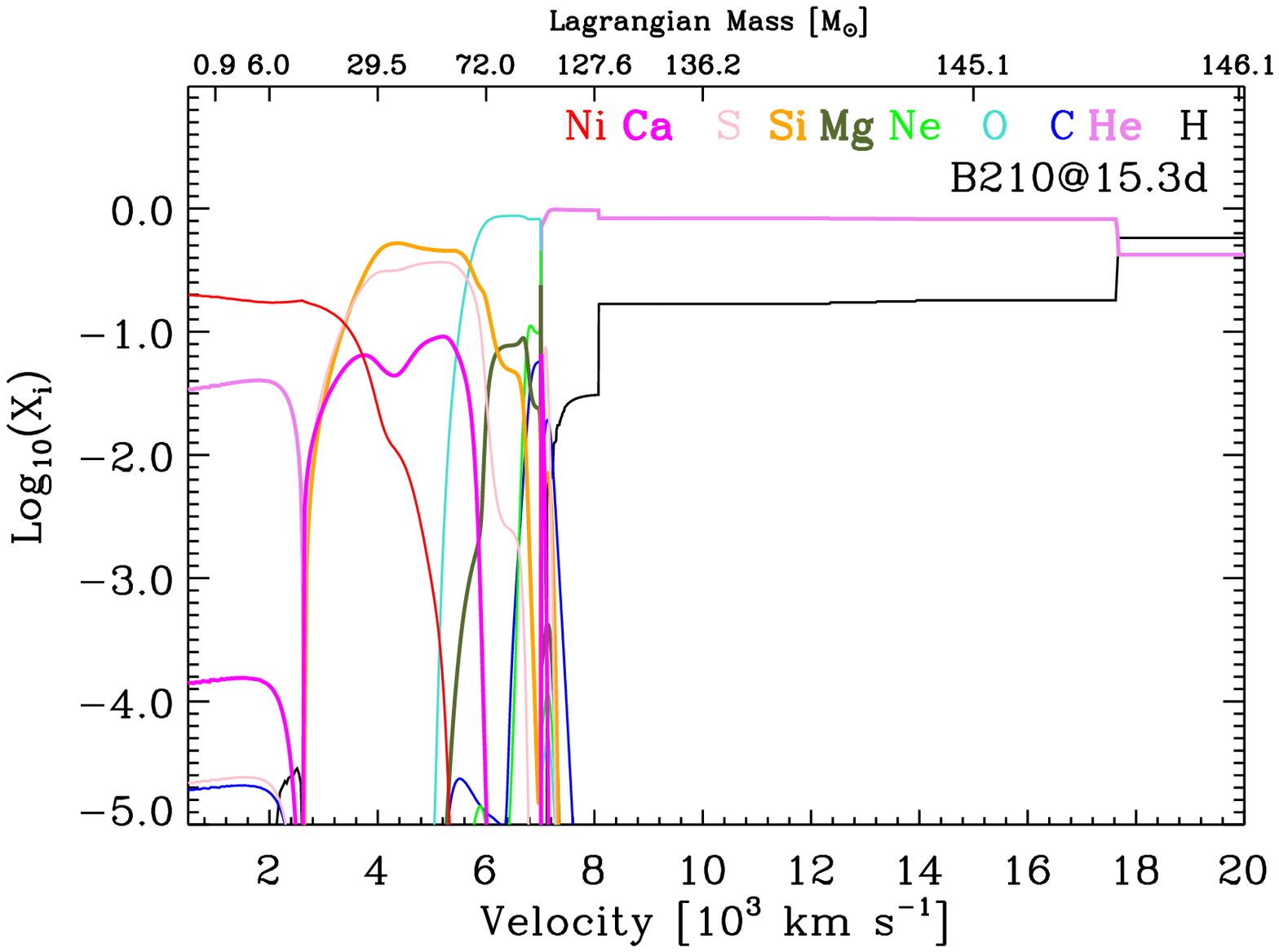,width=8.75cm}
\epsfig{file=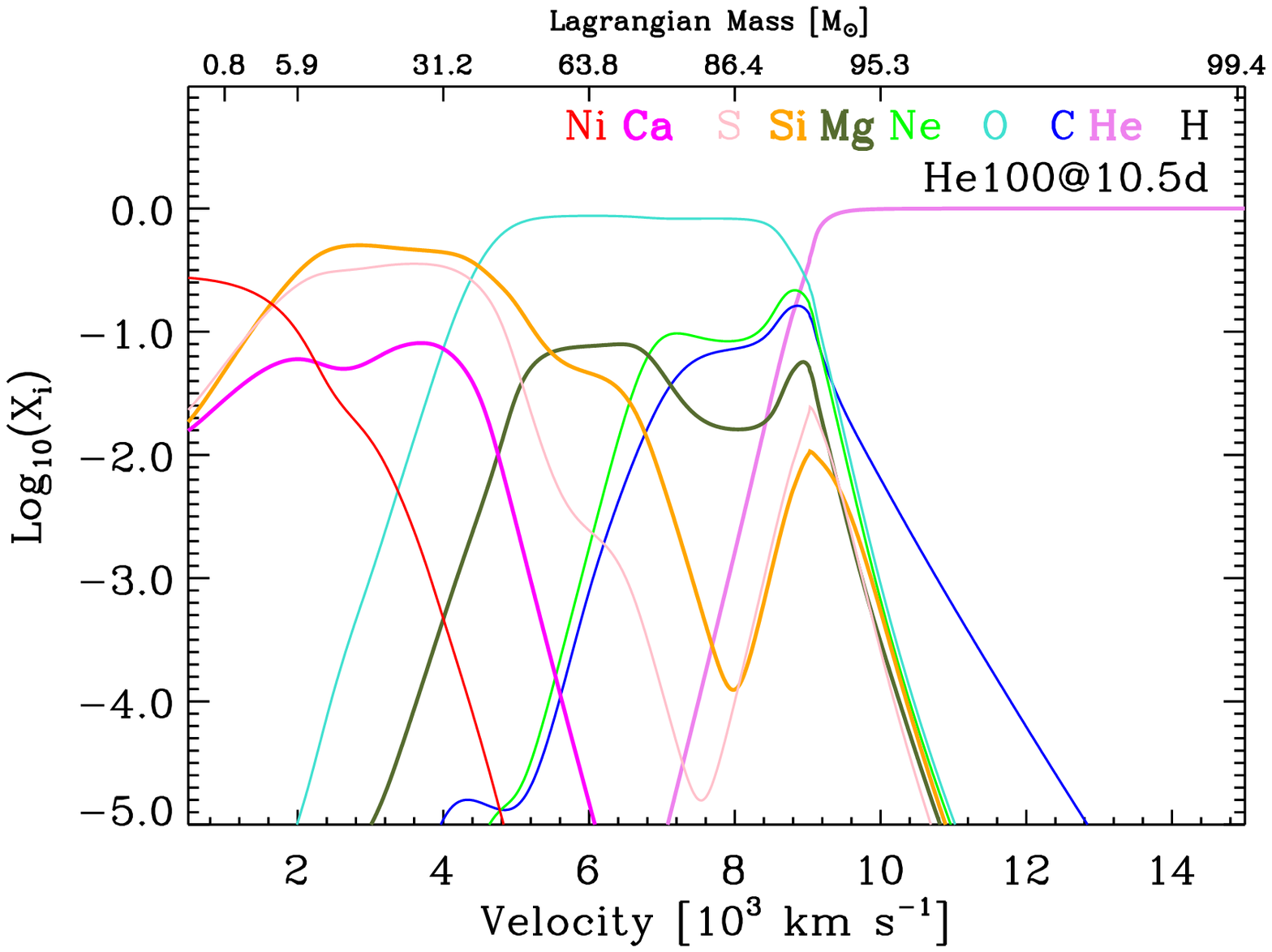,width=8.75cm}
\caption{Illustration of the chemical stratification in velocity space (bottom axis) and in mass space (top axis) for the four
models discussed in this study (B190: Top left; R190: Top right; B210: Bottom left; He100: Bottom right). The stratification
shown is not that of the original models -- rather we illustrate the models used as input for \cmfgen\ in
which the composition gradients were softened (except for B210).
For better visibility, we limit the range and thus do not show the outer (homogeneous) regions.
We also only show a selection of important species, rather than all species included in the calculations.
\label{fig_summary_presn}
}
\end{figure*}

\subsection{Explosion}
\label{sect_expl}

To model the explosion phase, we use a more extended reaction network, which in addition allows for $
(\alpha,n)$ reactions up to calcium, for $(n,\gamma)$ reactions up to nickel, and for the radioactive decay of
$\iso{56}Ni \rightarrow \iso{56}Co \rightarrow \iso{56}Fe$. In total, we include the isotopes: n, p, \iso{4}He,
\iso{12}C, \iso{14}N, \iso{16}O, \iso{20-22}Ne, \iso{23}Na, \iso{23-26}Mg, \iso{27}Al, \iso{27-30}Si,
\iso{31}P, \iso{31-34}S, \iso{35}Cl, \iso{35-38}Ar, \iso{39}K, \iso{39-42}Ca, \iso{43}Sc, \iso{44-46}Ti,
\iso{47}V, \iso{48-50}Cr, \iso{51}Mn, \iso{52-54,56}Fe, \iso{55-56}Co, \iso{56-58}Ni, \iso{59}Cu, \iso{60}Zn.

Following central neon exhaustion, oxygen is ignited, and consumed in less than a minute. At this stage
the typical central conditions are $T_{\rm c}\sim$\,3.5$\times$10$^9$\,K and $\rho_{\rm c}\sim$\,10$^6$\,g\,cm
$^{-3}$,  while the infall velocities reach several 1000\,\kms. In the subsequent $\sim$\,10\,s, central
silicon is exhausted, and the implosion turns into an explosion. The central conditions reached at
bounce depend on the mass of the model and the binding energy of the star, but are typically
$T_{\rm c}\sim$\,3.5--6$\times$\,10$^9$\,K and  $\rho_{\rm c}\sim$\,2-6$\times$10$^6$\,g\,cm$^{-3}$.

During this explosive-burning phase most of the energy is released by oxygen burning, and some by neon and
carbon burning. For example, in our 190\,\msun\ model, $\sim$\,40\,\msun\ of fuel are burnt,
split between 30\,\msun\ of oxygen, 7\,\msun\ of neon, and 4\,\msun\ of carbon, with a total energy release
of 44\,B. This fuel is converted primarily into silicon (20\,\msun)
and sulphur (13\,\msun), the remainder being argon (3\,\msun), calcium (2\,\msun) together
with $\sim$\,3\,\msun\ of \isoni. In contrast, our heavier model B210 burns about 70\%
more fuel, namely 50\,\msun\ of oxygen, 9\,\msun\ of neon, and 5\,\msun\ of carbon, releasing in the process
75\,B. This moderately increases the production of IMEs (23\,\msun\ of silicon, 16\,\msun\ of sulphur,
4\,\msun\ of argon, and 3\,\msun\ of calcium), but dramatically increases the \isoni\ production to 21\,\msun.

The energy liberated through explosive burning causes a huge increase in the ejecta internal energy (in the form
of radiation) and kinetic energy. As the envelope expands and cools, radiative pressure gradients accelerate
the material to its asymptotic velocity on timescales of hours to days. The energy lost to unbind the star
is considerable, on the order of 10\,B in all PISN models.
Asymptotically, the ejecta kinetic energy is 34.5\,B for model B190, 33.2\,B for model R190, 65.9\,B for model B210,
and 37.6\,B for model He100 (Table~\ref{tab_sum}). The mass-weighted average velocity for each model is, in the
same order, 4200, 4000, 6100, and 5500\,\kms\ --- this value can be used to infer a representative ejecta kinetic
energy (that differs  from the exact value by $\lesssim$\,30\%) of 23, 26, 54, and 30\,B.

\begin{figure}
\epsfig{file=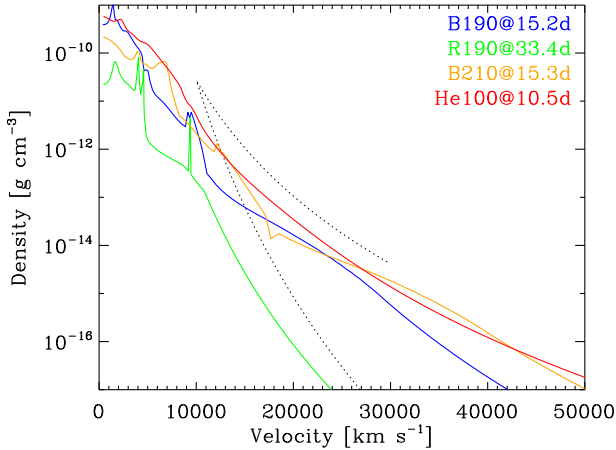,width=8.5cm}
\caption{Ejecta density distribution for the PISN Models B190, R190, B210, and He100
at the onset of our \cmfgen\ calculations at 15.2, 33.4, 15.3, and 10.5\,d after explosion.
For comparison, we overplot as a dotted line the slope for a power law with exponent 8 and 15.
\label{fig_rho_init}
}
\end{figure}

The yields and the chemical stratification are described in Table~\ref{tab_presn} and
Fig.~\ref{fig_summary_presn}. Although PISNe are thermonuclear explosions, the bulk of the burning takes
   place in the inner envelope of the progenitor, yielding a chemical stratification more reminiscent of core-collapse Type Ic SNe:
   The \isoni\ is produced at the base of the ejecta, remains confined to the lowest expanding material with velocities
   less than 4000\,\kms\ (Fig.~\ref{fig_summary_presn}), and the decay energy has to diffuse from there
   through the overlying mass to influence the outer ejecta and the light curve.

The density profiles for each model, at the start of the radiative-transfer simulations, are shown in Fig.~\ref{fig_rho_init}.
We stitch an outer density region to all models to ensure an optically-thin outer boundary. With this procedure,
the outer ejecta velocity for each model is 42000\,\kms\ (B190),  24000\,\kms\ (R190), 50000\,\kms\ (B210), and
52000\,\kms\ (He100). The structure for model He100 is quite smooth, with a density exponent of $\sim$\,8
beyond 5000\,\kms. There is a fair amount of structure in the density profiles for the BSG models and even
more so in the RSG model, in part associated with the formation and propagation of reverse shocks at shell interfaces
and exacerbated by the 1-D treatment.

Hydrodynamical instabilities in PISNe have been investigated by \citet{joggerst_whalen_11}.
They find that mixing in Pop {\sc iii} 150-250\,\msun\ SNe explosions is negligible compared to that obtained in 15-40\,\msun\
Pop {\sc ii} counterparts, although their investigation does not contain H-deficient compact  progenitors and
they treat nuclear burning through an initial phase performed in 1D. More recently, \citet{chen_etal_12} have
repeated such simulations and found that the treatment of nuclear burning in multi-D can enhance
the level of mixing significantly. Since this issue is not settled, we decided to
soften the composition discontinuities by mixing over adjacent mass shells, producing a mild smearing of the
(high-resolution) 1D input structures produced by \v1d. This is done more
as a convenience than aimed at describing the physical effect of mixing seen in multi-D simulations. The chemical stratification
of the models used as input to \cmfgen\ is shown in Fig.~\ref{fig_summary_presn}. Based on the present
work, we however anticipate that mixing will not significantly alter the observables we describe here.
Non-local energy deposition at very late times partly mimics the effect of mixing since \grays\ are then
able to travel from their deeply located emission site in the ejecta to the O-rich, the He-rich, and
possibly the H-rich outer ejecta (Sect.~\ref{sect_neb}).

\subsection{Radiative-transfer calculations}
\label{sect_setup_cmfgen}

   Apart from the different hydrodynamic inputs, the numerical procedure adopted in this work on PISN
radiative-transfer simulations is the same as that employed previously for the simulations of SN II-peculiar
and  in particular SN 1987A \citep{DH10,li_etal_12}, SNe II-Plateau \citep{DH11}, SNe IIb/Ib/Ic
\citep{dessart_etal_11,dessart_etal_12}. \citet{HD12} have recently given a full description of the code
\cmfgen\ for SN applications, detailing the important components for the radiative transfer in SN ejecta.
Since the code can naturally handle a wide range of composition mixtures (all important elements up to
Nickel can be treated), the extension to treat PISN ejecta requires no specific adjustment compared
to the modeling of standard CCSNe or SNe Ia.

  We perform time-dependent simulations for the full ejecta at all times, and start by re-mapping the ejecta
chemical stratification and structure (radius, density, temperature) computed with \v1d\ once homologous expansion
is  reached. In practice, we start the He100 model at 10.5\,d, the B190 model at 15.2\,d, the B210 model
at 15.3\,d, and the R190 model at 33.4\,d after explosion.
This time increases with progenitor radius because for larger stars, the corresponding
ejecta takes longer to significantly expand beyond its original spatial extent.\footnote{In the RSG explosion
model R190, the helium-rich shell, near 4000\,\kms, and sandwiched between the inner O-rich shell and
the outer extended H-rich envelope has still not settled into homology. Here, enforcing homology leads to an
artificial increase in velocity (and hence in radius) for the corresponding layer, producing a higher ejecta mass
by 7\%. The  \cmfgen\ model R190 is then in fact  176\,\msun. The BSG and He models are not affected by
this issue.\label{note_r190}} Given the long duration of the high-brightness phase for these events and
the low early SN luminosity for the  BSG and WR star explosions, starting this late is not a strong limitation here.
For a description of the shock breakout and early light-curve  signatures, the reader should
consult \citet{kasen_etal_11} or Waldman et al. (in prep.).\footnote{We note that given all the uncertainties
on the PISN progenitors and the extreme sensitivity of the breakout signature on the progenitor atmospheric
scale height, the presence of an optically-thick wind, or simply the stellar radius, the shock breakout  signal is far
from a clean signature to constrain the progenitor and explosion properties.}

   While the 1D Lagrangian hydrodynamical model computed with \v1d\ contains about 2000 mass shells, the
radiative transfer calculations are performed with a much smaller number of depth points. The critical scale
to resolve is not mass but optical depth and we find that we obtain converged results with typically 5--7 points
per optical-depth decade \citep{DH10}. In these PISN simulations, we use a fixed number of 150 depth points
at all times, providing a satisfactory resolution of $\sim$\,100\,\kms\ near the base of the ejecta, decreasing to
$\sim$\,1000\,\kms\ at the fast-expanding outer edge.

  The bulk of the ejecta is located at high optical depth at early times and its base remains optically-thick
  until 200-300\,d after bolometric maximum.
  However, for consistency and simplicity, we treat the entire ejecta in non-LTE at all times.
  This is not optimal in terms of
  memory requirements, but it turns out that the optically thick layers are close to LTE and converge very fast.
  It also allows us to smoothly handle the phase when the ejecta starts to thin out in the continuum --
  when this occurs some spectral regions may still be thick (e.g., in the UV where many lines overlap,
  blanket efficiently the radiation, and drive the photons towards a blackbody distribution locally) while other
  spectral regions are already transparent (e.g., in the IR where the distribution of line opacity is more sparse).
  In general, numerous lines will remain optically thick for weeks after the continuum has become optically thin.
  Because we model the full ejecta, we adopt a zero-flux inner boundary.

  We  explicitly treat non-thermal processes associated with radiatioactive decay. This is particularly
  important given the large mass of \isoni\ (we only consider the decay of \isoni\ and \isoco\  in this study) and
  the low ionization conditions we obtain in PISN ejecta. Through a solution of the Spencer-Fano
  equation \citep{spencer_fano_54}, we compute the contributions to non-thermal excitation and ionization
  for all elements as well as the contribution to heating, and incorporate this in the non-LTE rate equations
  and the energy  equation. A detailed presentation and application of this non-thermal treatment is given
  by \citet{li_etal_12,dessart_etal_12}.
  For the computation of non-thermal rates, we proceed by injecting all the decay energy as
  high-energy electrons of 2\,keV  and solve for the electron-degradation function at 2000 linearly-spaced
  energy bins down to 1\,eV.
  At each of the 150 ejecta depth points, we construct and solve an upper diagonal matrix of 2000x2000,
  which is trivial, but does require up to 30 minutes in these PISN simulations. We reduce the computational
  burden of our non-thermal solver by considering in the computation only the most important ions of each species.
  As the electron density is the main quantity modulating the magnitude of non-thermal processes in a given
  calculation, and since it varies slowly between consecutive iterations, we recompute the
  non-thermal electron-degradation  function every ten iterations.  With this procedure, the treatment of
  non-thermal processes increases the total computational time of a model typically by a few percent.

\begin{figure*}
\epsfig{file=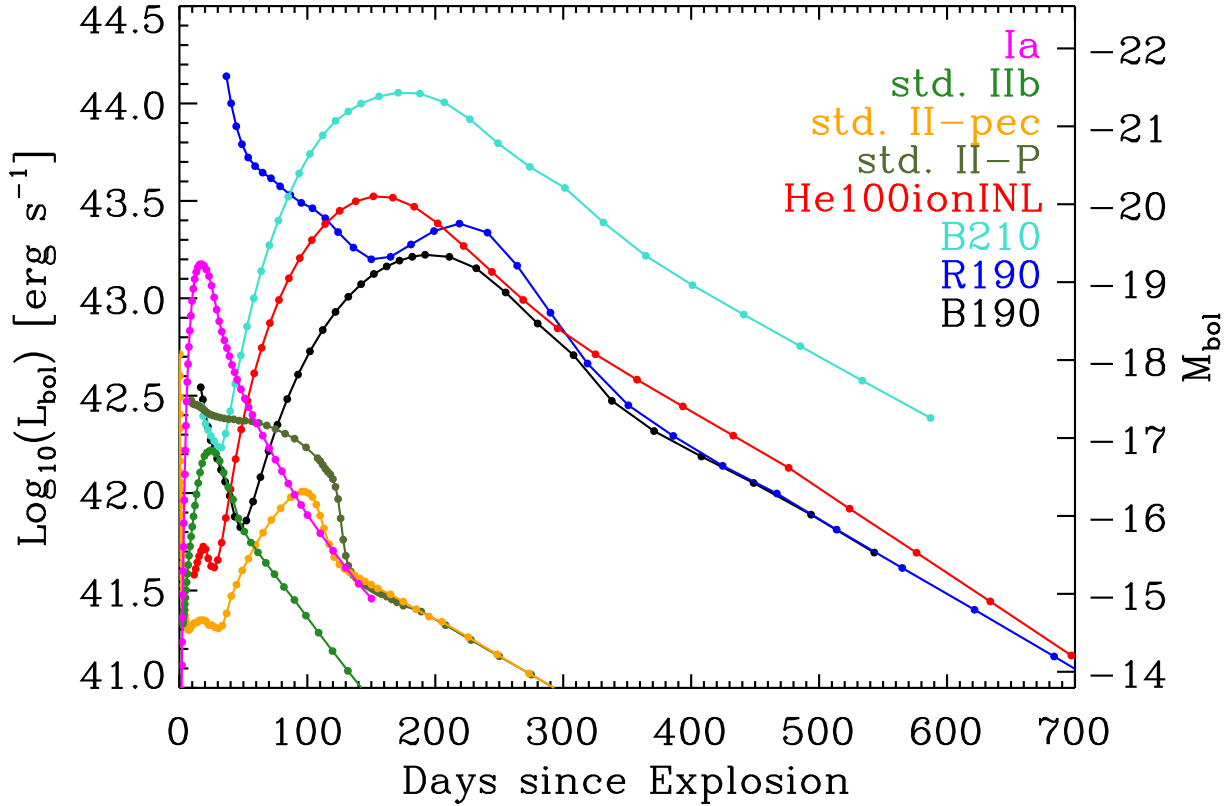,width=17cm}
\caption{Synthetic bolometric light curves extracted from our non-LTE time-dependent {\sc cmfgen}
simulations. Shown are our results for the PISN simulations discussed in this work:
B190 (black), R190 (blue), B210 (turquoise), and He100ionINL (red). To better reveal the extraordinary properties
of these PISN models, we also show our recent results  for a ``standard'' SN II-P (olive; evolved
at solar metallicity  from a single 15\,\msun\ main-sequence star; model  s15e12 of \citealt{DH11}), a ``standard"
II-pec (orange; evolved at the LMC metallicity  from a single 18\,\msun\ main-sequence star;
model  ``lm18a7Ad" of \citealt{DH10}), a ``standard" IIb SN (green; evolved at solar metallicity  from a binary 18\,\msun\
main-sequence star;  model 365A2 of Dessart et al., in prep),  and a SN Ia produced by the
delayed-detonation in a Chandrasekhar-mass white dwarf with 0.67\,\msun\ of \isoni\  (magenta; Dessart et al., in prep).
In these last two models, \gray\ transport is solved for, predicting the fast decline rate during the nebular phase
through  \gray\ escape.
Here, we use the term ``standard'' to stress that such II-P, II-pec, and IIb models match closely the bolometric
light curve for the representative SN of each type observed routinely in the local Universe.
Although B190(NL) and B210(NL) correspond spectroscopically to
a SN II-pec, R190(NL) to a SN II-P, and He100(NL)/He100ionI(NL) to a SN Ic,  their photometric properties
are obviously extraordinary.
For each sequence, small filled circles indicate the actual post-explosion times at which the computations are
performed. For reference, we give the bolometric magnitude on the right-hand-side ordinate axis.
\label{fig_lbol}
 }
\end{figure*}

  The \gray\ escape from the ejecta is minute for all times considered here, but the huge scale and mass
of the ejecta combined with the increasing \gray\ mean free path leads to non-local energy deposition within
the ejecta. We find that this effect leads to the appearance or strengthening of H\one\ or He\one\ lines at nebular
times, and more generally tends to cause the broadening of line profiles with time as the SN progresses into the
nebular phase.  Hence, we typically adopt local energy deposition up to 200\,d after explosion and employ
a \gray\ transport code \citep{HD12} to compute the non-local energy-deposition profile subsequently.
The energy deposition from \grays\ is calculated once at the start of the simulation since it depends
only on density, radius, time, and composition (i.e., quantities that are not affected by the radiative-transfer
solution or the ionization state of the gas) of the ejecta. Hence, \cmfgen\ models are run assuming
either local energy deposition, or non-local energy deposition (model name is then appended by ``NL'', e.g.,
models R190 and R190NL).

The model atoms adopted for all simulations in this work are comparable to that used in \citet{DH11} and are
presented in the appendix, in Table~B1, together with the relevant references. Not all simulations
were performed at the same time and not all have the same model atoms. Simulations B190, R190, B210, and
He100 were performed first, ran with the same model atoms, and neglected the neutral ions Mg\one, Si\one,
S\one, and Ca\one. We added these in the second-generation simulation He100ionI. We in fact use that
simulation to compare to observations of the Type Ic SN 2007bi, as well as to gauge the blanketing effects
from these neutral species. As we discuss in Appendix~C, they have a significant influence on
the spectra and colors, but hardly influence the bolometric light curve. Hence, most properties of model He100
apply to model He100ionI; when these differences are significant, we specifically address them.

The completeness of the model atoms is always an issue in our approach since we cannot blindly incorporate
millions of lines from an extended list. Our line opacity enters the computations through the specific treatment of
atomic/ionic levels and so we have to compromise between completeness and tractability. Each of the four
simulations we present here requires about 40 steps each taking about 3 days on 12 cores. The memory is
mostly taken by the matrix to invert \citep{HM98_lb,HD12}, which contains all the terms from the linearized
statistical and radiative equilibrium equations. The total memory allocated is typically
$NT^2 \times (N_{\rm bands}+1) \times ND$ where $NT$ is the total number of equations (variables),
$N_{\rm bands}$ is 1 (3) for a diagonal (tridiagonal) operator, and $ND$ is the number of depth points.
We typically use $NT=2000$ (as in Table~B1), $N_{\rm bands}=1$, and $ND=150$, corresponding
to a memory of $2000^2 \times 2 \times 150 \times 8 \sim 9.6$\,Gb.

Radiative-transfer simulations for PISNe have been performed in the past.
\citet{scannapieco_etal_03,scannapieco_etal_05} focused on RSG star progenitors giving rise to SNe II-P,
with approximations for the transfer such as flux-limited diffusion and blackbody spectra. Their bolometric light
curves should therefore be accurate but their spectral colors are  uncertain.
\citet{galyam_etal_09} performed similar simulations but focusing on a range of He cores and
explosion characteristics to model the observations of SN 2007bi (they model the light curve and
one nebular-phase spectrum). As we discuss in \citet{dessart_etal_12d}, non-LTE line-blanketed simulations
of  PISN  ejecta (which we present in detail in this work) yield a suitable match to the SN 2007bi light curve, but
are in  conflict  with numerous observed spectral characteristics of SN 2007bi.
\citet{kasen_etal_11} did an extended study of PISN light curves, including RSG, BSG, and WR star progenitors.
As in our grid of models, they do not have a well motivated mechanism for predicting PISN explosions from
BSG or WR stars  so the final masses and stellar types at death remain rather speculative and thus
represent exploratory material.  With their radiative transfer, they considered both the multi-band light curves
and spectra, highlighting for example the striking dependency on metallicity or  the signatures for shock breakout,
but their assumption of LTE for the gas, their approximate treatment of line opacity, and their focus on  early-time
photospheric-phase spectra limit the applicability of their models for confrontation to observations of PISNe
candidates like SN 2007bi. This SN, like most superluminous events today, are still discovered at or beyond
the peak of the LC, expected to occur weeks to months after explosion. At this phase, the SN ejecta has considerably
expanded and starts becoming optically thin in the continuum, while the potential presence of a large amount
of IGE may maintain a high optical depth in lines for many months still. This configuration requires a detailed
non-LTE line-blanketed treatment, which we adopt here.

The radiative-transfer calculations performed with \cmfgen\ yield the level populations,
electron density, and temperature as a function of depth and time. Summing over level populations gives
the ion and atom populations as well as the ionization state of the gas. Concurrently, the properties
of the radiation field are computed as a function of depth, frequency, and angle. In the next sections, we
describe both the gas and the radiation properties, in an attempt to identify
the photospheric-phase and nebular-phase signatures of PISNe.
In the next section, we start by discussing the bolometric light-curve properties.

\begin{figure*}
\epsfig{file=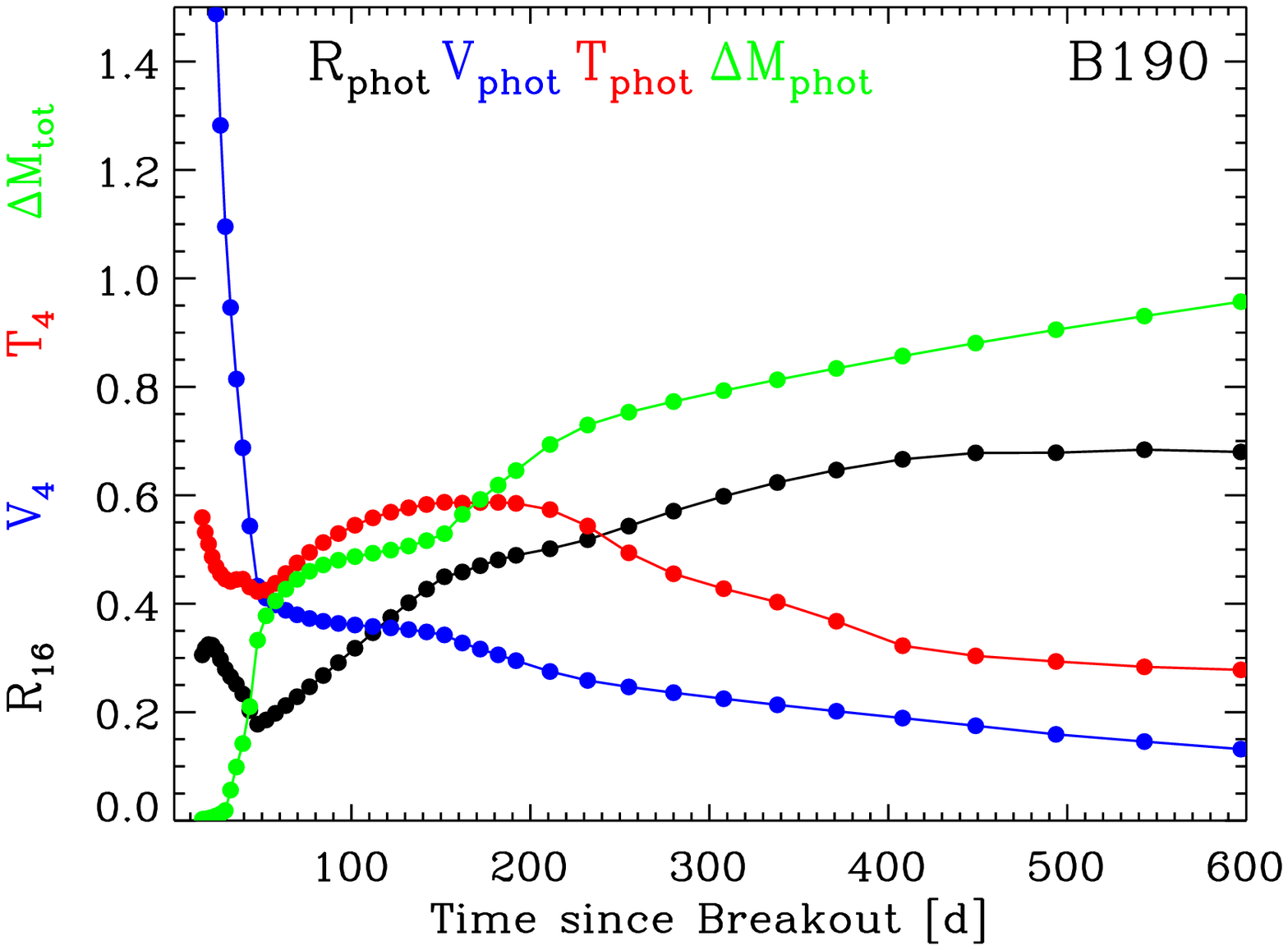,width=8.5cm}
\epsfig{file=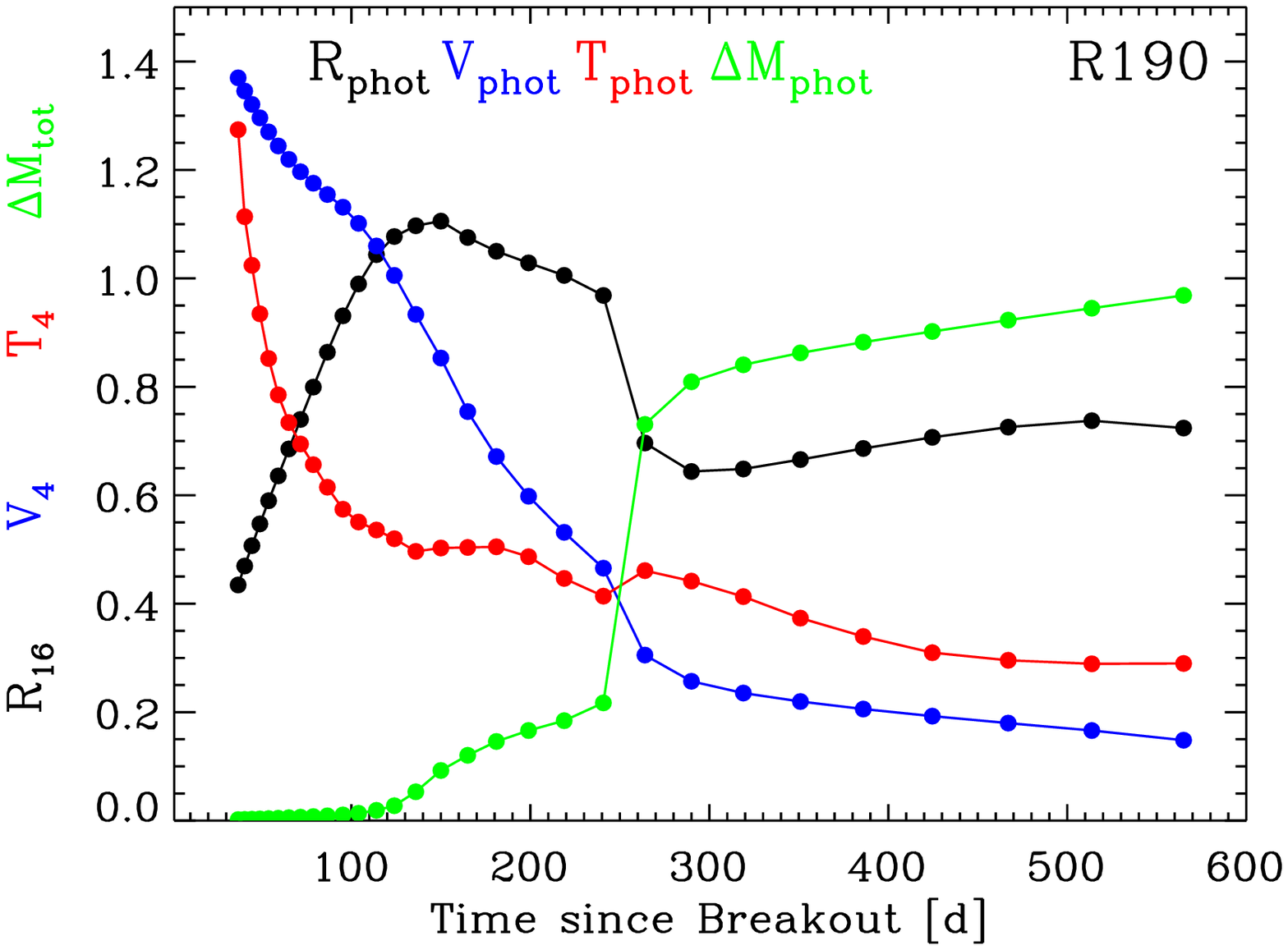,width=8.5cm}
\epsfig{file=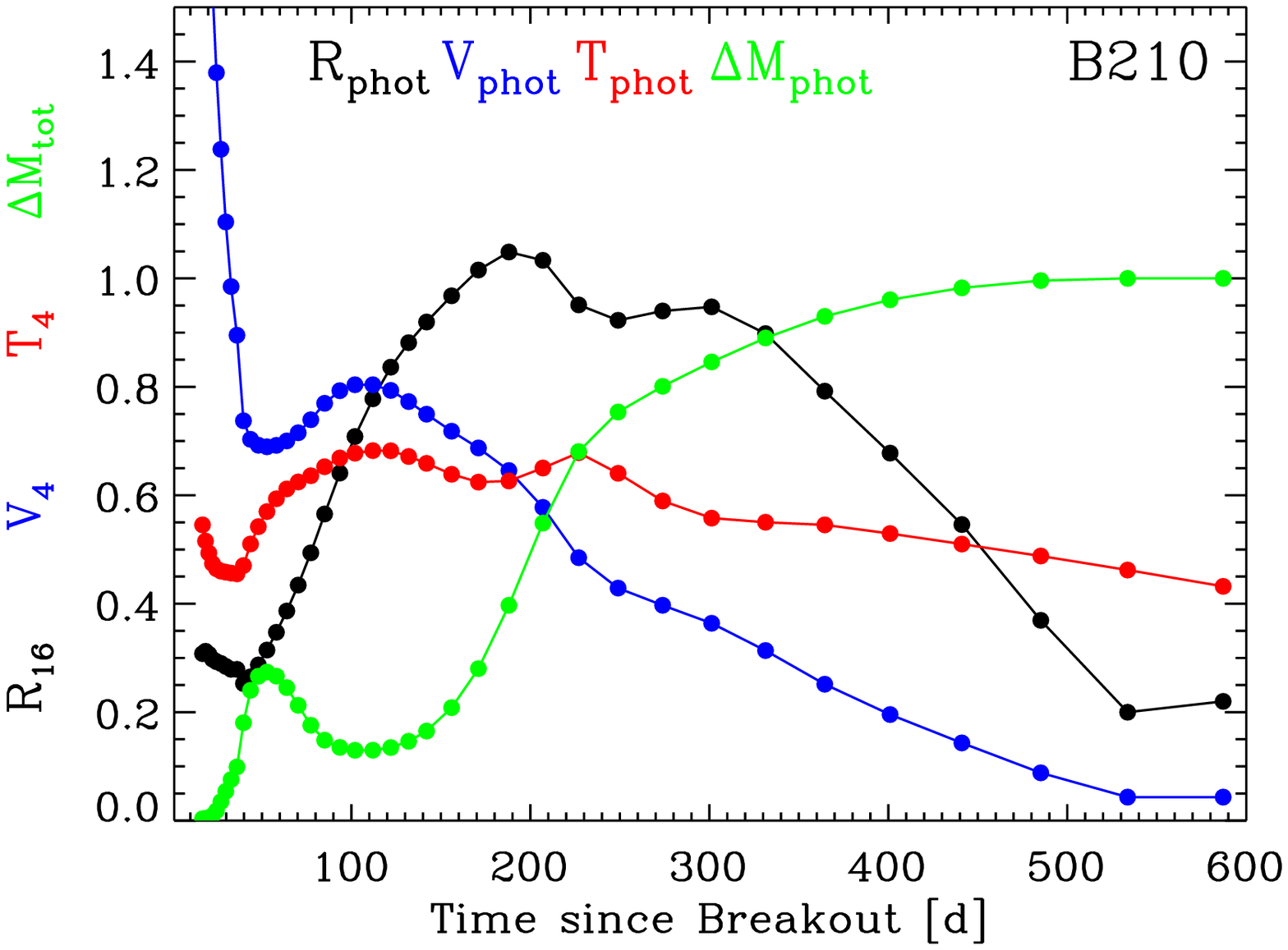,width=8.5cm}
\epsfig{file=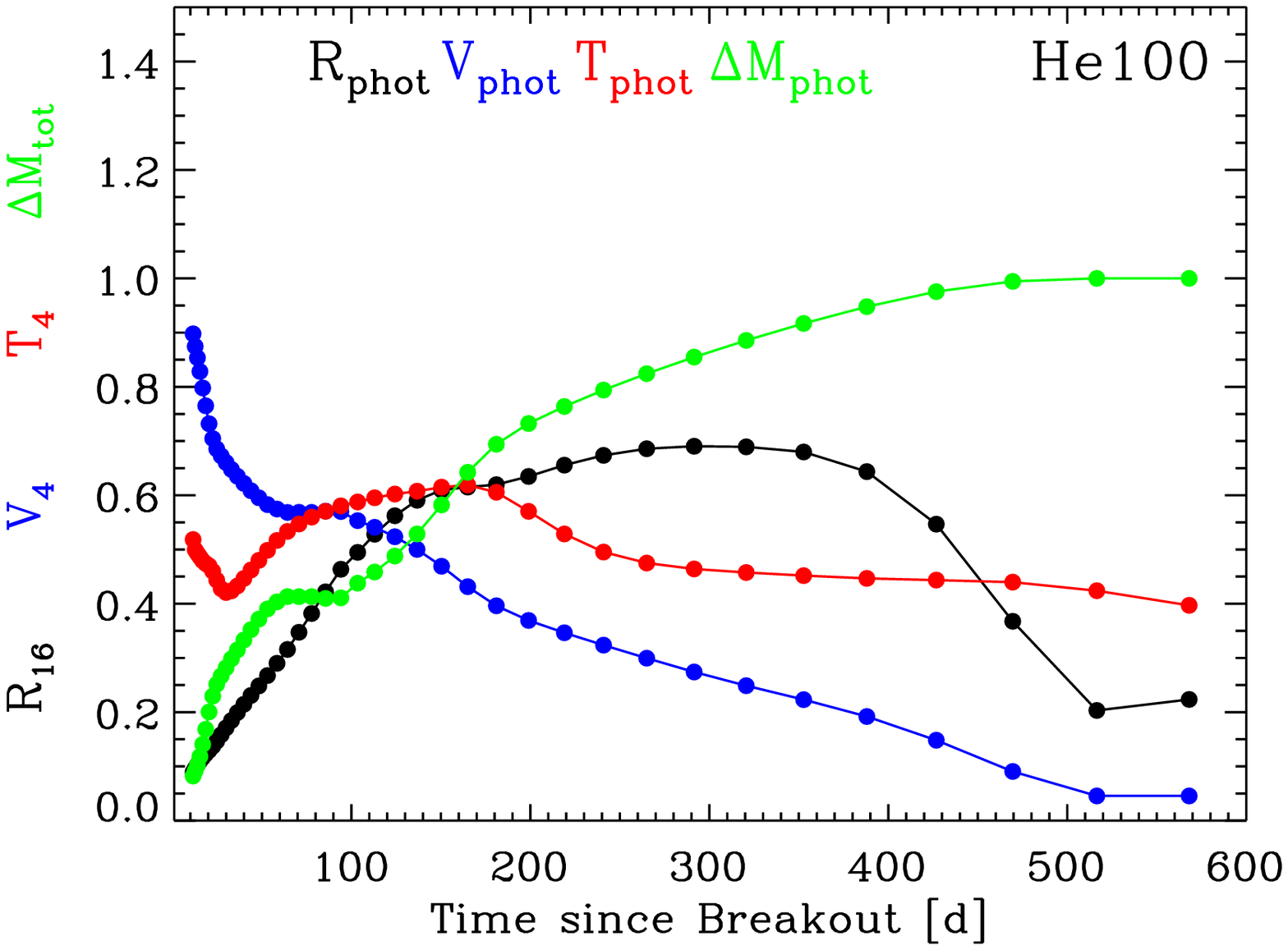,width=8.5cm}
\caption{Evolution of properties at the electron-scattering photosphere, and specifically the radius
(black; in units of 10$^{16}$\,cm), the velocity (blue; in units of 10000\,\kms), the temperature (red; in
units of 10000\,K), and the overlying mass (normalized to the total ejecta mass; green) for model B190 (top left),
R190 (top right), B210 (bottom left), and He100 (bottom right). The dots correspond to the actual times
computed with \cmfgen.
All ejecta models are optically-thin to electron scattering at $\gtrsim$\,550\,d (see Tables~A1--A4).
\label{fig_phot}
}
\end{figure*}


 \section{Light curve evolution}
\label{sect_lc}

   One important output of \cmfgen\ is the emergent spectrum in the observer's frame. By integrating this flux
   over frequency at each time step, we recover the bolometric light curve. We show the
   synthetic  bolometric luminosity for PISN models B190, R190, B210, and He100ionINL in Fig.~\ref{fig_lbol}.
   The fundamental signatures of  these PISN models is the very long duration of the high-brightness phase,
   typically of a few hundred days, reaching peak luminosities on the order of 10$^{10}$\,\lsun\ at 150--200\,d
   after explosion.
   Specifically, we find a bolometric maximum at 196, 220, 177, and 156\,d for models
   B190, R190, B210, and He100ionINL. In the same order, the bolometric maximum luminosities are
   4.4$\times$10$^9$, 6.3$\times$10$^9$, 3.0$\times$10$^{10}$, and 8.8$\times$10$^{9}$\,\lsun.
   Allowing for non-local energy deposition changes these values by a few percent over this timespan.
   In the lower mass model He100ionINL, \gray\ escape starts being visible at $\sim$\,500\,d after explosion,
   introducing a slight downward tilt of the fading rate below 0.01\,mag\,d$^{-1}$, while for all other models
   (with an ejecta mass in the range 130-160\,\msun), full \gray\ trapping holds exactly.
   Note that PISN explosions may not systematically be super-energetic and superluminous \citep{kasen_etal_11},
   but our set of models was selected to be so, with kinetic energies of a few tens of B and nucleosynthetic
   yields including a few \msun\ of \isoni.

 \begin{figure*}
\epsfig{file=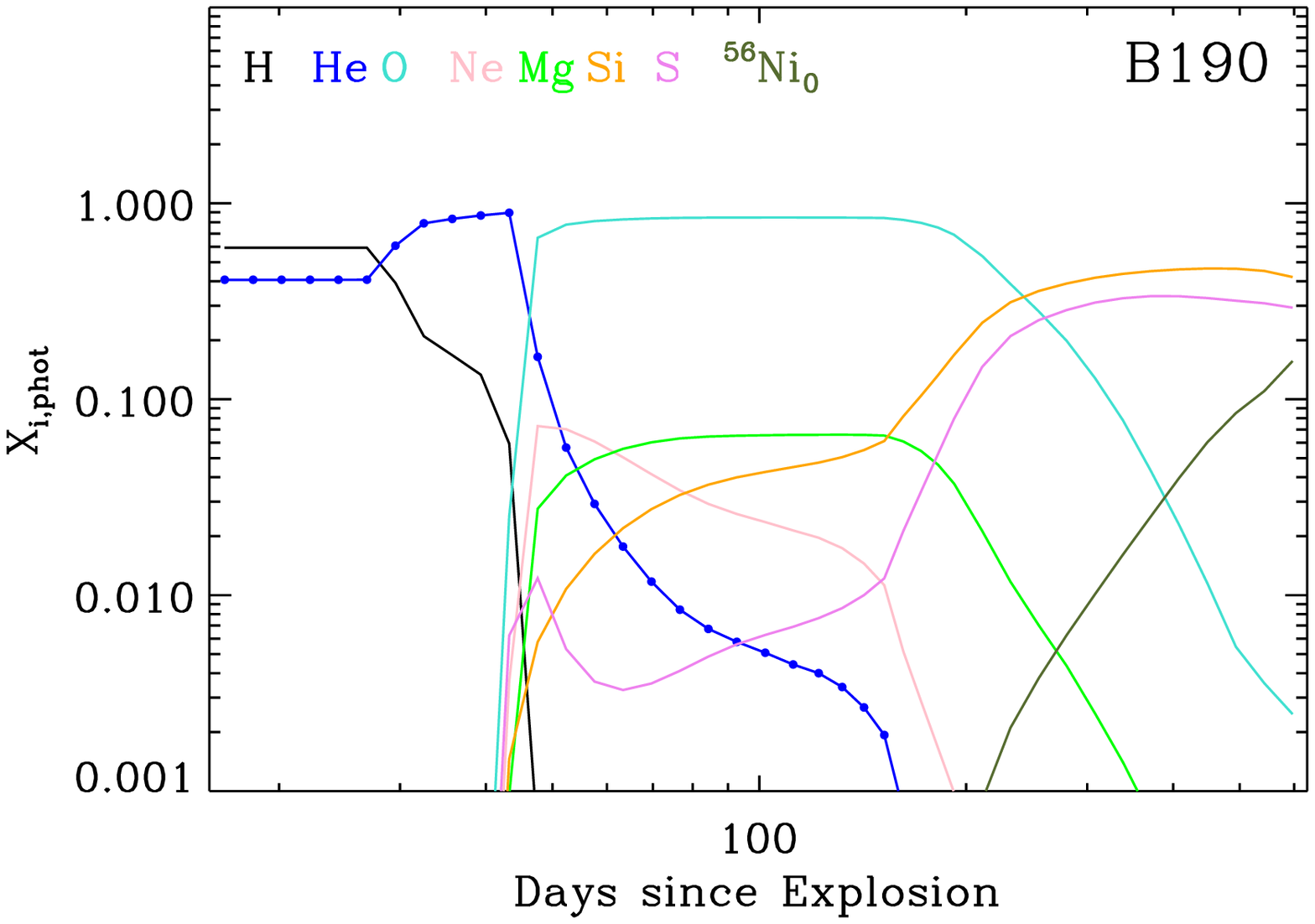,width=8.5cm}
\epsfig{file=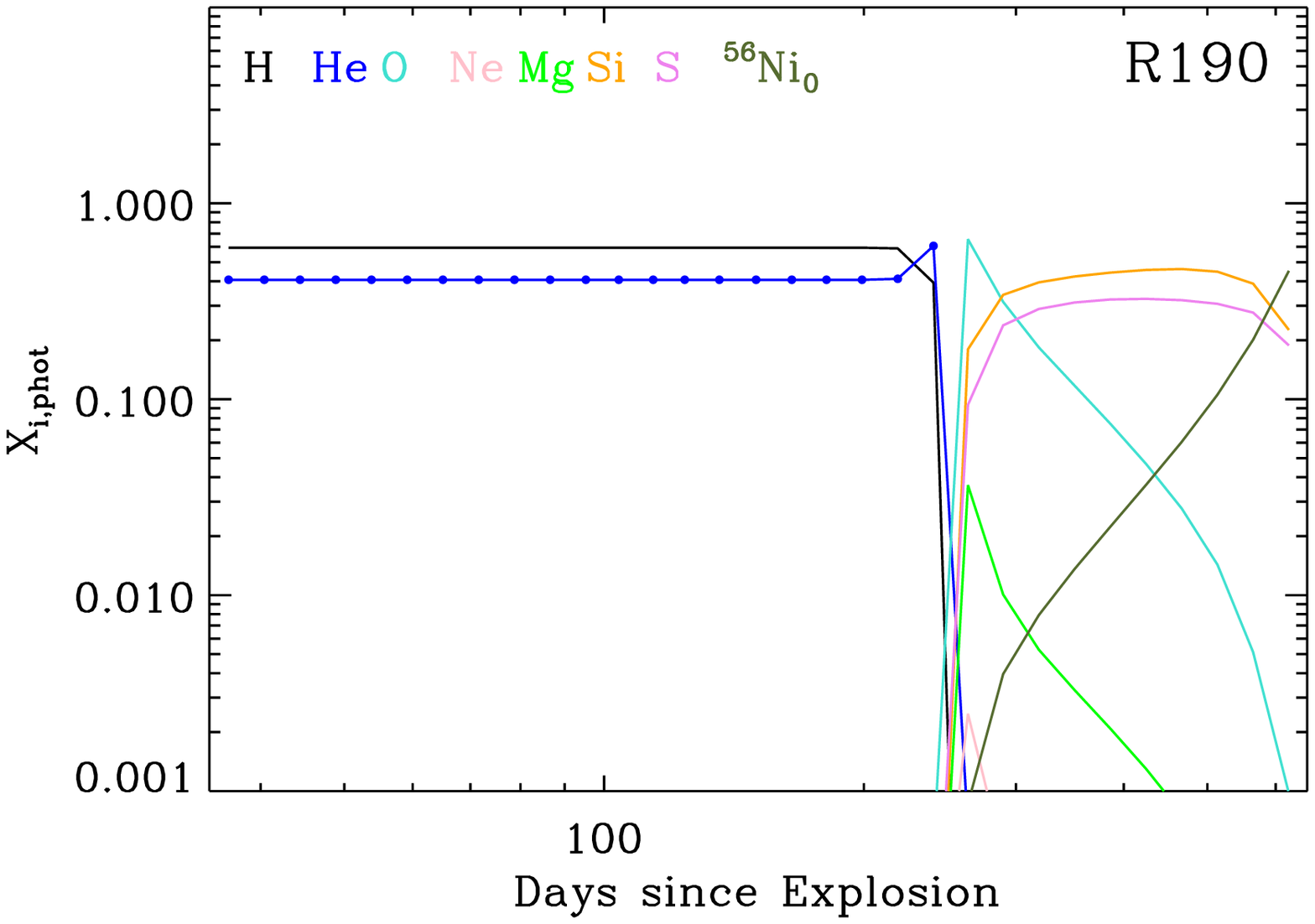,width=8.5cm}
\epsfig{file=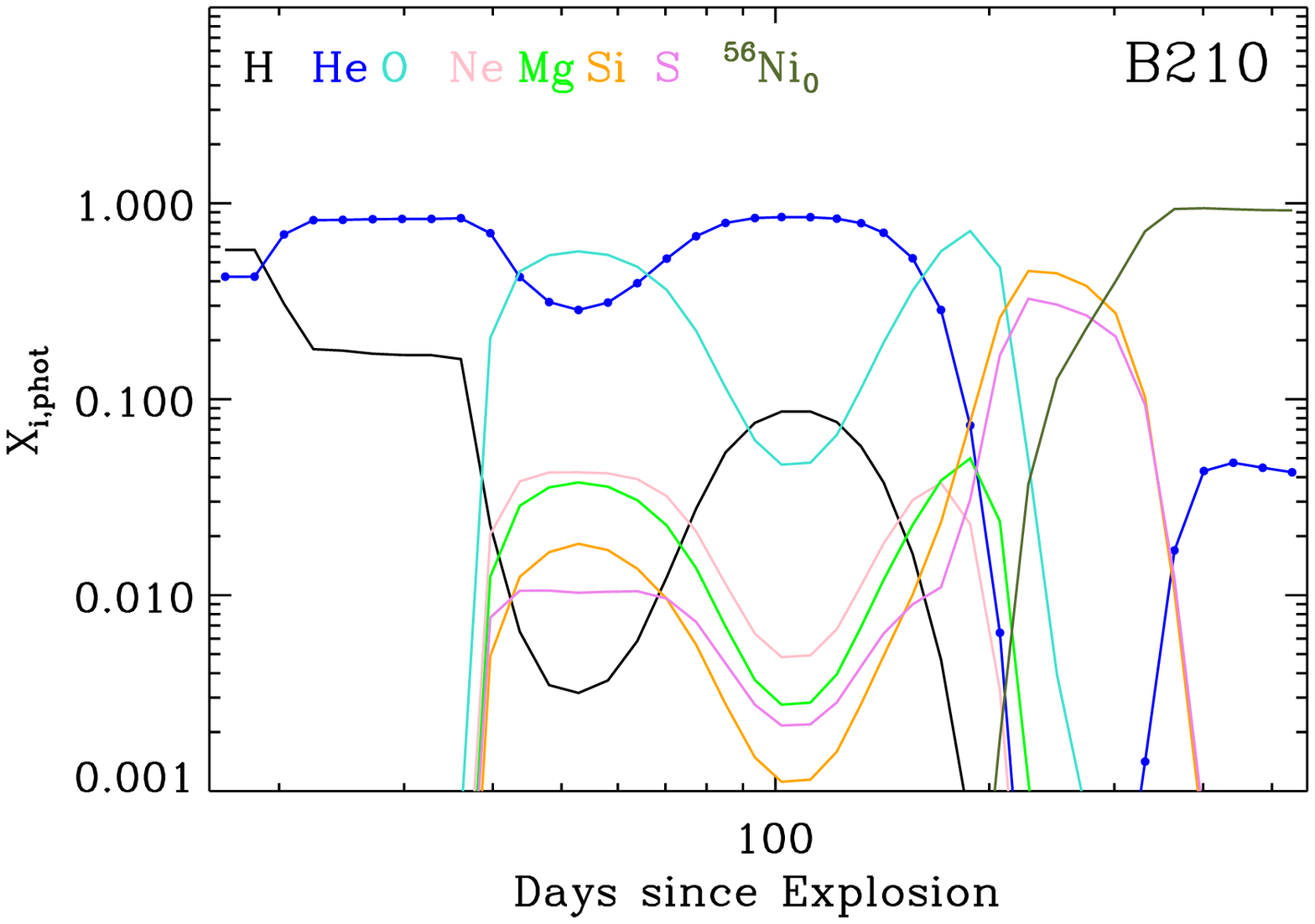,width=8.5cm}
\epsfig{file=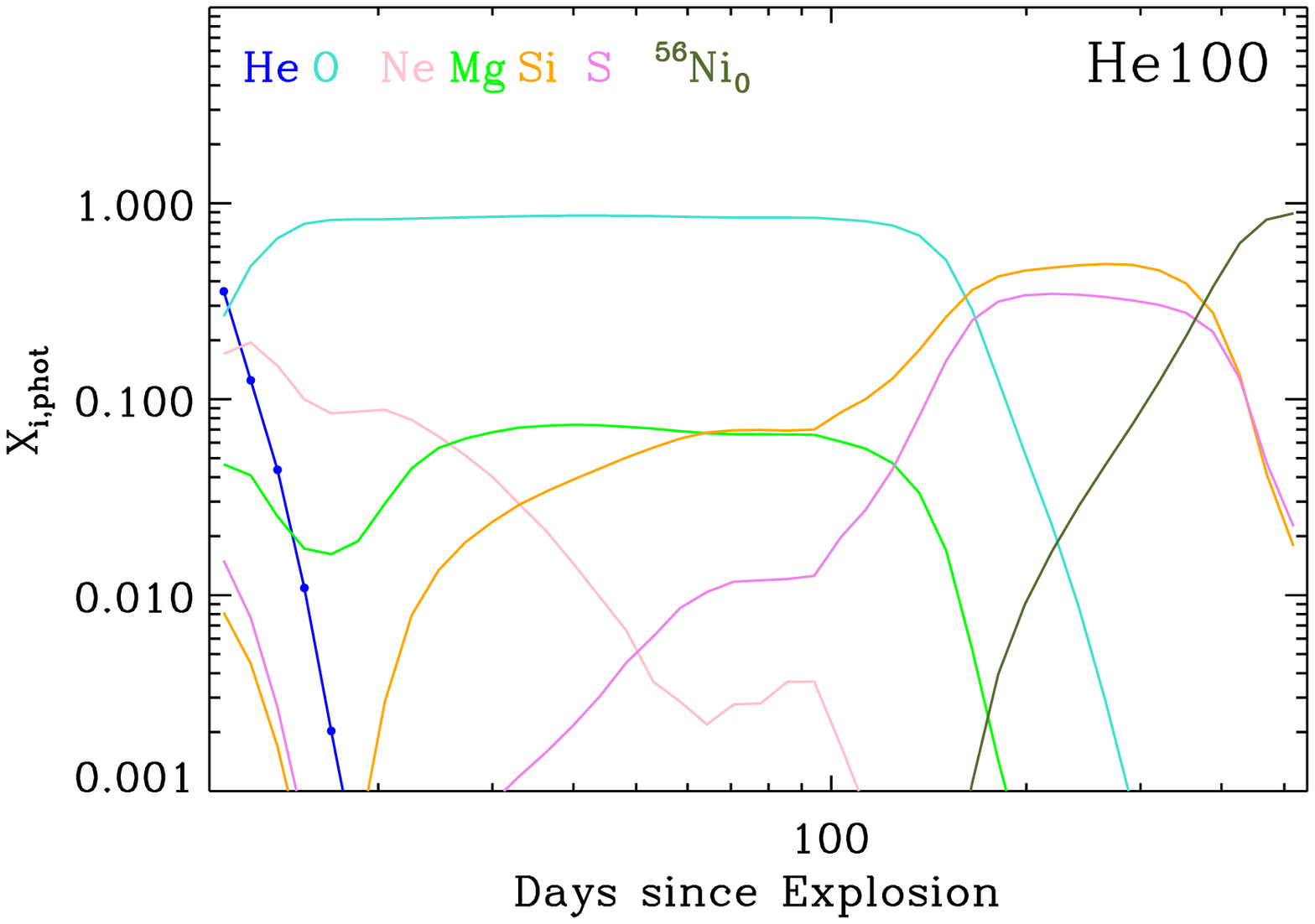,width=8.5cm}
\caption{Evolution of the composition for representative species at the electron-scattering photosphere
for model B190 (top left), R190 (top right), B210 (bottom left), and He100 (bottom right).
\isoni$_0$ refers to the sum of the mass fractions for the \isoni-decay products
(it thus corresponds to the initial \isoni\ mass fraction in the corresponding mass shell).
Most simulations have a short photosphere residence time in the H-rich envelope (B190 and B210),
or none at all for He100 in which the compact and H-deficient progenitor produces a photosphere in the O-rich
shell up to 200\,d. In model R190, up to 250\,d is spent in the H-rich extended progenitor envelope, but by the time
the photosphere reaches its base, the He-rich and O-rich shells are fully recombined and transparent so the photosphere
moves directly into the Si/S/IGE-rich inner ejecta.
\label{fig_phot_xfrac}
}
\end{figure*}

    The morphology of these light curves is analogous to the SNe we observe in the local Universe:
    the Type II-Plateau class (R190), arising from a RSG progenitor \citep{falk_arnett_77},
    the Type II-pec class (B190 and B210) arising from a BSG progenitor \citep{arnett_etal_89},
    and the Type Ib/c class (He100) arising from a WR star progenitor
    \citep{EW88,dessart_etal_11,dessart_etal_12}.

    The anomalously large initial radius and kinetic energy of
    model R190 yields a ``plateau'' brightness over 10 times more luminous than obtained in the SN II-P model
    resulting from a 1.2\,B 15\,\msun\ RSG progenitor (olive curve; model taken from \citealt{DH11}).
    The  2.63\,\msun\ of \isoni\ originally made in the  explosion causes a strong re-brightening at 200\,d after explosion,
    following the recession of the photosphere to deeper layers and the outward migration of a heat wave powered
    by \isoco\ decay. This hump is never seen in SNe II-P produced from lower mass RSG progenitors because
    they produce too little \isoni\ in their explosion. Beyond 350\,d, model R190 radiates at the instantaneous rate of
    decay energy deposition, even when allowance for \gray\ escape is made. The ejecta remains optically thick
    to \grays\ throughout the evolution modeled here, so that \grays\ may escape from the original site of emission
    but are eventually trapped somewhere else in the ejecta.\footnote{The use of a coarser grid and the successive
    re-mappings in the R190 time sequence  introduce a 10\% change in the total mass and the \isoni\ mass
    as time progresses. This causes model R190 to have the same nebular  luminosity as model B190,
    although the initial amount for \isoni\ is respectively 2.63 and 2.99\,\msun. This resolution problem affects
    negligibly models B190, B210, and He100. See footnote~\ref{note_r190}.}

   The light curve morphology for model B190 and B210 is explained in the same terms as for standard
   SNe II-pec like SN 1987A \citep{blinnikov_etal_00,DH10,li_etal_12}. However, as for R190,
   the light curve is much brighter and broader than for a standard II-pec model (orange; model evolved at
   the LMC metallicity  from a rotating single 18\,\msun\ main-sequence star; \citealt{DH10,li_etal_12}).
   The luminosity falloff after shock breakout is much more gradual and extended and the onset of re-brightening,
   due to heating by radioactive decay, occurs later at about 50\,d after explosion. The bright peak is much bigger
   due to the larger mass, optical depth, and hence diffusion time of the corresponding ejecta.
   In model B210, which synthesized 21.3\,\msun\ of \isoni, the peak luminosity is
   10$^{44}$\,erg\,s$^{-1}$ --- $\sim$\,7 times larger than in model  B190
    which synthesized only $\sim$\,3\,\msun\ of \isoni.
   As in model R190, the larger ejecta mass prevents the rapid luminosity evolution seen in lower mass Type II SNe.

   The light curve morphology for model He100ionI (and He100; see Appendix~C)
   is typical of a Type I CCSN, with a faint post-breakout plateau,
   followed by a steep rise due to decay heating at depth. The strong re-brightening starts $\sim$\,30\,d
   after explosion in this model, thus significantly later than in SNe IIb/Ib/Ic models \citep{EW88,dessart_etal_11}.
   Provided mixing is  weak in these PISNe (\citealt{joggerst_whalen_11}; but see \citealt{chen_etal_12}),  the
   rise to peak should not start much earlier than obtained here \citep{dessart_etal_12}. As for models B190 and B210,
   the peak of the light curve is broad and luminous and qualitatively analogous between all three models. Although B190
   and B210 correspond to a SN II-pec, R190 to a SN II-P, and He100ionI/He100 to a SN Ic, their photometric properties are obviously
   special and it makes sense to retain the PISN calling for these models.

   Figure~\ref{fig_lbol} comprises all the SN types studied so far with \cmfgen, and thus includes PISNe,
   SNe II-P,  SNe II-pec, and SN Ib/c. To complete the set, we add results from on-going SNe Ia simulations for
   a delayed detonation in a Chandrasekhar-mass white dwarf synthesizing 0.67\,\msun\ of \isoni\ (Dessart et al.,
   in prep). In terms of peak brightness, such a SN Ia rivals luminous PISNe but its much lower mass
   prevents a sustained luminosity beyond $\sim$\,30\,d after explosion.


\begin{figure*}
\epsfig{file=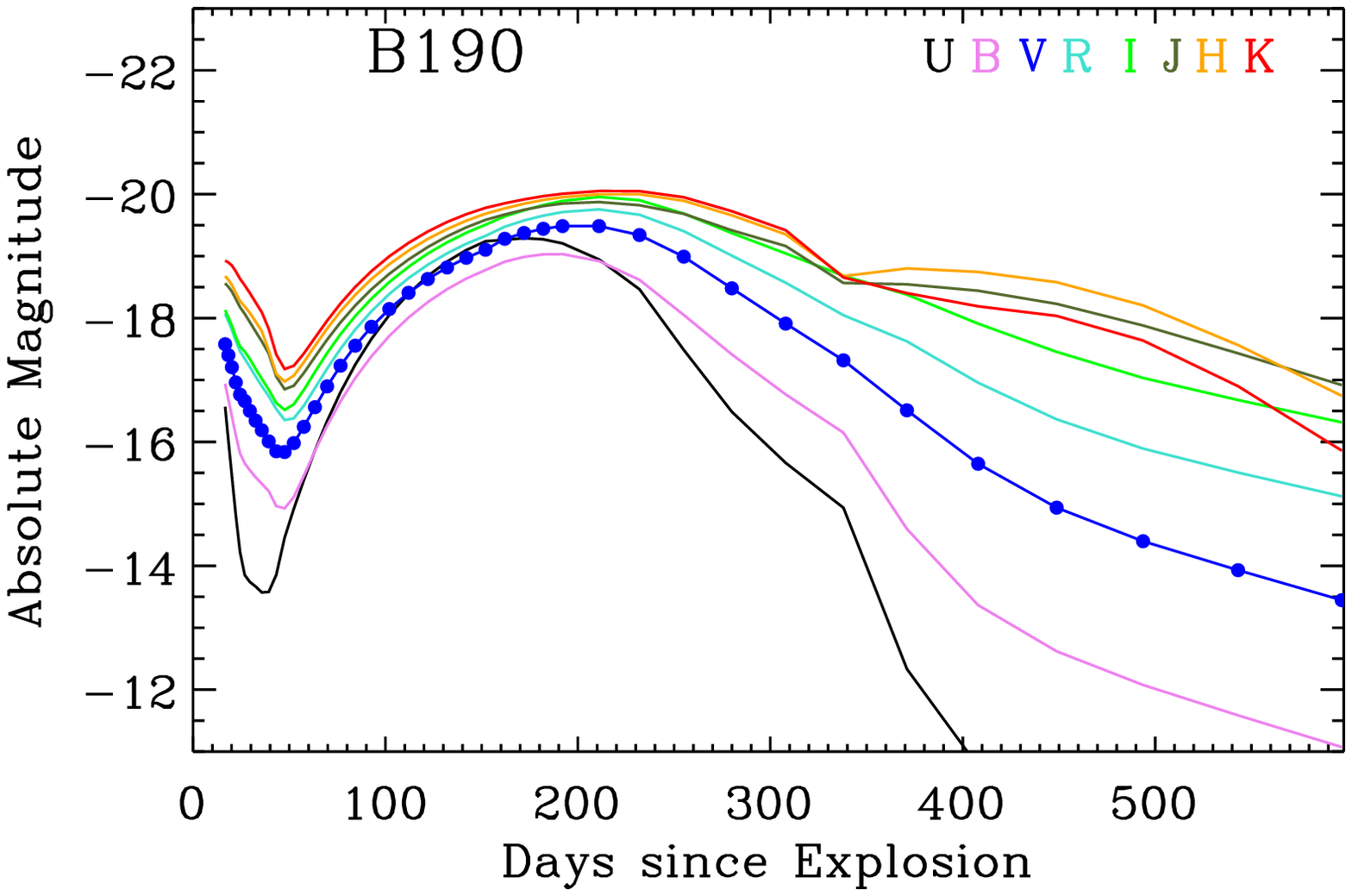,width=8.75cm}
\epsfig{file=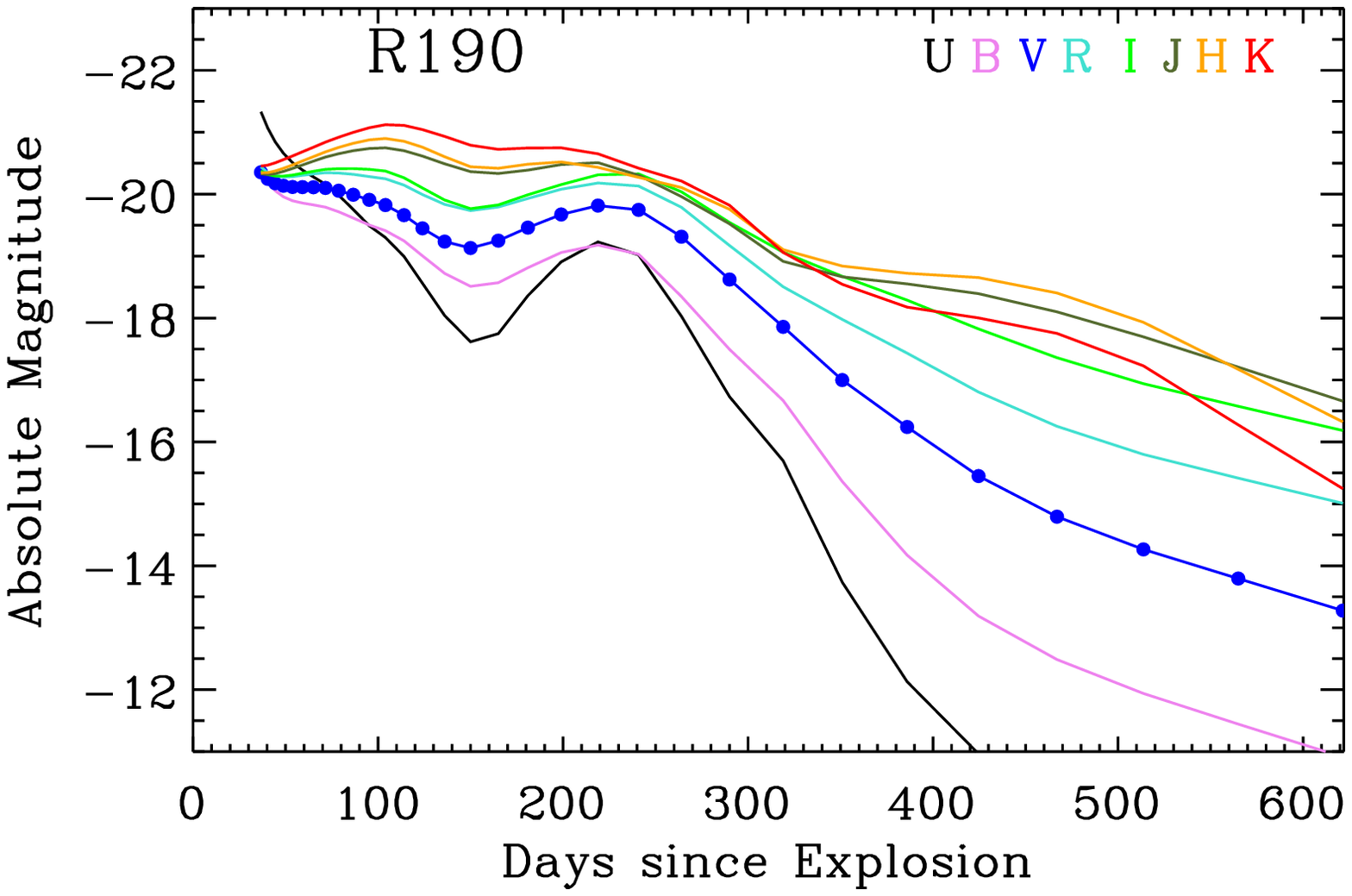,width=8.75cm}
\epsfig{file=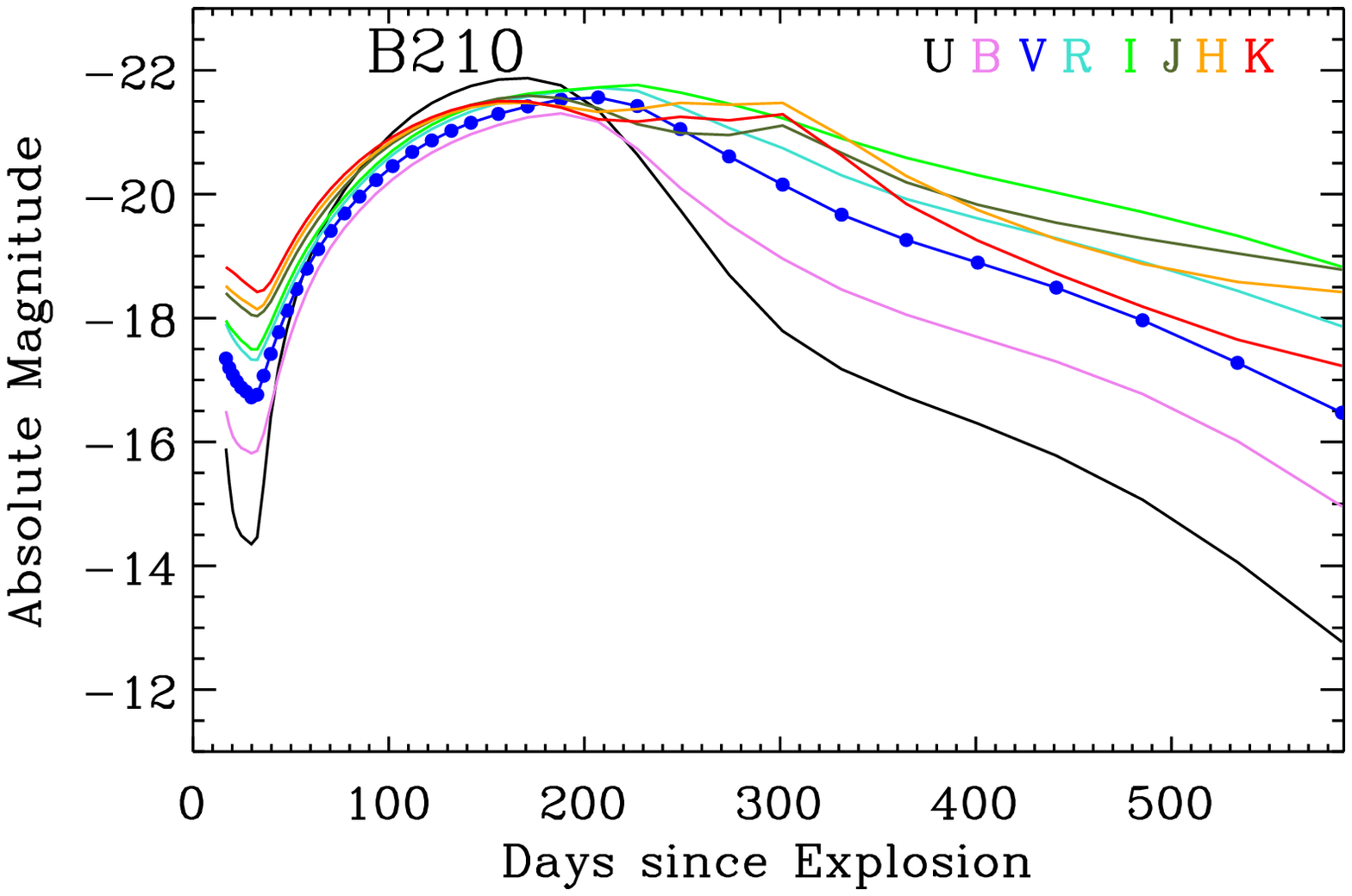,width=8.75cm}
\epsfig{file=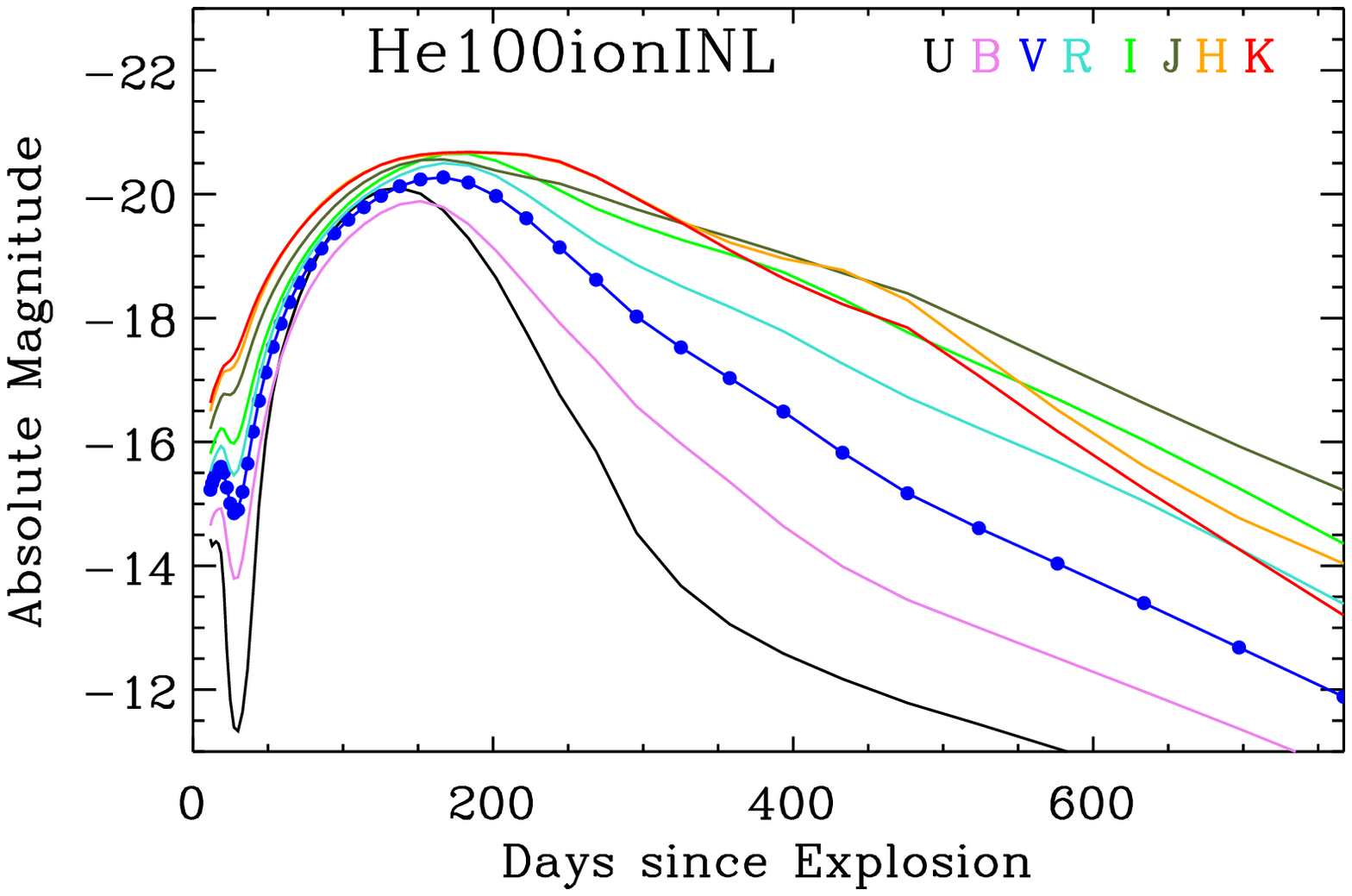,width=8.75cm}
\caption{Multi-band light curves  for our PISN simulations obtained by integrating at each epoch the synthetic
spectra over the band pass of filters U, B, V, R, I, J, H, and K. The absolute magnitude is shown.
Magnitudes plotted here are also logged into Tables~D1--D4.
In each sequence, we obtain slight kinks in the magnitudes (especially in bands U, J, H, and K, where
little flux is emitted) that do not occur in bolometric luminosity. These are tracked down to changes in parameters at the
corresponding time, e.g., as the medium turns optically thin (switch to nebular inner boundary), change in SL assignment (split the
lower 5 to splitting the lower 10 levels for all ions), or change in model atoms (e.g., switch to a big Fe\one\ model atom). While these
glitches are not physical, the ejecta relax to the new proper solution at the next time step.
\label{fig_mb_lc}
}
\end{figure*}

\section{Photospheric Properties}
\label{sect_phot}

   As can be seen in Fig.~\ref{fig_phot}, the photospheric\footnote{Here,
   we define the photosphere as the ejecta location where the inward integrated electron-scattering optical
   depth is 2/3. At early times, when there is little line opacity, most photons will originate (i.e., be created)
   below the ``electron-scattering'' photosphere. At later times, the ejecta is metal rich and will have considerable line
   opacity and so using the Rosseland-mean or the flux-mean opacity would yield a location for the  photosphere that
   is further out. Hence, our  discussion of the ``photosphere" is merely illustrative.} temperature is rather low in all
   models at all times, although a higher temperature is found for the R190 model because it starts from a much more extended
   configuration that dwarfs cooling from expansion (associated with the $PdV$ term in the energy
   equation; \citealt{DH11}). Since the photosphere is rather cool, PISN ejecta achieve a tremendous luminosity by
   remaining   optically thick out to large distances --  the maximum photospheric radii in models R190 and B210 exceed
   10$^{16}$\,cm, while the maximum photospheric radii for  B190 and He100 reach $5\times 10^{15}$\,cm.
   These radii are typically ten times larger than  predicted for SNe resulting from the explosion of lower-mass massive stars,
   whatever their type \citep{dessart_etal_08,DH10,DH11,dessart_etal_11}.

    Beyond 50\,d after explosion, and thus well before the light-curve peak for all models, the photosphere is
   at the recombination temperature for the dominant ion (H, C, or O), which happen to have a similar ionization
   potential of 11-14\,eV, and is on the order of 5000--6000\,K. We thus expect red colors and spectra
   dominated by neutral and once- and twice-ionized ions (Appendix~A \& Fig.~A1).

As the evolution proceeds the photosphere generally migrates to deeper layers -- this is best seen by looking at the
photospheric velocity, or the mass  that lies above the photosphere (Fig.~\ref{fig_phot}). The speed of recession is
connected to the ability of the SN to remain hot and ionized because free electrons can contribute significantly to the
opacity at the photosphere through electron scattering. Model R190 shows a slow progression of the photosphere
inwards initially, which remains in the outer 1\% of the ejecta (in mass coordinate), before showing a prompt
recession at the end of the plateau phase at $\sim$\,230\,d when it enters the core (this phase could have been
better resolved with our simulations). The photosphere then crosses 60\% of the ejecta mass in about 10\,d as the
ejecta becomes thin, essentially crossing the entire O-rich shell. In models B190 and He100, the recession to deeper
layers is fast early on and slows down as the decay heating becomes effective, halting temporarily the photospheric
recession. In B210 the heat wave is so fierce that it makes the photosphere move {\it outward} through
$\sim$\,27\,\msun\ of ejecta material between days 50 and 100 (Fig.~\ref{fig_phot}). As a consequence of this
phenomenon the photospheric velocity does not vary monotonically.

The migration of the photosphere to deeper layers leads to the complete probing of the ejecta, from
the progenitor surface to its core, over a time span of about 300\,d. Because of the large ejecta mass and
the presence of shells of distinct composition, exacerbated by the low-primordial metallicity, this migration reveals
an unprecedented chemical evolution at the photosphere (Fig.~\ref{fig_phot_xfrac}).

   The enormous energy of a few tens of B released through burning (Sect.~\ref{sect_expl})
   is compensated in PISN explosions by the huge ejecta mass. This is largely independent of the mass loss
   prescription since the pair-production instability can only occur if the core mass is greater than $\sim$\,60\,\msun\
   \citep{HW02}. No matter what, the ejecta mass will be huge, well above that of any CCSN progenitor identified
   or inferred so far. A representative ejecta velocity is $v_{\rm rep} \equiv \sqrt{2E/M} \sim 3200  \sqrt{E_1/M_{10}}$\,\kms,
   where $E_1$ is the kinetic energy in units of B and $M_{10}$ is the ejecta mass in units of 10\,\msun.
   A typical value for a RSG or BSG explosion in the local Universe is $E_1=1$ and $M_{10}=1$, i.e.,
   $v_{\rm rep} \sim 3200$\,\kms. For our PISN models, the increase in explosion energy more than compensates
   for the increase in ejecta mass and we have  $v_{\rm rep}  \sim 5000$\,\kms, comparable to SNe Ib/c
   \citep{dessart_etal_11} from intermediate-mass binary stars, but a factor of two lower than in a Chandrasekhar-mass
   white-dwarf explosion. Although both PISN and SN Ia are thermonuclear explosions, the bulk of the burning takes
   place in the inner envelope in a PISN, yielding a chemical stratification more typical of core-collapse Type Ic SNe:
   The \isoni\ is produced at the base of the ejecta, remains confined to the lowest expanding material with velocities
  less than 4000\,\kms\ (Fig.~\ref{fig_summary_presn}), and the decay energy has to diffuse from there
   through the overlying buffer of mass to influence the outer ejecta and the light curve.

    The various properties described above for the bolometric luminosity and the photospheric properties
assist the interpretation of the multi-band light-curves presented in Fig.~\ref{fig_mb_lc}.
All models reach absolute visual magnitudes on the order of $-$20 to $-$22\,mag.
Model R190 exhibits a quasi-plateau in the V band, together with a hump at 250\,d caused by decay heating.
The higher photospheric temperature makes the spectral-energy distribution (SED) peak short-ward of the
Balmer jump for about two months following explosion. This evolution is typical of RSG explosions,
with the magnitudes dropping fast in the blue but leveling for a long time in the red. After the secondary bolometric
peak, all magnitudes become fainter with time, but continue to ebb faster in the blue. The evolution of colors for models
B190, B210, and He100 are very different from model R190 at early times. The fading after shock emergence is more
obvious, and so is the re-brightening as the heat wave reaches the receding photosphere. The colors become bluer
on the rise to peak before fading again as the ejecta becomes thin. This fading is again faster in the blue.

\begin{figure}
\epsfig{file=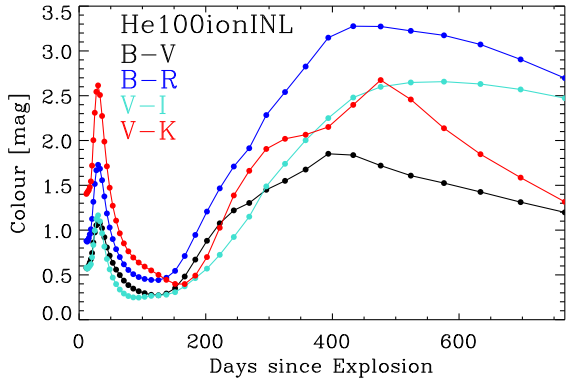,width=8.5cm}
\caption{Same as Fig.~\ref{fig_mb_lc}, but now showing selected colours for model He100ionINL.
An important feature of PISN model He100/He100ionI, shared with PISN models R190, B190, or B210 presented here, is
the red colour of their spectra, especially at and beyond peak.
\label{fig_he100_color}
}
\end{figure}

An important feature of these multi-band light curves is that the colors are systematically red after the peak of
the light curve. Figure~\ref{fig_he100_color} illustrates this color evolution
for model He100ionI, which is representative of BSG/WR star progenitors models at all times --- for
RSG star progenitor models, this similarity holds for a few 100\,d after explosion and beyond.
Before the peak, all models from BSG and WR star explosions
are getting bluer: This is because the photosphere is getting hotter due to the heating from decay {\it at depth}
but without the ``reddening" effects associated with high metal content at the photosphere (the photosphere still lies
outside of these metal rich layers before the peak). The switch from red to blue and back to red as the SN evolves
from the re-brightening to the peak, and eventually fading into the nebular phase is a distinctive
signature of \isoni\ powered PISNe. This trend is somewhat similar to SNe Ia, with an additional re-hardening of
the spectrum as the ejecta turn transparent.
We will return to this issue when comparing our PISN models to the observations of SN 2007bi (Sect.~\ref{sect_07bi}).

  \section{Spectral evolution: General properties}
\label{sect_spec_evol}

  From the preceding sections, we see that the four PISN models B190, R190, B210, and He100 show
  significant differences in their light curves (peak brightness, time to peak, presence or not of a
  post-breakout plateau), which are easily connected to the fundamental differences  between the progenitors.
  We also expect the early spectra to differ significantly between models because of the differences
  between the progenitor envelopes. However, despite the different mass loss histories
  these models have similar  super-massive helium cores  and systematically produce large amounts of IMEs and IGEs
  from O burning  (Sect.~\ref{sect_presn}--\ref{sect_expl}). Consequently, we expect the nebular phase properties
  to be rather similar due to the similarities of the progenitor cores. Thus we describe the spectral evolution
  during the photospheric phase for each progenitor type separately (RSG, BSG, WR) before grouping them to
  discuss their nebular-phase properties.

\begin{figure*}
\epsfig{file=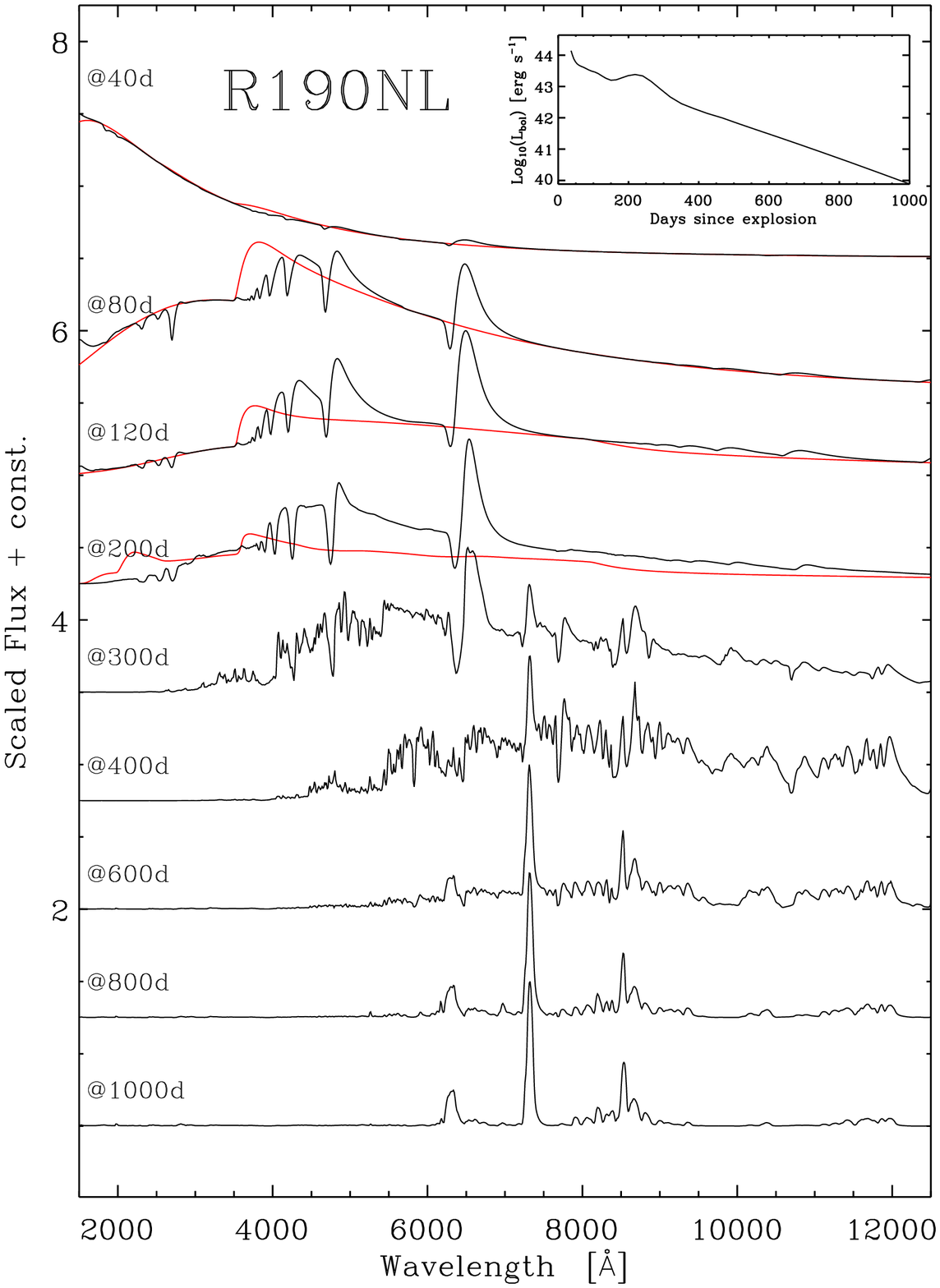,width=8.5cm}
\epsfig{file=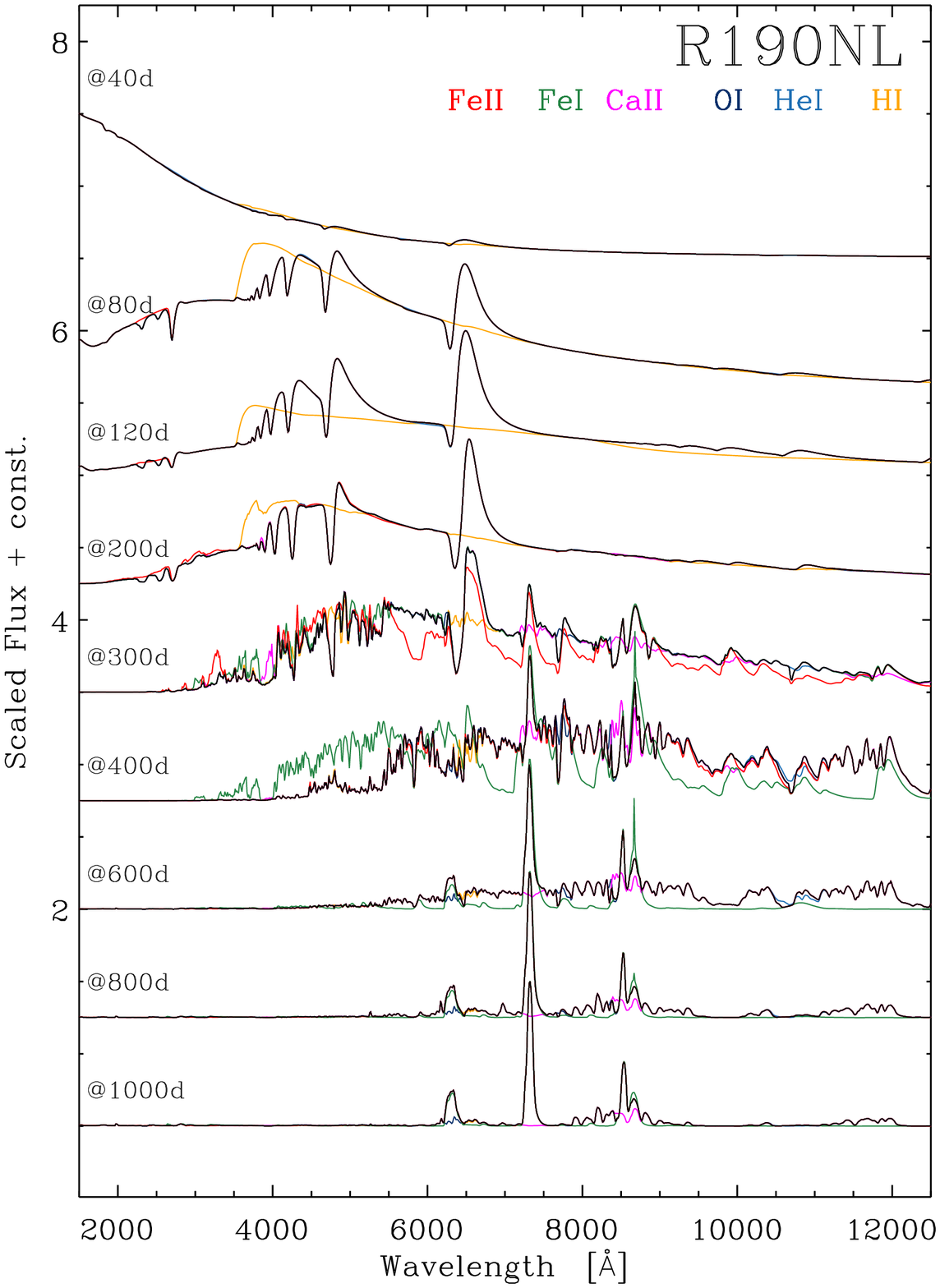,width=8.5cm}
\caption{
{\it Left:} Montage of synthetic spectra for the RSG model R190NL, showing the full spectrum (black) as well
as the contribution from continuum processes only (i.e., bound-free and free-free processes; red).
For completeness, we include an inset for the R190NL bolometric light curve.
The emergent flux, scaled for better visibility (see inset for the absolute bolometric
flux scale), is computed with \cmfgen\ by simulating the time-dependent non-LTE transport for the full ejecta from
36 until 1000\,d after explosions. The sequence thus covers from times when the ejecta is optical thick
until times when it is optically thin and nebular-line emission is strong.
{\it Right:} Same as left, but we now overplot the corresponding spectra obtained by {\it neglecting} an ion
(see color coding and labels at top-right). We illustrate the effect for ions that have a large impact on synthetic spectra.
\label{fig_R190}
 }
\end{figure*}

  \begin{figure*}
\epsfig{file=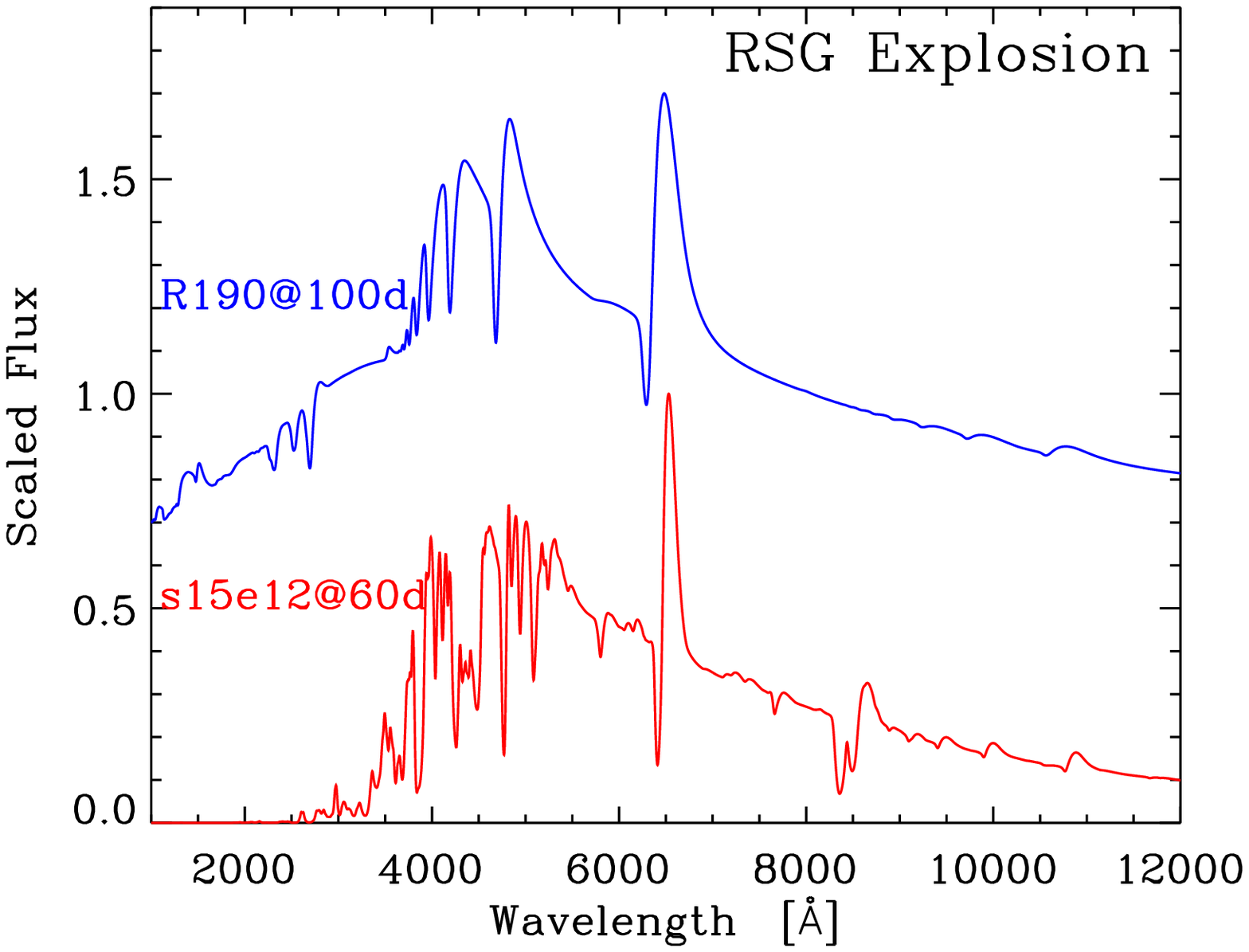,width=8.5cm}
\epsfig{file=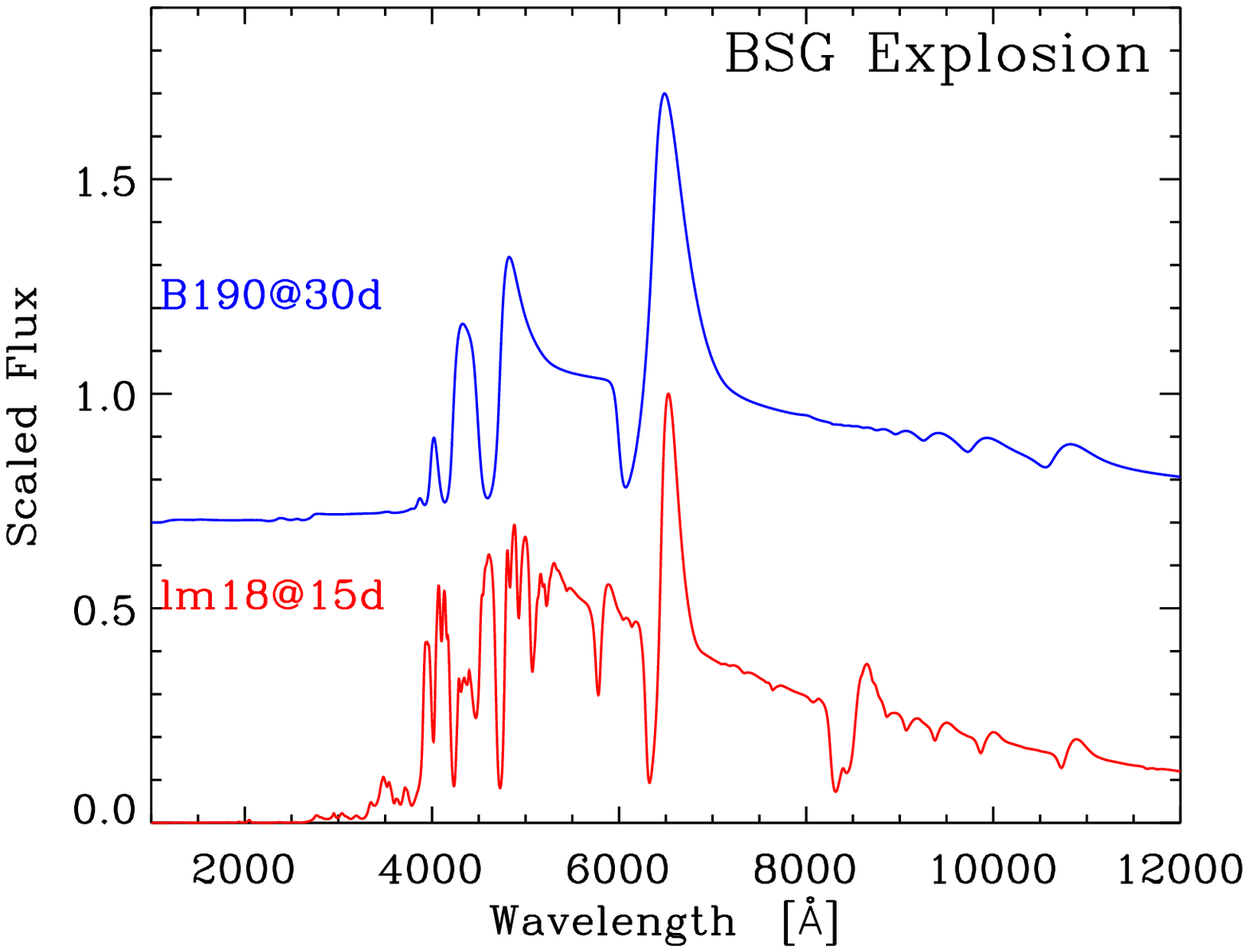,width=8.5cm}
\caption{The left panel shows a comparison between the RSG PISN model R190
and the equivalent explosion from a lighter massive star at \zsun\ (15\,\msun\ s15e12 SN II-P model), while the right
panel shows a comparison between the model B190 and the 18\,\msun\  ``lm18a7Ad"  SN II-pec model at the LMC metallicity.
The spectral morphology is similar, except that the PISN explosions at 10$^{-4}$\,\zsun\ show no metals lines
(not even Na\one\,D or the Ca\two\ triplet generally seen in all Type II SNe) at times when their equivalent
solar/LMC explosions show strong line blanketing throughout the UV and optical ranges.
\label{fig_comp_s15_lm18}
 }
\end{figure*}

All our PISN models have a very large oxygen mass (39.6 to 48.1\,\mo). However, as a consequence of the
large ionization energy (13.6 eV) and the cool photospheric temperatures, O lines generally only have a weak
direct influence on the spectra. The only exceptions are [O\,{\sc i}] 6303--6363\,\AA, which is seen in nebular-phase
spectra, and a few lines of O\,{\sc i} which weakly influence the early photospheric spectra of He100.

Initially all models present a  true continuum in the optical -- that is the observed spectra consist of (broad) lines
superimposed on an underlying  background which is formed via continuum (i.e., bound-free and free-free)
processes. However, eventually the photosphere recedes into layers where IMEs (primarily Si and S)
dominate the composition of the ejecta. At this, and later times, the true continuum flux weakens and
then disappears. However, before the true nebular phase,  a pseudo continuum, not generated by true
continuum processes, continues to be seen.  It is generated by the huge number of lines associated first with
IMEs  and later on with IGEs. In some cases this allows, for as long as 500-600\,d, the formation of
P-Cygni profiles associated with line transitions from ions present in overlying mass shells.

   \subsection{Spectral evolution during the photospheric phase of PISN model R190}
\label{sect_spec_RSG}

   The large progenitor radius ($\sim$\,4000\,\rsun) and explosion energy (33.2\,B) for model R190 yield a
   SN which is photometrically (Fig.~\ref{fig_lbol}) and spectroscopically (Fig.~\ref{fig_R190}) similar to a
   superluminous SN II-P.
   At early times, the outer ejecta and photospheric layers remain fully ionized (Fig.~\ref{fig_phot};
   Table~A2) producing a blue continuum (red curve in the montage of Fig.~\ref{fig_R190}).
   The relative high velocity and steep density profile of the photospheric layers (Fig.~\ref{fig_rho_init})
   yield broad but weak P-Cygni profiles \citep{DH05_qs_SN,dessart_etal_08}.
  At such times, the photosphere reside in the outer 1\% of the ejecta mass, which is essentially pure hydrogen
  and helium (the metallicity is 10$^{-4}$\,\zsun). Consequently, we find only H\one\ Balmer lines and a weak
  He\one\,5875\AA. These are in fact typical spectral signatures of early-time
  SNe II-P \citep{PST06_SN2005cs,QWH07_IIP_xrays,dessart_etal_08}, but there are some critical differences.
  To facilitate the analysis and line identifications we overplot in Fig.~\ref{fig_R190} (right panel)  the synthetic
  spectra computed by  {\it omitting} a given species.\footnote{The other option is to include only the selected species.
  However, this procedure becomes dicey when continuum processes weaken and/or when line opacity
  dominates --- both effects occur at nebular times in PISN explosions.
  In this case, taking out a dominant species completely alters the spectrum and the result is unusable.
  So, for consistency, we adopt the same procedure at all epochs and show these illustrations with the
  selected species {\it omitted}. This manipulation is done only for post-processing and analysis --- all \cmfgen\
  simulations are obviously performed with all species included.}

 As the ejecta expands helium quickly becomes neutral -- the inefficient mixing of \isoni\ from greater
 depths prevent any non-thermal effects at the photosphere \citep{dessart_etal_12} at these times in
 these PISN models. The adopted metallicity of 10$^{-4}$\,\zsun\  impacts  all species other than
 hydrogen and helium so that CNO, IMEs, and IGEs have little (direct) influence on the spectrum --
 up to 200\,d after explosion, we can only see (identify) H\one\ Balmer lines.  As the ejecta evolve,
 we do not see the appearance of Fe\two\ lines, Ca\two\,H\&K or Ca\two\ triplet at 8500\,\AA\ that are
 usually associated with Type II-P spectra.

   The lack of metals in the Hydrogen-rich envelope completely quenches the effect of line blanketing.
  As can be seen in Fig.~\ref{fig_R190}, the continuum (red curve) follows closely the total synthetic flux,
  except over spectral regions where the H\one\ Balmer lines reside. However, by 200\,d, a pseudo-continuum
  is visible in the full spectrum -- a large number of lines, with no noticeable signature in the full spectrum, are
  contributing to the total flux, even in regions apparently devoid of lines. At this time, the spectral
  formation region is receding into layers containing IMEs. However,  the photosphere is so cold  that the
  emergence of line blanketing on the very weak flux short ward of 4000\,\AA\ is difficult to see. By 300\,d
  after explosion, the spectrum has retained a similar color to that at 200\,d. However, the pseudo-continuum
  is no longer produced by genuine continuum processes (the red curve is down at zero flux) but instead
  by line opacity from Fe\two. By 400\,d, the main ion contributing to this background line-opacity switches
  from Fe\two\  to Fe\one\ and produces an even redder pseudo-continuum (red and green curves in
  Fig.~\ref{fig_R190}, right  panel). By 600\,d after explosion, the spectrum has visibly turned nebular.

   In Fig.~\ref{fig_comp_s15_lm18} (left panel), we illustrate the contrast between the spectrum  from the explosion
   of a 15\,\msun\ RSG progenitor at solar metallicity (model s15e12; \citealt{DH11}) and that from the R190 model
   at 10$^{-4}$\,\zsun. The line profiles are  broader in the PISN model, but the contrast with the low-mass SN II-P
   model is small due to the modest increase in $E/M$. Although the colors and H\one\ Balmer lines  appear similar,
   metals visibly affect the low-mass SN II-P spectrum in the UV (strong line blanketing) and optical
   (weaker blanketing, primarily due to Fe\two, Ti\two, and Sc\two) but have little effect on the PISN spectrum.
   This makes H$\beta$ easily recognized as a P-Cygni profile in the PISN spectrum while in the solar-metallicity
   spectrum, H$\beta$ overlaps with numerous metal lines (e.g., Fe\two),  and appears only as an absorption feature.
   A similar effect was discussed by \citet{kasen_etal_11}. Such RSG explosions in the early Universe could
   be used to constrain the environmental metallicity (or set an upper limit on the metallicity in case no
   metal lines are seen), either by means of the reduced blanketing in the UV or the weakness of metal lines
   (e.g. Fe, Ca) in the optical (see also Fig.~9 of \citealt{DH05_qs_SN}). The absence of Ca\two\ lines at the
   recombination epoch  is a unique feature of model R190, never seen in SNe II-P models, and never yet observed
   in SNe II-P.

\begin{figure*}
\epsfig{file=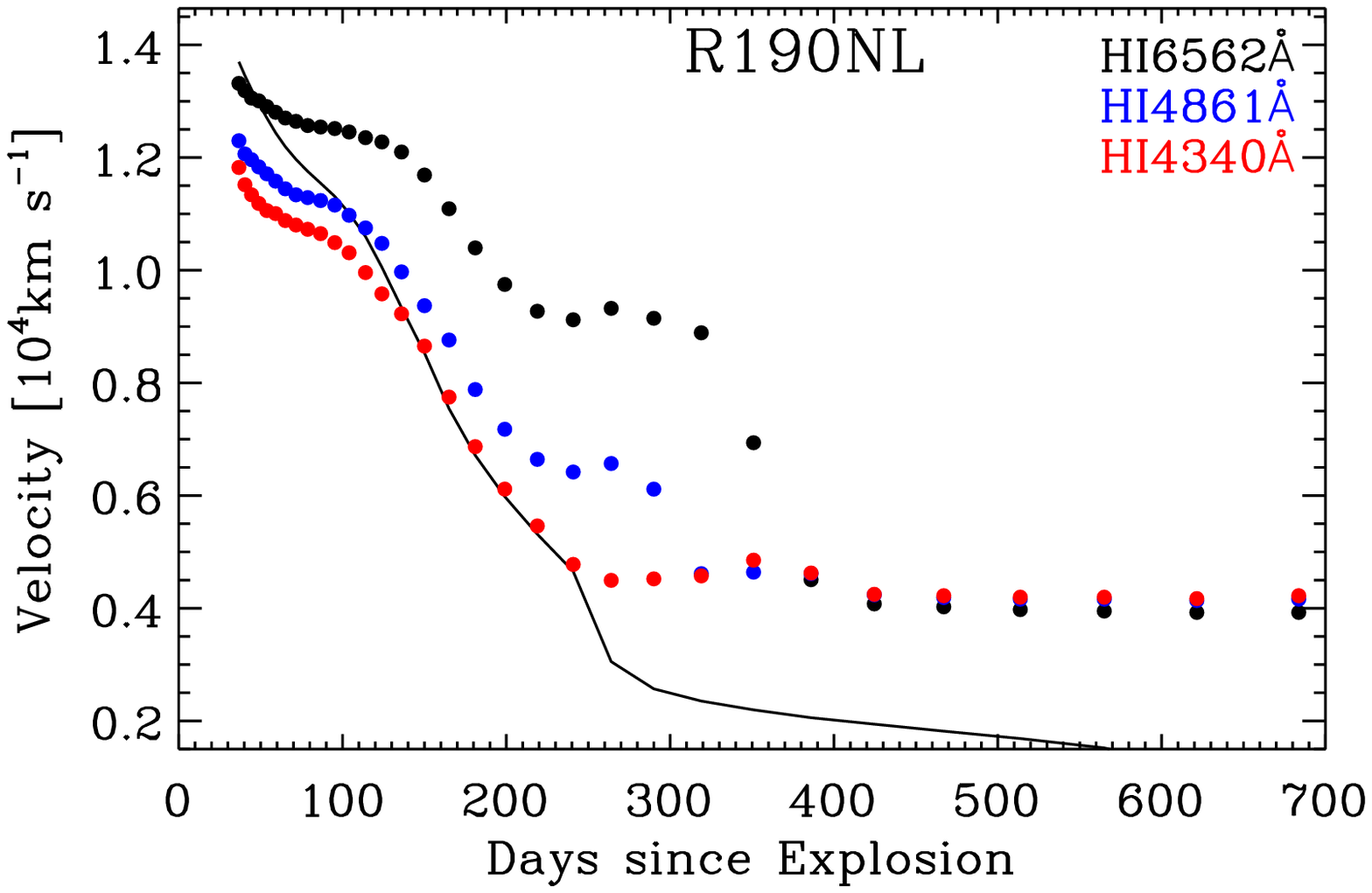,width=8.5cm}
\epsfig{file=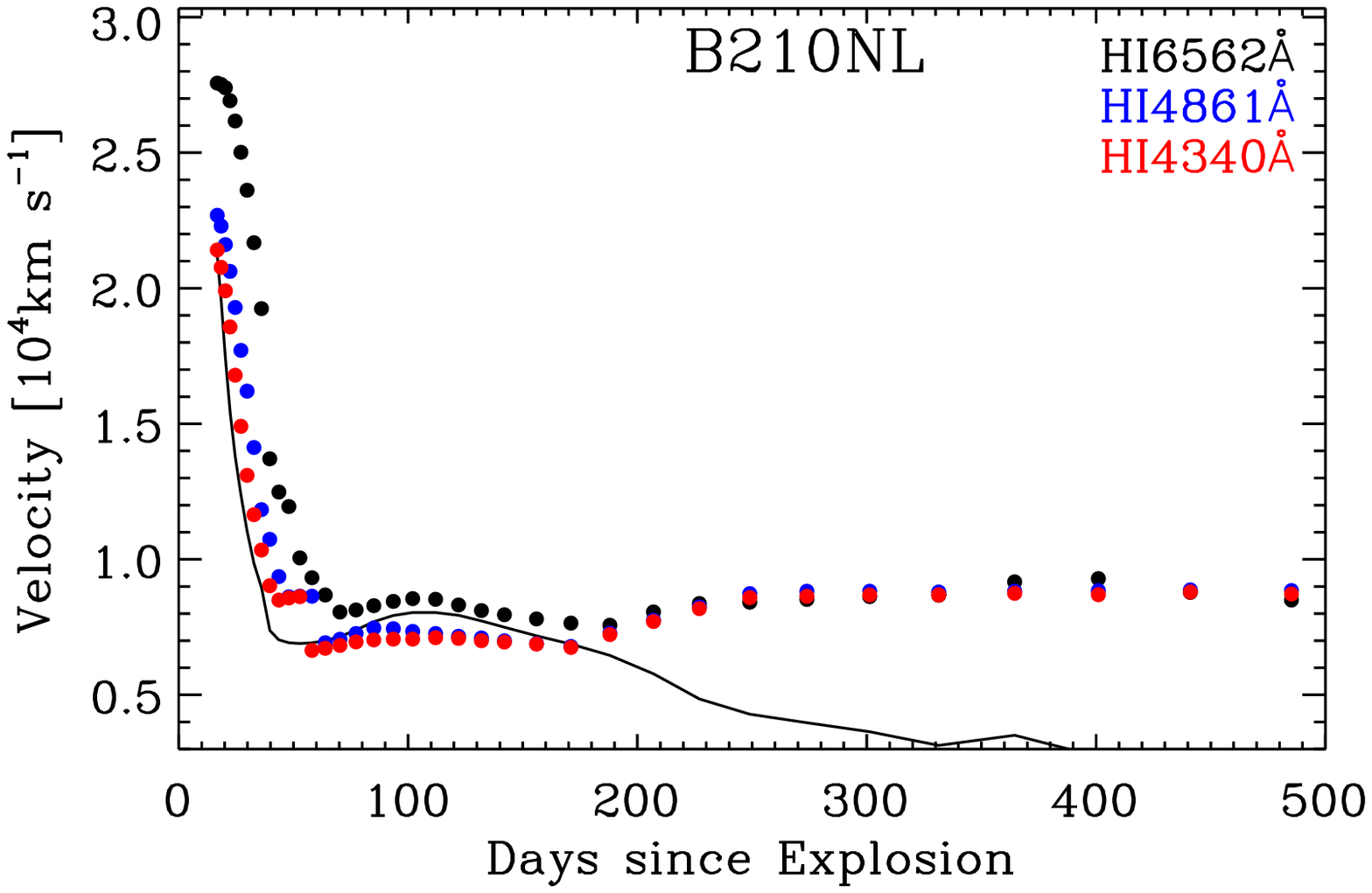,width=8.5cm}
\epsfig{file=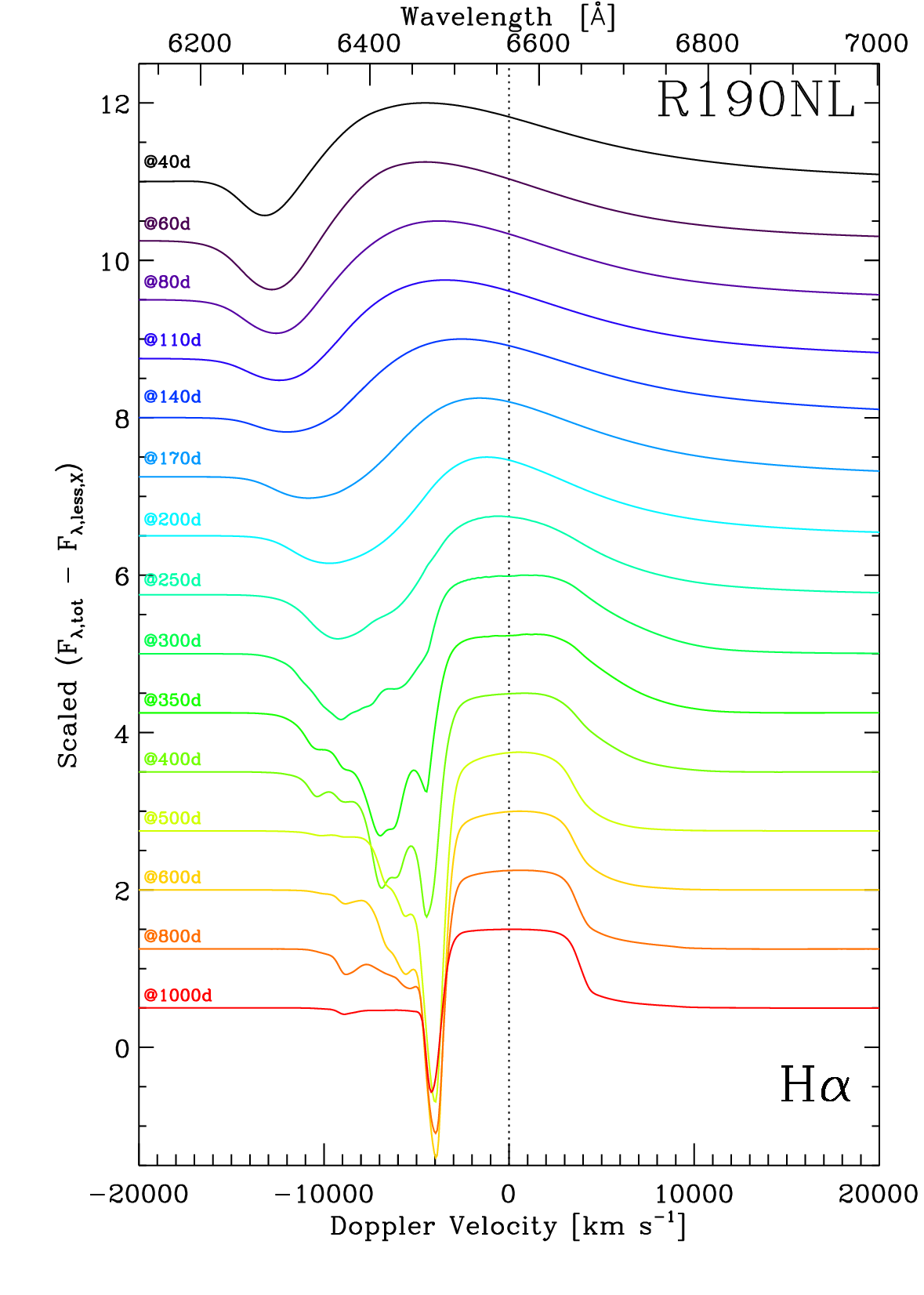,width=8.5cm}
\epsfig{file=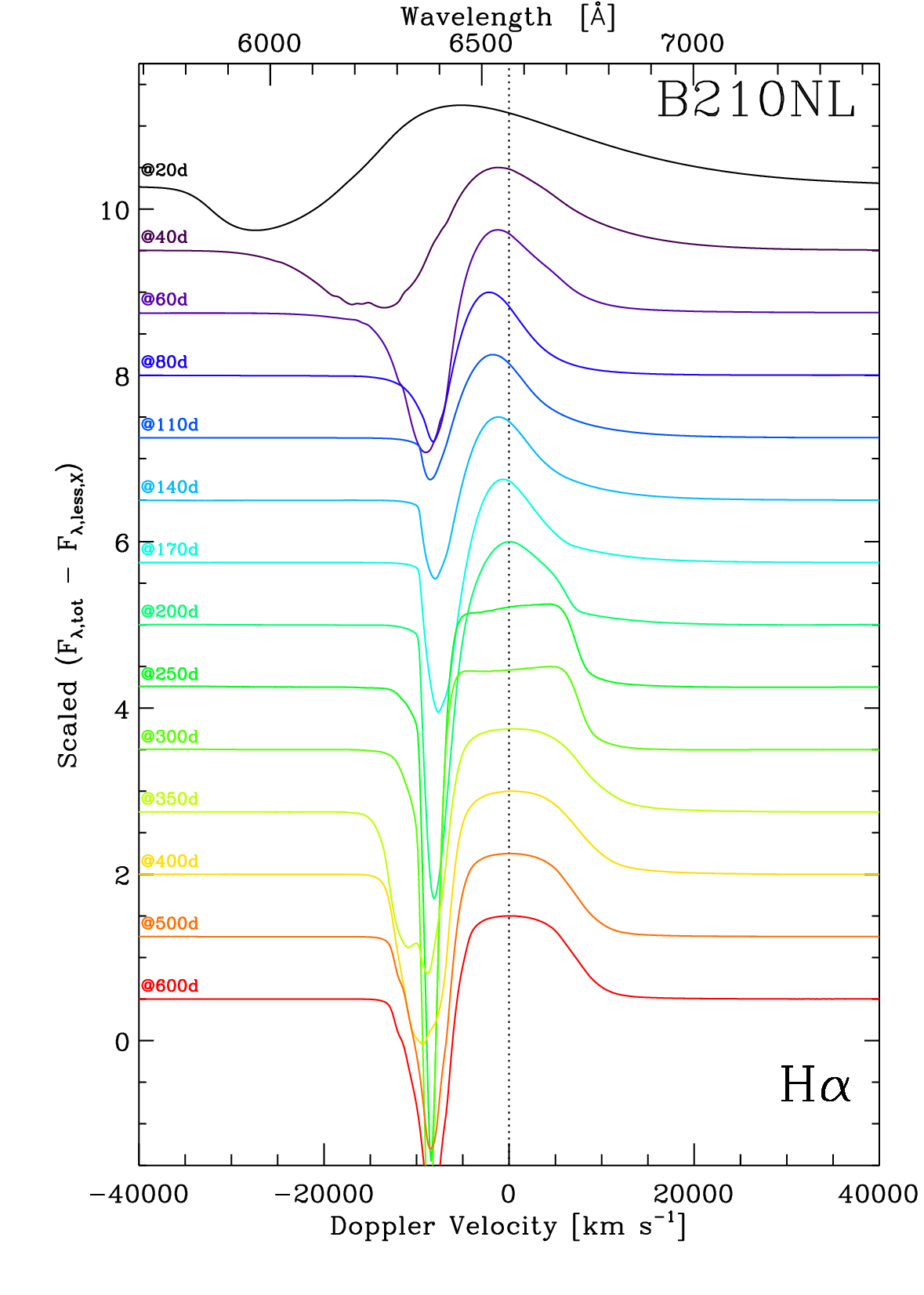,width=8.5cm}
\caption{{\it Top Row:}
Doppler velocity of the maximum absorption (filled dots) in H$\alpha$ (black), H$\beta$ (blue), and
H$\gamma$ (red) for model sequence R190NL (left) and B210NL (right).
For comparison, we overplot the evolution of the photospheric velocity (black line).
{\it Bottom Row:}
Evolution of the H$\alpha$ line profile in model R190NL (left) and B210NL (right).
Notice the typical P-Cygni profile morphology, already discussed in \citet{DH05_qs_SN},
with a maximum absorption that may underestimate or overestimate the photospheric
velocity, and the blue shifted peak emission.
At nebular times, the line profile evolves towards a flat topped emission with a P-Cygni profile absorption
fixed at $\sim$\,4000\,\kms\ (R190NL) and $\sim$\,8000\,\kms\ (B210NL), which in each case
corresponds to the velocity at the base of the hydrogen envelope (see Fig.~\ref{fig_summary_presn}).
\label{fig_vhi}
}
\end{figure*}

     The Doppler velocity at maximum absorption in the broad H\one\ Balmer lines tracks closely
     the decreasing photospheric velocity,  although weaker Balmer lines from higher up in the series may
     underestimate it (Fig.~\ref{fig_vhi}; left column).  This apparently peculiar finding is well explained by the
     line formation process in the homologously expanding ejecta of SNe \citep{DH05_epm}.
     Non-LTE and time-dependent  effects also contribute to maintain a significant line optical depth at large velocities,
     producing persistent, strong and broad lines for H$\alpha$ and H$\beta$ \citep{UC05_time_dep,DH08_time}.

   \subsection{Spectral evolution during the photospheric phase of PISN models B190 and B210}
   \label{sect_spec_BSG}

\begin{figure*}
\epsfig{file=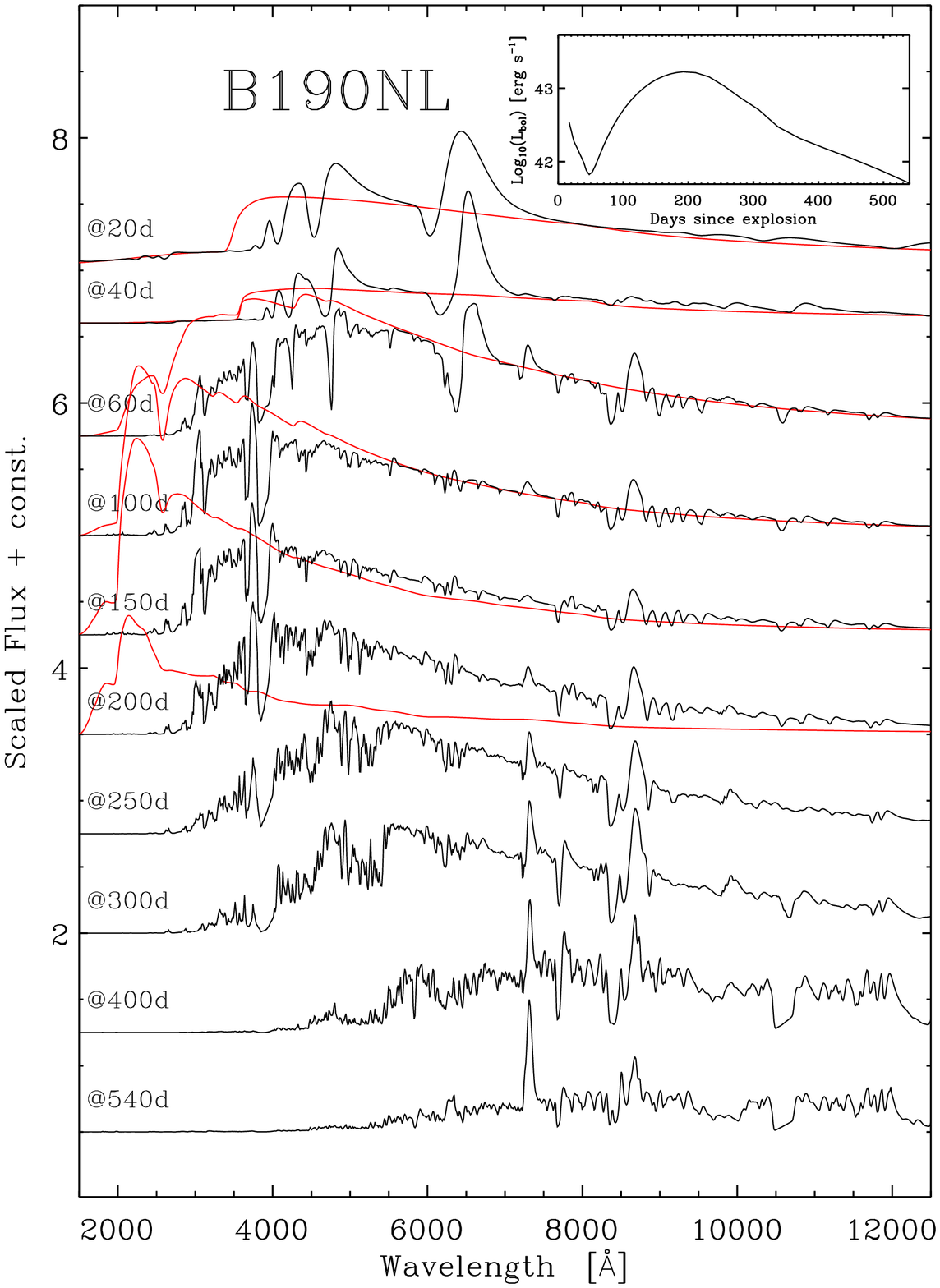,width=8.5cm}
\epsfig{file=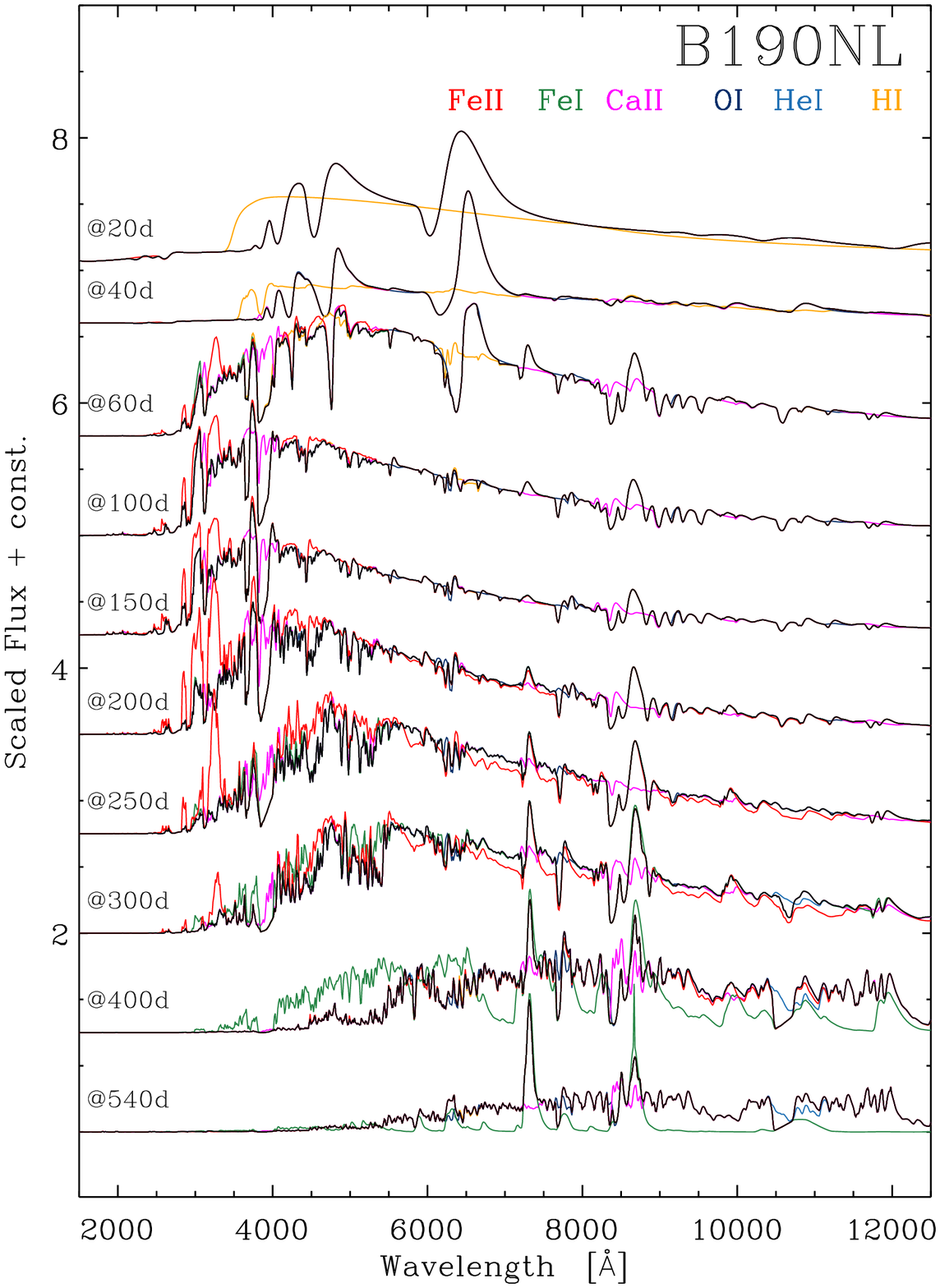,width=8.5cm}
\caption{Same as Fig.~\ref{fig_R190}, but now for BSG model B190NL with 190\,\msun\ on the main sequence.
\label{fig_B190}
 }
\end{figure*}

\begin{figure*}
\epsfig{file=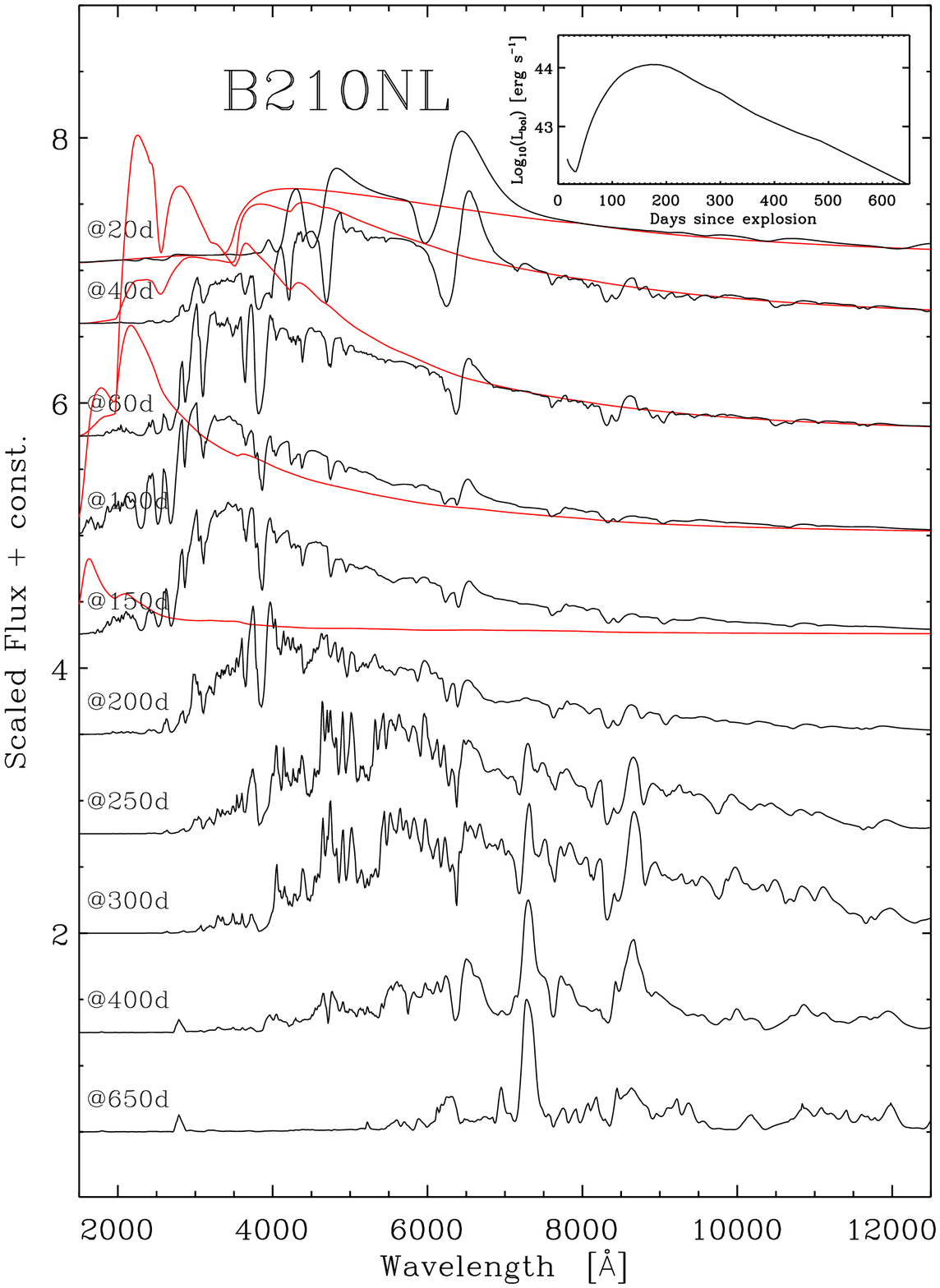,width=8.5cm}
\epsfig{file=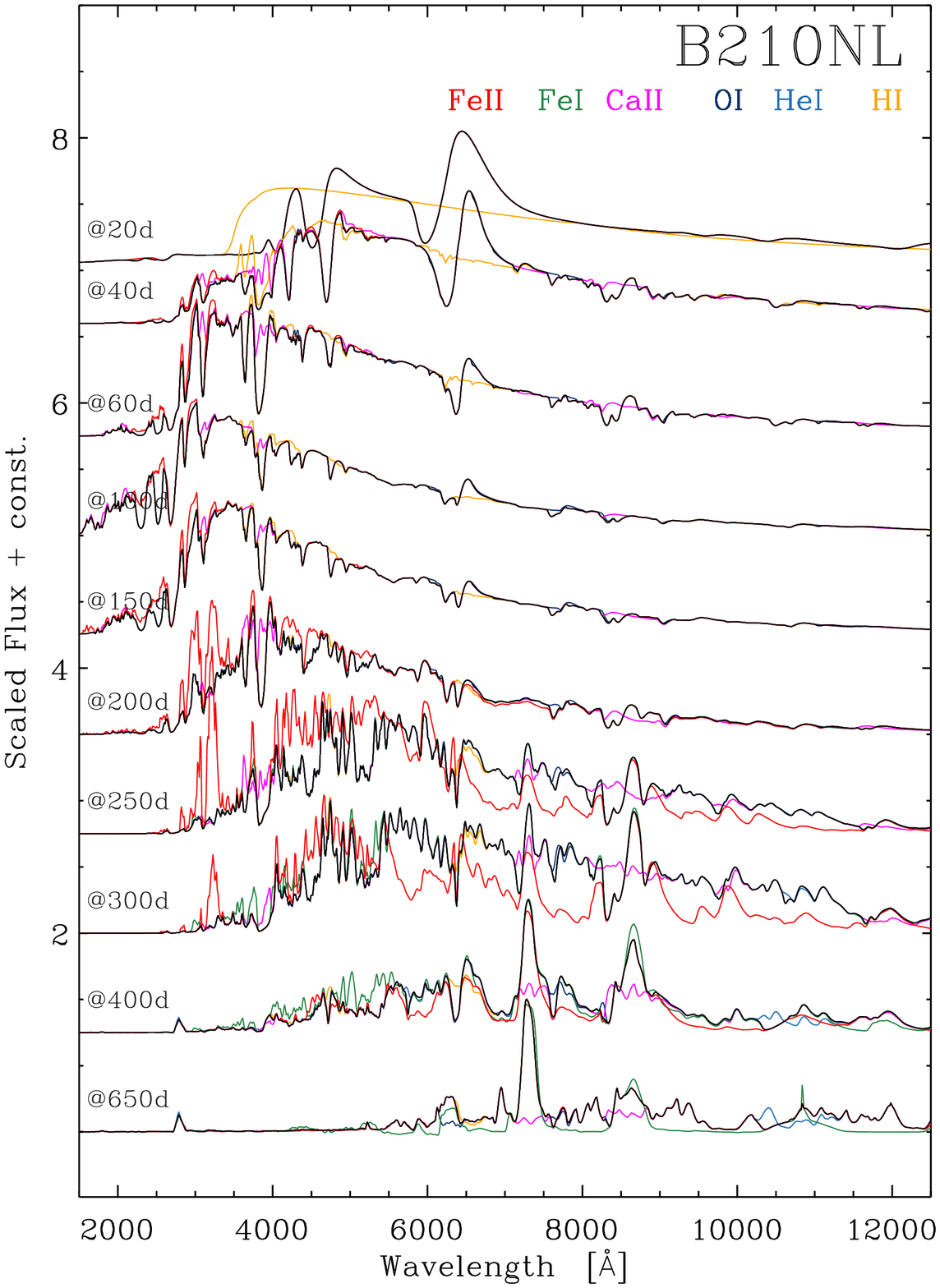,width=8.5cm}
\caption{Same as Fig.~\ref{fig_R190}, but now for BSG model B210NL with 210\,\msun\ on the main sequence.
Notice, in particular, the presence of He\one\,10830\,\AA\ at nebular times and the red colors at all post-peak epochs,
despite the huge \isoni\ mass synthesized in this explosion.
\label{fig_B210}
 }
\end{figure*}

   The spectral evolution of model B190 and B210 reflects the typical explosions from BSG stars, objects
characterized here by a surface radius of 186 and 146\,\rsun\ (Figs.~\ref{fig_B190}--\ref{fig_B210}).
As for model R190, the large explosion
energies of 34.5 and 65.9\,B, respectively, produce high representative ejecta velocities of $\sim$\,5100\,\kms\ and
6800\,\kms, well in excess of the value of  2800\,\kms\ inferred for SN 1987A \citep{arnett_etal_89}.
At early times, the spectral evolution for these two models is qualitatively similar to the SN II-pec model
of \citet{DH10,li_etal_12} and the observations of SN 1987A.

   Up to the phase of re-brightening ($\sim$\,50\,d), the optical spectra are comparable to those of model R190
   at the recombination epoch and are composed exclusively of H\one\ Balmer lines. The lack of metals in
   the primordial gas, and the absence of mixing into the outer ejecta layers by enriched material (arising from steady
   or  explosive burning), prevents any line blanketing in the UV or the optical. However, the smaller initial
   radius causes a much stronger cooling due to expansion, which induces early recombination at the photosphere
   and a fast recession of the photosphere (Fig.~\ref{fig_phot_xfrac}).
   While it took $\sim$\,250\,d to reach the helium core in model R190, the photosphere recedes to the base of the
   He-rich shell as early as $\sim$\,50\,d after explosion in models B190 and B210. As seen before
   for model R190, this corresponds to the appearance of new lines (from O\one\ or Ca\two) and the onset
   of line-blanketing effects, at this time primarily from IMEs.

   The photosphere recession to the helium core quenches the emission in H$\alpha$ and other Balmer lines --
   a consequence of the small $\mathcal{O}$(0.1) hydrogen mass fraction in the he-rich shell and the lack of non-local \gray\
   energy deposition into the H-rich envelope at such early times.

   In model B210, the larger \isoni\ mass causes
   a reversal of the photosphere, which moves back into the helium shell, but in model B190, decay heating merely
   stalls the recombination wave halfway through the ejecta (in mass), within the O-rich shell. It is around the
   light curve peak and beyond that the decoupling layers for the radiation recede to the IGE-rich layers,
   causing the appearance of iron line blanketing at $\sim$\,200\,d after explosion.

     In the right panel of Fig.~\ref{fig_comp_s15_lm18}, we show a comparison, at the recombination epoch,
     for models B190 at 30\,d after explosion and the SN II-pec model of \citet{DH10} at 15\,d after explosion.
     As for the R190/s15e12 comparison, the PISN models B190 and B210 have systematically broader Balmer
     lines and lack the strong lines from Ca\two, Na\one, Sc\two, and Fe\two, present in the ``lm18a7Ad"
     model at LMC metallicity. At such epochs these theoretical BSG explosions are relatively faint,
     and would be more difficult to detect than the much brighter R190 model.

     For models B190/B210, the Doppler velocity at maximum absorption in the H\one\ Balmer line is initially large,
    and declines as the photosphere recedes inward through the H-rich envelope (Fig.~\ref{fig_vhi}; right column,
    shown only for model B210). The decline is much more rapid than in R190, since the time scale for the
    photosphere to pass through the lower-mass H-rich envelope is $\sim$\,50\,d, much less than the $\sim$\,250\,d
    for model R190. In models B190/B210, the H\one\ Balmer lines present a maximum absorption at a Doppler
    velocity that strongly overestimates (by a factor of $\sim$\,2) the photospheric velocity at early times
    --  the same effect is seen in SN\,1987A \citep{DH10}. However, as the model re-brightens
    and its photosphere leaves the H-rich envelope, the maximum absorption of the H\one\ Balmer lines track
    a unique Doppler velocity, which corresponds to the velocity at the base of the hydrogen envelope.

    \subsection{Spectral evolution during the photospheric phase of PISN model He100}
\label{sect_spec_He100}

   In this section, we describe the spectral evolution of a PISN arising from a  WR progenitor using model He100ionI,
   which differs from model He100 by the treatment of additional neutral ions found to be important
   (Mg\one, Si\one, S\one, and Ca\one). We discuss specific  differences between He100 and He100ionI in
   Appendix~C, which are essentially confined to the colors  in the first few weeks following explosion, and
   additional line features and blanketing.

      At the start of the simulation, the He100ionI and He100 models are at 10.5\,d after explosion and
      the photosphere has already receded to the base of the $\sim$\,8\,\msun\ He-rich outer shell (Fig.~\ref{fig_phot_xfrac})
      at $\sim$\,9000\,\kms\ (Fig.~\ref{fig_phot}). The composition is still He-rich but the O, Ne, Mg and Ca
      are also abundant, with mass fraction for IMEs that are orders of magnitude larger than for a solar mixture
      at a metallicity of 10$^{-4}$\,\zsun. This explains the dominant role of line-blanketing from IMEs,
      including Si and S,  from such early times up to the peak of the light curve. It also explains
      the early appearance of strong Ca\two\ lines.

\begin{center}
\begin{figure*}
\epsfig{file=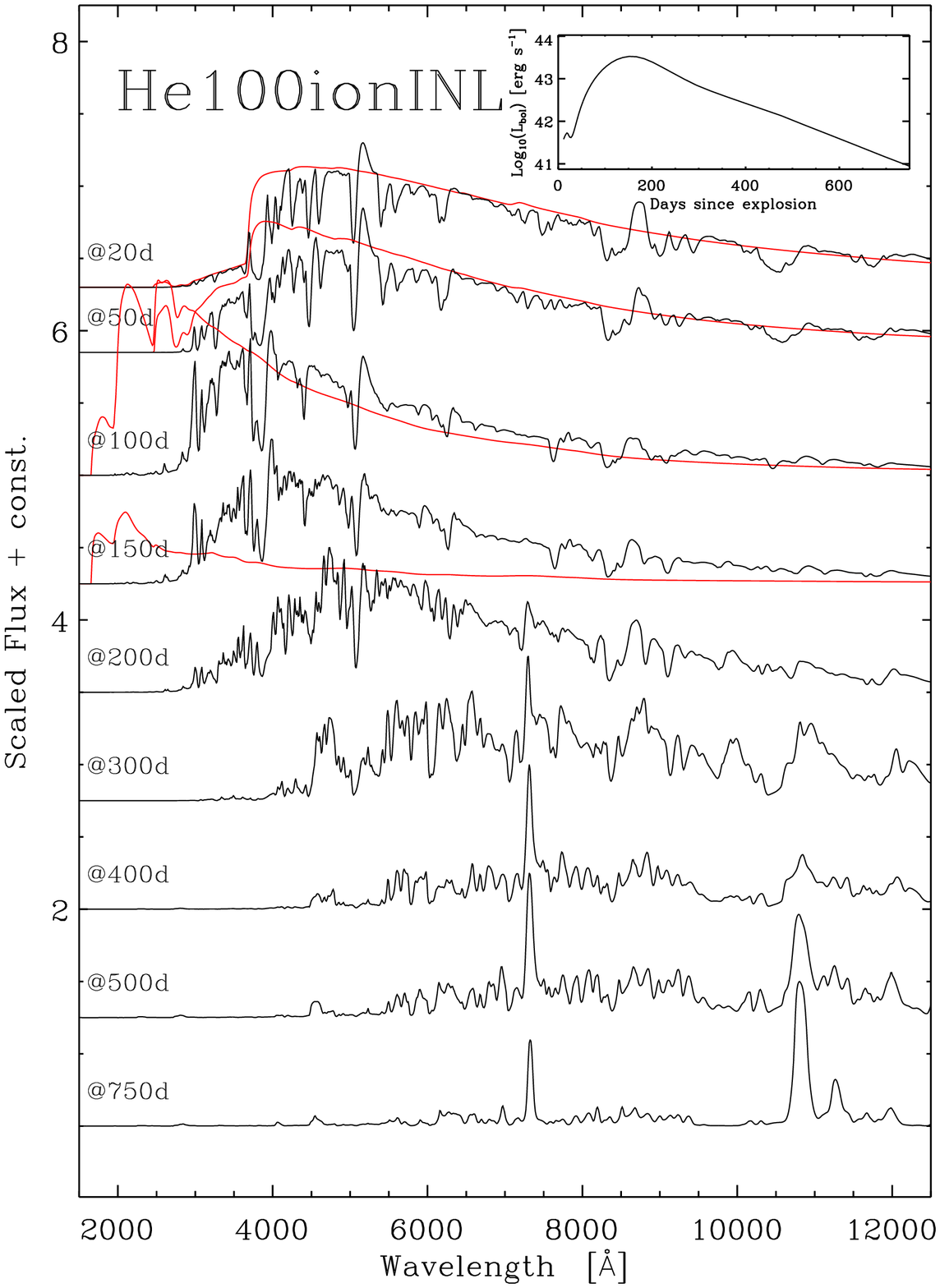,width=8.5cm}
\epsfig{file=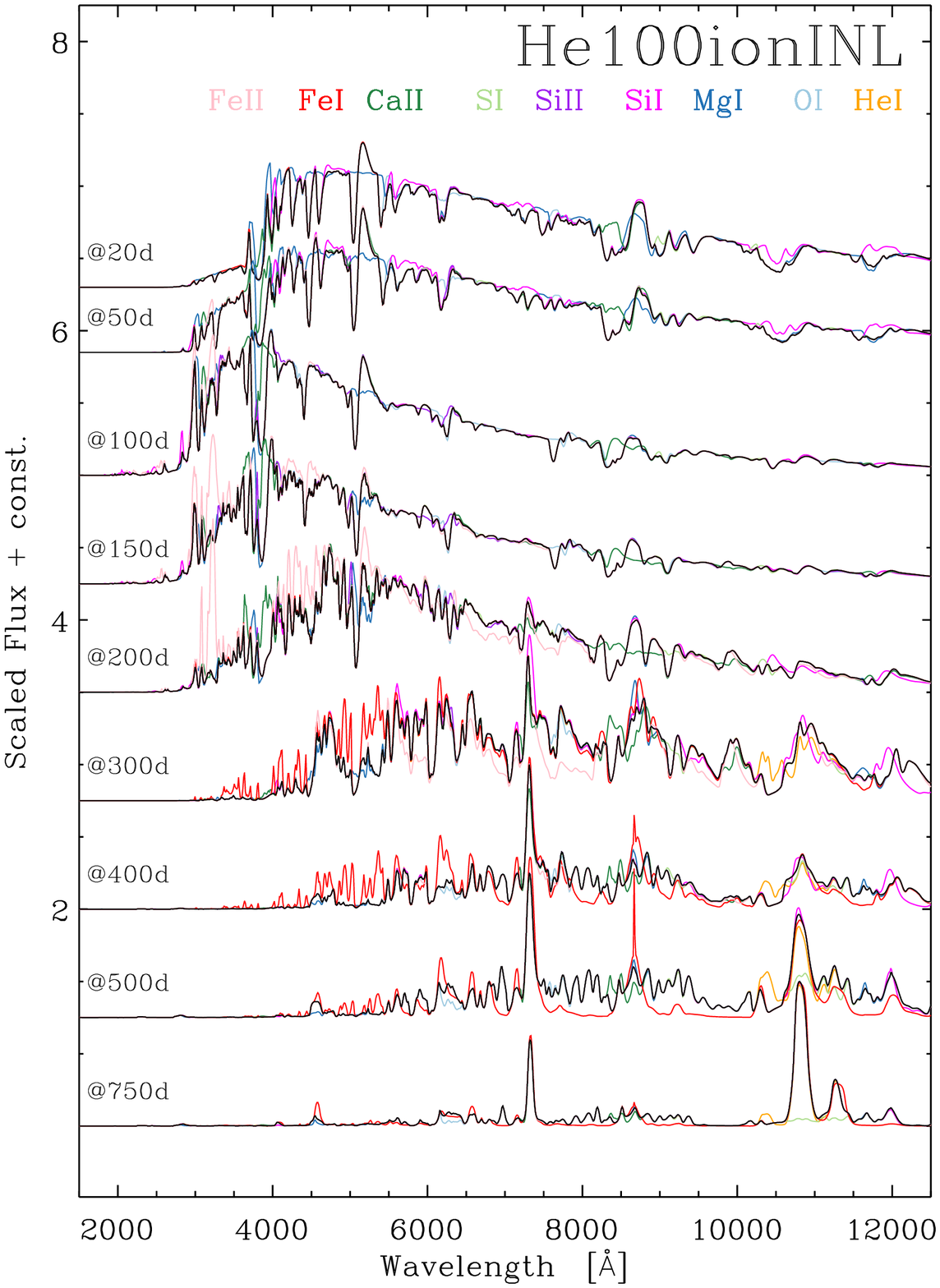,width=8.5cm}
\caption{Same as Fig.~\ref{fig_R190}, but now for 100\,\msun\ He-star model He100ionINL (same PISN model as He100,
but the radiative transfer is computed with allowance for Mg\one, Si\one, S\one, and Ca\one; we also allow for
non-local energy deposition).
\label{fig_he100}
 }
\end{figure*}
\end{center}

  This chemical stratification, although not as obvious as in the H-rich models discussed in the preceding
sections, is reflected in the spectral evolution of model He100ionI (Fig.~\ref{fig_he100}).
Despite representing $\sim$\,8\% of the total mass of the ejecta, He is only present in the outermost
ejecta shells. By 10\,d after explosion, the photosphere has already crossed these He-rich regions
so that  there is no visible signature of He during the remaining part of the photospheric phase,
This is consistent with the $\mathcal{O}$(0.1) mass fraction of He in the outer layers \citep{dessart_etal_11}
together with the negligible non-thermal excitation/ionization at such epochs in the outer ejecta \citep{dessart_etal_12}.
We find Si\one\ to be an important source of blanketing at the cool photosphere (for up to about 50\,d).
The continuum flux testifies for the presence of important photo-ionization cross sections (e.g., from Mg\one\
with an  important edge at 3757\,\AA) and the absence of strong line blanketing in the optical,
since the continuum flux closely follows the total flux for up to a month
(Fig.~\ref{fig_he100}). To be complete about these line identifications, we find O\one\ lines at 6157, 6454,
7002, 7254, 7777, 7990, and 9260\,\AA; Mg\one\ lines at 5167--5172--5183\,\AA; Mg\,\two\ lines at 4481 and
7890\,\AA; Si\one\ lines at 3905, 4102, 5500--5800 (numerous lines), 5948\,\AA, 7250, 7289,  7410, 7742--7799,
8648, 9413, 10585, 10694, 10827, 10870, 10885,  10982\,\AA; Si\two\ lines at 4129, 5056, 5957-5978, 6239,
6347--6371, 7849\,\AA; S\one\ lines at 6052, 6750, 8712.44,8874.48, 9228, 9421, 9650--9681,10457, 10633;
and Ca\two\ H\&K, 7103, 7162, 8133, 8201, 8248, 8498, 8542, 8662\,\AA. The strong Mg\one\,5167\,\AA\ is
a striking feature of He100 model around the peak of the LC.

\begin{figure}
\epsfig{file=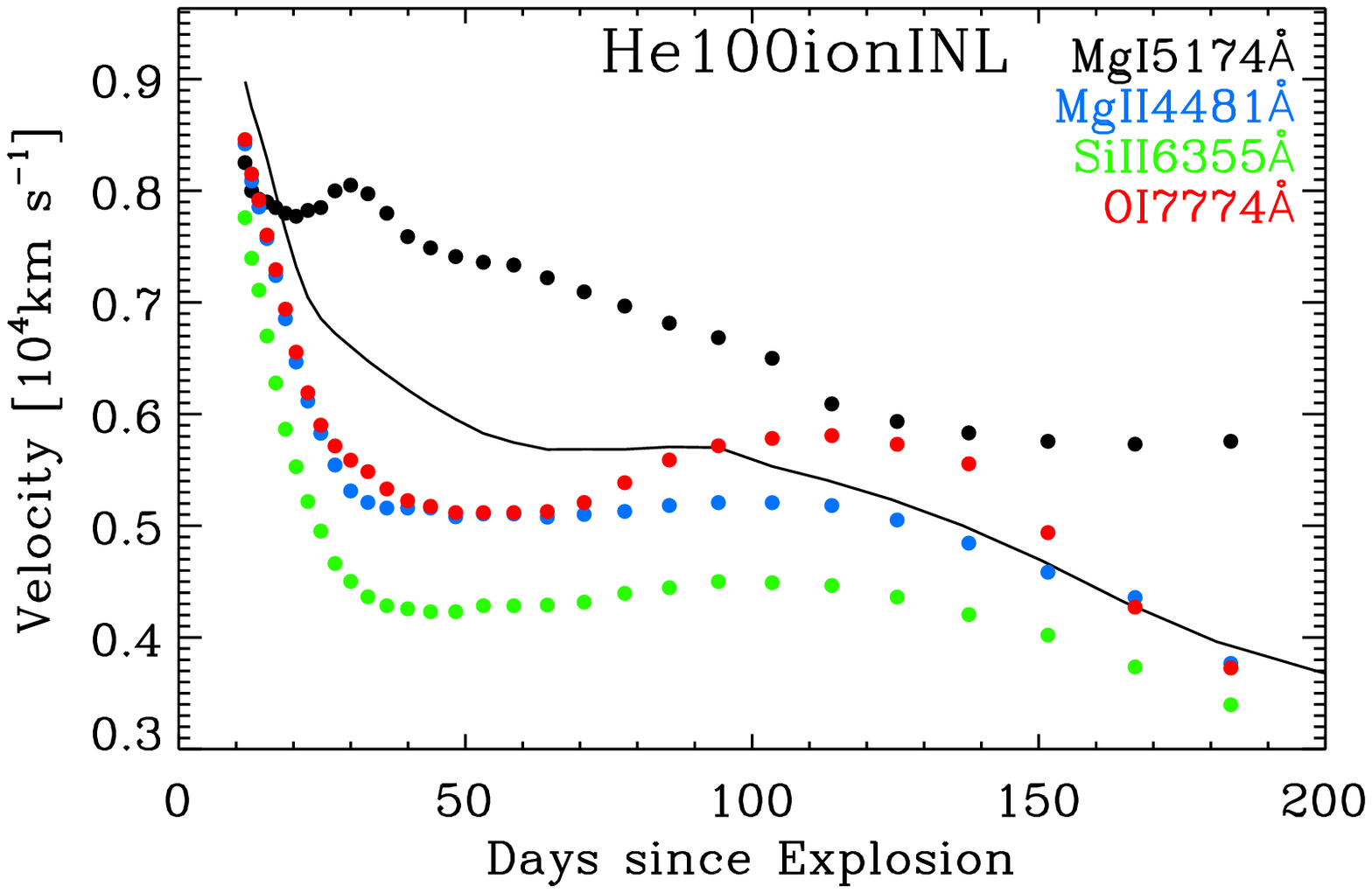,width=8.5cm}
\epsfig{file=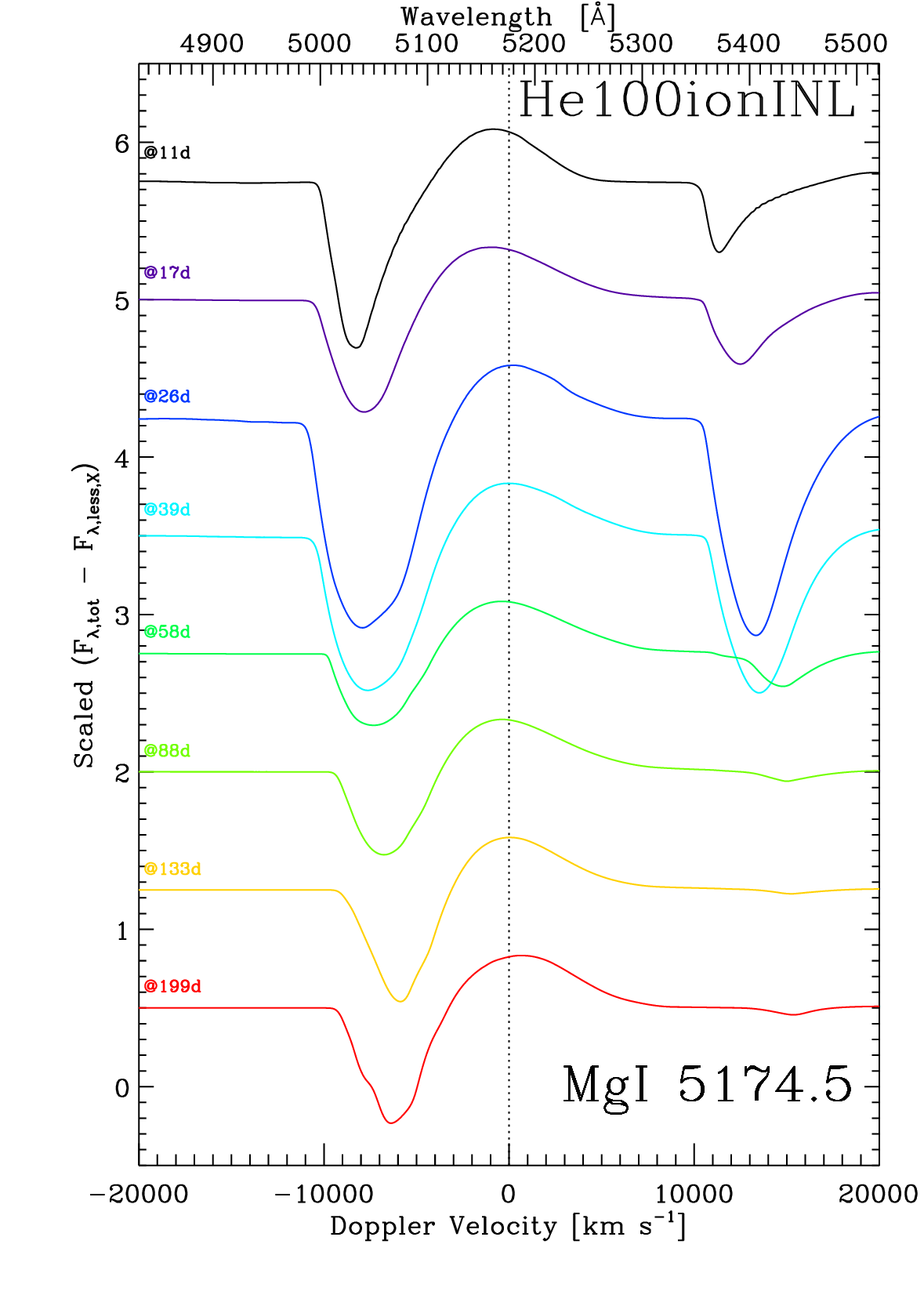,width=8.5cm}
\caption{Same as Fig.~\ref{fig_vhi}, but now showing the results for model He100ionINL.
We focus on early times when these lines remain optically thick.
\label{fig_vel_he100}
}
\end{figure}

   The larger photospheric temperature on the rise to bolometric maximum (Fig.~\ref{fig_phot}) is visible in the hardening of the
spectra (see also Fig.~\ref{fig_mb_lc}) and the temporary weakening of blanketing from neutral IMEs. During that phase, the photospheric
velocity is around 5000\,\kms, and the optical line profiles appear relatively narrow (i.e., compared to the
RSG/BSG progenitor models discussed above), apart from more optically-thick lines like Ca\two.
Numerous lines have a velocity at maximum absorption that
underestimates the photospheric velocity (Fig.~\ref{fig_vel_he100}). As noted before and in \citet{DH05_epm},
this is caused by a projection effect in homologous ejecta since for increasing impact parameters, the location of
maximum absorption tends to follow an iso-density contour, intersecting planes of ever smaller projected velocity.
During this brightening phase, the continuum synthetic flux is again a good diagnostic for this heating and increased
blanketing from earlier times.

   By $\sim$\,150\,d after explosion (i.e., the light curve peak), the photosphere has now receded halfway
through the ejecta, into the layers dominated by Si/S/IGEs, and its temperature goes through a maximum
(of $\sim$\,6000\,K).  From then onwards, the spectrum reddens, the blanketing from IGEs strengthens,
initially through the effects of Fe\two\ and eventually Fe\one. The spectrum starts again showing strong
lines of low-ionization and neutral species  (as it did at $\sim$\,50\,d), but with the additional contribution
from metal line blanketing. We also see the appearance of [Ca\two] at 7291--7323\,\AA. As seen before
for the B190 and B210 models, the spectra at and beyond light-curve peak are red, not blue, with the bulk
of the flux emerging at 5000\,\AA, and progressing to longer wavelength as time proceeds. Throughout this
phase, no He\one\ line is visible, which again confirms the notion that a Type Ic classification is not a guarantee
of helium deficiency \citep{dessart_etal_12}. The fundamental feature of these PISN spectra are the ubiquitous
and concomitant presence of lines from O, Mg, Si, and Ca, in a much more pronounced fashion than generally
seen in Type Ic SNe.   The low metallicity of the environment is not as obvious as for the R190 model, but
noticeable after inspection from the lack of Fe\two\ blanketing in the 5000\,\AA\ region; it is clear that the blanketing
occurs but it is not IGEs like Ti or Fe that cause it here but instead IMEs and in particular Si.

\begin{figure*}
\epsfig{file=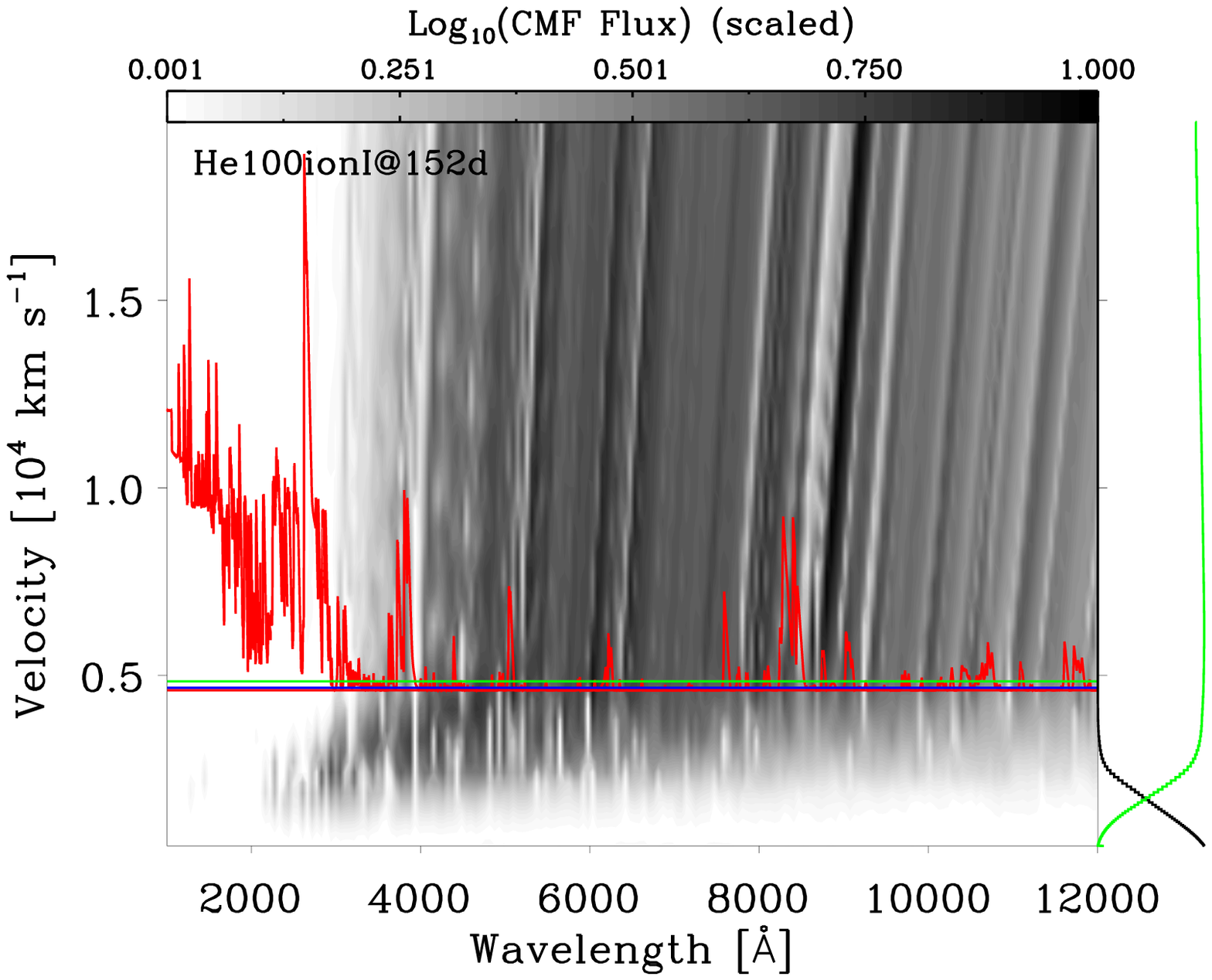,width=8.5cm}
\epsfig{file=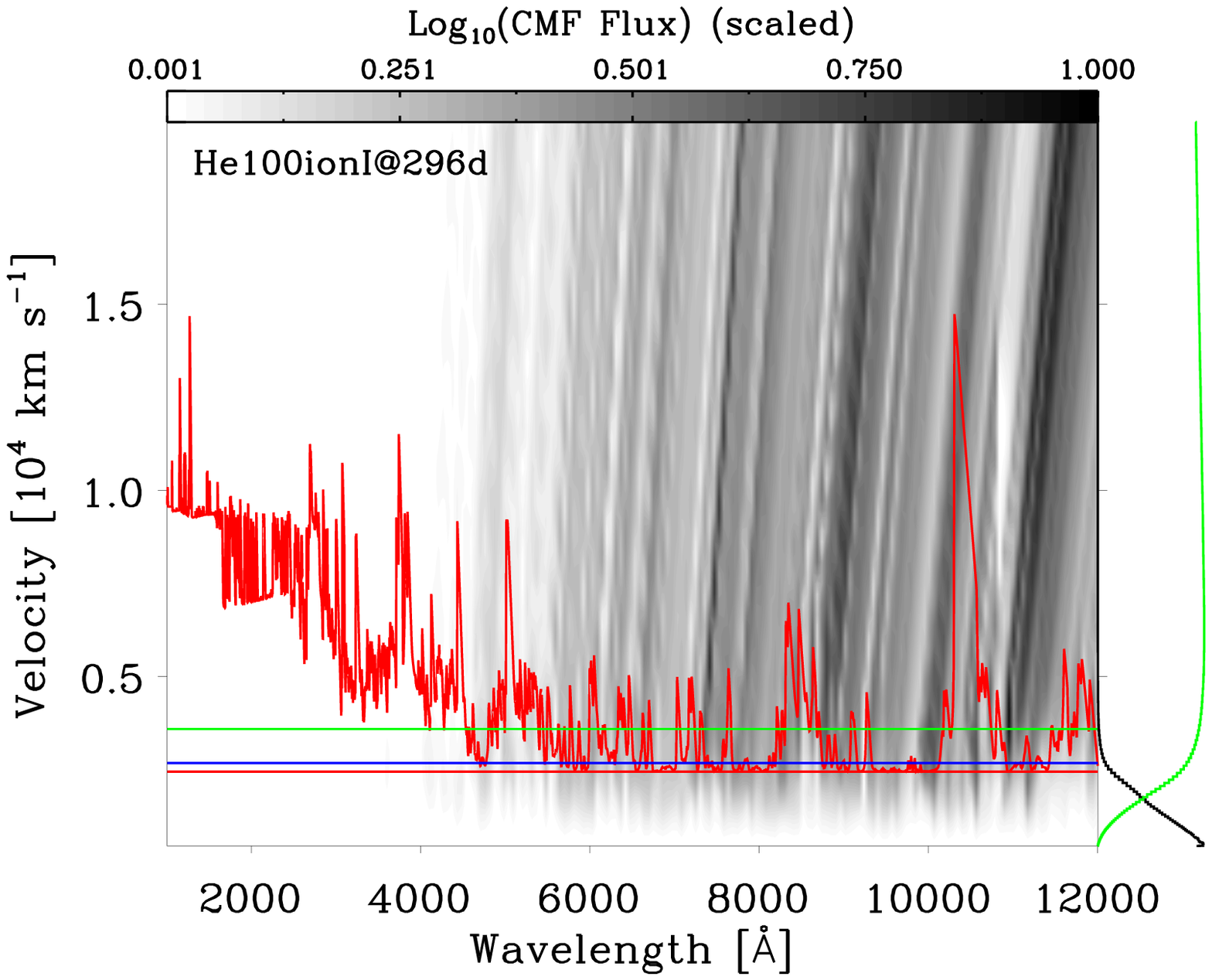,width=8.5cm}
\caption{
Variation of the co-moving frame flux versus wavelength and velocity (or depth in the ejecta)
for model He100ionI at 152 (left) and 296\,d after explosion (right).
We also overplot the location of the photosphere that results when including all sources of opacity (jagged red line),
electron scattering only (bottom red line), the Rosseland-mean opacity (blue line), or the flux-mean opacity
(green line).
Finally, for illustration, we show on the right side of each panel the local energy deposition from radioactive
decay of \isoni\ nuclei (black; in erg\,s$^{-1}$\,cm$^{-3}$; normalized to its maximum value which occurs in the innermost
shell) and the frequency-integrated CMF flux versus  velocity (green).
While the \isoni\ distribution is a given of the hydrodynamical input and thus fixed, the distribution of the energy
deposition extends farther outward as time progresses and as the photospheres recedes.
\label{fig_flux_vs_depth}
}
\end{figure*}

    To illustrate more vividly the blanketing caused by the numerous lines of metals, as well as its evolution
with time, we show in Fig.~\ref{fig_flux_vs_depth} the distribution of the flux versus wavelength and velocity
(equivalently the depth in the ejecta). Overplotted, we draw the photospheric location obtained for various
opacity sources (electron-scattering, Rosseland-mean opacity, and flux-mean opacity). These photospheric
velocities/radii are systematically smaller than obtained when accounting for the additional opacity of lines
(red curve). Moving from the light-curve peak to $\sim$\,300\,d, the blanketing is even stronger in the UV and
optical, and it continues to strengthen as the ejecta cools and recombines.
This blanketing enhances the likelihood of interaction with a line for a photon emitted short ward
of $\lesssim$\,5000\,\AA. Through fluorescence, this photon can give rise to multiple lower energy photons, which,
being subject to a smaller opacity, can escape. Alternatively, blanketing lengthens the photon-residence time at depth,
and thus enhances the probability for absorption.
Because of the decrease in temperature outwards,\footnote{In the optically-thin regions, the temperature distribution
can be quite complex in practice due to density and composition changes, as well as the effects of non-local energy
deposition.} this would translate into emission but now of lower
energy photons. Both effects tend to deplete the flux where blanketing is strong, ``reddening" the spectrum.
Indeed,  where the blanketing is strong, the flux is weak.

   \subsection{Nebular phase spectral evolution of all four PISN models}
\label{sect_neb}

The transition to the nebular phase is not sudden, but occurs gradually as the intensity of weaker lines become
less and less strong compared to the lines normally associated with the nebular phase of SN spectra
(e.g., [O\one]\,6303--6363\,\AA, [Ca\two]\,7290--7290\,\AA\, and the triplet lines of Ca\,\two\ at 8500\,\AA;
and H$\alpha$ when H is present).

This outer ejecta, which contains few tens of solar masses of IMEs, O, He, and in some cases H, is like
a nebula on top of a centrally illuminating source. This persistent incoming radiation is complemented
by non-local deposition of decay-energy from \grays\ escaping the core emission sites, which occurs
when the  \gray\ mean-free-path becomes sizable compared to the ejecta extent. This can resurrect
the outer ejecta which had earlier on become completely transparent and invisible. Hence, in PISN
models  that were faint initially, the nebular phase represents a means of probing the outer ejecta again.

The strongest nebular line is [Ca\two]\,7290--7290\,\AA\ -- it has a moderate critical density of $\sim$\,10$^6$\,cm$^{-3}$,
is easily excited by collisions with electrons, and for most of the SN evolution discussed in this paper Ca$^{+}$ is the
 dominant ionization stage of Calcium. [Ca\two]\,7290--7290\,\AA\ appears as a single broad feature
at late times in all models -- in fact it appears before the complete transition to the nebular phase, at 250$\pm$50\,d.
In models B190 and B210, it is even present for a few weeks near $\sim$\,50\,d. In R190NL (which is representative
of the whole set), we find that emission is limited to the regions below 4500\,\kms\ in that model (except for H$\alpha$),
thus limited at all times to the progenitor He-core. It peaks at 4000\,\kms\ at 200\,d and systematically recedes with time down
to $\sim$\,1500\,\kms\ at 1000\,d. In contrast to Type II-SNe at near solar metallicity, there is no Ca\two\
emission/absorption from the H envelope. This occurs because of the Ca deficiency outside of the core where
Ca has a mass fraction of only $6.44 \times 10^{-9}$. The low metallicity of the PISN progenitors in our set
quenches any outer-ejecta (i.e., the H or He envelope) emission/absorption from species other than H/He.
We note at nebular times the overlap of the [Ca\two]\,7300\,\AA\ doublet with Ca\one\ lines
at 7148 and 7326\,\AA\ (see Appendix~C).

The Ca\two\ triplet is initially absent during the photospheric phase, appears
as a relatively narrow P-Cygni profile as the photosphere recedes into the helium core, before it turns into
a set of relatively narrow emission lines at very late times. In model B210, the larger expansion rate causes  the
individual emissions to overlap and form a unique emission feature --- different ejecta kinetic energies and
stratification yield different velocities for this emission region but the principle holds the same in all.

     Interestingly, some lines appear or re-appear at the nebular
phase following the revival from non-local energy deposition. The most striking example of this is the
re-appearance and persistence of H$\alpha$ in model B210NL beyond 200\,d, after the line had been weakening
since the start of the simulations at
15\,d (Fig.~\ref{fig_B210}). Over the time span 200-500\,d, H$\alpha$ shows a strong P-Cygni profile, in particular
with a strong and broad absorption component. We show a montage for the complete evolution of H$\alpha$ in
Fig.~\ref{fig_vhi} for models R190NL (left column) and B210NL (right column). At nebular times, the Doppler
velocity at maximum absorption remains constant (see also the top panel in Fig.~\ref{fig_vhi}) and equal to the
velocity at the base of the H envelope. The emission component is flat topped and its half-width also matches
closely the velocity at maximum absorption. The H-deficiency in the inner ejecta (Fig.~\ref{fig_summary_presn})
together with the non-local energy deposition from \grays\ explains these properties. Note that due to strong line
overlap with metal lines, H$\alpha$ is not easily seen in all RSG/BSG models. Only B210 with its enhanced
heating shows it unambiguously.

\begin{figure*}
\epsfig{file=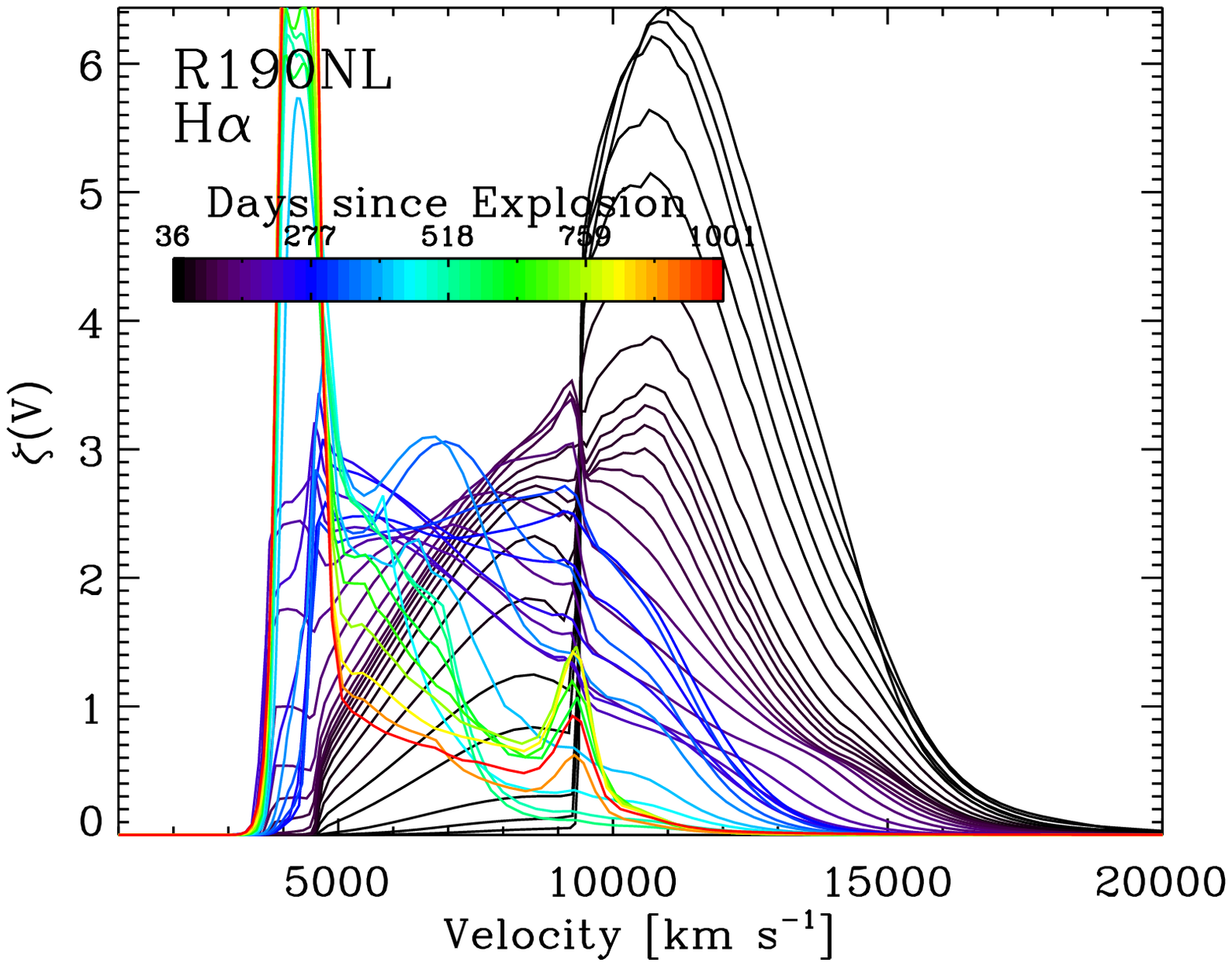,width=5.75cm}
\epsfig{file=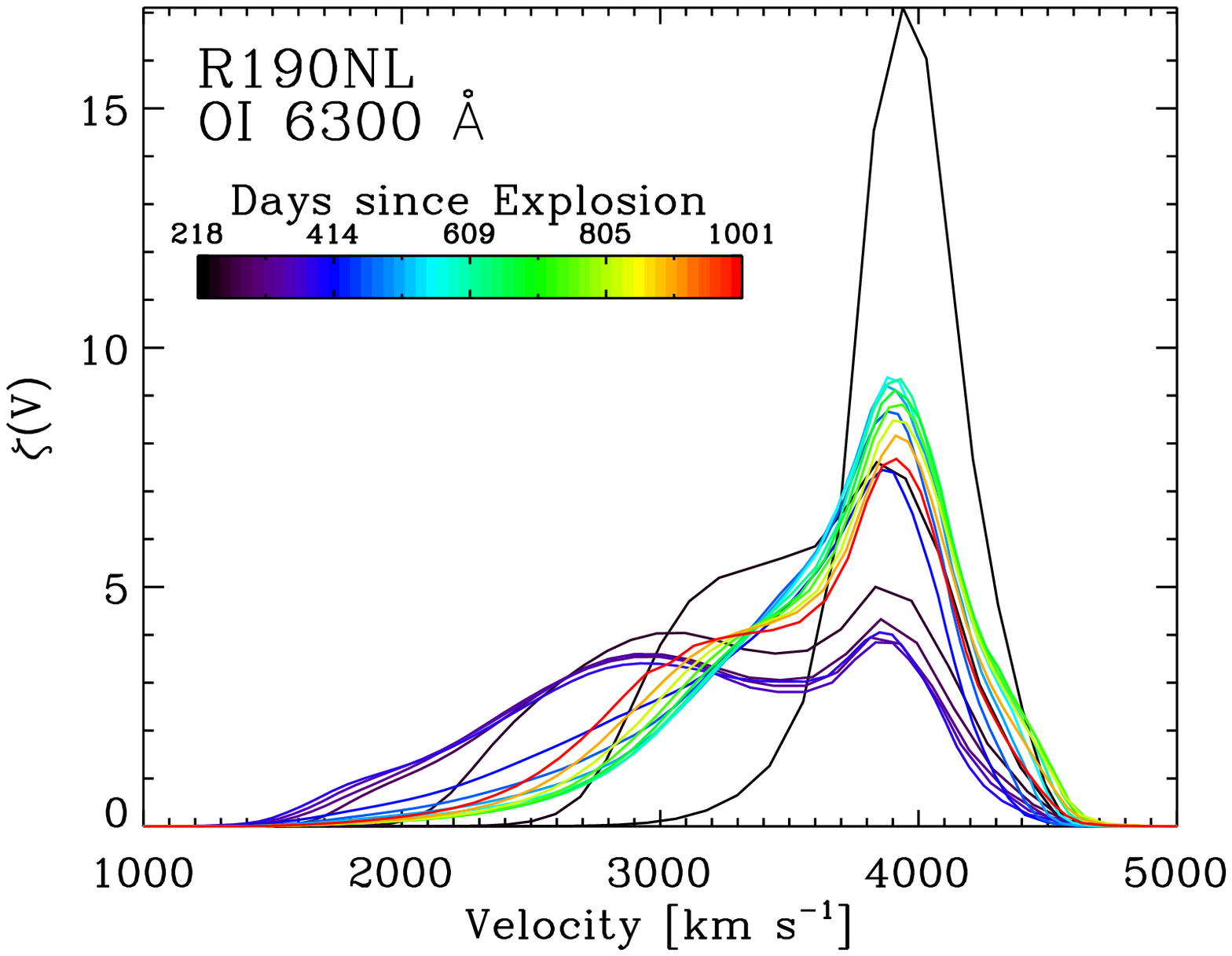,width=5.75cm}
\epsfig{file=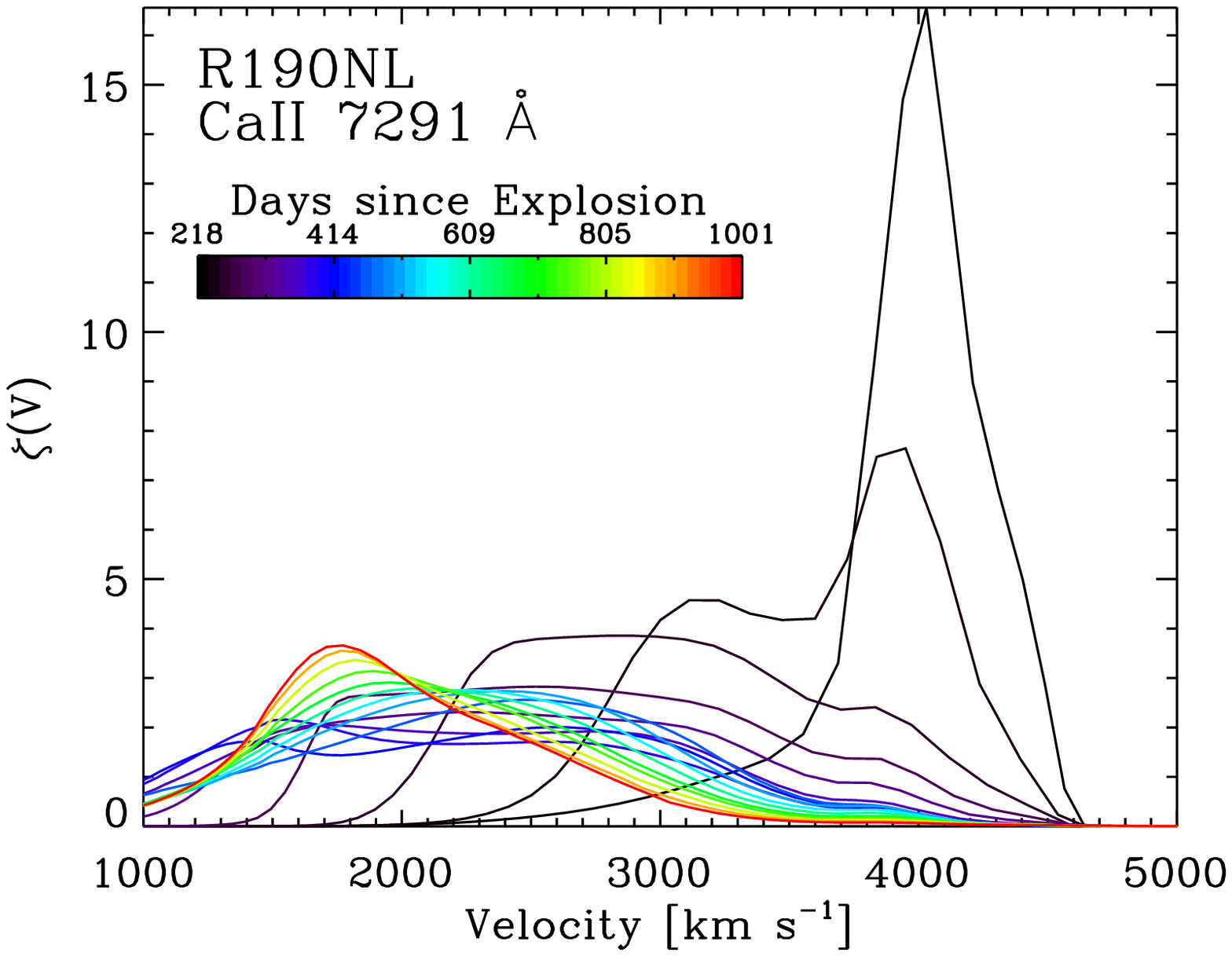,width=5.75cm}
\epsfig{file=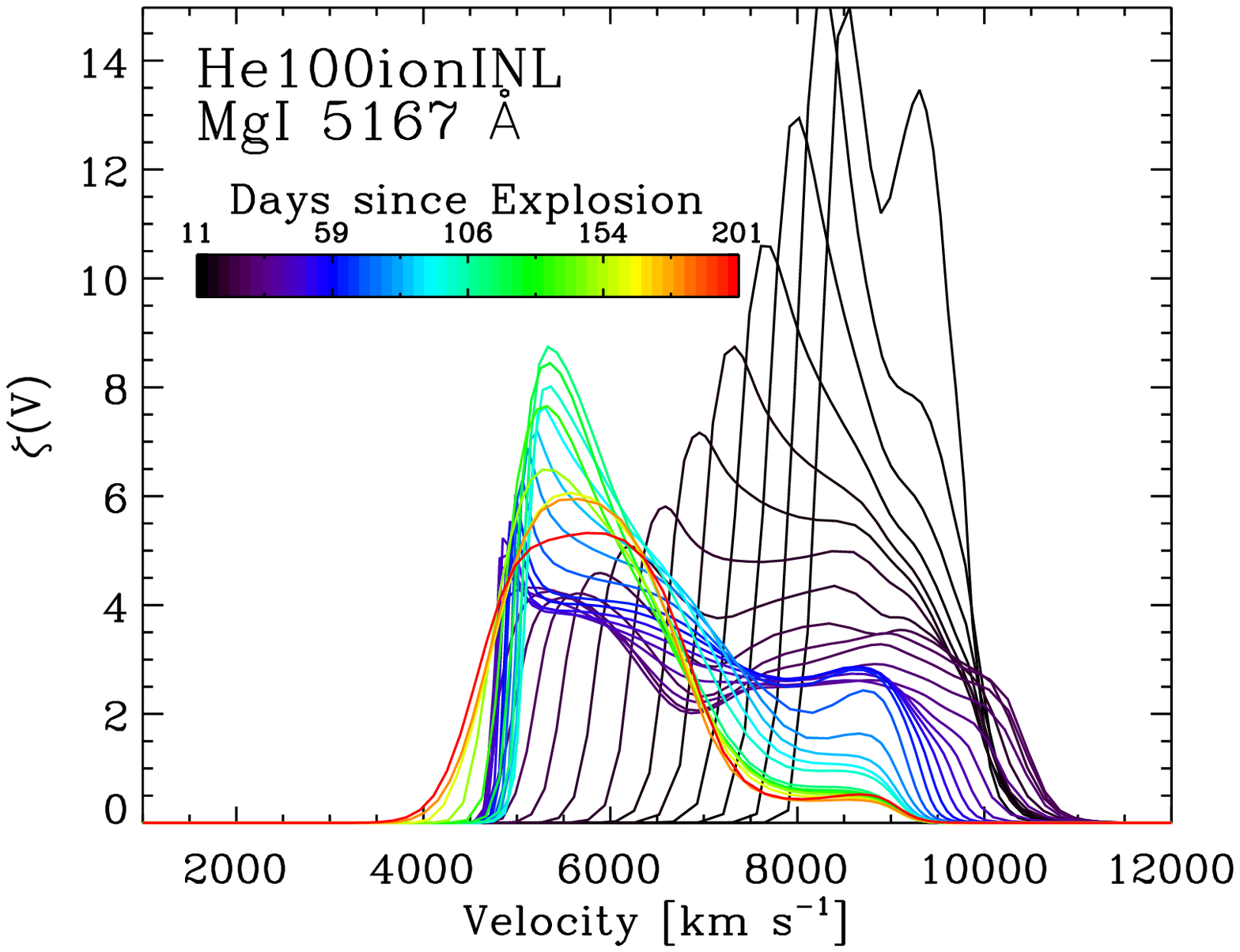,width=5.75cm}
\epsfig{file=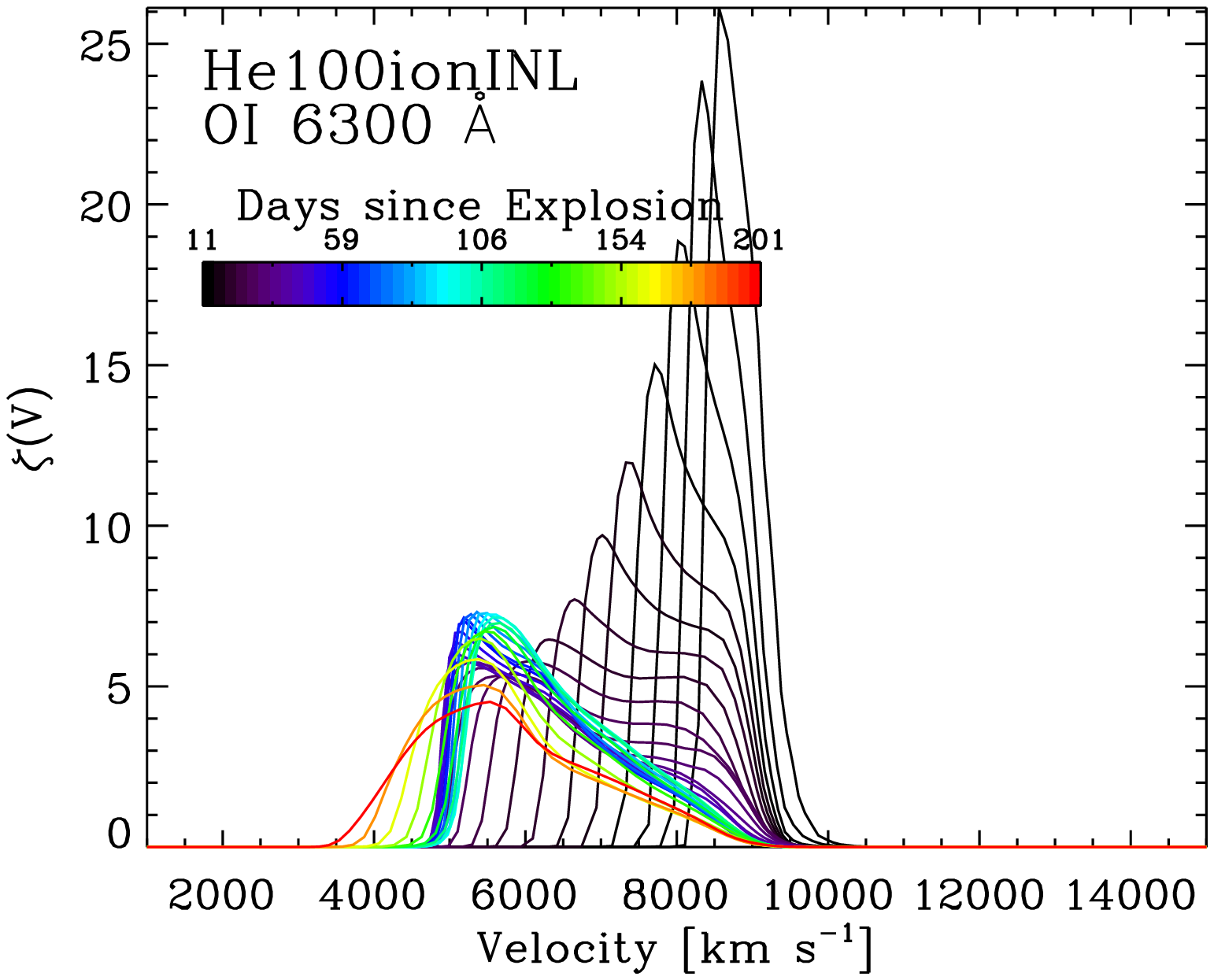,width=5.75cm}
\epsfig{file=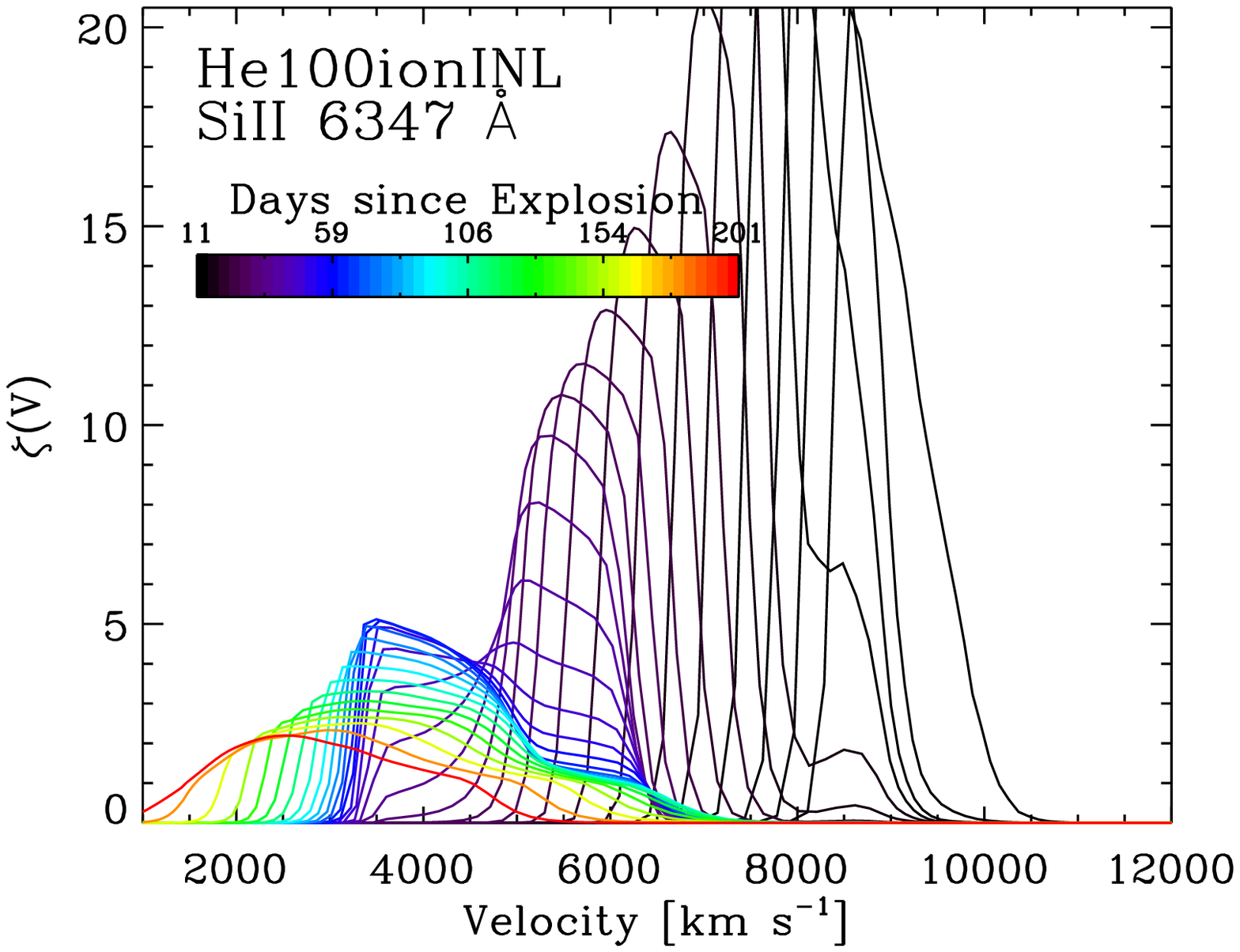,width=5.75cm}
\caption{Illustration of the spatial distribution of emission for representative lines (for multiplets,
we show the bluest component) of important ions in model R190NL (top row) and model He100ionINL (bottom row).
The quantity $\xi(v)$ is defined such that the line flux is proportional to $\int \xi \, d(\log v)$ \citep{hillier_87}.
The time coverage for each panel is adjusted to capture the time when the line is unambiguously present in the
synthetic spectrum.
We find that all lines are primarily segregated according to the chemical stratification.
\label{fig_ep_seq}
}
\end{figure*}

   In Fig.~\ref{fig_ep_seq}, we show the time evolution of the formation region for some of these lines
in models R190NL and He100ionINL. The emission regions are ultimately bounded by the abundance distribution
of the species under consideration, which reflect the pre-SN evolution,
the regions where explosive burning took place, and the additional smearing we apply to all models (except for B210)
--- see Fig.~\ref{fig_summary_presn} for details.
For example, [O\one]\,6303--6363\,\AA\ is  present in all models at very late times.
The lower the model $E/M$, the lower the expansion velocity of the O-rich shell. We obtain
representative velocities of 4000\,\kms\ (R190 and B190), 6000\,\kms\ (B210), and 7000\,\kms\ (He100).
Oxygen emission thus arises from regions where both the O mass fraction and the mass density are large.
This emission is also influenced by optical depth effects.
In model R190, O emission comes from regions with velocities extending from
1400 to 4500\,\kms, while in model He100 O emission is confined to regions with velocities between 3500
and 10,000\,\kms. At early times, the emission is weighted towards the outer regions because the high
opacity limits the emission from the inner regions. At later times emission tends to be weighted towards
the inner regions where the power source (the decay of \isoco) resides.

\begin{figure*}
\epsfig{file=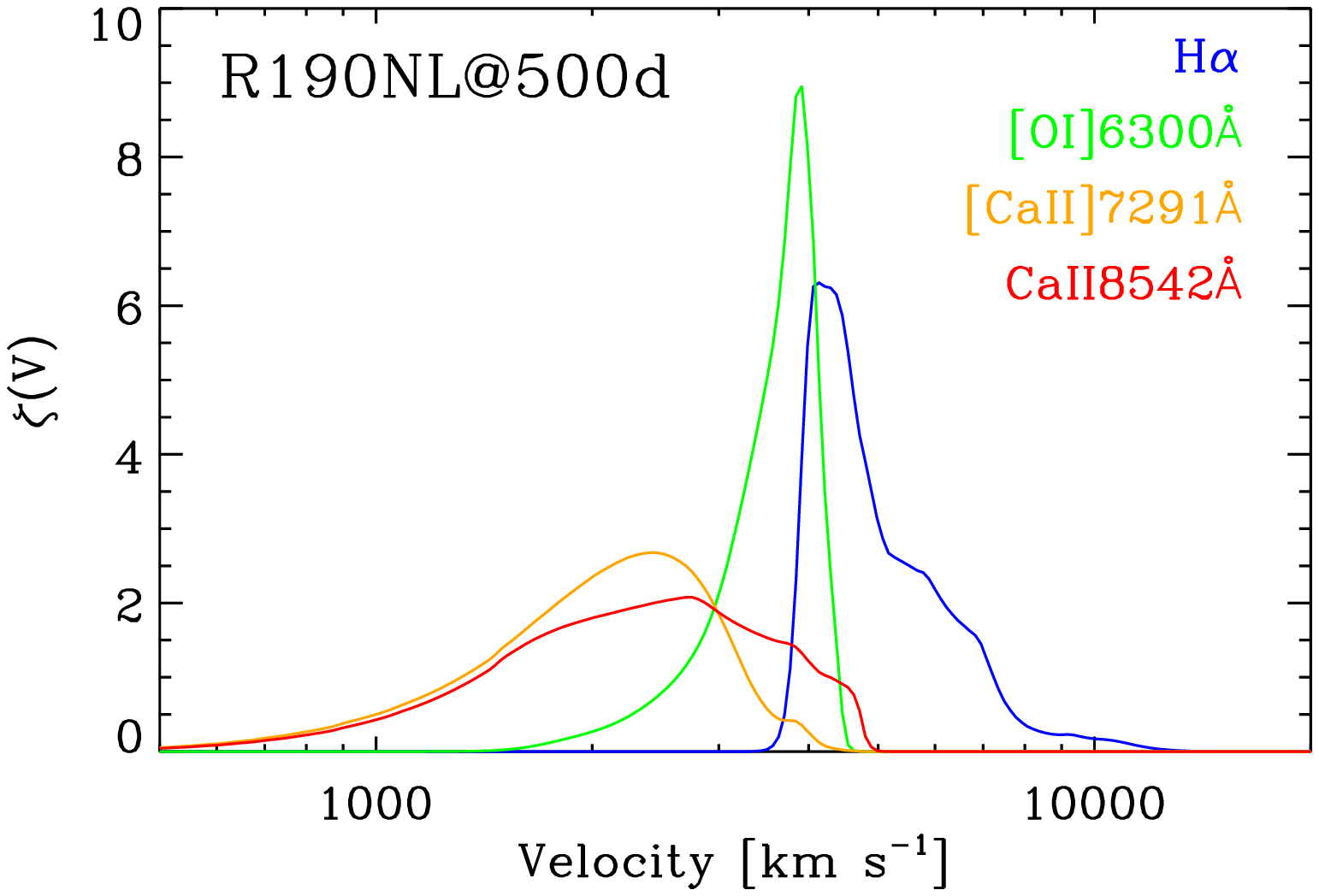,width=8.75cm}
\epsfig{file=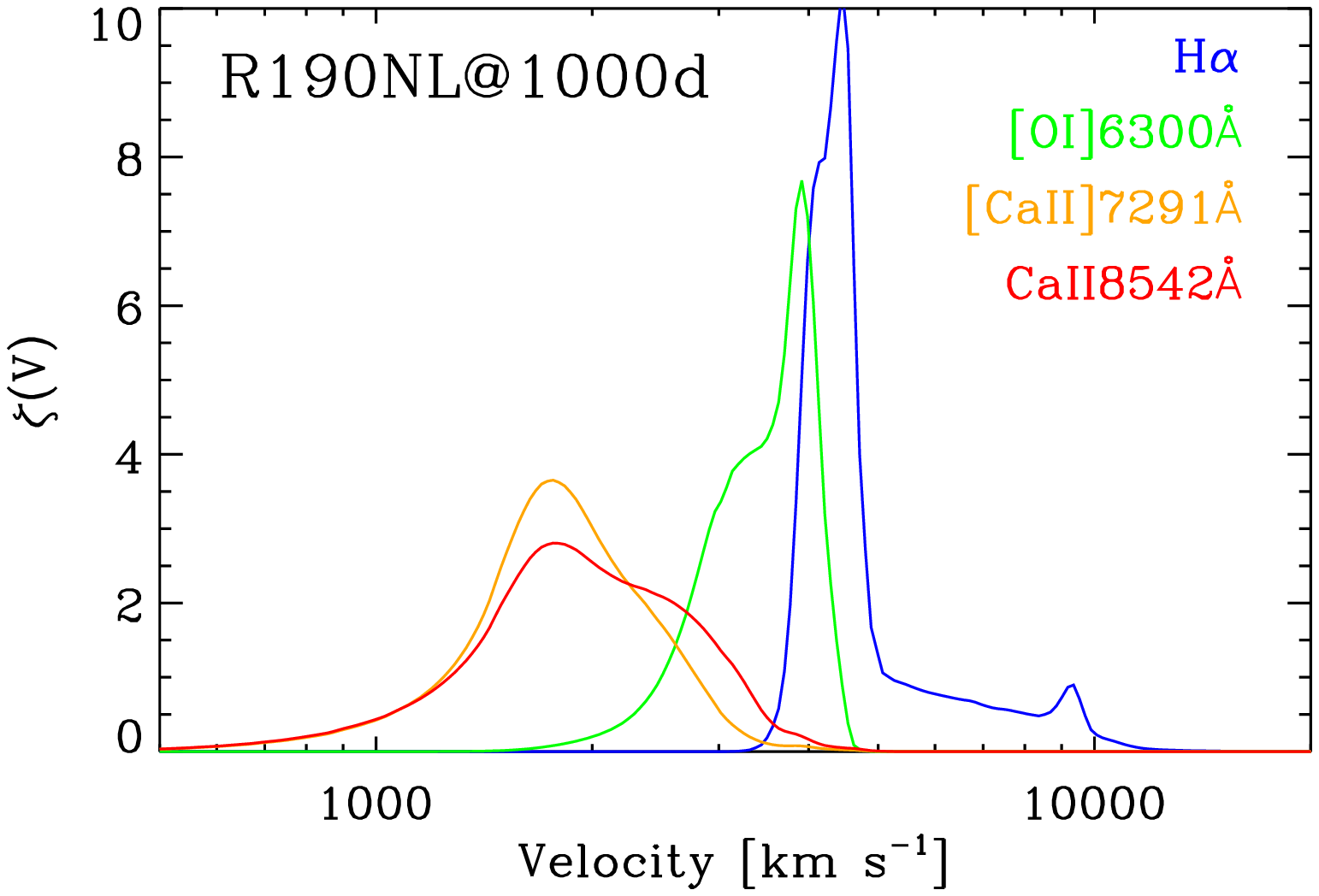,width=8.75cm}
\caption{Comparison of the formation regions for H$\alpha$, [O\one]\,6300\,\AA,
[Ca\two]\,7291\,\AA, and Ca\two\,8542\,\AA\ at 500\,d (left) and 1000\,d (right) after explosion
for model R190NL.
\label{fig_comp_ep}
}
\end{figure*}

       In Fig.~\ref{fig_comp_ep}, we compare this segregation of line emission at nebular times in model R190.
The regions of emission for the main lines evolve little with time and reflect directly the chemical stratification,
yielding broad H\one\ lines, intermediate witdh O\one, and narrow Ca\two. This is in contrast to solar-metallicity
CCSNe at nebular times in which a significant fraction of the nebular line flux stems from the envelope
unaffected by any nuclear burning, and thus at the primordial composition \citep{DH11,maguire_etal_12}.
Similarly, in nebular-phase spectra of Type II CCSNe, H$\alpha$ generally shows a narrow profile and requires mixing into the core \citep{utrobin_07,li_etal_12}, while here the H$\alpha$ profile stems entirely from the outer H-rich shell.
Mixing in those PISNe, which is expected to be inefficient (\citealt{joggerst_whalen_11}; but see
also \citealt{chen_etal_12}), should not alter this result.

\section{Comparison to observations of PISN candidates}
\label{sect_comp_obs}

\subsection{SN 2007\lowercase{bi}}
\label{sect_07bi}

Given the numerous uncertainties affecting the modeling of PISNe, our goal is to identify critical
   signatures from our simulations that will allow  PISN candidates  (e.g., SN 2007bi) to be confirmed
   or dismissed as a PISN. An important result of the present study has already been presented in the
   broader context of superluminous SNe, where we argue that PISN explosions
   have probably not yet been discovered \citep{dessart_etal_12d}.

\begin{figure}
\epsfig{file=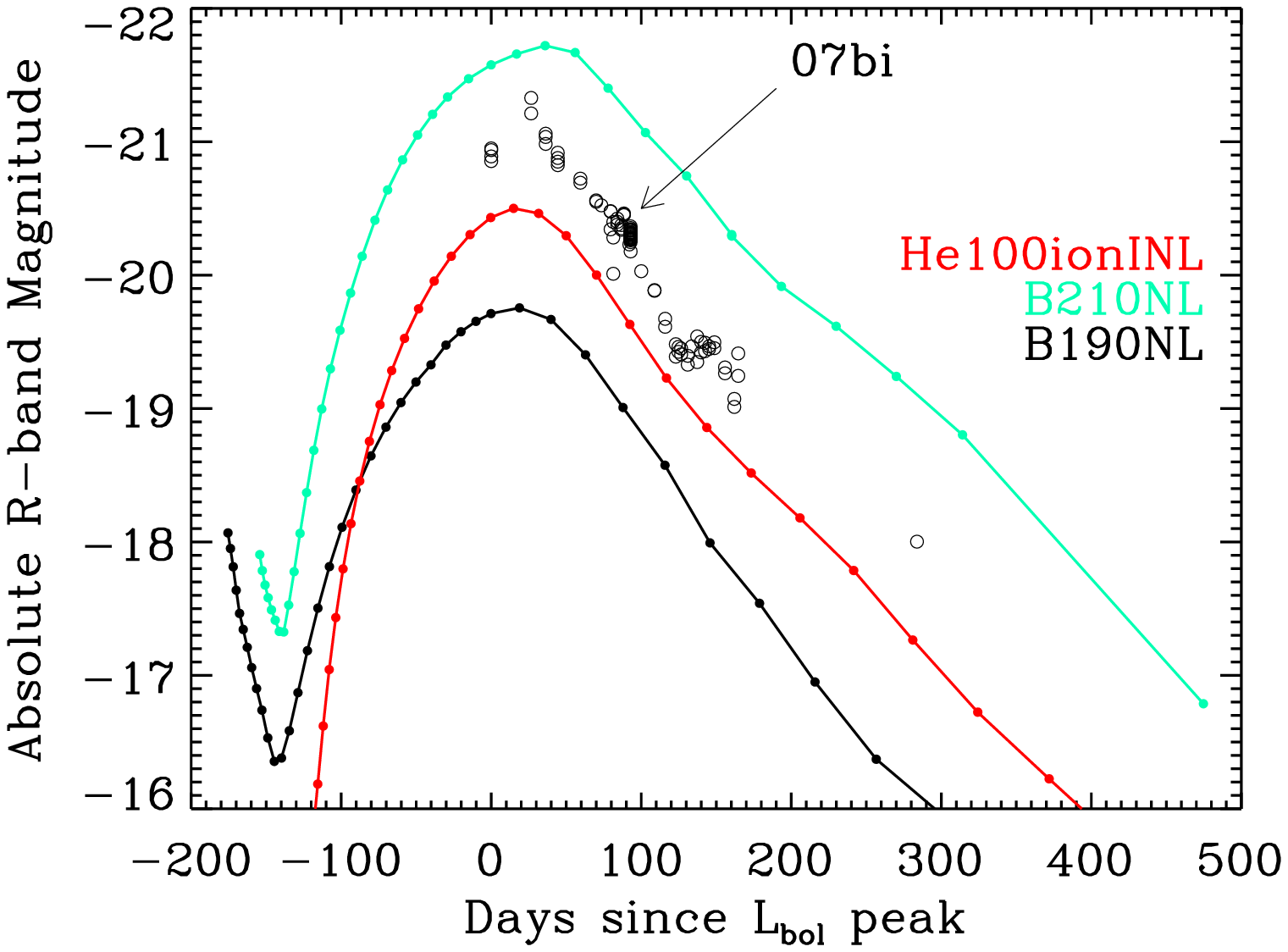,width=9.75cm}
\epsfig{file=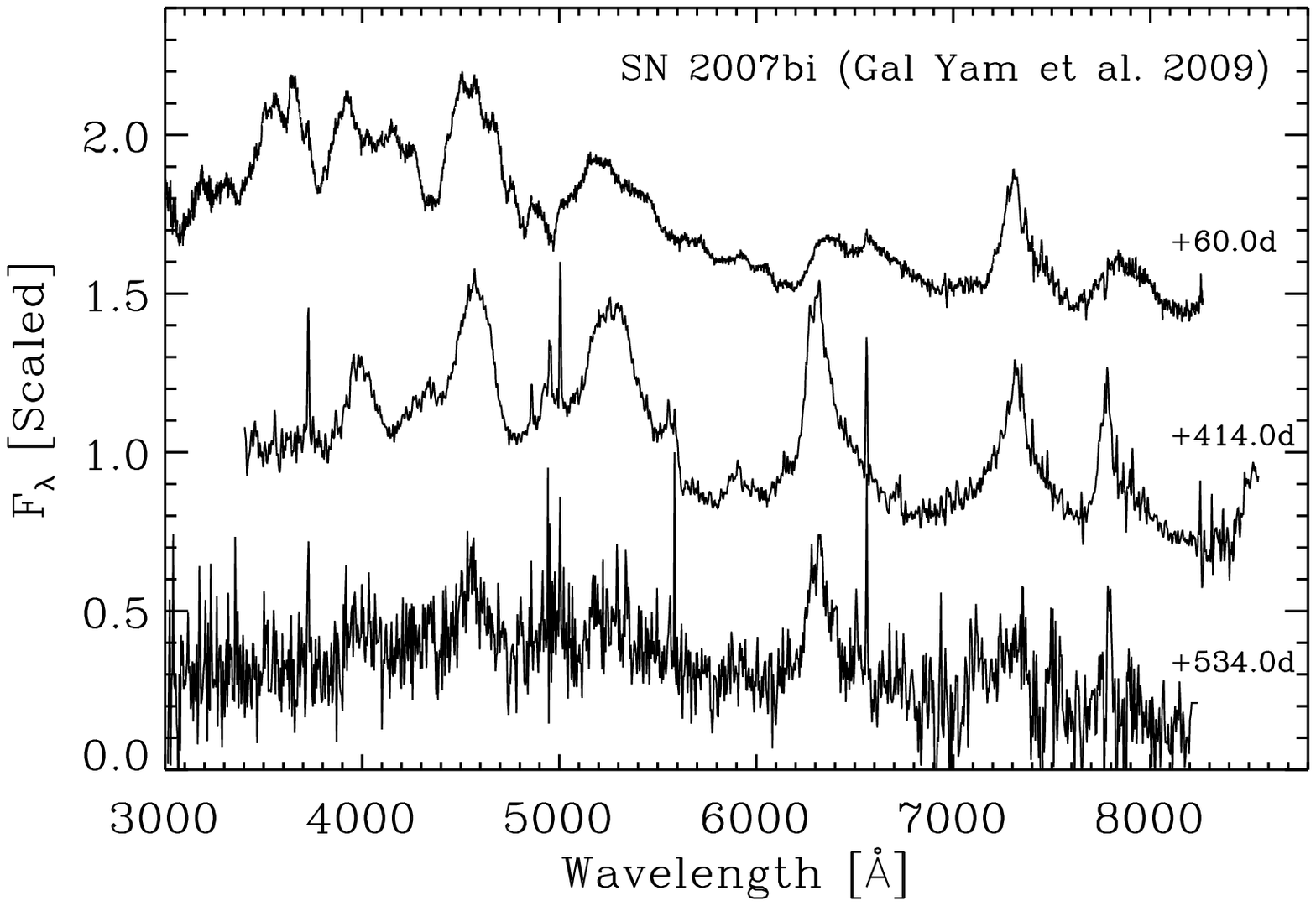,width=8.75cm}
\epsfig{file=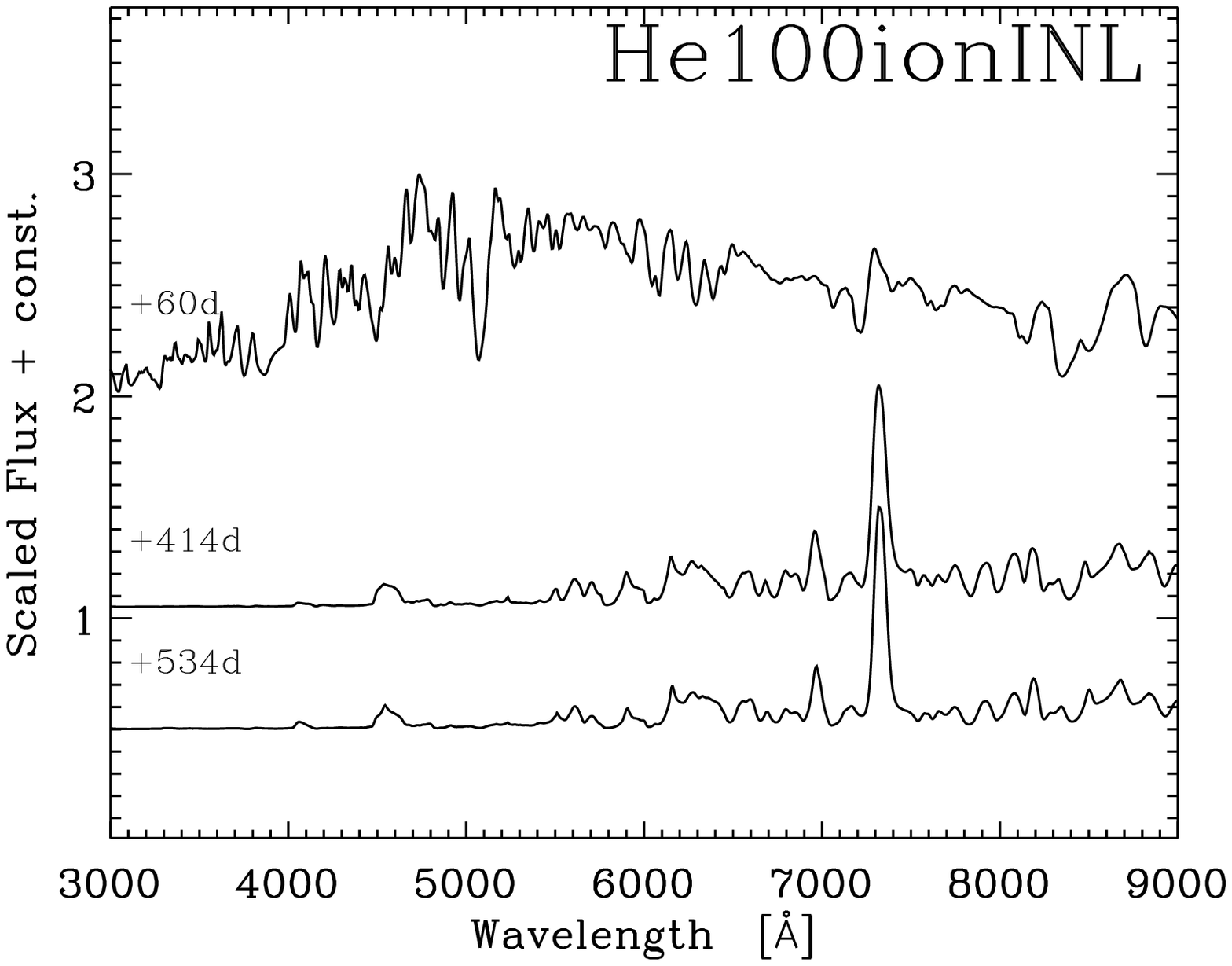,width=8.75cm}
\caption{
{\it Top:} R-band light curves for PISN models B190, B210, and He100, together with the observed R-band
light curve of SN\,2007bi (adjusted horizontally for convenience).  The abscissa is the time since
bolometric maximum --- it differs by merely a few days from the time of R-band
maximum used below. No correction for reddening is applied.  {\it Middle:} Montage of observed spectra of
SN\,2007bi \citep{galyam_etal_09}, stacked arbitrarily for visibility. Labels refer to the time since
R-band maximum. {\it Bottom: } Montage of synthetic spectra for model He100ionINL
at contemporaneous epochs.
\label{fig_07bi}
}
\end{figure}

   The observed light curve of SN 2007bi reveals a peak at $M_R\sim-$\,21.3\,mag and a slowly
   decreasing nebular flux compatible with full \gray\ trapping \citep{galyam_etal_09}. Our model
   He100 (or He100ionI; allowance for non-local energy deposition does not influence the results for
   this comparison) matches closely these light-curve properties, although our model is fainter
   by $\sim$\,0.8\,mag in the R band (Fig.~\ref{fig_07bi}). As discussed in \citet{dessart_etal_12d},
   the models He110 or He115 (corresponding to similar models to He100 but from 110 and 115\,\msun\ Helium
   cores) match better the luminosity of SN\,2007bi at and beyond the peak. However,
   their spectral properties are comparable to those of He100. Overall, the explosion and light-curve
   properties that we infer for model He100 are in agreement with the 100\,\msun\  Helium-core
   models computed by other groups \citep{galyam_etal_09,kasen_etal_11}.

   Matching the light curve is a necessary step to validate a model, but it is not a sufficient test.
   Light-curve degeneracy is epitomized by the good quality fit obtained by \citet{kasen_bildsten_10} and
   \citet{kasen_etal_11} using either a 20\,\msun\ ejecta powered by a magnetar or a 100\,\msun\ PISN
   powered  by 5\,\msun\ of \isoni. Modelling multi-epoch spectra is the critical next step to validate
   a proposed model.

   For a RSG progenitor, the high-brightness phase of the PISN is during the plateau, when the spectra
   are crammed with H\one\ lines. SN 2007bi is a Type Ic SN so it is unlikely  to result from the explosion
   of a RSG star, even one endowed with few solar masses of \isoni\ like our model R190. One may conceive
   of a contrived situation in which the SN 2007bi detection caught the \isoni\ bump that follows the plateau,
   at which time the H\one\ lines may be weak, but why would the SN not have been detected during
   the preceding 200\,d when the SN was even brighter in the optical?

   For the BSG progenitor B190, the light curve shape is in agreement with the observations of
   SN\,2007bi. The presence of hydrogen in the outer ejecta cannot be inferred from spectra at and
   beyond peak in this model (Fig.~\ref{fig_B190}) so it would  be at least in principle compatible with
   the observations of SN 2007bi and its Type Ic classification. Model B210 does show H$\alpha$
   unambiguously at nebular times (Fig.~\ref{fig_B210}),\footnote{This results from the huge \isoni\
   production combined with non-thermal ionization and excitation, as well as non-local \gray-energy deposition}
   but it produces a too high luminosity for SN\,2007bi. In model B190, the presence of an hydrogen
   envelope (not directly inferred from spectra at and beyond peak) leads to a smaller expansion rate of
   the He-core than in the He100 model. The broad line features seen in the optical spectra of SN\,2007bi
   suggest fast expansion of the emitting layers and so, in this PISN context, would favor the lowest mass
   progenitor capable of producing 3--5\,\msun\ of \isoni. Hence, our model He100 (He100ionI) seems
   the most suitable of all four for this PISN candidate and we therefore focus on this below.

  The main spectral signatures for model He100ionI are the dominance of absorption and emission
  processes from IMEs in the spectra up to the peak, and a hardening of the spectrum on the rise to
  peak (Fig.~\ref{fig_he100}). After the peak, the spectrum formation region is located within the
  Si/S/Fe-rich layers. The high metal content  together with the cooling of the photosphere, leads to strong blanketing
  in the blue and a severe reddening of the spectrum. This alone is the most fundamental
  disagreement with the observations of SN\,2007bi \citep[Fig~\ref{fig_07bi}; ][]{dessart_etal_12d},
  and in fact of other superluminous SNe which seem to be systematically blue after the
  peak \citep{quimby_etal_11}. Our PISN simulations are in contradiction with this observed property.
  Indeed, it would be surprising for PISN explosions to retain hot photospheres free of blanketing
  few hundred days after explosion.

  On the rise to peak, the photosphere feels the heat wave powered by decay
  and diffusing from greater depth, and consequently cannot feel the blanketing effects of these
  buried IGE; the spectra before peak are bluer. After the peak, the photosphere cools and recedes
  to IGE-rich layers, feeling the full effect of metal-line blanketing. The comparison made
  by \citet{Y10_full} between typical SN Ia spectra and those of SN\,2007bi reveals a striking
  similarity, but this similarity suggests SN 2007bi is not a PISN. In the best fitting model He100 of a
  100\,\msun\ ejecta, the \isoni\ mass is 5.02\,\msun, thus 5\% of the total mass. This is closer to
  the value of $\sim$\,1\% characterizing SNe II-P than the value of $\sim$\,50\% characterizing SNe Ia.
  As discussed in \citet{dessart_etal_12d}, models with increasing \isoni-to-ejecta-mass ratios have bluer colors
  but even in the most extreme case of PISN model He125, this ratio is 0.26 and the colors are still too red to
  match those of SN\,2007bi --- this He125 model is also much too bright with its \isoni\ mass of 32\,\msun.
  In the models He100---He115, which match quite well the SN\,2007bi light curve, the amount of energy
  released per unit mass is not so favorable to produce a hot emitting ejecta at and beyond the peak of
  the light curve, as suggested by SN 2007bi. And indeed, our nebular-phase synthetic spectra are
  systematically cool and suffer severe line-blanketing dwarfing the flux short ward of $\sim$\,5000\,\AA.
  This color mismatch can be inferred by comparing the model He100 color $B-R$ with the observations
  of SN\,2007bi. In the model, it is $\sim$\,0.55 at the light curve peak and subsequently, it steadily
  rises to reach 2.5--3 after 200\,d (Fig.~\ref{fig_he100_color}). In contrast, the observations of SN\,2007bi
  show a $B-R$ of $\sim$\,0.23\,mag (0.9\,mag) 54\,d (150\,d) after peak \citep{Y10_full}, hence a much bluer colour.

  A further problem with the similarity between SNe Ia spectra and SN 2007bi illustrated by \citet{Y10_full}
is that in a PISN explosion, the \isoni\ is produced in the hottest and densest regions of the ejecta.
In model He100, these regions are ejected at a speed $\lesssim$\,4000\,\kms, while in SNe Ia these layers
reach up to 15000\,\kms. This strong contrast in expansion rate for the \isoco/\isofe-rich layers
conflicts with this spectral similarity: metal lines in SNe Ia are broad while those in PISNe are narrow.
It is also illustrated by the inconsistency of the PISN model
presented in \citet{galyam_etal_09}: inferring a mass-weighted mean velocity of 8000\,\kms\ and an ejecta
mass of 100\,\msun, the explosion has to deliver $\sim$\,100\,B (note that
$\sim$\,10\,B is needed to just unbind such a super-massive compact star), which is then
incompatible with the inferred 3--10\,\msun\ of \isoni\ synthesized in the explosion. For comparison,
the extreme model B210 presented here releases ``only'' 75\,B through burning, producing an
ejecta with a kinetic energy of 65\,B.

   \citet{kasen_etal_11} have compared one synthetic spectrum for their He100 model at 50\,d before peak
   with the spectrum of SN 2007bi at 54\,d after peak. They do reproduce the blue spectrum but this stems from
   the fact that they have a 100-d mismatch. At 50\,d after explosion, the color is no longer blue and the spectra
   are strongly line blanketed, in contradiction with the observations.

   It thus seems unlikely that SN\,2007bi is a PISN explosion. In addition with the issues raised by star formation
   \citep{hosokawa_etal_11} and stellar evolution \citep{langer_etal_07} for such super-massive star
   progenitors,\footnote{On possible way to overcome the evolution argument is through mergers in a dense
   star cluster, as recently discussed by \cite{pan_etal_12b}.} we have provided independent arguments that
   suggest that even if  PISN ejecta existed, their radiative properties would conflict in numerous ways
   with the observations of SN\,2007bi. As we propose in \citet{dessart_etal_12d}, we instead favor a magnetar
   scenario for SN\,2007bi, as well as for superluminous SNe characterized by blue colors.

\subsection{SN\,2006gy}
\label{sect_06gy}

The spectral evolution of our H-rich PISN models (i.e., R190, B190, B210) can also be compared to that
of PISN candidate SN\,2006gy, whose superluminous display was associated with \isoni-power or CSM interaction
\citep{ofek_etal_07,smith_etal_07}. This superluminous Type IIn SN reveals a spectrum dominated noticeably by
H\one\ Balmer and Fe\two\ lines with narrow and broad velocity components \citep{smith_etal_10}.
Its spectral morphology suggests a slow cooling of the emitting region, with little signs of line blanketing,
and, surprisingly, no emission of the usual [Ca\two] and [O\one] nebular lines seen in massive-star explosions.

  The \isoni-power vs. CSM-interaction models can be discussed further in light of our simulations.
  The explosion of a RSG progenitor is directly rejected from the bell-shape light-curve morphology of
SN\,2006gy, while the presence of hydrogen would require a blue or a yellow supergiant
(in agreement with the expectations from stellar evolutionary calculations; \citealt{langer_etal_07}).
However, our simulations, wherein the luminosity is powered by radioactive decay primarily,
indicate that by the time the SN reaches its peak brightness, the photosphere
is in the Helium core and radiates a spectrum that contains little sign for the presence of hydrogen (models
B190 and B210). The bulk of that energy arises from the core and the emission is channeled for a sizable
part into those fine structure lines, which should thus be observable.
The fact that SN\,2006gy retained a Type II spectral morphology for hundreds of days with no signs
of blanketing short ward of 5000\,\AA\ and no forbidden line emission from [Ca\two] and [O\one]
invalidates the \isoni-power model for SN\,2006gy.


   \section{PISN as metallicity indicators in the Universe}
\label{sect_z_indic}

   Being so luminous, PISN explosions represent attractive probes of the young Universe where
   the first stars formed. In particular, those that die as RSG stars would sustain a high luminosity in the optical
from
   early after breakout until the transition to the nebular phase about a year later (in the SN frame).
   As discussed earlier (Sect.~\ref{sect_spec_RSG}), the representative model R190 has a nearly
   pure H\one\ spectrum (with small contributions from He\one) with no sign of line blanketing from either
   IMEs or IGEs for $\gtrsim$\,250\,d after explosion. During that extended time, which may last a few years
   for an observer on earth depending on the redshift,
   spectroscopic observations could reveal the environmental metallicity out of which the PISN progenitor formed.
   At large redshift, this determination is generally done from analysis of nebular lines
   \citep{osterbrock_89}, but doing this task using the SN spectrum offers a very interesting alternative.
   In the present case reported, the metallicity is so low that no metal lines are seen at early times, which would
   be an unambiguous signature that the metallicity is extremely low. For higher metal abundances, one expects
   a gradual rise in associated line strengths \citep{kasen_etal_11}.

   As we discussed in detail, in particular for model He100ionI, on the rise to peak the photosphere probes the
   He-rich and O-rich envelope above the IGE-rich inner ejecta and thus still reflects the primordial abundances
   for IGEs. After the peak, this information is all lost since we see emission/absorption from the ashes
   produced by steady and explosive burning. This still leaves an extended window for determining iron
   abundances, for example, for the primordial gas out of which the star formed.

\begin{figure}
\epsfig{file=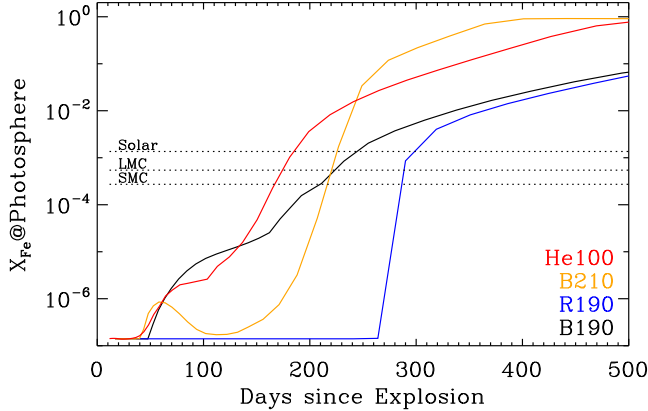,width=9.3cm}
\caption{Illustration of the iron mass fraction at the electron-scattering photosphere versus
time since explosion for each model. Two competing effects cause this quantity to generally
increase with time, i.e., the photosphere recession to IGE-rich layers as time proceeds
and the decay with time of \isoni\ and \isoco\ isotopes. We draw a dotted horizontal line
to represent the iron mass fraction in the SMC, LMC, and in the solar neighborhood.
More generally, RSG-star explosions constitute an attractive probe to constrain the metallicity,
especially those occurring in the early Universe.
\label{fig_z}
}
\end{figure}

   In Fig.~\ref{fig_z}, we show how the iron mass fraction varies at the photosphere for all models, with horizontal
   lines indicating the levels corresponding to the SMC, the LMC, and the solar iron mass fraction.
   Over a few hundred days, the iron mass fraction varies in all cases over 7 orders of magnitude, i.e. from
   the iron-deficient outer ejecta to the pure iron layers of the inner ejecta. Note however that the primordial
   metallicity is preserved at the photosphere of model R190 for $\sim$\,200\,d (because the thermalization
   region is much deeper than the electron-scattering photosphere, metals influence the emergent
   spectrum of model R190 before 250\,d; see Fig.~\ref{fig_R190}).

\section{Conclusion}
\label{sect_disc}

   In this paper, we have explored the radiative properties of PISNe arising from
the explosion of RSG, BSG, and WR star progenitors at a very low metallicity of 10$^{-4}$\,\zsun\
and without rotation.
Our approach is based on physically-consistent modeling of the evolution from the main sequence until the
onset of the pair-production instability using the stellar evolution code \mesa,
the explosion phase and the evolution of the ejecta with
the radiation-hydrodynamics code \v1d\ (with allowance for explosive burning), and the non-LTE
time-dependent radiative-transfer modeling with \cmfgen\ for the multi-band light-curves and
spectra from the photospheric to the nebular phase. In this paper, we have focused on this last step,
while the former two steps will be described in Waldman et al. (in prep.).  In practice, we focus
on a 190\,\msun\ RSG star (model R190), a 190 and 210\,\msun\ BSG stars (models B190 and B210),
and a 100\,\msun\ WR star (model He100).

   The exceptional progenitor and explosion properties of the PISNe we modeled set them apart from
the SNe we routinely observe in the local Universe. With ejecta masses of $\gtrsim$\,100\,\msun,
ejecta kinetic energies of few tens of B, \isoni\ yields of a few \msun, their luminosity peaks at
$\sim$\,10$^{10}$\,\lsun\ about 150-200\,d after explosion,
remaining above $\sim$\,10$^{9}$\,\lsun\ for $\gtrsim$\,100\,d. Their large
ejecta mass ensures full \gray-trapping so that the \isoco\ decay rate gives the distinctive
$\sim$\,0.01\,mag\,d$^{-1}$ nebular decline rate up to $\sim$\,500\,d after explosion
($\sim$\,300\,d after bolometric maximum). Over the next 500\,d, the fading rate slowly steepens in
the lower ejecta-mass model He100 but stays the same in the more massive ejecta models.

   The ingredients ruling the evolution of ``standard'' SNe in the local Universe apply to PISNe
directly. The influence of the progenitor radius is key in yielding the high-luminosity plateau
of the RSG model R190, while the more compact BSG and WR progenitors are characterized
by a single highly luminous \isoni-powered peak. Furthermore, the large progenitor radius of
model R190 yields a hot ionized photosphere for weeks after explosion. In the BSG model B190,
recombination turns on much more quickly, as in SN\,1987A. In the helium-rich model He100,
the re-brightening phase to the bolometric peak corresponds to a photospheric residence in
the O-rich core, with inefficient non-thermal excitation of the outer helium. Hence, our model
R190 would be classified unambiguously as a SN II-P. Models B190/B210, despite having a sizable
amount of H and He, would be Type IIc if observed early on \citep{dessart_etal_12} but most
probably a Ic if discovered at peak or beyond. Finally, model He100 would definitely be a Type Ic
despite the $\sim$\,8\,\msun\ of helium in its outer ejecta.

   Despite these significant spectral and light-curve differences up to the bolometric peak,
   all models reveal very similar light-curve and spectral properties beyond it.
   This naturally stems from the comparable core
properties of the progenitor stars --- here, differences in cumulative mass for O or \isoni\
change little the temperature or ionization controlling the radiative properties from the core
and conspire to produce this degeneracy.

   Two fundamental properties emerge from
PISN simulations (see also \citealt{kasen_etal_11}).
First, homologous expansion in such massive ejecta implies that while the
outer layers may be traveling fast, as witnessed early on from the width of the Balmer lines in
the SN II-P model R190, the inner ejecta travels at relatively small speeds and should therefore
lead to rather modest nebular line widths. The \isoni\ rich core travels at $\lesssim$\,2000\,\kms\
in model B190/R190 and at $\lesssim$\,4000\,\kms\ in models B210/He100. Hence, although these
objects are thermonuclear explosions, their chemical stratification is closer to that of core-collapse SNe
than SNe Ia. In practice, we find all our PISN simulations to follow this principle and reveal
nebular lines with a narrow width, in particular arising from the O-rich and Ca-rich shell
on top of the \isoni-rich core.

  The second, perhaps counterintuitive, property of these PISNe is the relatively modest ejecta and
photospheric temperatures of $\lesssim$\,5000\,K at and beyond the bolometric maximum.
What controls the ejecta ionization and temperature is not just the mass of \isoni, which is
admittedly large, but instead the ratio of \isoni\ mass to ejecta mass, which is typically quite
small in these PISNe, again more typical of core-collapse SNe than SNe Ia.
Indeed, the large ejecta mass means that the energy released is to be shared. This is an inevitable
property of PISNe since pair-production can only occur in stars with super-massive He/CO cores.
Indirectly, this increases the diffusion time of the ejecta, which becomes transparent late,
at $\lesssim$\,200\,d in this set. By then, the ejecta has expanded to 10$^{16}$\,cm, corresponding
to an increase in radius of 10$^5$ for the core material at that time. At such late times, \isoco\ decay
dominates, with a weaker heating rate compared to \isoni.
Consequently, the PISN spectra at and beyond bolometric maximum are red, i.e., little flux is
emitted shortward of $\sim$\,5000\,\AA,  and strongly blanketed by Fe\two\ and eventually Fe\one.
This is in stark contrast with any known superluminous SN today.
As we emphasize in \citet{dessart_etal_12d}, these properties are at odds with the
observations of the PISN candidate SN\,2007bi \citep{galyam_etal_09}.
An attractive alternative, which needs further study, is the magnetar model whereby energy is
added to a SN ejecta but without the compromising
blanketing effects of metals.

The current status of PISNe today is rather grim.
Recent numerical simulations indicate that super-massive stars are hard to form \citep{hosokawa_etal_11}.
If they did form, we do not currently expect them to explode as Type Ic PISNe unless at very low Z \citep{langer_etal_07}.
If they did form and exploded at \zsun/3 as a Type Ic SN, we do not expect them to resemble SN\,2007bi at all,
as demonstrated here, but rather display red colors and lines with a modest width at all times.

A more attractive scenario for producing blue weakly-blanketed broad-lined superluminous SNe
is by the magnetar scenario, or any similar setup in which a large energy is fed into the ejecta once
it has expanded to a SN size \citep{dessart_etal_12d}. Two effects suggest that this scenario
is more amenable to explain these events. First, the formation of a magnetar involves the fast rotation
of the progenitor Fe core prior to collapse \citep{hirschi_etal_04,WH06,georgy_etal_09},
which can be produced in a wide
range of massive stars at LMC-like metallicities or less. We expect a diversity both in the magnetar
properties and in the associated SN ejecta. Importantly, a large magnetar radiation is function of the
magnetar properties and can therefore occur in combination with low- or moderate-mass ejecta. In contrast,
a large \isoni\ mass in a  PISN requires by essence a larger mass progenitor. The snow-plow effect
from magnetar energy injection can thus easily alter the dynamics of the inner ejecta and lead to
the formation of broad lines at nebular times \citep{kasen_bildsten_10,woosley_10,dessart_etal_12d}.
Mistaking such a magnetar-powered ejecta for a PISN (in homologous expansion and showing a
\isoni-powered LC) would lead to a poor estimate of the ejecta mass and kinetic energy.
Such inconsistencies have arisen for SN\,2005bf and SN\,2007bi if one assumes a super-massive
\isoni-powered SN \citep{folatelli_etal_06,galyam_etal_09,kasen_etal_11} rather than a magnetar-powered SN
\citep{maeda_etal_07,kasen_bildsten_10}. What we demonstrate in this work is that distinguishing
the two models should be done through detailed spectroscopic modeling.

Our simulations give the fundamental properties we expect for PISNe in the early
Universe and stemming from non-rotating progenitor stars.
Proposed PISN candidates have all been observed in the nearby Universe,
at metallicities close to the solar value. At smaller metallicities, the evolution of the progenitor
massive star will be altered, in ways that depend considerably on the mass loss
rate. Exploratory calculations by \citet{langer_etal_07} suggest that H-rich PISNe are
possible at metallicities as high as \zsun/3, but that H-deficient (Type Ic) PISNe may
be limited to lower-metallicity environments. Besides the impact on the progenitor
properties (and its propensity to lead to core collapse or pair production), variations in the
metallicity will tend to modulate the effects of line blanketing. Based on our results, we
anticipate this will play a role at early times (plateau phase in SNe II-P, pre-peak
phase in SNe II-pec and Ib/c), before the photosphere recedes to the inner ejecta
rich in explosively-synthesized IGEs.

A potential caveat of our simulations, to be remedied in the future, is the neglect of stellar rotation
for the pre-SN evolution.
The PISN channel for rotating massive Pop\,{\sc iiii} stars has been recently studied by \citet{CW12}
and \citet{yoon_etal_12}. By increasing the stellar core mass, fast rotation can lower the minimum
main-sequence mass needed to encounter the pair-production instability. Neglecting mass loss entirely,
\citet{CW12} suggest a lower mass limit of $\sim$\,65\,\msun; allowing for radiation-driven mass loss
from optically-thick lines of CNO elements, \citet{yoon_etal_12} give a lower main-sequence
mass limit of $\sim$\,90\,\msun, corresponding to a final mass of $\sim$\,70\,\msun.
Rotation can thus permit to produce lower mass PISN ejecta, about 30\% less massive
than our He100 model.
Lowering the ejecta mass reduces the rise time to light-curve peak, although this shift
will depend on the explosion energy and \isoni\ mass produced.
This will  diversify the PISN light-curve and spectral properties, introducing a range of peak luminosities
and spectral properties. In \citet{dessart_etal_12d}, we presented results for PISNe from He cores
between 100 and 125\,\msun\ (evolved without mass loss). These covered a huge range of \isoni\ mass
between 5.0 and 32.4\,\msun, ejecta kinetic energy between 37.6 and 74.2\,B. More energetic models
produced higher peak luminosities with broader spectral lines at any given epoch. However, as
emphasized in \citet{dessart_etal_12d},  the ejecta richer in \isoni\ have larger temperatures but suffer
stronger blanketing from IGEs so that the range of colors is limited and the resulting PISN spectra
are rather red. To conclude, the variations of a few tens of percent in ejecta mass permitted by the inclusion
of progenitor rotation will likely have only a modest impact on PISN radiative properties.

Ongoing and future deep systematic surveys of the sky may eventually detect unambiguously
the explosion of the first stars. Given their longer high-brightness light curve and bluer colours,
the explosion of RSG stars should be prime targets. Their extended and massive hydrogen-envelope would
also allow the inference of the environmental metallicity in which they form. Furthermore, such
RSG-star explosions could be used to determine distances out to large redshifts using the
Expanding Photosphere Method \citep{KK74_EPM,E96,DH05_epm}.

\section*{Acknowledgments}

LD and SB acknowledge financial support from the European Community through an
International Re-integration Grant, under grant number PIRG04-GA-2008-239184,
and from ``Agence Nationale de la Recherche" grant ANR-2011-Blanc-SIMI-5-6-007-01.
DJH acknowledges support from STScI theory grants HST-AR-11756.01.A and HST-AR-12640.01,
and NASA theory grant NNX10AC80G.
This work was granted access to the HPC resources of CINES under the allocation 2011--c2011046608 and
c2012046608 made by GENCI (Grand Equipement National de Calcul Intensif).
A subset of the computations were also performed at Caltech's Center for Advanced Computing Research
on the cluster Zwicky funded through NSF grant no. PHY-0960291 and the Sherman Fairchild Foundation.

\appendix

\section{Ejecta properties}
\label{sect_ejecta_prop}

   In this section, we present more quantitatively the photospheric and ejecta properties.
   We give tabulated values (Tables~\ref{tab_phot_B190}---\ref{tab_phot_He100}) corresponding to
   Fig.~5 discussed in Section~4. Specifically, we give
   the time evolution for the radius, velocity, temperature, and overlying mass
   at the electron-scattering photosphere, together with the base ejecta electron-scattering optical depth.

   Concerning the latter, we note that ejecta recombination induces a decline of the optical
   depth that is steeper than the $1/t^2$ scaling corresponding to homologous expansion.
   Nonetheless,  the electron-scattering optical depth remains $\gtrsim$\,1 for about 600\,d
   in all four simulations. This is a rather unique property of PISN explosions, inherent to the huge
   progenitor/ejecta masses involved. This property may be misleading because by the time the ejecta base is
optically thin,
   the outer ejecta has a much lower density and may thus be in a nebular regime.
   This hybrid configuration is reflected spectroscopically
   by the concomitant presence of P-Cygni profiles (associated with transitions that remain optically
   thick for years) and forbidden emission lines (such as the [Ca\two]\,7300\,\AA\ doublet).
   There is thus continued interaction between the radiation and the gas, but the depletion of the radiation field at
short wavelength
   (in the bound-free continua of H/He/CNO/IMEs) is very weak and thus unable to cause much photo-ionization.
   At late times, \grays\ travel some distance before being absorbed, inducing non-local energy deposition.
   The corresponding thermal energy, and the associated non-thermal excitation and ionization, may cause
   the ionization to rise again.

   We note that the huge energy release from decay (and the diffusion of that heat)
   prevent the formation of a sharp drop in optical depth at any time, something that is most dramatically seen
   in ``standard" SNe II-P simulations at the end of the plateau phase \citep{DH11}. In that case, the transition
   to this nebular regime often coincides with a sharp rise in polarization \citep{leonard_etal_06,DH11b}.

\setcounter{footnote}{9}

\begin{table}
\centering
\caption{Evolution of important quantities at the electron-scattering photosphere for model B190.
By definition, these are limited to so-called photospheric epochs. They are also primarily illustrative
since electron scattering becomes irrelevant as metal-line opacity increase and eventually dominate
at late times.
\label{tab_phot_B190}
}
\begin{tabular}{l@{\hspace{1.6mm}}c@{\hspace{1.6mm}}c@{\hspace{1.6mm}}c@{\hspace{1.6mm}}
c@{\hspace{1.6mm}}
c@{\hspace{1.6mm}}}
\multicolumn{6}{c}{B190} \\
\hline
 Age       & $\tau_{\rm base,es}$ &    $R_{\rm phot}$      &   $V_{\rm phot}$   &   $T_{\rm phot}$  &   $\Delta
M_{\rm phot}$   \\
 \hline
 [d]      &                    &     [$10^{15}$\,cm]   &  [\kms]  &   [K]  & [\msun] \\
\hline
        16.7 &      8456.35 &         3.06 &        21224 &         5589 &        0.313 \\
        18.3 &      6805.03 &         3.18 &        20058 &         5318 &        0.412 \\
        20.2 &      5436.70 &         3.25 &        18627 &         5101 &        0.563 \\
        22.2 &      4317.84 &         3.23 &        16870 &         4863 &        0.801 \\
        24.4 &      3436.37 &         3.14 &        14874 &         4680 &        1.152 \\
        26.9 &      2754.18 &         2.97 &        12821 &         4549 &        1.650 \\
        29.5 &      2226.68 &         2.80 &        10954 &         4456 &        2.485 \\
        32.5 &      1810.37 &         2.66 &         9462 &         4411 &        7.520 \\
        35.7 &      1471.41 &         2.51 &         8143 &         4446 &       13.200 \\
        39.3 &      1180.57 &         2.34 &         6876 &         4453 &       18.964 \\
        43.2 &       940.03 &         2.03 &         5432 &         4308 &       28.061 \\
        47.6 &       750.98 &         1.78 &         4325 &         4225 &       44.386 \\
        52.3 &       605.75 &         1.86 &         4104 &         4250 &       50.383 \\
        57.6 &       493.08 &         1.98 &         3978 &         4376 &       54.026 \\
        63.3 &       403.21 &         2.12 &         3878 &         4557 &       56.901 \\
        69.7 &       329.64 &         2.28 &         3795 &         4752 &       59.300 \\
        76.6 &       268.23 &         2.47 &         3727 &         4946 &       61.280 \\
        84.3 &       219.36 &         2.68 &         3675 &         5128 &       62.839 \\
        92.7 &       179.38 &         2.91 &         3637 &         5294 &       63.989 \\
       102.0 &       146.02 &         3.18 &         3607 &         5447 &       64.867 \\
       112.0 &       118.34 &         3.46 &         3579 &         5583 &       65.699 \\
       122.0 &        96.86 &         3.75 &         3554 &         5688 &       66.461 \\
       132.0 &        80.21 &         4.02 &         3522 &         5769 &       67.447 \\
       142.0 &        66.94 &         4.27 &         3479 &         5830 &       68.769 \\
       152.0 &        56.06 &         4.50 &         3424 &         5871 &       70.493 \\
       162.0 &        46.49 &         4.59 &         3276 &         5855 &       75.231 \\
       172.0 &        38.57 &         4.70 &         3163 &         5866 &       78.905 \\
       182.0 &        31.97 &         4.81 &         3057 &         5869 &       82.450 \\
       192.0 &        26.73 &         4.89 &         2949 &         5850 &       86.025 \\
       211.0 &        19.27 &         5.01 &         2750 &         5735 &       92.428 \\
       232.0 &        13.32 &         5.18 &         2584 &         5432 &       97.182 \\
       255.0 &         9.19 &         5.43 &         2465 &         4938 &      100.306 \\
       280.0 &         6.41 &         5.71 &         2359 &         4552 &      102.933 \\
       308.0 &         4.52 &         5.98 &         2247 &         4277 &      105.600 \\
       338.0 &         3.43 &         6.24 &         2135 &         4029 &      108.257 \\
       371.0 &         2.82 &         6.47 &         2017 &         3677 &      110.939 \\
       408.0 &         2.31 &         6.66 &         1890 &         3225 &      113.995 \\
       448.8 &         1.88 &         6.78 &         1748 &         3037 &      117.161 \\
       493.7 &         1.51 &         6.78 &         1590 &         2934 &      120.450 \\
       543.0 &         1.19 &         6.84 &         1458 &         2835 &      123.780 \\
       597.0 &         0.96 &         6.80 &         1318 &         2780 &      127.295 \\
       657.0 &         0.78 &         5.87 &         1034 &         2891 &      130.894 \\
\hline
\end{tabular}
\end{table}

\begin{table}
\centering
\caption{Same as Table~\ref{tab_phot_B190}, but now showing the photospheric properties for model R190.
Note that in the radiative-transfer model R190 the total ejecta mass is 7\% more massive than the
original non-homologously expanding \v1d\ input at 30\,d (see Footnote~2 for details).
\label{tab_phot_R190}
}
\begin{tabular}{l@{\hspace{1.6mm}}c@{\hspace{1.6mm}}c@{\hspace{1.6mm}}c@{\hspace{1.6mm}}
c@{\hspace{1.6mm}}
c@{\hspace{1.6mm}}}
\multicolumn{6}{c}{R190} \\
\hline
 Age       & $\tau_{\rm base,es}$ &    $R_{\rm phot}$      &   $V_{\rm phot}$   &   $T_{\rm phot}$  &   $\Delta
M_{\rm phot}$   \\
 \hline
 [d]      &                    &     [$10^{15}$\,cm]   &  [\kms]  &   [K]  & [\msun] \\
\hline
        36.7 &      2568.10 &         4.35 &        13698 &        12742 &        0.348 \\
        40.4 &      2087.00 &         4.70 &        13454 &        11137 &        0.409 \\
        44.4 &      1699.22 &         5.07 &        13209 &        10240 &        0.483 \\
        48.9 &      1357.41 &         5.47 &        12960 &         9346 &        0.573 \\
        53.8 &      1060.76 &         5.90 &        12700 &         8522 &        0.688 \\
        59.1 &       827.13 &         6.36 &        12442 &         7851 &        0.829 \\
        65.0 &       661.70 &         6.85 &        12195 &         7340 &        0.994 \\
        71.6 &       541.43 &         7.40 &        11966 &         6945 &        1.177 \\
        78.7 &       438.99 &         8.00 &        11755 &         6563 &        1.383 \\
        86.6 &       348.97 &         8.64 &        11546 &         6149 &        1.627 \\
        95.2 &       276.88 &         9.31 &        11312 &         5742 &        1.957 \\
       104.0 &       217.71 &         9.90 &        11014 &         5507 &        2.488 \\
       114.0 &       164.08 &        10.44 &        10597 &         5359 &        3.469 \\
       124.0 &       127.94 &        10.77 &        10054 &         5198 &        5.015 \\
       136.0 &        99.02 &        10.97 &         9336 &         4965 &        9.677 \\
       150.0 &        73.83 &        11.06 &         8530 &         5028 &       16.744 \\
       165.0 &        56.21 &        10.75 &         7543 &         5038 &       21.750 \\
       181.0 &        40.87 &        10.50 &         6714 &         5049 &       26.307 \\
       199.0 &        29.22 &        10.29 &         5982 &         4867 &       29.889 \\
       218.9 &        19.41 &        10.06 &         5316 &         4466 &       33.147 \\
       240.8 &        11.69 &         9.69 &         4655 &         4139 &       38.840 \\
       264.0 &         7.30 &         6.96 &         3052 &         4610 &      130.108 \\
       290.0 &         4.70 &         6.44 &         2569 &         4416 &      144.765 \\
       319.0 &         3.31 &         6.49 &         2353 &         4131 &      150.075 \\
       350.9 &         2.67 &         6.66 &         2196 &         3734 &      153.387 \\
       386.0 &         2.18 &         6.86 &         2057 &         3395 &      156.586 \\
       424.6 &         1.78 &         7.07 &         1926 &         3097 &      159.586 \\
       467.0 &         1.44 &         7.26 &         1798 &         2956 &      163.258 \\
       513.7 &         1.15 &         7.38 &         1661 &         2891 &      166.955 \\
       565.0 &         0.92 &         7.24 &         1483 &         2897 &      170.969 \\
       621.5 &         0.75 &         5.68 &         1058 &         3060 &      175.209 \\
\hline
\end{tabular}
\end{table}

\begin{table}
\centering
\caption{Same as Table~\ref{tab_phot_B190}, but now showing the photospheric properties for model B210.
\label{tab_phot_B210}
}
\begin{tabular}{l@{\hspace{1.6mm}}c@{\hspace{1.6mm}}c@{\hspace{1.6mm}}c@{\hspace{1.6mm}}
c@{\hspace{1.6mm}}
c@{\hspace{1.6mm}}}
\multicolumn{6}{c}{B210} \\
\hline
 Age       & $\tau_{\rm base,es}$ &    $R_{\rm phot}$      &   $V_{\rm phot}$   &   $T_{\rm phot}$  &   $\Delta
M_{\rm phot}$   \\
 \hline
 [d]      &                    &     [$10^{15}$\,cm]   &  [\kms]  &   [K]  & [\msun] \\
\hline
        16.8 &      3806.45 &         3.08 &        21214 &         5453 &        0.491 \\
        18.5 &      3090.62 &         3.12 &        19555 &         5155 &        0.605 \\
        20.3 &      2496.34 &         3.07 &        17476 &         4934 &        0.786 \\
        22.4 &      2009.13 &         2.98 &        15414 &         4741 &        1.362 \\
        24.6 &      1631.11 &         2.93 &        13793 &         4645 &        2.656 \\
        27.1 &      1329.80 &         2.90 &        12380 &         4601 &        5.131 \\
        29.8 &      1085.34 &         2.84 &        11041 &         4582 &        7.947 \\
        32.8 &       876.82 &         2.79 &         9849 &         4565 &       11.126 \\
        36.1 &       707.42 &         2.79 &         8951 &         4550 &       14.555 \\
        39.7 &       572.21 &         2.53 &         7373 &         4704 &       26.563 \\
        43.6 &       461.90 &         2.65 &         7031 &         5102 &       35.364 \\
        48.0 &       373.02 &         2.87 &         6922 &         5421 &       39.163 \\
        52.8 &       302.94 &         3.15 &         6894 &         5699 &       40.280 \\
        58.1 &       246.63 &         3.47 &         6920 &         5935 &       39.177 \\
        63.9 &       199.83 &         3.87 &         7002 &         6117 &       36.066 \\
        70.3 &       160.92 &         4.34 &         7152 &         6245 &       31.235 \\
        77.3 &       129.92 &         4.94 &         7393 &         6360 &       25.803 \\
        85.0 &       105.40 &         5.66 &         7695 &         6526 &       21.755 \\
        93.6 &        85.15 &         6.41 &         7929 &         6687 &       19.797 \\
       102.0 &        69.46 &         7.08 &         8038 &         6776 &       19.035 \\
       112.0 &        55.20 &         7.78 &         8038 &         6823 &       19.055 \\
       122.0 &        44.78 &         8.36 &         7933 &         6820 &       19.792 \\
       132.0 &        36.76 &         8.81 &         7728 &         6715 &       21.504 \\
       142.0 &        30.26 &         9.20 &         7495 &         6589 &       24.283 \\
       156.0 &        22.26 &         9.68 &         7181 &         6387 &       30.617 \\
       171.0 &        17.00 &        10.15 &         6873 &         6242 &       41.190 \\
       188.0 &        13.15 &        10.49 &         6455 &         6264 &       58.370 \\
       207.0 &         9.81 &        10.33 &         5777 &         6502 &       80.717 \\
       227.0 &         6.83 &         9.51 &         4849 &         6785 &      100.010 \\
       249.0 &         4.90 &         9.23 &         4288 &         6405 &      110.797 \\
       273.9 &         3.79 &         9.40 &         3971 &         5895 &      117.728 \\
       301.3 &         2.81 &         9.47 &         3639 &         5578 &      124.344 \\
       331.4 &         1.77 &         8.98 &         3136 &         5502 &      130.862 \\
       364.5 &         1.42 &         7.92 &         2515 &         5454 &      136.697 \\
       401.0 &         1.16 &         6.78 &         1956 &         5294 &      141.143 \\
       441.1 &         0.95 &         5.46 &         1432 &         5099 &      144.326 \\
       485.2 &         0.78 &         3.69 &          881 &         4880 &      146.294 \\
       533.7 &         0.63 &         2.00 &          433 &         4621 &      146.849 \\
       587.1 &         0.51 &         2.20 &          433 &         4320 &      146.833 \\
\hline
\end{tabular}
\end{table}

\begin{table}
\centering
\caption{Same as Table~\ref{tab_phot_B190}, but now showing the photospheric properties for model He100.
\label{tab_phot_He100}
}
\begin{tabular}{l@{\hspace{1.6mm}}c@{\hspace{1.6mm}}c@{\hspace{1.6mm}}c@{\hspace{1.6mm}}
c@{\hspace{1.6mm}}
c@{\hspace{1.6mm}}}
\multicolumn{6}{c}{He100} \\
\hline
 Age       & $\tau_{\rm base,es}$ &    $R_{\rm phot}$      &   $V_{\rm phot}$   &   $T_{\rm phot}$  &   $\Delta
M_{\rm phot}$   \\
 \hline
 [d]      &                    &     [$10^{15}$\,cm]   &  [\kms]  &   [K]  & [\msun] \\
\hline
        11.6 &      6433.32 &         0.90 &         8977 &         5183 &        8.295 \\
        12.7 &      5215.37 &         0.96 &         8746 &         4992 &        9.262 \\
        14.0 &      4236.04 &         1.03 &         8535 &         4926 &       10.345 \\
        15.4 &      3449.59 &         1.10 &         8284 &         4861 &       11.927 \\
        16.9 &      2816.45 &         1.17 &         7979 &         4791 &       14.178 \\
        18.6 &      2300.95 &         1.23 &         7650 &         4742 &       16.979 \\
        20.5 &      1875.86 &         1.29 &         7319 &         4699 &       20.203 \\
        22.5 &      1524.64 &         1.37 &         7046 &         4598 &       23.147 \\
        24.8 &      1237.52 &         1.47 &         6855 &         4435 &       25.342 \\
        27.2 &      1005.06 &         1.58 &         6726 &         4275 &       26.909 \\
        30.0 &       818.25 &         1.71 &         6604 &         4207 &       28.425 \\
        33.0 &       666.34 &         1.85 &         6475 &         4237 &       30.085 \\
        36.3 &       541.71 &         1.99 &         6351 &         4327 &       31.746 \\
        39.9 &       438.67 &         2.14 &         6218 &         4465 &       33.583 \\
        43.9 &       355.25 &         2.31 &         6084 &         4621 &       35.513 \\
        48.3 &       288.98 &         2.48 &         5953 &         4800 &       37.467 \\
        53.1 &       235.24 &         2.68 &         5828 &         4984 &       39.364 \\
        58.4 &       190.65 &         2.90 &         5746 &         5166 &       40.677 \\
        64.3 &       153.97 &         3.16 &         5683 &         5333 &       41.662 \\
        70.7 &       124.85 &         3.47 &         5685 &         5470 &       41.650 \\
        77.8 &       101.46 &         3.82 &         5684 &         5597 &       41.661 \\
        85.6 &        81.86 &         4.22 &         5706 &         5704 &       41.316 \\
        94.1 &        65.21 &         4.64 &         5700 &         5804 &       41.418 \\
       103.5 &        51.21 &         4.95 &         5532 &         5876 &       44.158 \\
       113.0 &        41.30 &         5.28 &         5409 &         5954 &       46.228 \\
       124.3 &        32.20 &         5.62 &         5235 &         6022 &       49.186 \\
       136.7 &        24.34 &         5.91 &         5002 &         6075 &       53.297 \\
       150.4 &        17.93 &         6.10 &         4691 &         6150 &       58.696 \\
       165.0 &        13.51 &         6.15 &         4314 &         6189 &       64.763 \\
       181.0 &        10.05 &         6.20 &         3962 &         6055 &       69.968 \\
       199.0 &         7.04 &         6.35 &         3691 &         5704 &       73.820 \\
       219.0 &         5.10 &         6.56 &         3464 &         5286 &       76.981 \\
       240.9 &         3.80 &         6.74 &         3236 &         4953 &       80.037 \\
       265.0 &         2.79 &         6.86 &         2995 &         4750 &       83.083 \\
       291.5 &         2.05 &         6.90 &         2741 &         4641 &       86.162 \\
       320.6 &         1.68 &         6.89 &         2488 &         4575 &       89.257 \\
       352.7 &         1.38 &         6.80 &         2230 &         4518 &       92.417 \\
       388.0 &         1.13 &         6.44 &         1919 &         4466 &       95.531 \\
       426.8 &         0.93 &         5.47 &         1482 &         4435 &       98.325 \\
       469.5 &         0.76 &         3.68 &          906 &         4398 &      100.230 \\
       516.5 &         0.62 &         2.03 &          455 &         4240 &      100.789 \\
       568.1 &         0.50 &         2.23 &          455 &         3970 &      100.789 \\
       624.9 &         0.40 &         2.46 &          455 &         3701 &      100.789 \\
\hline
\end{tabular}
\end{table}

\begin{figure*}
\epsfig{file=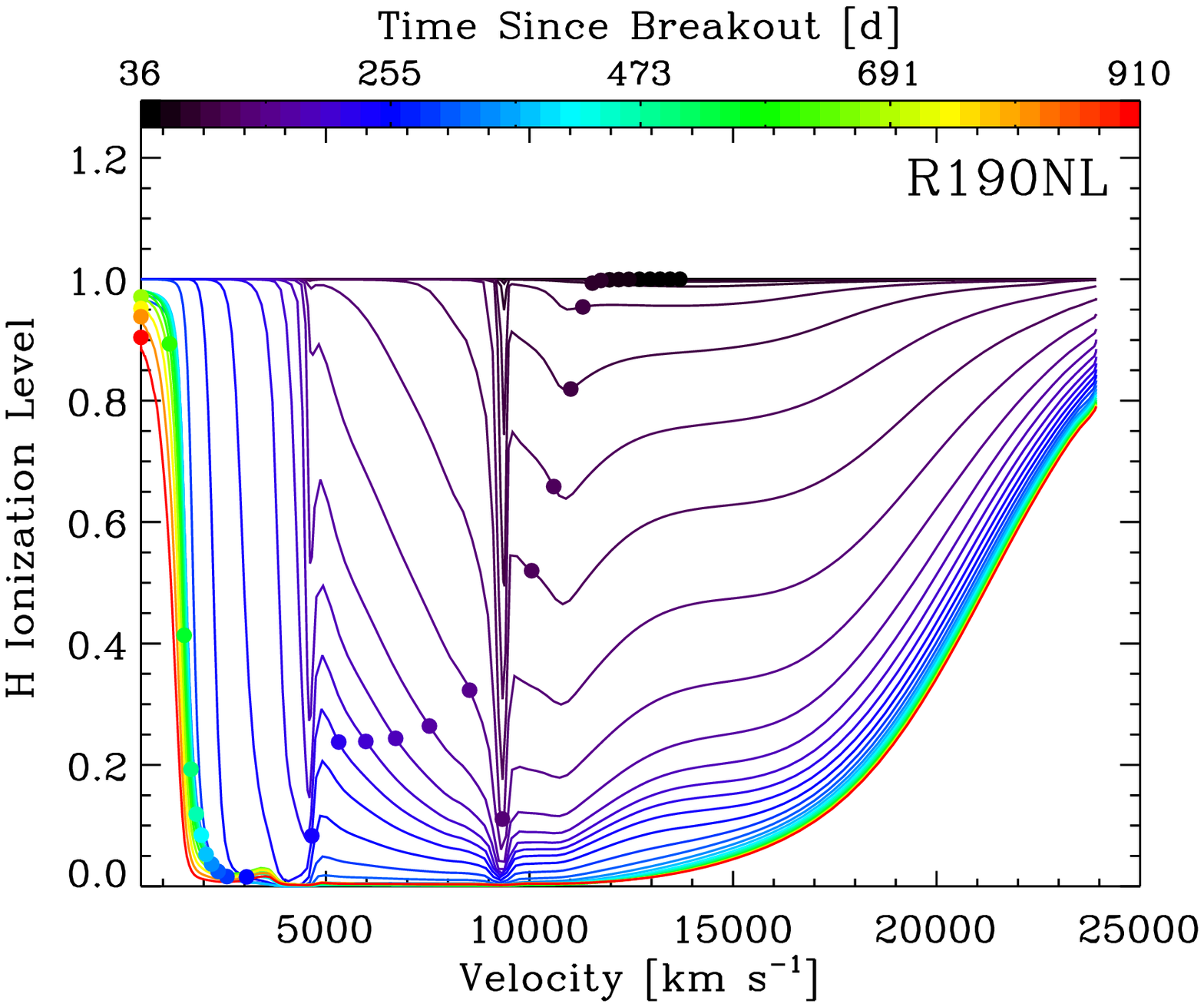,width=5.5cm}
\epsfig{file=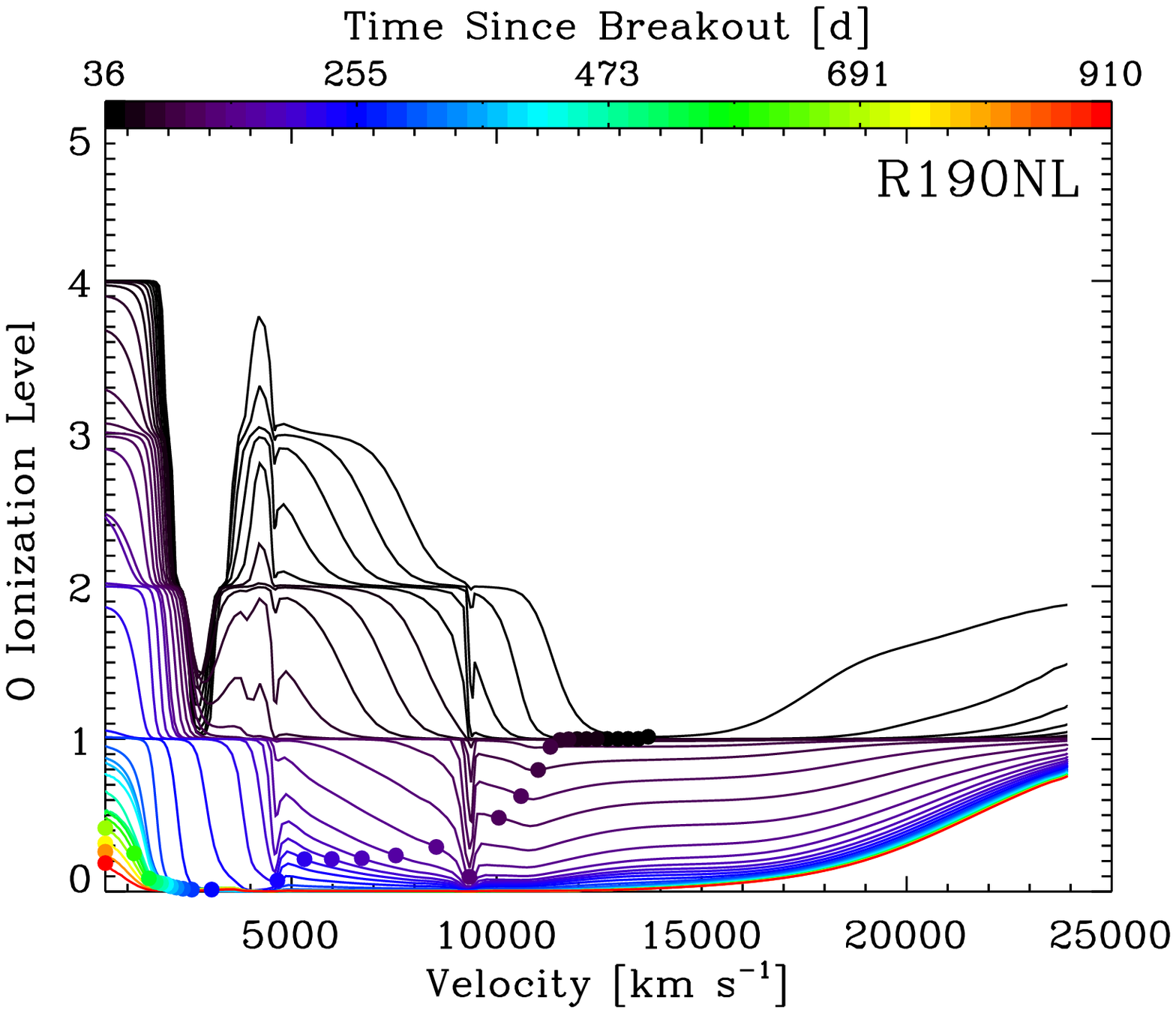,width=5.5cm}
\epsfig{file=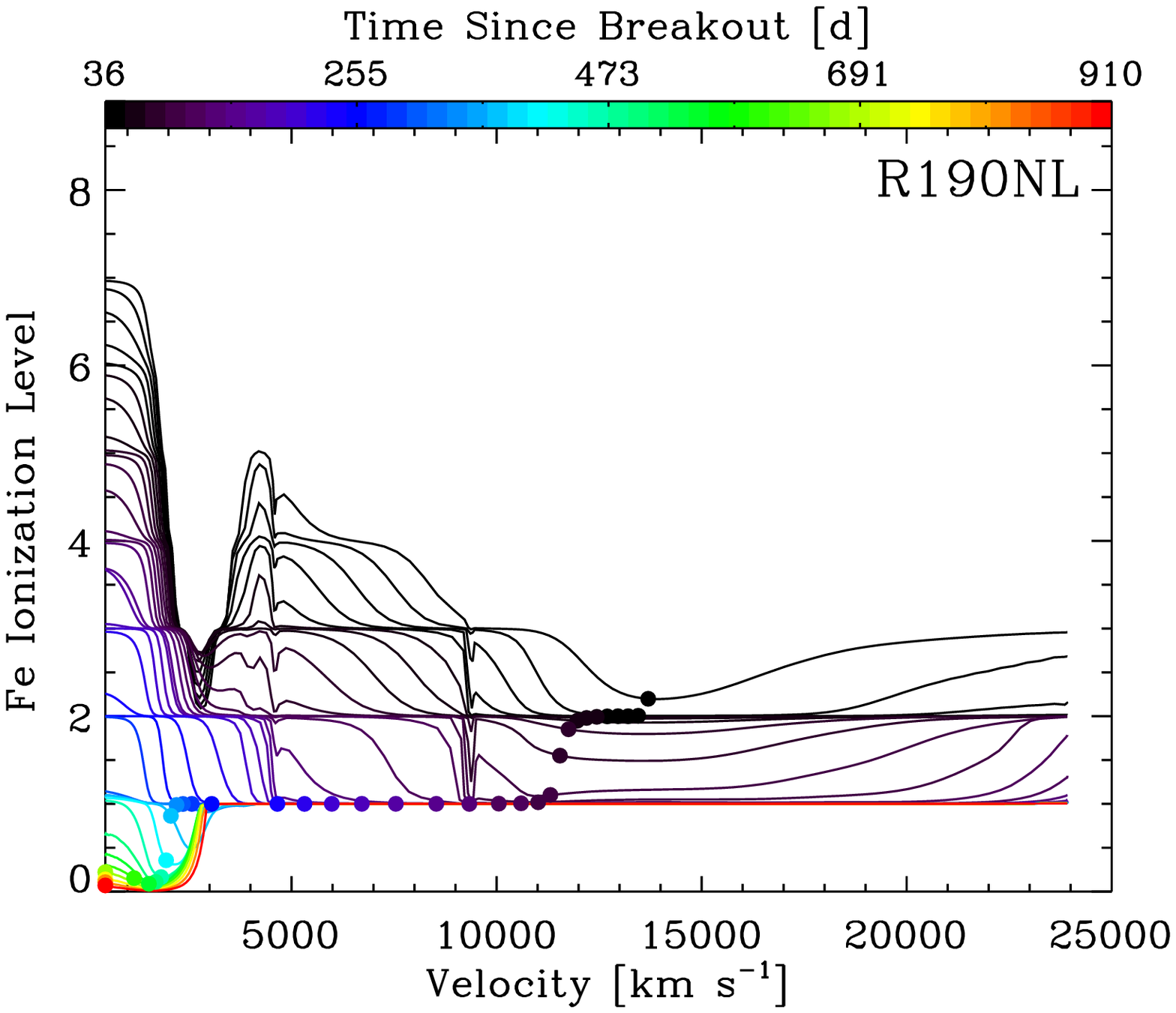,width=5.5cm}
\vspace{0.5cm}\\
\epsfig{file=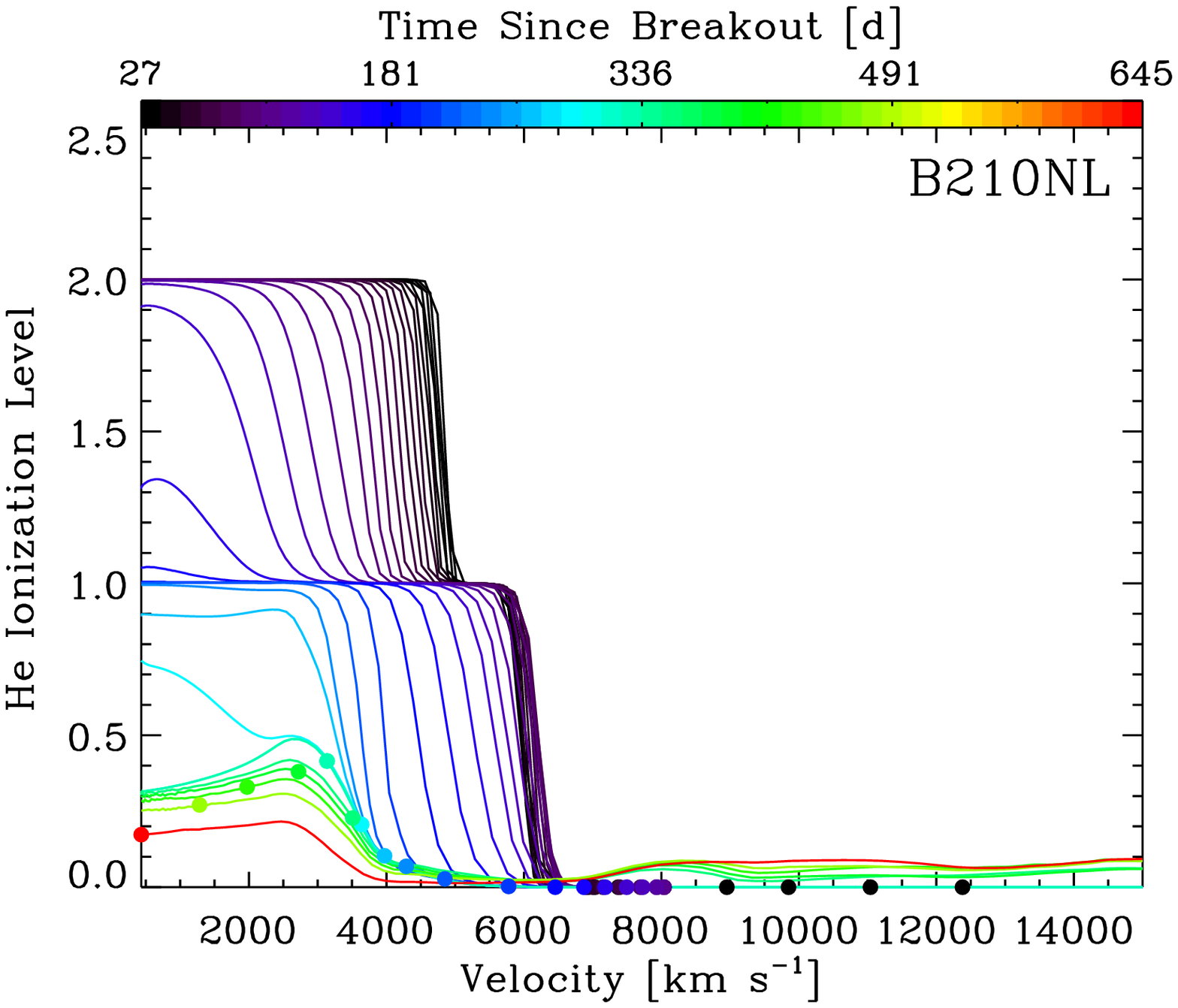,width=5.5cm}
\epsfig{file=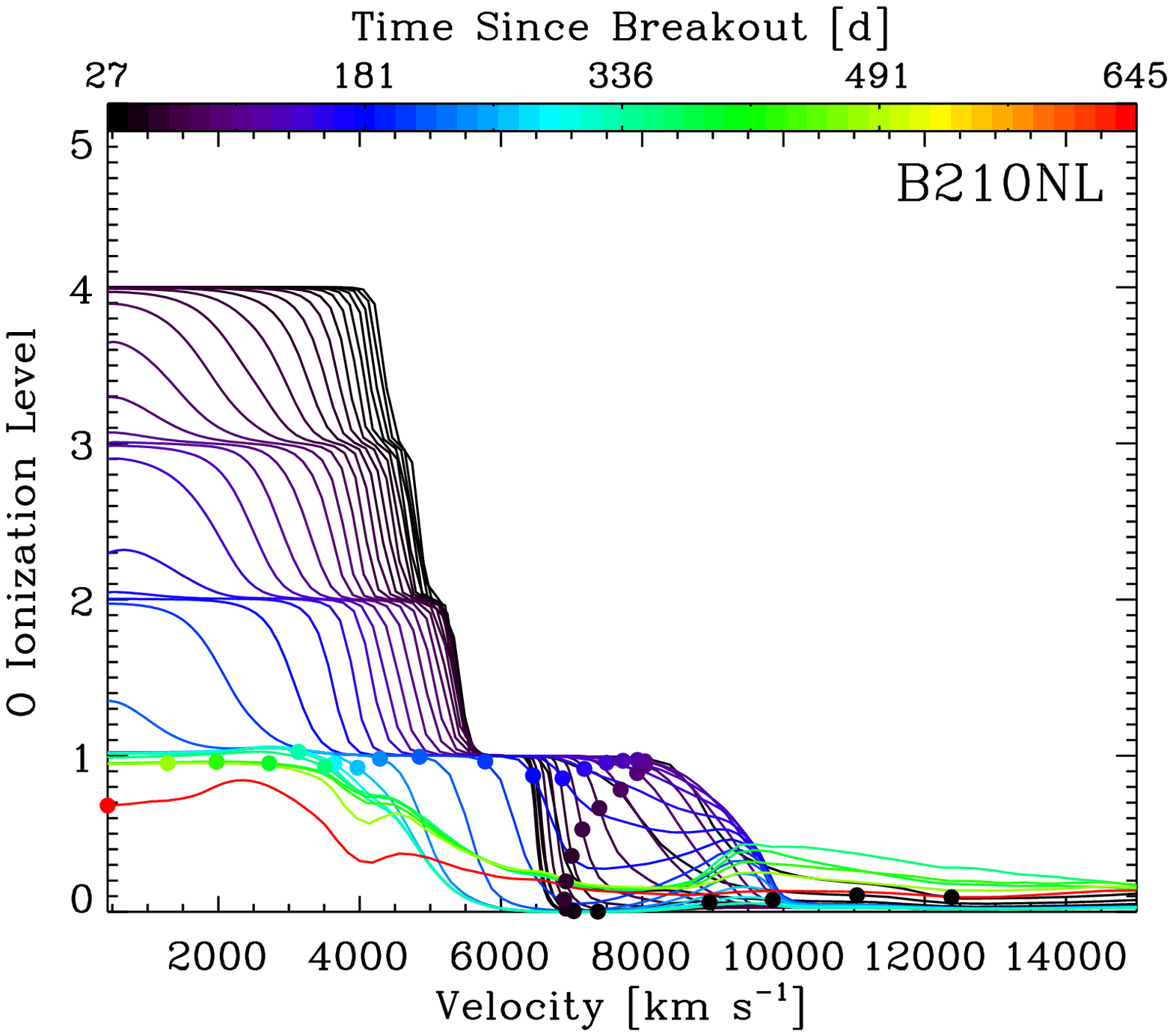,width=5.5cm}
\epsfig{file=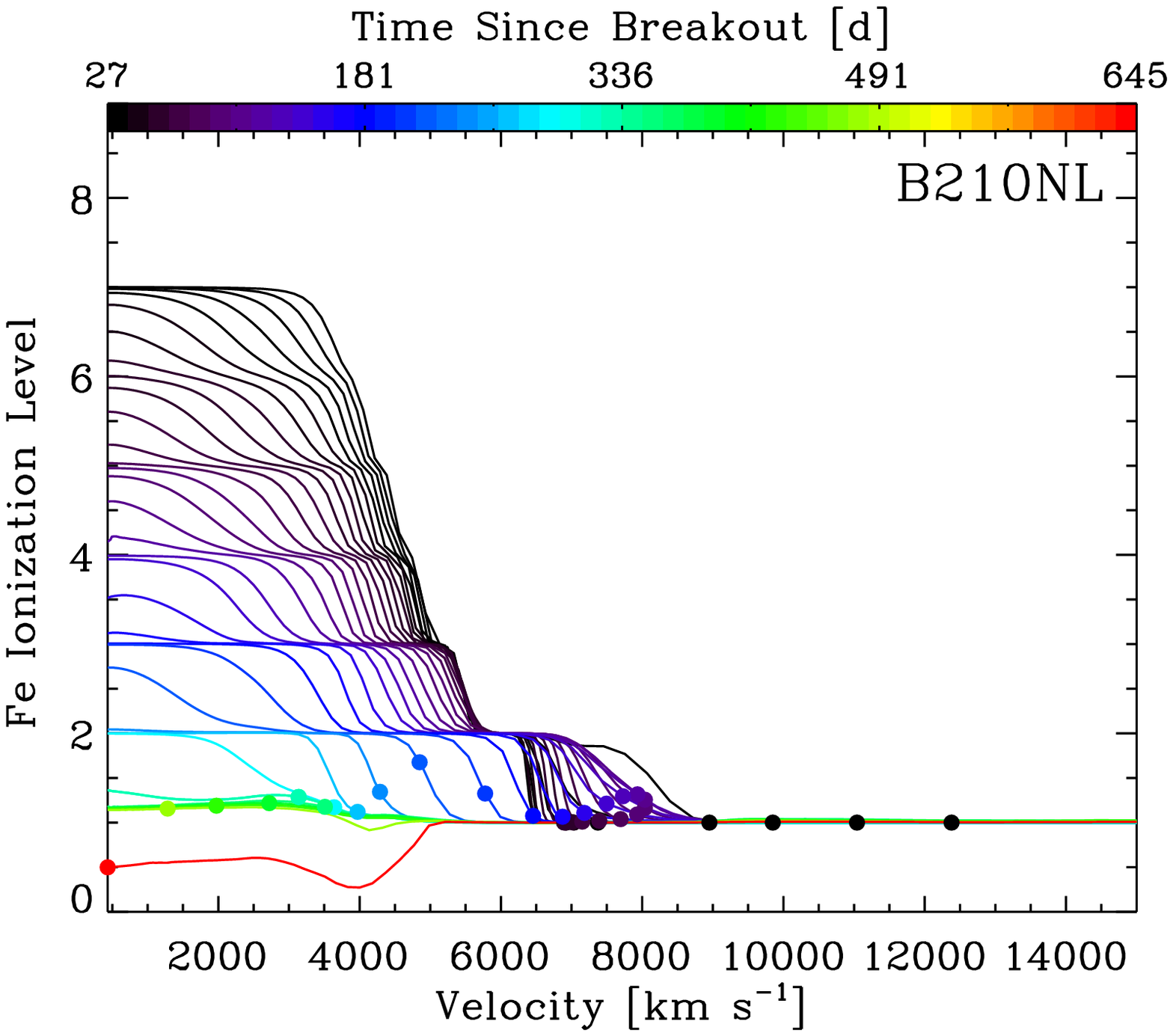,width=5.5cm}
\vspace{0.5cm}\\
\epsfig{file=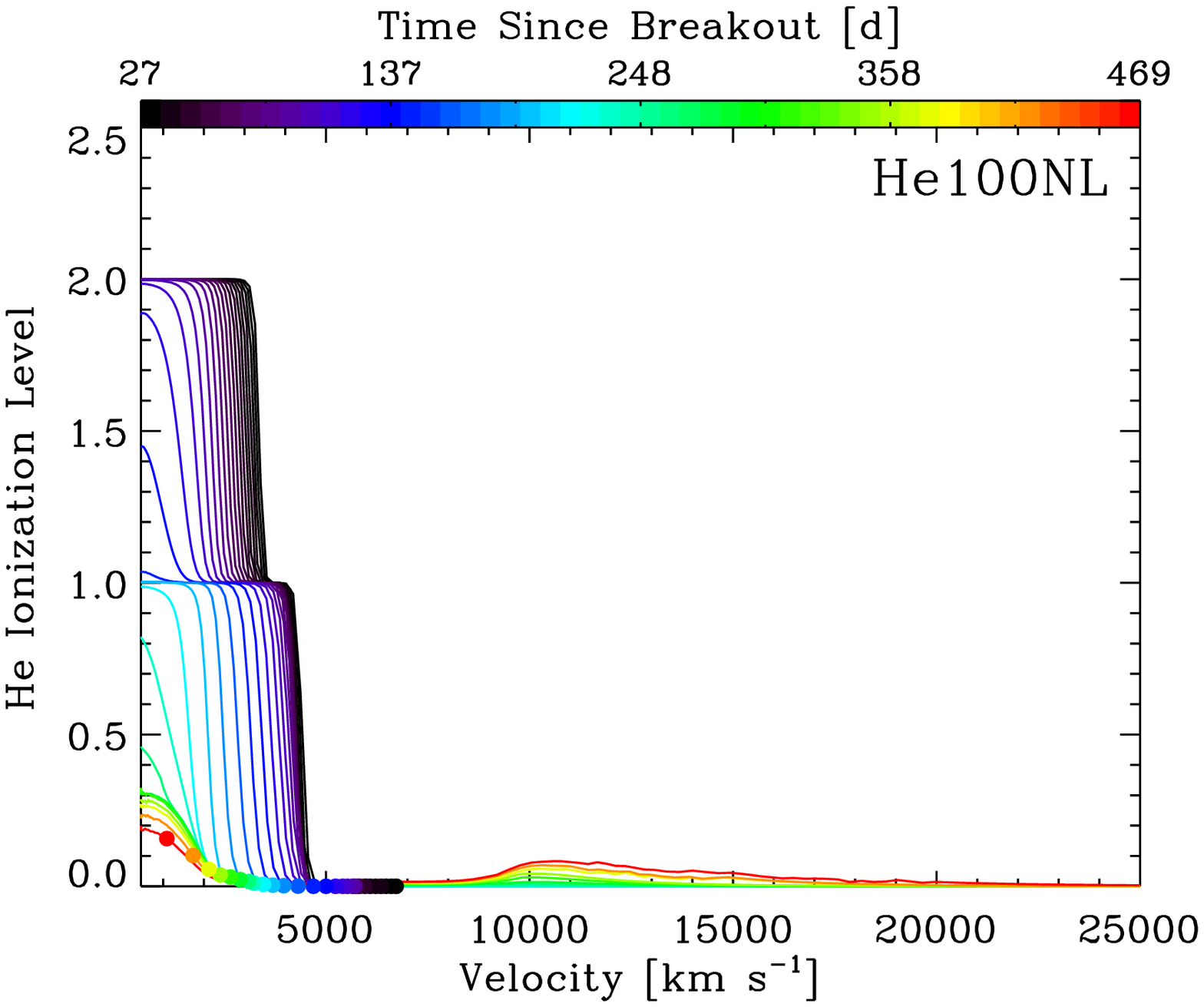,width=5.5cm}
\epsfig{file=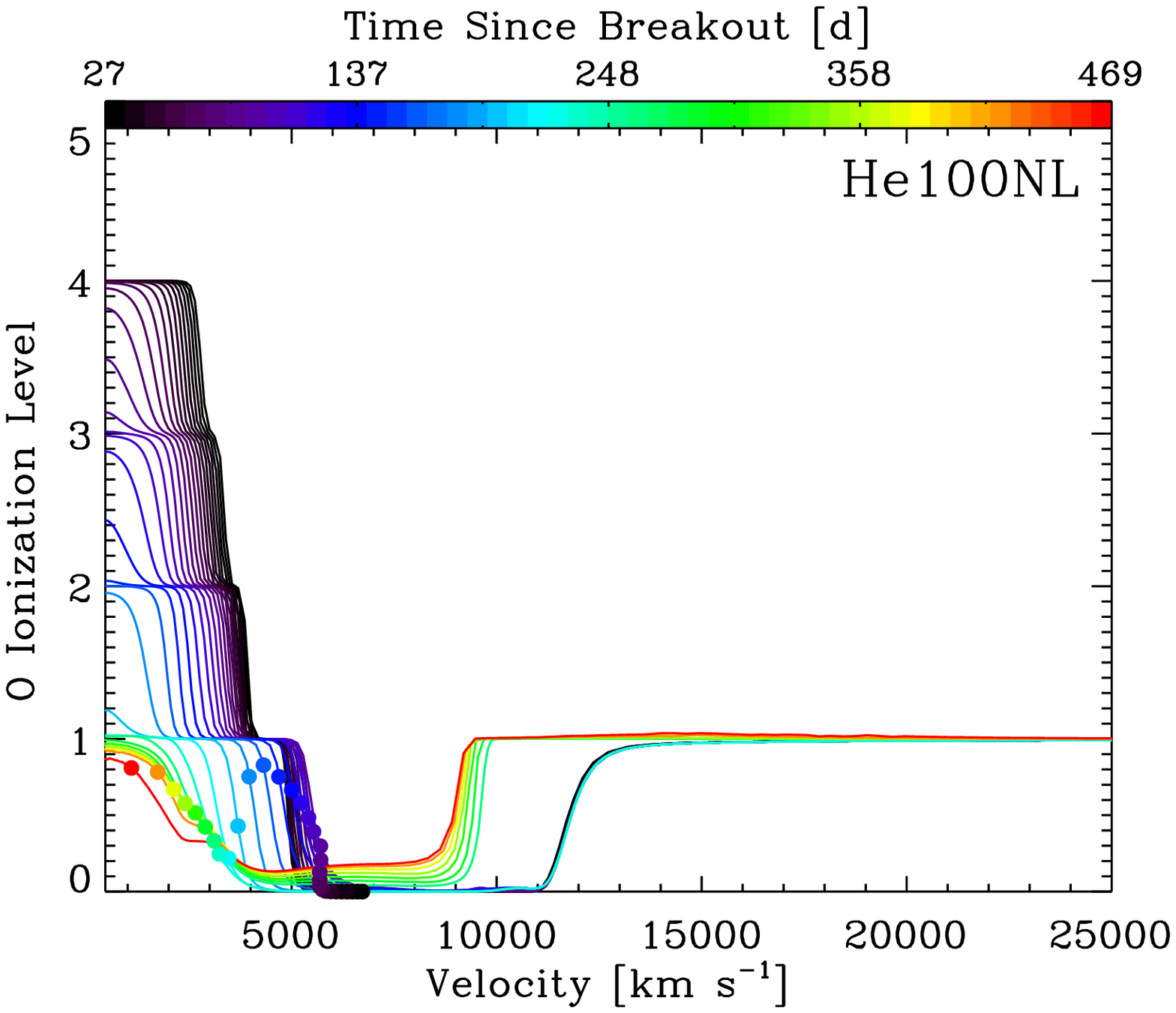,width=5.5cm}
\epsfig{file=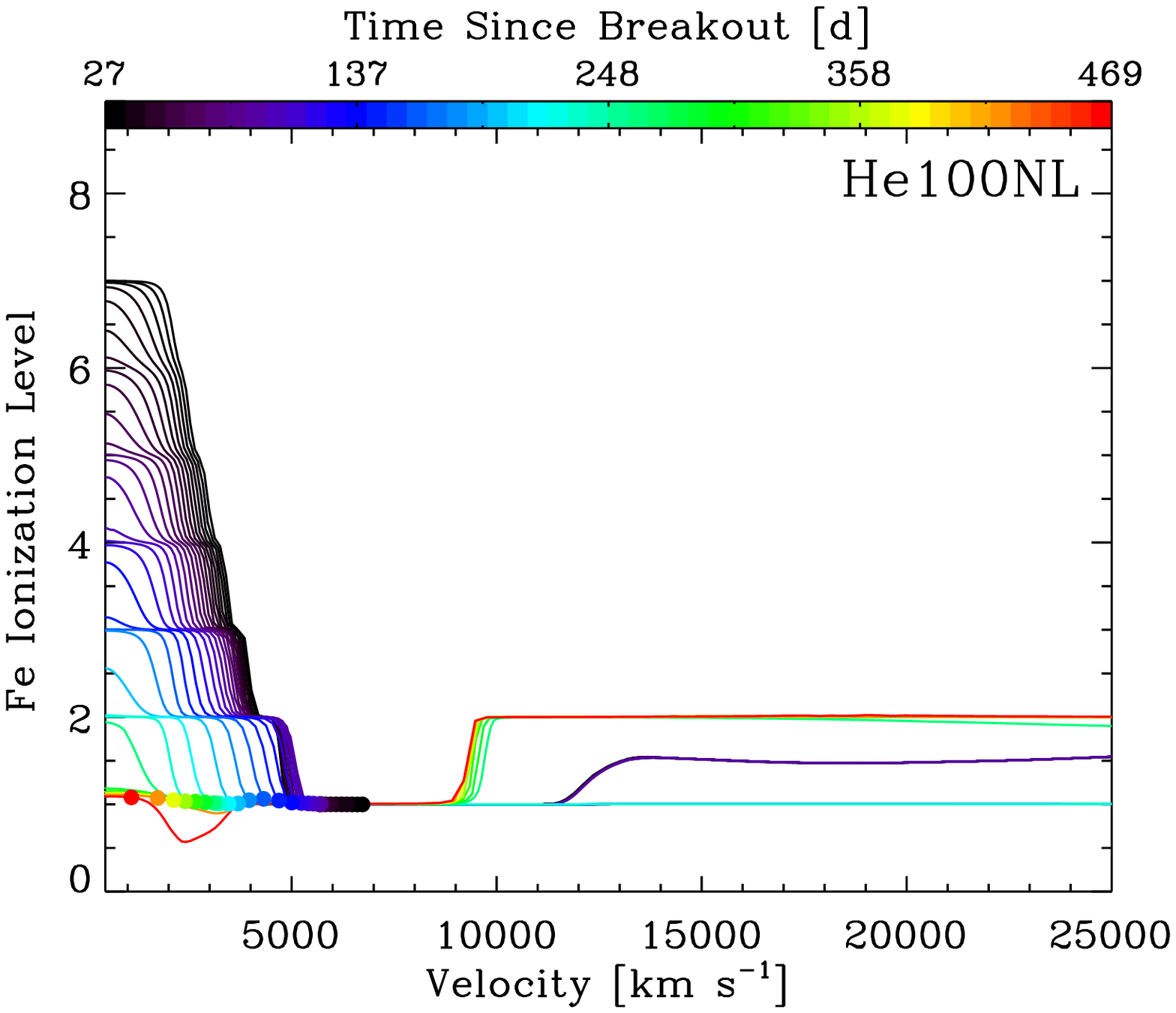,width=5.5cm}

\caption{Evolution of the ionization state of selected species versus time (color coding) and velocity (equivalent to
the depth in the ejecta)
for models R190NL (top row; we show H, O, and Fe), B210NL (middle row; He, O, and Fe), and He100NL (bottom row;
He, O, and Fe).
An ionization level $n$, where $n$ is a positive real number,
for a species X means that this species X is found at that location primarily in its $n^{\rm th}$-time ionized
form (i.e., the ionization level is equal to $\sum_i i {\rm X}^{i+}/ \sum_i {\rm X}^{i+})$.
Along each curve, a dot represents the photospheric location at each epoch.
\label{fig_ionization}
 }
\end{figure*}

   We also show the ionization state of the gas in  Fig.~\ref{fig_ionization}, as it indicates which species
   contribute opacity at depth (thus controling the diffusion of heat) and at the photosphere (affecting
   the spectrum formation). We restrict the illustration to models R190, B210, and He100ionI, and the dominant
   species H and/or He, O, and Fe.
   A generic property of these ejecta is their low ionization. This result may seem counter intuitive given the
   large amount of \isoni\ but it stems from the large ejecta mass which increases the diffusion time, making the
   SN evolve on very long time scales and turning transparent after few hundred days when the decay
   energy rate is small. This large mass also means this decay energy is to be shared, so that the energy
   released per unit mass is in fact quite comparable to what obtains in standard core-collapse SNe.
   Combined with the huge explosion energies, the small initial radius implies a very large expansion,
   associated with strong cooling from $PdV$ work.

   These curves reproduce the general morphology of ionization profiles in ``standard" SN ejecta  (e.g., \citealt{DH10}).
   The fast-expanding low-density outer regions maintain a high ionization --  this ionization freeze-out is in a large
    part a time-dependent effect \citep{DH08_time}.
   At low velocity, radioactive decay contributes significant
   heating, weakly affected by radiative cooling due to the larger optical depth of the inner ejecta layers.
   The photosphere (marked as a filled circle in the figure) is by essence a tracer of the region bridging
   thick and thin conditions, and thus delimits the higher ionization ejecta at depth from the recombined
   conditions immediately above it.
   The minimum ionization is  found at the photosphere up to the peak of the light curve because
   it suffers strong radiative cooling, fast expansion cooling, and weaker heating.\footnote{We note that species
tend to be over-ionized
   when subdominant so these plots have to be interpreted with the chemical stratification in mind
   (Fig.~2). We also note a peculiar dip in ionization at 2000-4000\,\kms\ in model
   R190, which is associated with the O-rich shell: This shell is highly-bound but \isoni\ deficient so that at
   early times when diffusion is inhibited the corresponding layers evolve adiabatically and cool tremendously
   by expanding from their small original radius.}

\begin{figure}
\smallskip
\smallskip
\epsfig{file=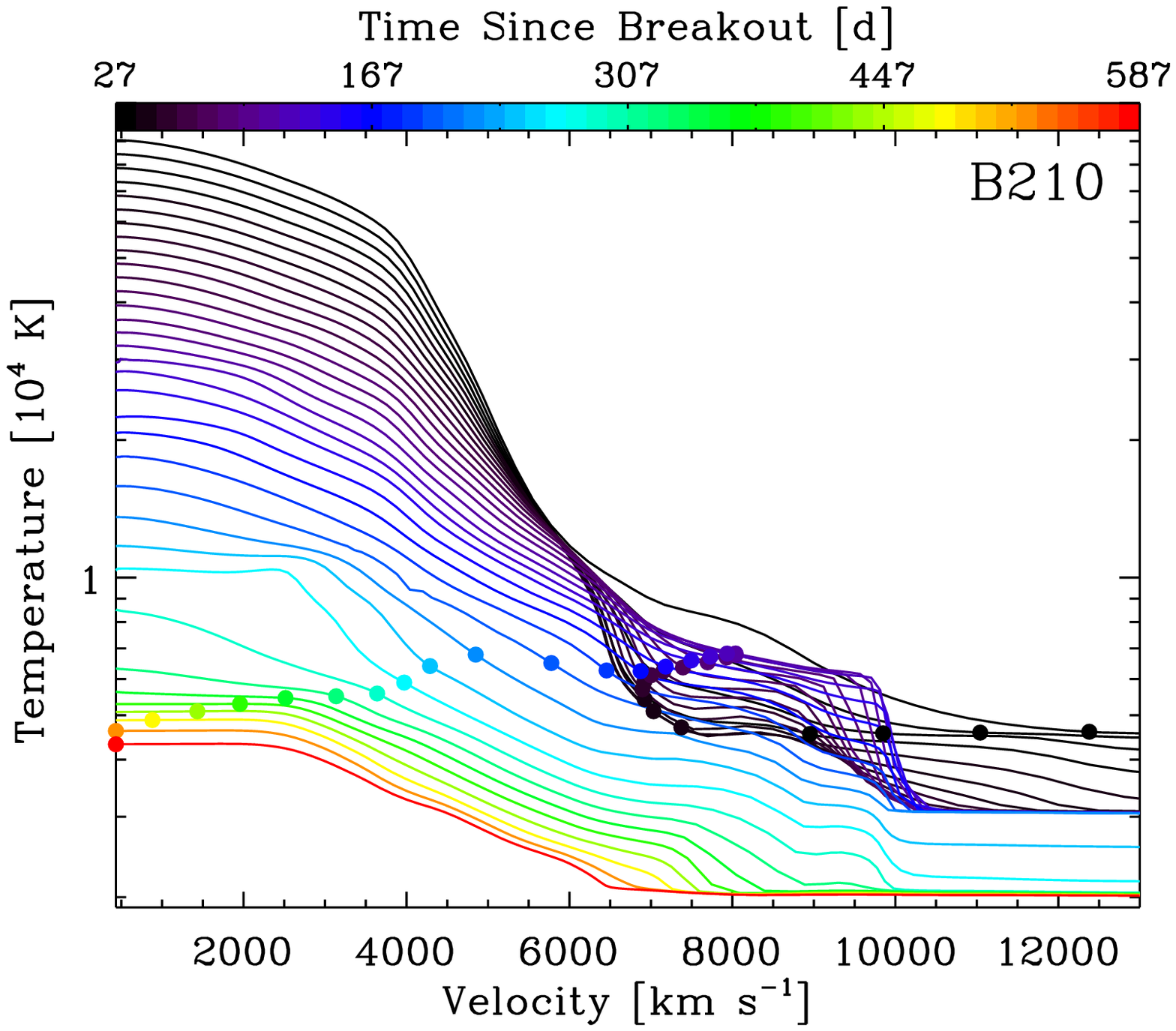,width=8.5cm}
\caption{Evolution of the ejecta temperature for model B210
from 25 until $\sim$\,600\,d after explosion. The dots refer to the location of the electron-scattering photosphere
and highlight the non-monotonic behavior of the photospheric location within the ejecta. Indeed, the strong heat
wave powered by the decay energy from 21.3\,\msun\ of \isoni\ causes the photosphere to migrate outwards in
mass (or velocity) between 50 and 100\,d after explosion. This behavior, unique to model B210, is seen in both
mass and velocity at the photosphere in Fig.~5. Note that despite the huge \isoni\ mass,
the temperature at a few hundred days after explosion is $\lesssim$\,5000\,K throughout the ejecta.
\label{fig_m210_temp}
}
\end{figure}

Finally, we illustrate in Fig.~\ref{fig_m210_temp} the temperature evolution of the B210 model ejecta,
and in particular the presence of a reversal in the trajectory of the photosphere in velocity/mass space.
This reversal is also seen in the ionization of oxygen, for example (Fig.~\ref{fig_ionization}).
This occurs here because of the strong heating from decay. Early on, the photosphere recedes in mass
until the heat wave from greater depth halts its recession, and in the present (but rare) situation, reverses
that recession to outward migration. We obtained such an effect in SNe Ib/Ic models in which no macroscopic
mixing of \isoni\ was applied \citep{dessart_etal_11}. Observationally, this may occur if a strong heat source
is turned on, as in a magnetar. SN 2005bf may well have been influenced by such a delayed heating,
given the observed double-peaked light curve, the reversal of the velocity at maximum absorption
in He\,\one\,5875\,\AA, and the blue colors of the spectra at late times \citep{maeda_etal_07,dessart_etal_12d}.

\section{Atomic data sources and model atoms}
\label{sect_atom_data}

The model atoms adopted for all simulations in this work are identical to that used in \citet{DH11}.
The sources of atomic data are varied, and in many cases multiple data sets for a given ion are available.
In some cases these multiple data sets represent an evolution in data quality and/or quantity, while in
other cases they represent different sources and/or computational methods. Comparisons of models
calculated with different data sets and atomic models potentially provide insights into the sensitivity
of our results to the adopted model atoms and hydrodynamical inputs (although such calculations have yet to be
undertaken for SNe).

Oscillator strengths for CNO elements were originally taken from
\citet{NS83_LTDR, NS84_CNO_LTDR}. These authors also provide transition probabilities to
states in the ion
continuum. The largest source of oscillator data is from \citet{Kur09_ATD,Kur_web}; its principal advantage
over many other sources (e.g., Opacity Project) is that LS coupling is not assumed. More recently, non-LS
oscillator strengths have become available through the Iron Project \citep{HBE93_IP}, and work done by
the atomic-data group at Ohio State University \citep{Nahar_OSU}. Other important
sources of radiative data for Fe include \citet{BB92_FeV,  BB95_FeVI, BB95_FeIV}, \cite{Nahar95_FeII}.
Atomic data from the opacity project comes from TOPBASE \citep{Topbase93}.
Energy levels have generally been obtained from National Institute of Standards and Technology.
Collisional data is sparse, particularly for
states far from the ground state. The principal source for collisional data among low lying states
for a variety of species is the tabulation by \citet{Men83_col}; other sources include
\citet{BBD85_col}, \citet{LDH85_CII_col}, \citet{LB94_N2}, \citet{SL74},
\citet{T97_SII_col,T97_SIII_col}, Zhang \& Pradhan (\citeyear{ZP95_FeII_col,ZP95_FeIII_col,ZP97_FeIV_col}).
Photoionization data is taken from the Opacity Project \citep{Sea87_OP,Topbase93}, the
Iron Project \citep{HBE93_IP,NP96_FeIII}, and \citet{NP93_SiI}. Unfortunately Ni and Co
photoionization data is generally unavailable,
and we have utilized crude approximations. Charge exchange cross-sections are from the tabulation
by \citet{KF96_chg}.
Atomic data for C\four\  was obtained from \citet{Lei72_CIV,PSS88_LI_seq},
and for the carbon isoelectronic sequence from \citet{LP89_C_seq}.
Collision strengths for Ar\two\ are from \citet{TH96_ArII_col}.
The LS Ne\,{\sc i} photoionization cross-sections were modified according to \cite{Sea98_NeI_phot}.
The same procedure was applied to using Ar\,{\sc i} mixing coefficients computed at
http://aphysics2.lanl.gov/tempweb/lanl.
Additional data for Ne\,\one\ was obtained from the MCHF/MCDHF web site: http://nlte.nist.gov/MCHF.

\begin{table}
\begin{center}
\caption[]{Summary of the model atoms used in our PISN calculations with \cmfgen\ (H is not included in model He100).
N$_{\rm f}$ refers to the number of full levels, N$_{\rm s}$ to the number of super levels, and N$_{\rm trans}$
to the corresponding number of bound-bound transitions. The last column refers to the upper level for each ion
treated.
At late times, we exclude the high ionization stages and simultaneously increase the number of levels
for Fe\one\ and Fe\two. After a few hundred days, we also split the lower 5, 10, or 15 super-levels to account
explicitly for more processes between low-lying states of IGEs.
$n$w\,$^2$W refers to a state with principal quantum number $n$ (all $l$ states combined into
a single state), and spin 2. Similarly, 8z\,$^1$Z refers to the $n=8$ state with high l states (usually $l-4$ and above)
combined and spin 1.
\label{tab_atom}}
\begin{tabular}{l@{\hspace{3mm}}r@{\hspace{3mm}}r@{\hspace{3mm}}r@{\hspace{3mm}}l}
\hline
 Species        &  N$_{\rm f}$  &  N$_{\rm s}$ & N$_{\rm trans}$ & Upper Level \\
\hline
    H\,{\sc i}      & 30  &  20   &     435    & $n=30$\\
    He\,{\sc i}     & 51  &  40   &     374    & $n=11$\\
    He\,{\sc ii}    & 30  &  13   &     435    & $n=30$\\
     C{\,\sc i}\,    &  26  &   14 &    120 & 2s2p$^3$$\,^3$P\opar\                  \\
     C{\,\sc ii}\,   &  26  &   14 &     87 & 2s$^2$4d\,$^2$D$_{5/2}$                \\
     C{\,\sc iii}\,  & 112  &   62 &    891 & 2s8f\,$^1$Fo                       \\
     C{\,\sc iv}\,   &  64  &   59 &   1446 & $n=30$                         \\
     O{\,\sc i}\,    &  51  &   19 &    214 & 2s$^2$2p$^3$($^4$S\opar)4f\,$^3$F$_{3}$         \\
     O{\,\sc ii}\,   & 111  &   30 &   1157 & 2s$^2$2p$^2$($^3$P)4d\,$^2$D$_{5/2}$       \\
     O{\,\sc iii}\,  &  86  &   50 &    646 & 2p4f\,$^1$D                       \\
     O{\,\sc iv}\,   &  72  &   53 &    835 & 2p$^2$($^3$P)3p\,$^2$P\opar\                   \\
     Ne\one\,     & 139 &   70 & 	1587 & 2s$^2$2p$^5$($^2$P\oparsub{3/2})6d\,$^2$[5/2]\oparsub{3}  \\
     Ne{\,\sc ii}\,  &  91  &   22 &   1106 & 2s$^2$2p$^4$($^3$P)4d\,$^2$P$_{3/2}$       \\
     Ne{\,\sc iii}\, &  71  &   23 &    460 & 2s$^2$2p$^3$($^2$D\opar)3d\,$^3$S$_{1}$         \\
     Na{\,\sc i}\,   &  71  &   22 &   1614 & 30w\,$^2$W                         \\
     Mg\one\   & 122  	& 39 & 1486 & 3s15w\,$^1$W \\
     Mg{\,\sc ii}\,  &  65  &   22 &   1452 & 30w\,$^2$W                         \\
     Mg{\,\sc iii}\, &  99  &   31 &    775 & 2p$^5$7s$\,^1$P                    \\
       Si\one\    & 187 & 100 & 4329  & 3s$^2$3p($^2$P\oparsub{3/2})6g\,$^2$[5/2]\oparsub{2}  \\
     Si{\,\sc ii}\,  &  59  &   31 &    354 & 3s$^2$7g\,$^2$G$_{7/2}$             \\
     Si{\,\sc iii}\, &  61  &   33 &    310 & 3s5g\,$^1$G$_{4}$                    \\
     Si{\,\sc iv}\,  &  48  &   37 &    405 & 10f\,$^2$Fo                        \\
     S\one\       & 322 & 106  & 8540 & 3s$^2$3p$^3$($^4$S)10f\,$^3$F$_4$   \\
     S{\,\sc ii}\,   & 324  &   56 &   8208 & 3s3p$^3$($^5$S\opar)4p\,$^6$P             \\
     S{\,\sc iii}\,  &  98  &   48 &    837 & 3s3p$^2$($^2$D)3d$\,^3$P             \\
     S{\,\sc iv}\,   &  67  &   27 &    396 & 3s3p($^3$P\opar)4p$^2$D$_{5/2}$         \\
     Ar\one\      & 110 & 56  & 1541        & 3s$^2$3p$^5$($^2$P\oparsub{3/2})7p\,$^2$[3/2]$_2$  \\
     Ar\two\       & 415 & 134   & 20197  &   	3s$^2$3p$^4$($^3$P$_1$)7i\,$^2$[6]$_{11/2}$  \\
     Ar{\,\sc iii}\, & 346  &   32 &   6898 & 3s$^2$3p$^3$($^2$D\opar)8s\,$^1$D\opar\            \\
     Ca\one\      & 98 & 76 & 688  & 4s8z\,$^1$Z \\
     Ca{\,\sc ii}\,  &  77  &   21 &   1736 & 3p$^6$30w\,$^2$W                     \\
     Ca{\,\sc iii}\, &  40  &   16 &    108 & 3s$^2$3p$^5$5s\,$^1$P                \\
     Ca{\,\sc iv}\,  &  69  &   18 &    335 & 3s3p$^5$($^3$P\opar)3d\,$^4$D\oparsub{1/2}        \\
     Fe{\,\sc i}\,   & 136  &   44 &   1900 & 3d$^6$($^5$D)4s4p\,$^5$F\opar$_{3}$           \\
     Fe{\,\sc ii}\,  & 827  &  275 &  44\,831 & 3d$^5$($^6$S)4p$^2$($^3$P)\,$^4$P$_{1/2}$        \\
     Fe{\,\sc iii}\, & 607  &   69 &   9794 & 3d$^5$($^4$D)6s$^3$D$_{2}$               \\
     Fe{\,\sc iv}\,  &1000  &  100 &  72\,223 & 3d$^4$($^3$G)4f\,$^4$P$_{5/2}$            \\
     Fe{\,\sc v}\,   & 191  &   47 &   3977 & 3d$^3$($^4$F)4d\,$^5$F$_{3}$              \\
     Fe{\,\sc vi}\,  & 433  &   44 &  14\,103 & 3p$^5$($^2$P)3d$^4$($^1$S)$^2$Pc$_{3/2}$      \\
     Fe{\,\sc vii}\  & 153 & 29 & 1753 & 3p5($^2$P)3d$^3$(b$^2$D)\,$^1$P\opar$_1$ \\
     Co{\,\sc ii}\,  &1000  &   81 &  61\,986 & 3d$^7$($^4$P)4f\,$^5$F\opar$_{4}$              \\
     Co{\,\sc iii}\, &1000  &   72 &  68\,462 & 3d$^6$($^5$D)5f\,$^4$F$_{9/2}$            \\
     Co{\,\sc iv}\,  &1000  &   56 &  69\,425 & 3d$^5$($^2$D)5s\,$^1$D$_{2}$              \\
     Co{\,\sc v}\,   & 387  &   32 &  13\,605 & 3d$^4$($^3$F)4d\,$^2$H$_{9/2}$            \\
     Co{\,\sc vi}\,  & 323  &   28 &   9608 & 3d$^3$($^2$D)4d$\,^1$S$_{0}$              \\
     Co{\,\sc vii}\  & 319 & 31 & 9096  & 3p5($^2$P\opar)3d$^4$(3F)\,$^2$D\opar$_{3/2}$\\
     Ni{\,\sc ii}\,  &1000  &   59 &  51\,707 & 3d$^8$($^3$F)7f\,4I\opar$_{9/2}$            \\
     Ni{\,\sc iii}\, &1000  &   47 &  66\,486 & 3d$^7$($^2$D)4d\,$^3$Sb$_{1}$             \\
     Ni{\,\sc iv}\,  &1000  &   54 &  72\,898 & 3d$^6$($^5$D)6p$^6$F$_{11/2}$           \\
     Ni{\,\sc v}\,   & 183  &   46 &   3065 & 3d$^5$($^2$D3)4p$\,^3$F\opar$_{3}$             \\
     Ni{\,\sc vi}\,  & 314  &   37 &   9569 & 3d$^4$($^5$D)4d\,$^4$F$_{9/2}$            \\
     Ni{\,\sc vii}\  & 308 & 37 & 9225 & 3d$^3$($^2$D)4d\,$^3$P$_2$  \\
\hline
\end{tabular}
\end{center}
\end{table}

\section{Dependency on model atoms}
\label{sect_atom_dep}

The main impact of increasing the size of
the model atoms in our simulations, in particular for metals, is to enhance the magnitude of line blanketing,
which tends to make the SED redder. This effect is generally weaker than obtained
here
through variations in composition between models or versus time (i.e., H vs. IMEs, IMEs vs. IGEs etc.),
although it noticeably alters the colors after the light-curve peak.

For the early-time simulations of models R190, B190, and B210, the photosphere is located in
essentially a pure Hydrogen/Helium plasma. Line blanketing effects are negligible and there
is no concern with the opacity of IMEs and IGEs and the completeness of our model atoms at such
early times.

However, as the photosphere recedes to deeper layers in the ejecta where IMEs and IGEs are abundant
the completeness of model atoms becomes a central concern for the reliability of our predictions.
After much experimentation in past simulations \citep{DH10,DH11,li_etal_12,dessart_etal_12b},  we have
converged
to an adequate assignment for the number of full and super levels to include (for a discussion on our super-level
approach, see \citealt{HM98_lb}). Recently, we emphasized the critical need of including Fe\one\ in simulations
of Type II SNe \citep{li_etal_12}, while earlier on we found that Sc and Ti play a critical role in optical spectra
of SNe II-P \citep{DH11}.

\begin{figure*}
\epsfig{file=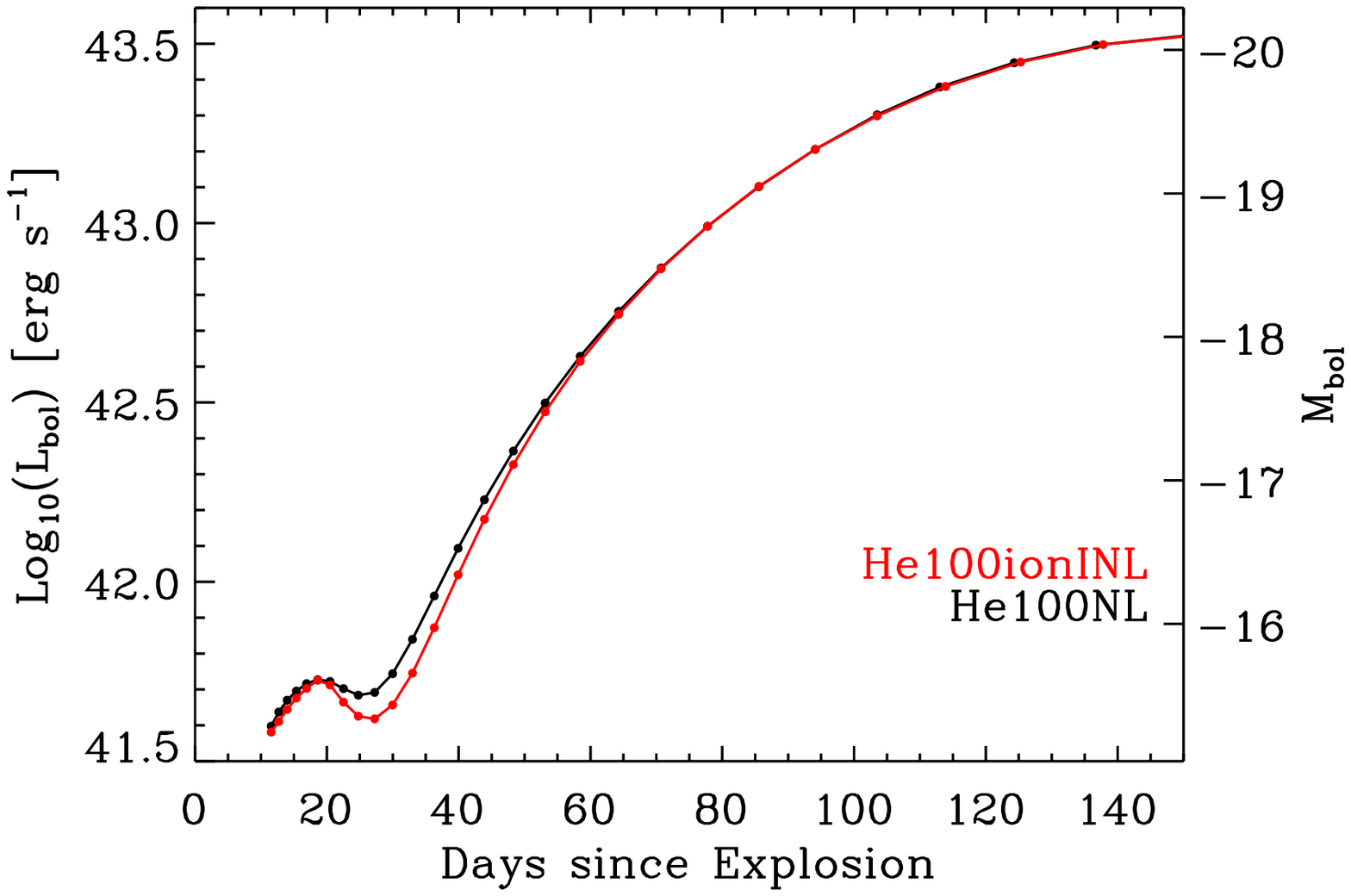,width=6.5cm}
\epsfig{file=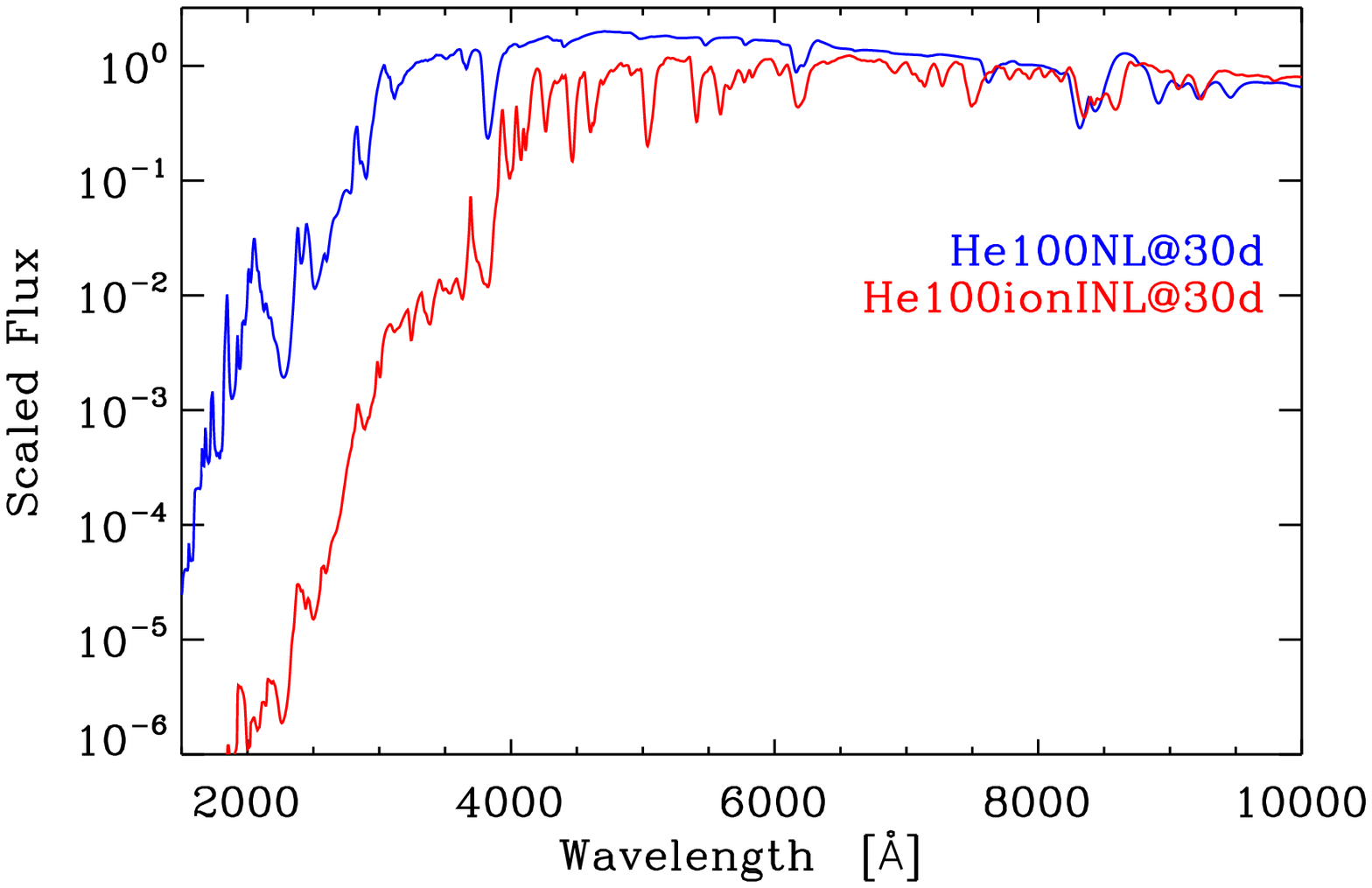,width=6.5cm}
\epsfig{file=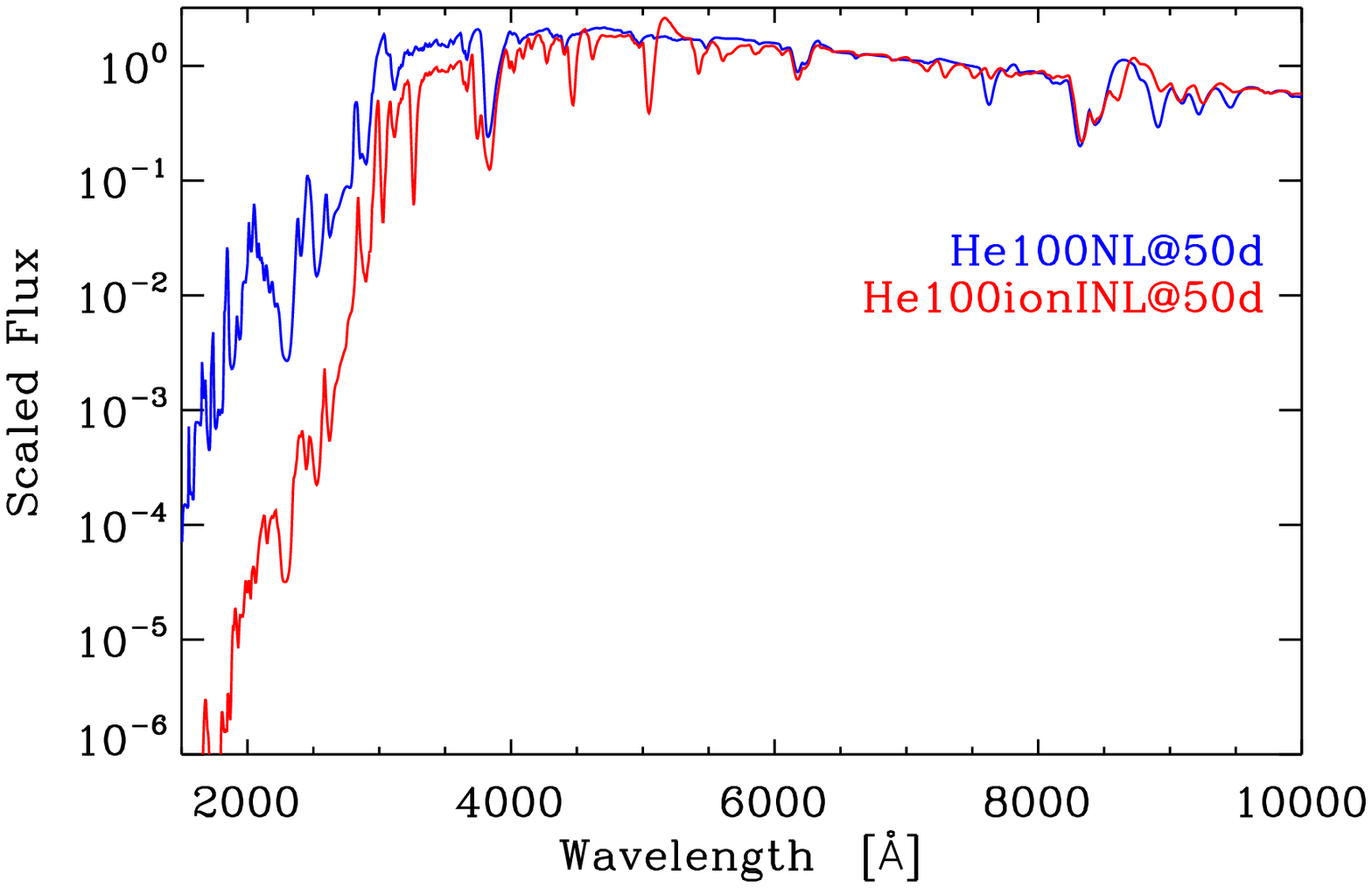,width=6.5cm}
\epsfig{file=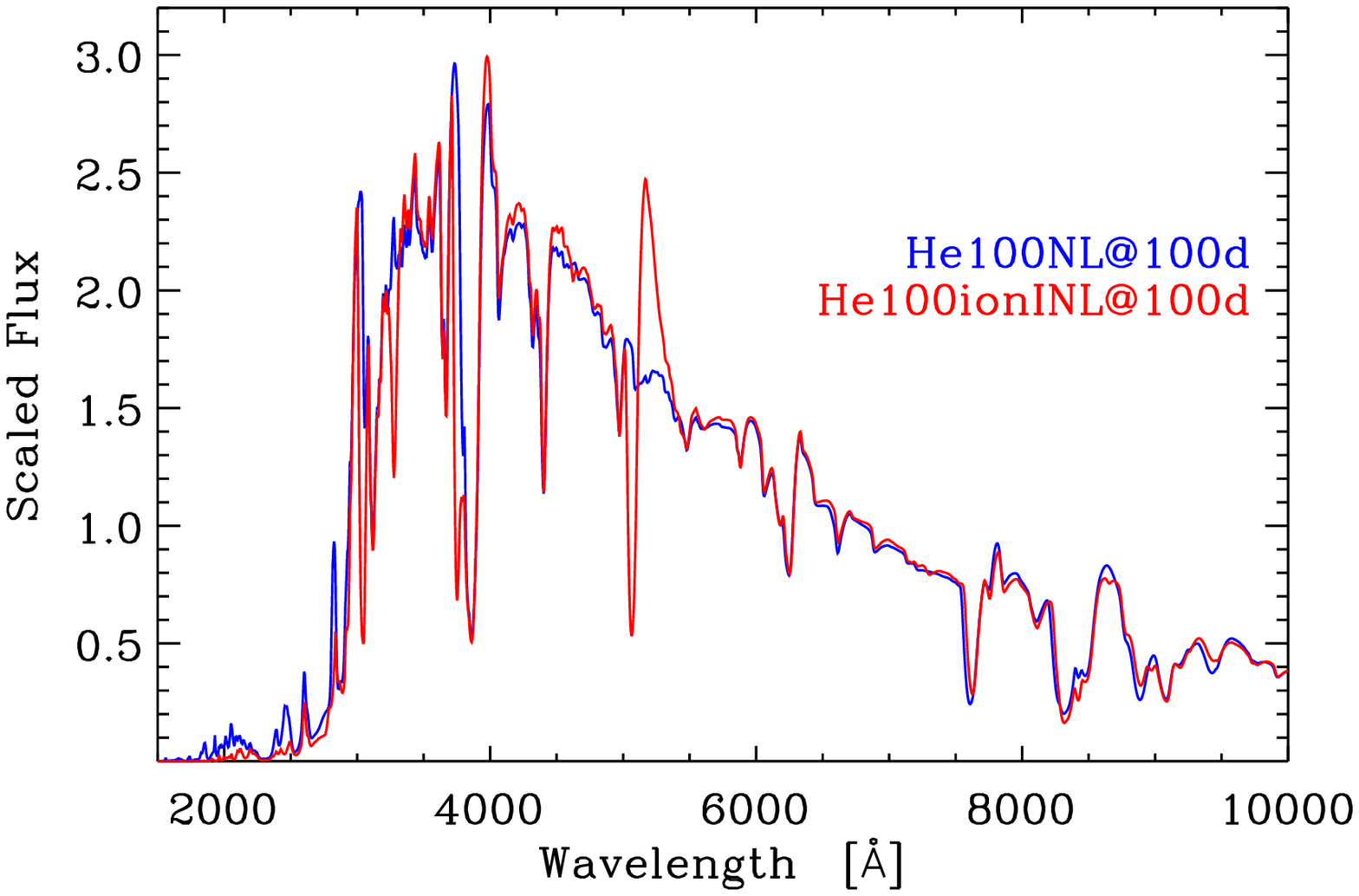,width=6.5cm}
\epsfig{file=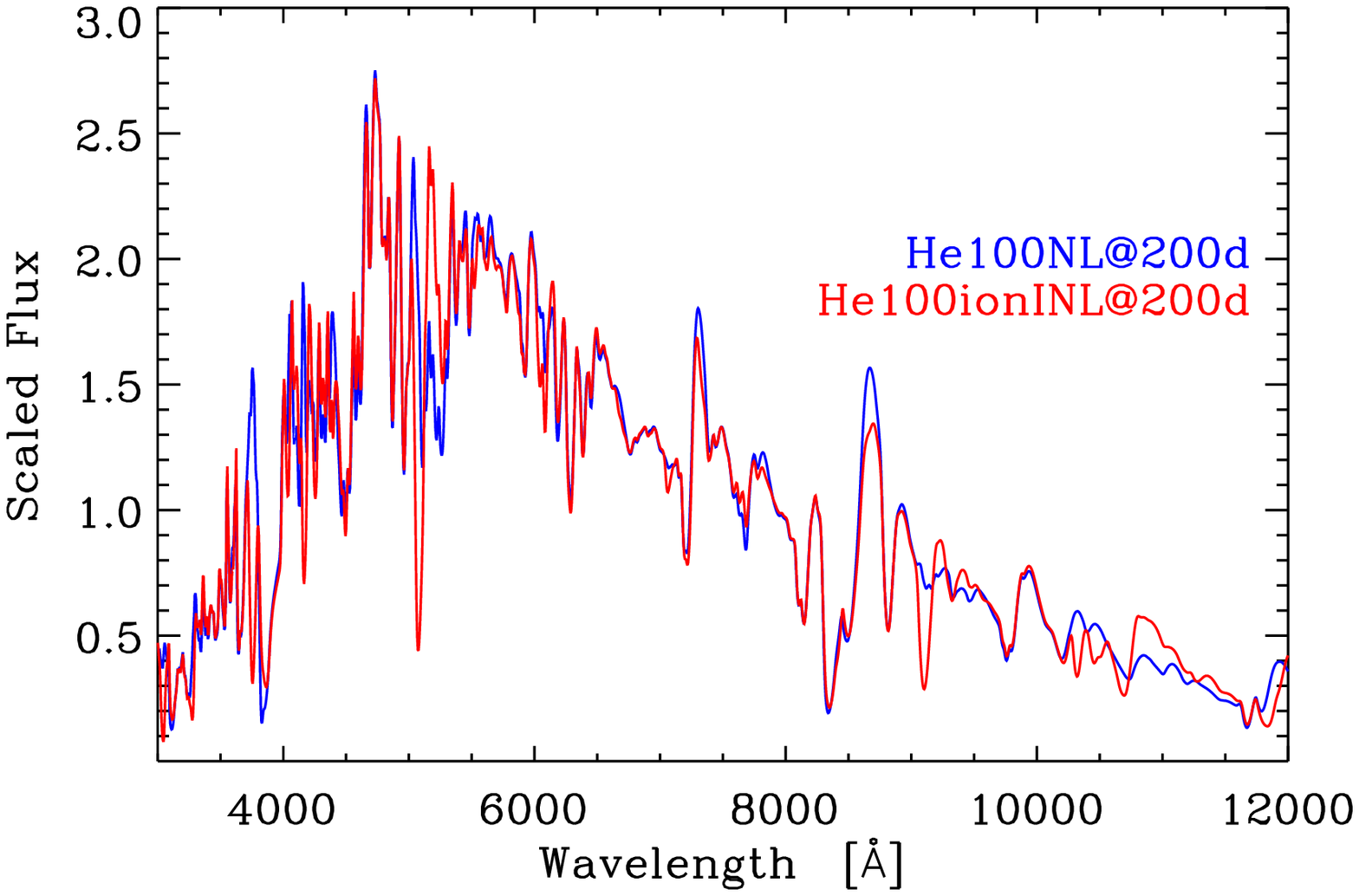,width=6.5cm}
\epsfig{file=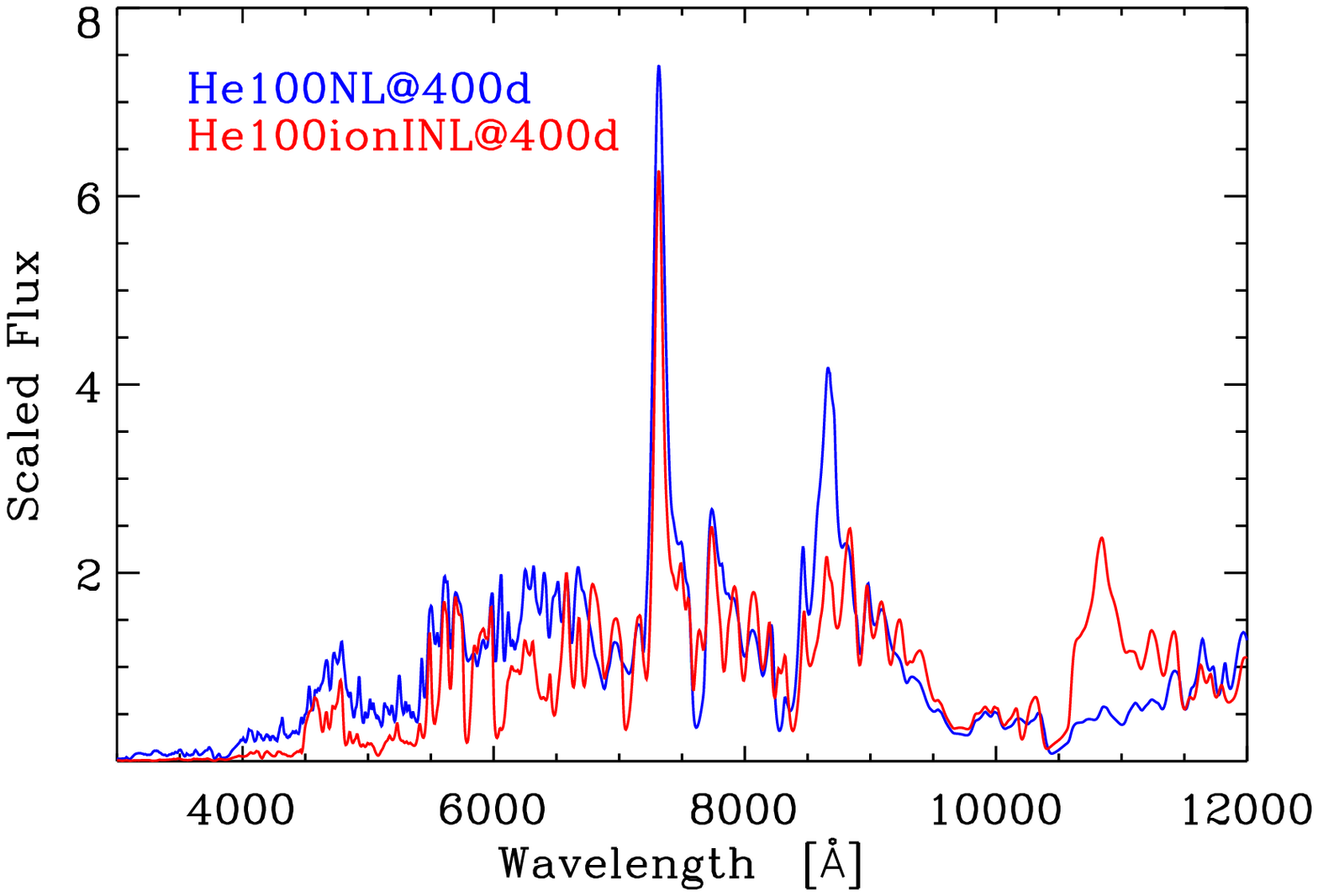,width=6.5cm}
\epsfig{file=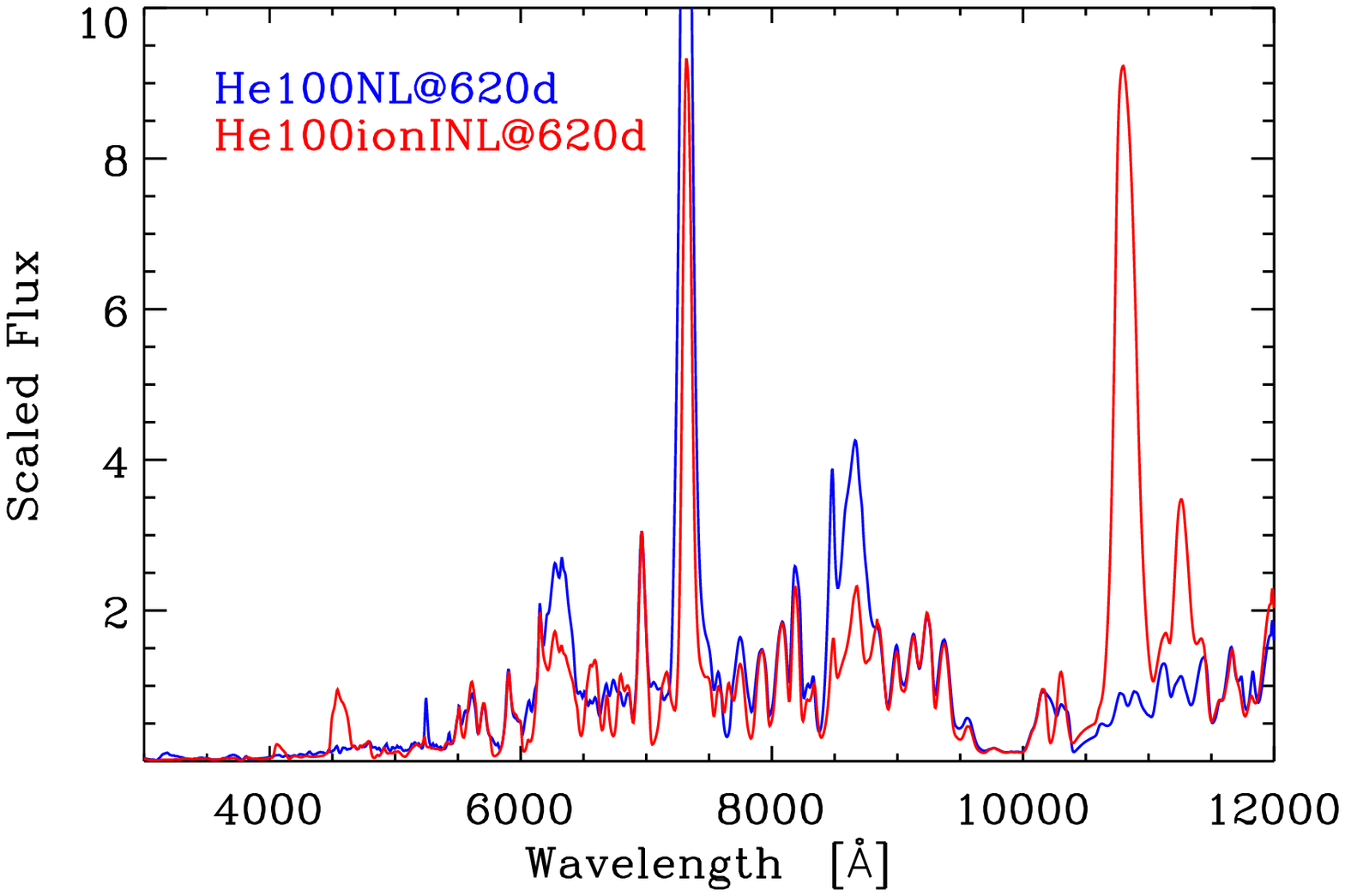,width=12cm}
 \caption{
Illustration of the changes to the bolometric light curve (top left) and the spectra (subsequent panels, showing
snapshots at 30, 50, 100, 200, 400, and 620\,d after explosion) for the ejecta model He100ionINL,
which treats the neutral species Mg\one, Ca\one, Si\one, and S\one, with model He100NL, which
ignores them. Under favorable ionization conditions, these additional neutral states can increase the
magnitude of line blanketing and bound-free opacity. Notice at nebular times the impact on the strength of
the [Ca\two]\,7300\,\AA\ doublet (He100NL), reduced by the overlap with Ca\one\ lines at 7148 and 7326\,\AA\
(He100ionNL). Opacity effects are associated with line blanketing and photo-ionization cross sections,
and are strongly modulated by the ionization state at the photosphere.
\label{fig_comp_He100}
 }
\end{figure*}

In ejecta dominated by metals, the situation is more complicated than in type II SNe. To reach a reliable
radiative-transfer result,
we first find that more levels need to be included. Secondly, while we typically include in the non-LTE solver all
the metal line transitions
with a statistical weight greater than 0.002, we find that we now need to go down to 10$^{-4}$.\footnote{This cut
only applies to elements beyond Ne in the periodic table, does not apply to the lowest $n$ levels
($n$ is typically 9), and we only cut a transition when there is at least $m$ ($m$ is typically 9) stronger
downward transitions from the level. Thus, this procedures does not cut important transitions to ground levels,
and forbidden and semi-forbidden transitions among low-lying states.} This means
treating many more lines
in total, which increases the computation time. Finally, in the course of this study started with models R190,
B190, B210, and He100,
we found that the ejecta were rather cold at the nebular phase, and for the compact progenitors, these
conditions were
also cold prior to re-brightening at early times. Because of the dominance of IMEs in the corresponding emitting
layers, we have run
a companion model, named He100ionI, identical to He100 (which has the same model atoms as R190 etc), but
now including the neutral states for species Mg, Si, S, and Ca.
We compare these two sequences at selected post-explosion times in Fig.~\ref{fig_comp_He100}.

The bolometric light curve is largely insensitive to the addition of these neutral atoms. We understand this from
the
fact that the light curve is primarily conditioned by the diffusion of heat through the optically-thick layers of the
ejecta.
These shells tend to be ionized and hot, and thus dominated by more ionized species. The only slight change
occurs at
very early  times, prior to re-brightening, when the photosphere is very cold (i.e., $\sim$\,4000\,K; Fig.~5).
At such times, the blanketing
from neutral species is actually huge, because of the IME-rich photosphere, but also because the IGEs are
under abundant compared to their solar-metallicity value. Hence, these neutral species act as an overlying
blanketing
layer, modifying the blocking power of the last-scattering layer. The blanketing is in part due to lines, especially in
the optical and
the near-IR, but more importantly stems from bound-free opacity from the ground and excited states
(note in particular the photo-ionization edge of Mg\one\ at 3757\,\AA)
of the corresponding ions (see also the continuum curve in Fig.~14). As the photosphere heats up
on the rise to peak, species
become once-ionized and the difference between models He100 and He100ionI weakens, essentially limited to
the presence
of the strong Mg\one\,5173\,\AA\ line.

However, as the SN light curve passes its peak, the color reddens again and the photosphere cools down to
4000\,K within 100-200\,d.
The SN becomes nebular and blanketing from neutral species turns back on. At nebular times, additional
blanketing reddens the SED
below 5500\,\AA, but also impacts the long wavelength range. This  stems indirectly from the impact on the
blanketed radiation field. It also arises directly from additional optically-thick lines from Mg\one, Si\one\ or S\one.

As discussed above, metal line blanketing is irrelevant up to the peak because the composition really becomes
IGE dominated at the photosphere only then. So, to expedite our very time consuming simulations, we have
used modest-sized iron model atoms up to the peak (Fe\one\,[136,44] and Fe\two\,[115,50]), improved it early
after the peak
(Fe\one\,[136,44] and Fe\two\,[827,275]), and finally used huge model atoms for both at later times (Fe\one\,
[1142,413]
and Fe\two\,[827,275]) --- the numbers in square brackets represent the number of full and super levels.
This choice of model atoms yields converged results in our SNe Ia simulations \citep{dessart_etal_12c} and we
expect
the same level of accuracy to be reached in those similarly metal-rich PISN ejecta at late times.

\section{Log of magnitudes for each PISN model}
\label{sect_tab_mag}

To complement the discussion on the colors of our PISN simulations, we present in
Tables~\ref{tab_mag_B190NL}--\ref{tab_mag_he100NL} the magnitudes and luminosities (bolometric
and UVOIR, i.e., integrated flux from the blue edge of the U band to the red edge of the I band).
From these, one can easily infer various bolometric corrections and compare to observations.

\begin{table*}
\caption{Bolometric and UVOIR luminosities as well as synthetic magnitudes for PISN model B190NL.
The UVOIR luminosity corresponds to the integrated flux over the wavelength interval 3000--9200\,\AA,
i.e., from the blue edge of the U band to the red edge of the I band.
\label{tab_mag_B190NL}
}
\begin{tabular}{l@{\hspace{1.6mm}}c@{\hspace{1.6mm}}c@{\hspace{1.6mm}}
c@{\hspace{1.6mm}}c@{\hspace{1.6mm}}c@{\hspace{1.6mm}}
c@{\hspace{1.6mm}}c@{\hspace{1.6mm}}c@{\hspace{1.6mm}}
c@{\hspace{1.6mm}}c@{\hspace{1.6mm}}c@{\hspace{1.6mm}}} \\
\hline
Age   & $L_{\rm bol}$ & $L_{\rm UVOIR}$ & $M_{\rm bol}$ & $M_{\rm U}$ & $M_{\rm B}$
& $M_{\rm V}$ & $M_{\rm R}$ & $M_{\rm I}$ & $M_{\rm J}$ & $M_{\rm H}$ & $M_{\rm K}$ \\
   (d)   & (erg\,s$^{-1}$) & (erg\,s$^{-1}$) & (mag) & (mag)& (mag)& (mag)& (mag)& (mag)& (mag)& (mag)& (mag)\\
\hline
 16.67 & 3.4929e+42 & 2.2666e+42 &  -17.651 & -16.568 & -16.938 & -17.578 & -18.068 & -18.129 & -18.560 & -18.677 & -18.926\\
 18.34 & 3.0239e+42 & 1.9194e+42 &  -17.495 & -16.028 & -16.691 & -17.401 & -17.951 & -18.011 & -18.506 & -18.618 & -18.900\\
 20.17 & 2.6104e+42 & 1.6095e+42 &  -17.335 & -15.449 & -16.419 & -17.207 & -17.814 & -17.876 & -18.429 & -18.534 & -18.849\\
 22.19 & 2.1978e+42 & 1.3107e+42 &  -17.148 & -14.824 & -16.111 & -16.960 & -17.638 & -17.701 & -18.310 & -18.408 & -18.755\\
 24.41 & 1.8688e+42 & 1.0867e+42 &  -16.972 & -14.222 & -15.816 & -16.764 & -17.464 & -17.540 & -18.176 & -18.274 & -18.633\\
 26.85 & 1.6860e+42 & 9.7437e+41 &  -16.860 & -13.851 & -15.654 & -16.661 & -17.344 & -17.458 & -18.068 & -18.187 & -18.528\\
 29.54 & 1.5034e+42 & 8.6590e+41 &  -16.736 & -13.739 & -15.543 & -16.501 & -17.210 & -17.338 & -17.935 & -18.081 & -18.403\\
 32.49 & 1.3165e+42 & 7.5756e+41 &  -16.592 & -13.666 & -15.430 & -16.344 & -17.058 & -17.181 & -17.784 & -17.944 & -18.254\\
 35.74 & 1.1436e+42 & 6.6064e+41 &  -16.439 & -13.570 & -15.324 & -16.189 & -16.901 & -17.018 & -17.624 & -17.788 & -18.093\\
 39.32 & 9.6780e+41 & 5.7329e+41 &  -16.258 & -13.577 & -15.206 & -16.009 & -16.739 & -16.842 & -17.428 & -17.480 & -17.826\\
 43.25 & 7.5661e+41 & 4.7465e+41 &  -15.990 & -13.855 & -14.965 & -15.849 & -16.531 & -16.629 & -17.056 & -17.097 & -17.410\\
 47.58 & 6.6828e+41 & 4.3700e+41 &  -15.856 & -14.454 & -14.928 & -15.837 & -16.355 & -16.518 & -16.850 & -16.976 & -17.177\\
 52.34 & 7.2292e+41 & 4.8332e+41 &  -15.941 & -14.920 & -15.111 & -15.981 & -16.380 & -16.608 & -16.907 & -17.071 & -17.229\\
 57.57 & 9.0472e+41 & 6.1748e+41 &  -16.184 & -15.382 & -15.452 & -16.244 & -16.584 & -16.840 & -17.113 & -17.286 & -17.426\\
 63.33 & 1.2072e+42 & 8.4140e+41 &  -16.498 & -15.860 & -15.854 & -16.561 & -16.870 & -17.130 & -17.373 & -17.551 & -17.682\\
 69.66 & 1.6449e+42 & 1.1691e+42 &  -16.833 & -16.338 & -16.261 & -16.897 & -17.185 & -17.436 & -17.652 & -17.830 & -17.958\\
 76.63 & 2.2447e+42 & 1.6236e+42 &  -17.171 & -16.804 & -16.656 & -17.231 & -17.504 & -17.740 & -17.931 & -18.108 & -18.235\\
 84.29 & 3.0375e+42 & 2.2306e+42 &  -17.499 & -17.248 & -17.031 & -17.554 & -17.815 & -18.033 & -18.203 & -18.375 & -18.503\\
 92.72 & 4.0579e+42 & 3.0184e+42 &  -17.814 & -17.664 & -17.383 & -17.859 & -18.110 & -18.314 & -18.464 & -18.630 & -18.758\\
102.00 & 5.3415e+42 & 4.0149e+42 &  -18.112 & -18.050 & -17.712 & -18.146 & -18.389 & -18.581 & -18.715 & -18.870 & -18.997\\
112.00 & 6.8734e+42 & 5.2077e+42 &  -18.386 & -18.394 & -18.009 & -18.408 & -18.645 & -18.829 & -18.947 & -19.090 & -19.215\\
122.00 & 8.5088e+42 & 6.4832e+42 &  -18.618 & -18.679 & -18.258 & -18.630 & -18.861 & -19.042 & -19.147 & -19.278 & -19.398\\
132.00 & 1.0182e+43 & 7.7896e+42 &  -18.813 & -18.911 & -18.466 & -18.817 & -19.045 & -19.223 & -19.319 & -19.437 & -19.551\\
142.00 & 1.1816e+43 & 9.0680e+42 &  -18.974 & -19.098 & -18.637 & -18.973 & -19.199 & -19.378 & -19.464 & -19.572 & -19.677\\
152.00 & 1.3333e+43 & 1.0257e+43 &  -19.105 & -19.241 & -18.776 & -19.103 & -19.328 & -19.508 & -19.585 & -19.685 & -19.780\\
162.00 & 1.4574e+43 & 1.1338e+43 &  -19.202 & -19.264 & -18.910 & -19.280 & -19.476 & -19.642 & -19.672 & -19.772 & -19.853\\
172.00 & 1.5608e+43 & 1.2171e+43 &  -19.277 & -19.290 & -18.988 & -19.374 & -19.576 & -19.740 & -19.748 & -19.847 & -19.916\\
182.00 & 1.6339e+43 & 1.2731e+43 &  -19.326 & -19.271 & -19.029 & -19.442 & -19.654 & -19.824 & -19.808 & -19.908 & -19.968\\
192.00 & 1.6723e+43 & 1.2992e+43 &  -19.351 & -19.206 & -19.033 & -19.485 & -19.712 & -19.891 & -19.849 & -19.952 & -20.007\\
211.00 & 1.6321e+43 & 1.2504e+43 &  -19.325 & -18.946 & -18.920 & -19.485 & -19.755 & -19.956 & -19.873 & -19.999 & -20.051\\
232.00 & 1.4259e+43 & 1.0593e+43 &  -19.178 & -18.465 & -18.617 & -19.341 & -19.668 & -19.902 & -19.822 & -19.998 & -20.049\\
255.00 & 1.0714e+43 & 7.5083e+42 &  -18.868 & -17.487 & -18.050 & -18.991 & -19.403 & -19.687 & -19.684 & -19.894 & -19.951\\
280.00 & 7.4220e+42 & 4.9189e+42 &  -18.469 & -16.488 & -17.413 & -18.480 & -19.008 & -19.372 & -19.417 & -19.659 & -19.726\\
308.00 & 5.1085e+42 & 3.2255e+42 &  -18.064 & -15.741 & -16.787 & -17.925 & -18.575 & -19.050 & -19.170 & -19.355 & -19.420\\
338.00 & 2.9760e+42 & 1.9662e+42 &  -17.477 & -15.067 & -16.157 & -17.303 & -17.993 & -18.651 & -18.520 & -18.585 & -18.589\\
371.00 & 2.0821e+42 & 1.1506e+42 &  -17.089 & -12.730 & -14.573 & -16.445 & -17.540 & -18.301 & -18.504 & -18.796 & -18.364\\
408.00 & 1.5414e+42 & 7.0200e+41 &  -16.763 & -11.737 & -13.546 & -15.679 & -16.950 & -17.896 & -18.429 & -18.752 & -18.164\\
448.80 & 1.1270e+42 & 4.3687e+41 &  -16.423 & -11.322 & -12.889 & -15.005 & -16.371 & -17.447 & -18.214 & -18.598 & -18.050\\
493.70 & 7.7498e+41 & 3.0059e+41 &  -16.016 & -11.055 & -12.425 & -14.518 & -15.943 & -17.060 & -17.850 & -18.163 & -17.612\\
543.10 & 4.9505e+41 & 2.1863e+41 &  -15.530 & -10.741 & -11.976 & -14.087 & -15.596 & -16.734 & -17.380 & -17.374 & -16.675\\
\hline
\end{tabular}
\end{table*}

\begin{table*}
\caption{Bolometric and UVOIR luminosities as well as synthetic magnitudes for PISN model R190NL.
\label{tab_mag_R190NL}
}
\begin{tabular}{l@{\hspace{1.6mm}}c@{\hspace{1.6mm}}c@{\hspace{1.6mm}}
c@{\hspace{1.6mm}}c@{\hspace{1.6mm}}c@{\hspace{1.6mm}}
c@{\hspace{1.6mm}}c@{\hspace{1.6mm}}c@{\hspace{1.6mm}}
c@{\hspace{1.6mm}}c@{\hspace{1.6mm}}c@{\hspace{1.6mm}}} \\
\hline
Age   & $L_{\rm bol}$ & $L_{\rm UVOIR}$ & $M_{\rm bol}$ & $M_{\rm U}$
& $M_{\rm B}$ & $M_{\rm V}$ & $M_{\rm R}$ & $M_{\rm I}$ &
$M_{\rm J}$ & $M_{\rm H}$ & $M_{\rm K}$ \\
   (d)   & (erg\,s$^{-1}$) & (erg\,s$^{-1}$) & (mag) & (mag)& (mag)& (mag)& (mag)& (mag)& (mag)& (mag)& (mag)\\
\hline
   36.72 & 1.3795e+44 & 4.3057e+43 & -21.642 &-21.330 & -20.380 & -20.354 & -20.440 & -20.374 & -20.329 & -20.346 & -20.454\\
   40.39 & 1.0017e+44 & 3.6363e+43 & -21.295 &-21.066 & -20.193 & -20.247 & -20.348 & -20.317 & -20.317 & -20.348 & -20.469\\
   44.43 & 7.6372e+43 & 3.2061e+43 & -21.000 &-20.843 & -20.055 & -20.176 & -20.297 & -20.292 & -20.333 & -20.375 & -20.507\\
   48.87 & 6.1774e+43 & 2.9380e+43 & -20.770 &-20.660 & -19.958 & -20.136 & -20.279 & -20.292 & -20.368 & -20.418 & -20.558\\
   53.76 & 5.2790e+43 & 2.7743e+43 & -20.600 &-20.509 & -19.894 & -20.116 & -20.283 & -20.308 & -20.415 & -20.472 & -20.617\\
   59.14 & 4.7693e+43 & 2.6861e+43 & -20.489 &-20.390 & -19.857 & -20.114 & -20.304 & -20.338 & -20.472 & -20.536 & -20.685\\
   65.05 & 4.4163e+43 & 2.6210e+43 & -20.406 &-20.278 & -19.827 & -20.112 & -20.328 & -20.370 & -20.534 & -20.606 & -20.758\\
   71.56 & 4.1327e+43 & 2.5494e+43 & -20.334 &-20.160 & -19.790 & -20.101 & -20.347 & -20.401 & -20.599 & -20.684 & -20.841\\
   78.72 & 3.7587e+43 & 2.4079e+43 & -20.231 &-19.984 & -19.716 & -20.056 & -20.341 & -20.411 & -20.654 & -20.755 & -20.920\\
   86.59 & 3.3955e+43 & 2.2317e+43 & -20.120 &-19.754 & -19.616 & -19.991 & -20.320 & -20.413 & -20.702 & -20.822 & -20.999\\
   95.25 & 3.0894e+43 & 2.0488e+43 & -20.018 &-19.491 & -19.502 & -19.909 & -20.284 & -20.401 & -20.738 & -20.878 & -21.073\\
  104.00 & 2.8994e+43 & 1.9156e+43 & -19.949 &-19.300 & -19.412 & -19.825 & -20.248 & -20.374 & -20.749 & -20.900 & -21.122\\
  114.00 & 2.5779e+43 & 1.6759e+43 & -19.821 &-18.996 & -19.247 & -19.662 & -20.147 & -20.265 & -20.703 & -20.855 & -21.111\\
  124.00 & 2.1881e+43 & 1.3864e+43 & -19.643 &-18.562 & -19.006 & -19.449 & -19.995 & -20.101 & -20.616 & -20.759 & -21.044\\
  136.00 & 1.8206e+43 & 1.1264e+43 & -19.444 &-18.040 & -18.722 & -19.235 & -19.835 & -19.906 & -20.491 & -20.605 & -20.933\\
  150.00 & 1.5858e+43 & 9.8104e+42 & -19.294 &-17.617 & -18.512 & -19.131 & -19.735 & -19.766 & -20.364 & -20.443 & -20.791\\
  165.00 & 1.6326e+43 & 1.0444e+43 & -19.325 &-17.749 & -18.571 & -19.250 & -19.793 & -19.828 & -20.334 & -20.419 & -20.724\\
  181.00 & 1.8889e+43 & 1.2586e+43 & -19.484 &-18.362 & -18.813 & -19.461 & -19.931 & -19.990 & -20.388 & -20.483 & -20.746\\
  199.00 & 2.2135e+43 & 1.5332e+43 & -19.656 &-18.908 & -19.058 & -19.673 & -20.080 & -20.155 & -20.478 & -20.520 & -20.749\\
  218.90 & 2.4176e+43 & 1.7416e+43 & -19.752 &-19.229 & -19.179 & -19.816 & -20.181 & -20.312 & -20.509 & -20.434 & -20.652\\
  240.80 & 2.1726e+43 & 1.6053e+43 & -19.636 &-19.020 & -19.031 & -19.747 & -20.132 & -20.327 & -20.306 & -20.272 & -20.420\\
  264.00 & 1.4704e+43 & 1.0458e+43 & -19.212 &-18.029 & -18.347 & -19.314 & -19.788 & -20.038 & -19.960 & -20.107 & -20.212\\
  290.00 & 8.4332e+42 & 5.6610e+42 & -18.608 &-16.730 & -17.494 & -18.622 & -19.171 & -19.543 & -19.522 & -19.757 & -19.820\\
  319.00 & 4.6253e+42 & 3.0791e+42 & -17.956 &-15.695 & -16.669 & -17.858 & -18.506 & -19.064 & -18.915 & -19.104 & -19.060\\
  350.90 & 2.8228e+42 & 1.7351e+42 & -17.420 &-13.735 & -15.362 & -16.999 & -17.978 & -18.674 & -18.665 & -18.840 & -18.545\\
  386.00 & 1.9654e+42 & 1.0575e+42 & -17.027 &-12.129 & -14.173 & -16.242 & -17.434 & -18.289 & -18.551 & -18.723 & -18.176\\
  424.60 & 1.3801e+42 & 6.3219e+41 & -16.643 &-11.844 & -13.372 & -15.471 & -16.804 & -17.823 & -18.352 & -18.624 & -17.880\\
  467.00 & 9.9485e+41 & 3.9880e+41 & -16.288 &-11.553 & -12.780 & -14.843 & -16.263 & -17.359 & -18.089 & -18.417 & -17.823\\
  513.70 & 6.4897e+41 & 2.8621e+41 & -15.824 &-11.402 & -12.400 & -14.435 & -15.894 & -17.012 & -17.635 & -17.752 & -17.030\\
  565.00 & 4.1263e+41 & 2.0683e+41 & -15.332 &-11.179 & -11.974 & -13.990 & -15.546 & -16.680 & -17.128 & -16.889 & -15.982\\
  621.50 & 2.5179e+41 & 1.4186e+41 & -14.796 &-10.926 & -11.484 & -13.454 & -15.148 & -16.295 & -16.542 & -15.925 & -14.811\\
  683.70 & 1.4502e+41 & 9.0849e+40 & -14.197 &-10.580 & -10.909 & -12.810 & -14.678 & -15.842 & -15.859 & -14.876 & -13.560\\
  752.10 & 7.8231e+40 & 5.5235e+40 & -13.527 &-10.144 & -10.213 & -12.202 & -14.167 & -15.324 & -15.121 & -13.415 & -11.947\\
  827.30 & 3.9231e+40 & 2.9709e+40 & -12.777 & -9.623 &  -9.482 & -11.362 & -13.492 & -14.689 & -14.243 & -12.345 & -10.831\\
  910.00 & 1.8103e+40 & 1.4445e+40 & -11.938 & -8.989 &  -8.676 & -10.413 & -12.699 & -13.940 & -13.244 & -11.334 & -9.784\\
 1001.00 & 7.6570e+39 & 6.2713e+39 & -11.003 & -8.277 &  -7.828 &  -9.424 & -11.785 & -13.052 & -12.178 & -10.366 & -8.765\\
\hline
\end{tabular}
\end{table*}

\begin{table*}
\caption{Bolometric and UVOIR luminosities as well as synthetic magnitudes for PISN model B210NL
\label{tab_mag_B210NL}
}
\begin{tabular}{l@{\hspace{1.6mm}}c@{\hspace{1.6mm}}c@{\hspace{1.6mm}}
c@{\hspace{1.6mm}}c@{\hspace{1.6mm}}c@{\hspace{1.6mm}}
c@{\hspace{1.6mm}}c@{\hspace{1.6mm}}c@{\hspace{1.6mm}}
c@{\hspace{1.6mm}}c@{\hspace{1.6mm}}c@{\hspace{1.6mm}}} \\
\hline
Age   & $L_{\rm bol}$ & $L_{\rm UVOIR}$ & $M_{\rm bol}$ & $M_{\rm U}$ & $M_{\rm B}$
& $M_{\rm V}$ & $M_{\rm R}$ & $M_{\rm I}$ & $M_{\rm J}$ & $M_{\rm H}$ & $M_{\rm K}$ \\
   (d)   & (erg\,s$^{-1}$) & (erg\,s$^{-1}$) & (mag) & (mag)& (mag)& (mag)& (mag)& (mag)& (mag)& (mag)& (mag)\\
\hline
   16.80 & 2.8082e+42 & 1.7870e+42 &  -17.414 & -15.893 & -16.499 & -17.346 & -17.905 & -17.961 & -18.404 & -18.519 & -18.822\\
   18.48 & 2.4868e+42 & 1.5465e+42 &  -17.282 & -15.344 & -16.264 & -17.196 & -17.785 & -17.864 & -18.346 & -18.464 & -18.783\\
   20.33 & 2.2713e+42 & 1.3845e+42 &  -17.184 & -14.879 & -16.090 & -17.080 & -17.677 & -17.792 & -18.295 & -18.416 & -18.740\\
   22.36 & 2.1097e+42 & 1.2687e+42 &  -17.104 & -14.625 & -15.982 & -16.974 & -17.582 & -17.719 & -18.240 & -18.365 & -18.688\\
   24.60 & 1.9644e+42 & 1.1713e+42 &  -17.026 & -14.490 & -15.904 & -16.881 & -17.491 & -17.639 & -18.178 & -18.308 & -18.622\\
   27.10 & 1.8524e+42 & 1.1017e+42 &  -16.962 & -14.421 & -15.865 & -16.813 & -17.413 & -17.574 & -18.121 & -18.259 & -18.559\\
   29.81 & 1.7287e+42 & 1.0242e+42 &  -16.887 & -14.350 & -15.817 & -16.721 & -17.329 & -17.495 & -18.051 & -18.202 & -18.491\\
   32.79 & 1.7152e+42 & 1.0362e+42 &  -16.879 & -14.460 & -15.856 & -16.762 & -17.325 & -17.494 & -18.031 & -18.141 & -18.421\\
   36.07 & 2.0198e+42 & 1.2960e+42 &  -17.056 & -15.322 & -16.131 & -17.068 & -17.527 & -17.684 & -18.110 & -18.214 & -18.452\\
   39.68 & 2.6317e+42 & 1.7885e+42 &  -17.344 & -16.396 & -16.587 & -17.421 & -17.777 & -17.924 & -18.263 & -18.408 & -18.585\\
   43.65 & 3.6298e+42 & 2.5640e+42 &  -17.693 & -17.200 & -17.073 & -17.773 & -18.064 & -18.213 & -18.499 & -18.659 & -18.807\\
   48.01 & 5.0858e+42 & 3.6894e+42 &  -18.059 & -17.824 & -17.547 & -18.121 & -18.370 & -18.517 & -18.768 & -18.931 & -19.063\\
   52.81 & 7.1644e+42 & 5.2956e+42 &  -18.431 & -18.370 & -17.999 & -18.467 & -18.687 & -18.825 & -19.048 & -19.208 & -19.329\\
   58.09 & 9.9973e+42 & 7.4758e+42 &  -18.793 & -18.861 & -18.417 & -18.797 & -18.997 & -19.121 & -19.325 & -19.475 & -19.590\\
   63.90 & 1.3780e+43 & 1.0351e+43 &  -19.141 & -19.307 & -18.797 & -19.111 & -19.298 & -19.407 & -19.597 & -19.732 & -19.843\\
   70.29 & 1.8721e+43 & 1.4037e+43 &  -19.474 & -19.711 & -19.141 & -19.408 & -19.587 & -19.686 & -19.864 & -19.980 & -20.085\\
   77.32 & 2.5052e+43 & 1.8614e+43 &  -19.790 & -20.067 & -19.452 & -19.690 & -19.866 & -19.954 & -20.126 & -20.223 & -20.319\\
   85.05 & 3.3223e+43 & 2.4331e+43 &  -20.097 & -20.394 & -19.736 & -19.962 & -20.142 & -20.219 & -20.382 & -20.460 & -20.542\\
   93.56 & 4.3745e+43 & 3.1614e+43 &  -20.395 & -20.719 & -20.009 & -20.228 & -20.412 & -20.480 & -20.624 & -20.683 & -20.751\\
  102.00 & 5.5049e+43 & 3.9495e+43 &  -20.645 & -20.995 & -20.241 & -20.454 & -20.639 & -20.706 & -20.827 & -20.868 & -20.927\\
  112.00 & 6.8640e+43 & 4.9270e+43 &  -20.885 & -21.263 & -20.477 & -20.682 & -20.865 & -20.934 & -21.026 & -21.051 & -21.101\\
  122.00 & 8.1394e+43 & 5.8866e+43 &  -21.070 & -21.471 & -20.671 & -20.868 & -21.051 & -21.123 & -21.193 & -21.200 & -21.242\\
  132.00 & 9.0999e+43 & 6.7721e+43 &  -21.191 & -21.626 & -20.831 & -21.024 & -21.206 & -21.278 & -21.330 & -21.316 & -21.356\\
  142.00 & 9.9900e+43 & 7.5971e+43 &  -21.292 & -21.747 & -20.968 & -21.154 & -21.336 & -21.404 & -21.440 & -21.403 & -21.439\\
  156.00 & 1.0871e+44 & 8.5119e+43 &  -21.384 & -21.850 & -21.118 & -21.296 & -21.473 & -21.532 & -21.544 & -21.468 & -21.503\\
  171.00 & 1.1360e+44 & 9.1407e+43 &  -21.432 & -21.873 & -21.240 & -21.418 & -21.577 & -21.621 & -21.591 & -21.476 & -21.497\\
  188.00 & 1.1249e+44 & 9.2770e+43 &  -21.421 & -21.757 & -21.304 & -21.530 & -21.658 & -21.674 & -21.552 & -21.420 & -21.399\\
  207.00 & 1.0145e+44 & 8.4987e+43 &  -21.309 & -21.362 & -21.173 & -21.563 & -21.721 & -21.725 & -21.387 & -21.329 & -21.209\\
  227.00 & 8.2998e+43 & 6.8282e+43 &  -21.091 & -20.638 & -20.732 & -21.424 & -21.669 & -21.764 & -21.131 & -21.375 & -21.172\\
  249.00 & 6.2501e+43 & 4.8175e+43 &  -20.783 & -19.744 & -20.100 & -21.052 & -21.402 & -21.641 & -20.986 & -21.474 & -21.250\\
  273.90 & 4.7246e+43 & 3.3740e+43 &  -20.479 & -18.700 & -19.510 & -20.610 & -21.069 & -21.460 & -20.955 & -21.445 & -21.193\\
  301.30 & 3.6861e+43 & 2.4050e+43 &  -20.210 & -17.791 & -18.959 & -20.155 & -20.744 & -21.227 & -21.109 & -21.474 & -21.290\\
  331.40 & 2.4510e+43 & 1.6278e+43 &  -19.766 & -17.175 & -18.460 & -19.670 & -20.304 & -20.901 & -20.665 & -20.942 & -20.625\\
  331.40 & 2.4371e+43 & 1.6453e+43 &  -19.760 & -17.516 & -18.467 & -19.627 & -20.293 & -20.940 & -20.685 & -20.864 & -20.575\\
  364.50 & 1.6297e+43 & 1.1526e+43 &  -19.323 & -16.990 & -17.944 & -19.133 & -19.916 & -20.659 & -20.243 & -20.270 & -19.774\\
  401.00 & 1.1467e+43 & 8.4434e+42 &  -18.942 & -16.434 & -17.419 & -18.704 & -19.617 & -20.388 & -19.953 & -19.815 & -19.169\\
  441.10 & 7.9475e+42 & 5.7846e+42 &  -18.544 & -15.721 & -16.647 & -18.145 & -19.241 & -20.065 & -19.759 & -19.545 & -18.684\\
  485.20 & 5.6956e+42 & 3.8594e+42 &  -18.182 & -15.033 & -15.792 & -17.542 & -18.803 & -19.720 & -19.604 & -19.458 & -18.379\\
  645.80 & 1.0517e+42 & 5.9927e+41 &  -16.348 & -12.727 & -13.115 & -15.085 & -16.787 & -17.795 & -18.048 & -17.577 & -16.151\\
\hline
\end{tabular}
\end{table*}

\begin{table*}
\caption{Bolometric and UVOIR luminosities as well as synthetic magnitudes for PISN model He100ionINL.
\label{tab_mag_he100NL}
}
\begin{tabular}{c@{\hspace{1.6mm}}c@{\hspace{1.6mm}}c@{\hspace{1.6mm}}
c@{\hspace{1.6mm}}c@{\hspace{1.6mm}}c@{\hspace{1.6mm}}c@{\hspace{1.6mm}}
c@{\hspace{1.6mm}}c@{\hspace{1.6mm}}c@{\hspace{1.6mm}}c@{\hspace{1.6mm}}
c@{\hspace{1.6mm}}} \\
\hline
Age   & $L_{\rm bol}$ & $L_{\rm UVOIR}$ & $M_{\rm bol}$ & $M_{\rm U}$ & $M_{\rm B}$
& $M_{\rm V}$ & $M_{\rm R}$ & $M_{\rm I}$ & $M_{\rm J}$ & $M_{\rm H}$ & $M_{\rm K}$ \\
   (d)   & (erg\,s$^{-1}$) & (erg\,s$^{-1}$) & (mag) & (mag)& (mag)& (mag)& (mag)& (mag)& (mag)& (mag)& (mag)\\
\hline
   11.55 & 3.8088e+41 & 2.4709e+41 &  -15.245 & -14.431 & -14.649 & -15.228 & -15.529 & -15.811 & -16.213 & -16.496 & -16.632 \\
   12.71 & 4.0804e+41 & 2.6687e+41 &  -15.320 & -14.324 & -14.760 & -15.332 & -15.631 & -15.902 & -16.317 & -16.611 & -16.750 \\
   13.98 & 4.4089e+41 & 2.8803e+41 &  -15.404 & -14.359 & -14.834 & -15.418 & -15.717 & -15.996 & -16.415 & -16.720 & -16.850 \\
   15.38 & 4.7409e+41 & 3.0781e+41 &  -15.483 & -14.392 & -14.887 & -15.495 & -15.798 & -16.083 & -16.514 & -16.827 & -16.942 \\
   16.92 & 5.0436e+41 & 3.2326e+41 &  -15.550 & -14.355 & -14.917 & -15.556 & -15.870 & -16.157 & -16.613 & -16.935 & -17.036 \\
   18.61 & 5.3148e+41 & 3.3305e+41 &  -15.607 & -14.207 & -14.922 & -15.598 & -15.931 & -16.218 & -16.716 & -17.051 & -17.142 \\
   20.47 & 5.1565e+41 & 3.0544e+41 &  -15.574 & -13.646 & -14.761 & -15.505 & -15.886 & -16.202 & -16.774 & -17.130 & -17.225 \\
   22.52 & 4.6197e+41 & 2.4974e+41 &  -15.455 & -12.646 & -14.403 & -15.264 & -15.719 & -16.098 & -16.766 & -17.151 & -17.268 \\
   24.77 & 4.2201e+41 & 2.0684e+41 &  -15.356 & -11.833 & -14.031 & -15.005 & -15.546 & -16.000 & -16.760 & -17.167 & -17.315 \\
   27.25 & 4.1484e+41 & 1.8889e+41 &  -15.338 & -11.393 & -13.795 & -14.849 & -15.463 & -15.975 & -16.800 & -17.217 & -17.393 \\
   29.98 & 4.5362e+41 & 2.0219e+41 &  -15.435 & -11.333 & -13.814 & -14.903 & -15.543 & -16.068 & -16.913 & -17.331 & -17.518 \\
   32.99 & 5.5622e+41 & 2.5671e+41 &  -15.656 & -11.642 & -14.116 & -15.193 & -15.801 & -16.291 & -17.106 & -17.521 & -17.701 \\
   36.29 & 7.4467e+41 & 3.7020e+41 &  -15.973 & -12.327 & -14.628 & -15.649 & -16.185 & -16.613 & -17.358 & -17.764 & -17.923 \\
   39.92 & 1.0462e+42 & 5.6987e+41 &  -16.342 & -13.543 & -15.244 & -16.165 & -16.620 & -16.981 & -17.634 & -18.026 & -18.153 \\
   43.91 & 1.4940e+42 & 8.8947e+41 &  -16.729 & -14.972 & -15.857 & -16.664 & -17.043 & -17.342 & -17.907 & -18.280 & -18.376 \\
   48.30 & 2.1214e+42 & 1.3572e+42 &  -17.110 & -16.011 & -16.404 & -17.119 & -17.433 & -17.681 & -18.168 & -18.524 & -18.592 \\
   53.13 & 2.9774e+42 & 2.0116e+42 &  -17.478 & -16.779 & -16.898 & -17.533 & -17.798 & -18.003 & -18.419 & -18.765 & -18.805 \\
   58.44 & 4.1165e+42 & 2.8960e+42 &  -17.829 & -17.396 & -17.350 & -17.909 & -18.137 & -18.306 & -18.661 & -18.999 & -19.015 \\
   64.28 & 5.5664e+42 & 4.0347e+42 &  -18.157 & -17.873 & -17.758 & -18.256 & -18.457 & -18.592 & -18.899 & -19.216 & -19.220 \\
   70.71 & 7.4557e+42 & 5.5287e+42 &  -18.474 & -18.337 & -18.136 & -18.572 & -18.754 & -18.864 & -19.130 & -19.430 & -19.424 \\
   77.78 & 9.8045e+42 & 7.3869e+42 &  -18.772 & -18.748 & -18.474 & -18.860 & -19.029 & -19.124 & -19.359 & -19.639 & -19.623 \\
   85.56 & 1.2666e+43 & 9.6485e+42 &  -19.050 & -19.115 & -18.777 & -19.124 & -19.285 & -19.373 & -19.584 & -19.841 & -19.818 \\
   94.12 & 1.6074e+43 & 1.2345e+43 &  -19.308 & -19.438 & -19.050 & -19.368 & -19.526 & -19.615 & -19.803 & -20.032 & -20.006 \\
  103.50 & 1.9896e+43 & 1.5376e+43 &  -19.540 & -19.705 & -19.293 & -19.589 & -19.748 & -19.844 & -20.006 & -20.205 & -20.182 \\
  113.90 & 2.4051e+43 & 1.8705e+43 &  -19.746 & -19.922 & -19.512 & -19.790 & -19.956 & -20.056 & -20.191 & -20.349 & -20.340 \\
  125.30 & 2.8126e+43 & 2.1992e+43 &  -19.916 & -20.065 & -19.700 & -19.973 & -20.142 & -20.244 & -20.351 & -20.473 & -20.474 \\
  137.80 & 3.1465e+43 & 2.4699e+43 &  -20.038 & -20.101 & -19.836 & -20.128 & -20.304 & -20.407 & -20.474 & -20.561 & -20.573 \\
  151.60 & 3.3367e+43 & 2.6215e+43 &  -20.101 & -20.008 & -19.887 & -20.238 & -20.432 & -20.545 & -20.550 & -20.616 & -20.637 \\
  166.80 & 3.2851e+43 & 2.5632e+43 &  -20.085 & -19.741 & -19.790 & -20.272 & -20.501 & -20.645 & -20.561 & -20.651 & -20.670 \\
  183.50 & 2.9521e+43 & 2.2540e+43 &  -19.968 & -19.291 & -19.517 & -20.188 & -20.463 & -20.652 & -20.504 & -20.670 & -20.680 \\
  201.90 & 2.4249e+43 & 1.7783e+43 &  -19.755 & -18.653 & -19.089 & -19.970 & -20.296 & -20.540 & -20.382 & -20.662 & -20.669 \\
  222.10 & 1.8562e+43 & 1.2748e+43 &  -19.465 & -17.772 & -18.535 & -19.611 & -20.001 & -20.335 & -20.276 & -20.625 & -20.637 \\
  244.30 & 1.3665e+43 & 8.6887e+42 &  -19.132 & -16.763 & -17.921 & -19.141 & -19.631 & -20.064 & -20.171 & -20.515 & -20.528 \\
  268.70 & 9.8105e+42 & 5.8724e+42 &  -18.772 & -15.845 & -17.315 & -18.618 & -19.229 & -19.768 & -19.980 & -20.275 & -20.280 \\
  295.60 & 7.0062e+42 & 4.0058e+42 &  -18.407 & -14.530 & -16.574 & -18.024 & -18.858 & -19.512 & -19.756 & -19.943 & -19.930 \\
  325.20 & 5.1579e+42 & 2.9173e+42 &  -18.074 & -13.678 & -15.976 & -17.526 & -18.517 & -19.266 & -19.536 & -19.565 & -19.545 \\
  357.70 & 3.8206e+42 & 2.1646e+42 &  -17.748 & -13.057 & -15.354 & -17.030 & -18.180 & -19.033 & -19.311 & -19.220 & -19.095 \\
  393.50 & 2.7860e+42 & 1.5224e+42 &  -17.406 & -12.583 & -14.638 & -16.490 & -17.786 & -18.740 & -19.044 & -18.961 & -18.641 \\
  432.90 & 1.9716e+42 & 9.6298e+41 &  -17.030 & -12.173 & -13.989 & -15.826 & -17.265 & -18.306 & -18.727 & -18.776 & -18.224 \\
  476.20 & 1.3480e+42 & 5.8865e+41 &  -16.617 & -11.785 & -13.451 & -15.171 & -16.724 & -17.770 & -18.399 & -18.288 & -17.846 \\
  523.80 & 8.3169e+41 & 3.6854e+41 &  -16.093 & -11.441 & -13.000 & -14.608 & -16.224 & -17.256 & -17.869 & -17.442 & -17.066 \\
  576.20 & 4.9389e+41 & 2.2191e+41 &  -15.527 & -11.048 & -12.511 & -14.036 & -15.685 & -16.693 & -17.273 & -16.518 & -16.173 \\
  633.80 & 2.7777e+41 & 1.2231e+41 &  -14.902 & -10.618 & -11.972 & -13.398 & -15.044 & -16.029 & -16.626 & -15.610 & -15.245 \\
  697.20 & 1.4668e+41 & 6.0678e+40 &  -14.209 & -10.131 & -11.369 & -12.681 & -14.274 & -15.251 & -15.931 & -14.773 & -14.267 \\
  766.90 & 7.4843e+40 & 2.7223e+40 &  -13.479 &  -9.560 & -10.688 & -11.886 & -13.385 & -14.359 & -15.217 & -14.035 & -13.202 \\
\hline
\end{tabular}
\end{table*}


\label{lastpage}

\end{document}